\documentclass[
  11pt          % 
  ]{book}

\usepackage{epsfig,amsmath,amssymb,verbatim}
\usepackage[dvips]{color}
\title{Hadronic Effects in Parity Violating Electron Scattering}
\author
{
	Candidate Gian Franco Sacco \\ \\
	Advisor Michael J. Ramsey-Musolf \\\\
    Associate Advisor Gerald V. Dunne \\\\
    Associate Advisor Richard T. Jones 	
}
\date{University of Connecticut 2004}
\begin{document}
\pagestyle{plain}
\def\simge{\hspace*{0.2em}\raisebox{0.5ex}{$>$}
     \hspace{-0.8em}\raisebox{-0.3em}{$\sim$}\hspace*{0.2em}}
\def\simle{\hspace*{0.2em}\raisebox{0.5ex}{$<$}
     \hspace{-0.8em}\raisebox{-0.3em}{$\sim$}\hspace*{0.2em}}
\def\bra#1{{\langle#1\vert}}
\def\ket#1{{\vert#1\rangle}}
\def\coeff#1#2{{\scriptstyle{#1\over #2}}}
\def\undertext#1{{$\underline{\hbox{#1}}$}}
\def\hcal#1{{\hbox{\cal #1}}}
\def\sst#1{{\scriptscriptstyle #1}}
\def\eexp#1{{\hbox{e}^{#1}}}
\def\rbra#1{{\langle #1 \vert\!\vert}}
\def\rket#1{{\vert\!\vert #1\rangle}}
\def\lsim{{ <\atop\sim}}
\def\gsim{{ >\atop\sim}}
\def\nubar{{\bar\nu}}
\def\psibar{{\bar\psi}}
\def\Gmu{{G_\mu}}
\def\alr{{A_\sst{LR}}}
\def\wpv{{W^\sst{PV}}}
\def\evec{{\vec e}}
\def\notq{{\not\! q}}
\def\notl{{\not\! \ell}}
\def\notk{{\not\! k}}
\def\notp{{\not\! p}}
\def\notpp{{\not\! p'}}
\def\notder{{\not\! \partial}}
\def\notcder{{\not\!\! D}}
\def\notA{{\not\!\! A}}
\def\notv{{\not\! v}}
\def\Jem{{J_\mu^{em}}}
\def\Jana{{J_{\mu 5}^{anapole}}}
\def\nue{{\nu_e}}
\def\mn{{m_\sst{N}}}
\def\mns{{m^2_\sst{N}}}
\def\me{{m_e}}
\def\mes{{m^2_e}}
\def\mq{{m_q}}
\def\mqs{{m_q^2}}
\def\mz{{M_\sst{Z}}}
\def\mzs{{M^2_\sst{Z}}}
\def\ubar{{\bar u}}
\def\dbar{{\bar d}}
\def\sbar{{\bar s}}
\def\qbar{{\bar q}}
\def\sstw{{\sin^2\theta_\sst{W}}}
\def\sstwo{{\sin^2\theta_\sst{W}^0}}
\def\gv{{g_\sst{V}}}
\def\ga{{g_\sst{A}}}
\def\pv{{\vec p}}
\def\pvs{{{\vec p}^{\>2}}}
\def\ppv{{{\vec p}^{\>\prime}}}
\def\ppvs{{{\vec p}^{\>\prime\>2}}}
\def\qv{{\vec q}}
\def\qvs{{{\vec q}^{\>2}}}
\def\xv{{\vec x}}
\def\xpv{{{\vec x}^{\>\prime}}}
\def\yv{{\vec y}}
\def\tauv{{\vec\tau}}
\def\sigv{{\vec\sigma}}

\def\sst#1{{\scriptscriptstyle #1}}
\def\gpnn{{g_{\sst{NN}\pi}}}
\def\grnn{{g_{\sst{NN}\rho}}}
\def\gnnm{{g_\sst{NNM}}}
\def\hnnm{{h_\sst{NNM}}}

\def\xivz{{\xi_\sst{V}^{(0)}}}
\def\xivt{{\xi_\sst{V}^{(3)}}}
\def\xive{{\xi_\sst{V}^{(8)}}}
\def\xiaz{{\xi_\sst{A}^{(0)}}}
\def\xiat{{\xi_\sst{A}^{(3)}}}
\def\xiae{{\xi_\sst{A}^{(8)}}}
\def\xivtez{{\xi_\sst{V}^{T=0}}}
\def\xivteo{{\xi_\sst{V}^{T=1}}}
\def\xiatez{{\xi_\sst{A}^{T=0}}}
\def\xiateo{{\xi_\sst{A}^{T=1}}}
\def\xiva{{\xi_\sst{V,A}}}

\def\rvz{{R_\sst{V}^{(0)}}}
\def\rvt{{R_\sst{V}^{(3)}}}
\def\rve{{R_\sst{V}^{(8)}}}
\def\raz{{R_\sst{A}^{(0)}}}
\def\rat{{R_\sst{A}^{(3)}}}
\def\rae{{R_\sst{A}^{(8)}}}
\def\rvtez{{R_\sst{V}^{T=0}}}
\def\rvteo{{R_\sst{V}^{T=1}}}
\def\ratez{{R_\sst{A}^{T=0}}}
\def\rateo{{R_\sst{A}^{T=1}}}

\def\mk{{m_\sst{K}}}
\def\mro{{m_\rho}}
\def\mks{{m_\sst{K}^2}}
\def\mpi{{m_\pi}}
\def\mpis{{m_\pi^2}}
\def\mom{{m_\omega}}
\def\mphi{{m_\phi}}
\def\Qhat{{\hat Q}}

\def\FOS{{F_1^{(s)}}}
\def\FTS{{F_2^{(s)}}}
\def\GAS{{G_\sst{A}^{(s)}}}
\def\GES{{G_\sst{E}^{(s)}}}
\def\GMS{{G_\sst{M}^{(s)}}}
\def\GATEZ{{G_\sst{A}^{\sst{T}=0}}}
\def\GATEO{{G_\sst{A}^{\sst{T}=1}}}
\def\mdax{{M_\sst{A}}}
\def\mustr{{\mu_s}}
\def\rsstr{{r^2_s}}
\def\rhostr{{\rho_s}}
\def\GEG{{G_\sst{E}^\gamma}}
\def\GEZ{{G_\sst{E}^\sst{Z}}}
\def\GMG{{G_\sst{M}^\gamma}}
\def\GMZ{{G_\sst{M}^\sst{Z}}}
\def\GEn{{G_\sst{E}^n}}
\def\GEp{{G_\sst{E}^p}}
\def\GMn{{G_\sst{M}^n}}
\def\GMp{{G_\sst{M}^p}}
\def\GAp{{G_\sst{A}^p}}
\def\GAn{{G_\sst{A}^n}}
\def\GA{{G_\sst{A}}}
\def\GETEZ{{G_\sst{E}^{\sst{T}=0}}}
\def\GETEO{{G_\sst{E}^{\sst{T}=1}}}
\def\GMTEZ{{G_\sst{M}^{\sst{T}=0}}}
\def\GMTEO{{G_\sst{M}^{\sst{T}=1}}}
\def\lamd{{\lambda_\sst{D}^\sst{V}}}
\def\lamn{{\lambda_n}}
\def\lams{{\lambda_\sst{E}^{(s)}}}
\def\bvz{{\beta_\sst{V}^0}}
\def\bvo{{\beta_\sst{V}^1}}
\def\Gdip{{G_\sst{D}^\sst{V}}}
\def\GdipA{{G_\sst{D}^\sst{A}}}
\def\fks{{F_\sst{K}^{(s)}}}
\def\FIS{{F_i^{(s)}}}
\def\fpi{{F_\pi}}
\def\fk{{F_\sst{K}}}

\def\RAp{{R_\sst{A}^p}}
\def\RAn{{R_\sst{A}^n}}

\def\RVp{{R_\sst{V}^p}}
\def\RVn{{R_\sst{V}^n}}
\def\rva{{R_\sst{V,A}}}
\def\xbb{{x_B}}

\def\mlq{{M_\sst{LQ}}}
\def\mlqs{{M_\sst{LQ}^2}}
\def\lscal{{\lambda_\sst{S}}}
\def\lvect{{\lambda_\sst{V}}}

\def\PR#1{{{\em   Phys. Rev.} {\bf #1} }}
\def\PRC#1{{{\em   Phys. Rev.} {\bf C#1} }}
\def\PRD#1{{{\em   Phys. Rev.} {\bf D#1} }}
\def\PRL#1{{{\em   Phys. Rev. Lett.} {\bf #1} }}
\def\NPA#1{{{\em   Nucl. Phys.} {\bf A#1} }}
\def\NPB#1{{{\em   Nucl. Phys.} {\bf B#1} }}
\def\AoP#1{{{\em   Ann. of Phys.} {\bf #1} }}
\def\PRp#1{{{\em   Phys. Reports} {\bf #1} }}
\def\PLB#1{{{\em   Phys. Lett.} {\bf B#1} }}
\def\ZPA#1{{{\em   Z. f\"ur Phys.} {\bf A#1} }}
\def\ZPC#1{{{\em   Z. f\"ur Phys.} {\bf C#1} }}
\def\etal{{{\em   et al.}}}

\def\delalr{{{delta\alr\over\alr}}}
\def\pbar{{\bar{p}}}
\def\lamchi{{\Lambda_\chi}}

\def\qw0{{Q_\sst{W}^0}}
\def\qwp{{Q_\sst{W}^P}}
\def\qwn{{Q_\sst{W}^N}}
\def\qwe{{Q_\sst{W}^e}}
\def\qem{{Q_\sst{EM}}}

\def\gae{{g_\sst{A}^e}}
\def\gve{{g_\sst{V}^e}}
\def\gvf{{g_\sst{V}^f}}
\def\gaf{{g_\sst{A}^f}}
\def\gvu{{g_\sst{V}^u}}
\def\gau{{g_\sst{A}^u}}
\def\gvd{{g_\sst{V}^d}}
\def\gad{{g_\sst{A}^d}}

\def\gvftil{{\tilde g_\sst{V}^f}}
\def\gaftil{{\tilde g_\sst{A}^f}}
\def\gvetil{{\tilde g_\sst{V}^e}}
\def\gaetil{{\tilde g_\sst{A}^e}}
\def\gvqtil{{\tilde g_\sst{V}^e}}
\def\gaqtil{{\tilde g_\sst{A}^e}}
\def\gvutil{{\tilde g_\sst{V}^e}}
\def\gautil{{\tilde g_\sst{A}^e}}
\def\gvdtil{{\tilde g_\sst{V}^e}}
\def\gadtil{{\tilde g_\sst{A}^e}}

\def\hvf{{h_\sst{V}^f}}
\def\hvu{{h_\sst{V}^u}}
\def\hvd{{h_\sst{V}^d}}
\def\hve{{h_\sst{V}^e}}
\def\hvq{{h_\sst{V}^q}}

\def\delp{{\delta_P}}
\def\delzp{{\delta_{00}}}
\def\deld{{\delta_\Delta}}
\def\dele{{\delta_e}}

\def\apv{{A_\sst{PV}}}
\def\apvnsid{{A_\sst{PV}^\sst{NSID}}}
\def\apvnsd{{A_\sst{PV}^\sst{NSD}}}
\def\qpv{{Q_\sst{W}}}

\def\RAd{{R_\sst{A}^\Delta}}
\def\RAs{{R_\sst{A}^{\mbox{Siegert}}}}
\def\RAa{{R_\sst{A}^{\mbox{anapole}}}}
\def\RAewk{{R_\sst{A}^{\mbox{ewk}}}}
\def\RAdw{{R_\sst{A}^{\mbox{d-wave}}}}

\def\lamchi{{\Lambda_\chi}}
\def\lamchis{{\Lambda_\chi^2}}

\def\alrd{{A_\sst{LR}}}
\def\alrds{{A_\sst{LR}^{{\mbox{Siegert}}}}}
\def\alrda{{A_\sst{LR}^{{\mbox{anapole}}}}}
\newcommand{\sem}{\sqrt{E^2-m^2} }
 \newcommand{\sd}{\!  \cdot \! }
\newcommand{\bge}{\begin{equation}}
  \newcommand{\ee}{\end{equation}}
  \newcommand{\la}{\langle}   \newcommand{\ra}{\rangle}  
\newcommand{\bga}{\begin{eqnarray}}
  \newcommand{\ea}{\end{eqnarray}}  \addtolength{\hoffset}{-.5cm}  \addtolength{\textwidth}{2cm}
  \newcommand{\bgd}{\begin{displaymath}}
  \newcommand{\ed}{\end{displaymath}} 
 \newcommand{\nn}{\ensuremath{\nonumber}} 
 \newcommand{\lp}{\left(} 
 \newcommand{\rp}{\right)} 
 \newcommand{\sla}{\hspace{-1.8mm}/} 
 \newcommand{\slab}{\hspace{-3.4mm}/}

%\authorspreviousdegreelong{

%  B.S., Universita' di Torino, 1996\\
%  M.S., University of Connecticut, 2001
 % }
%\yearofpublication{2004}

%\MajorAdvisor{Michael J. Ramsey-Musolf}
%\AssociateAdvisorA{Gerald V. Dunne}
%\AssociateAdvisorB{Richard T. Jones}
\frontmatter

\maketitle

\tableofcontents
\listoffigures
\listoftables

\mainmatter

% chapters of the thesis come here.
%\input{ch0.tex}
\chapter{Introduction}
The study  of the parity violating electron scattering (PVES) has played a  fundamental role in corroborating 
the Standard Model (SM) of electroweak (EW) interactions and understanding  the detailed structure of the weak-neutral current. 
The first experiment of this kind was performed at the Stanford Linear Accelerator Center (SLAC) in the late 1970's~\cite{prescot} where the  parity violating asymmetry
\bga
A_{RL}\equiv \frac{\sigma_R-\sigma_L}{\sigma_R+\sigma_L}\,,
\ea
where $\sigma_R$ and $\sigma_L$ are the cross sections for a right handed and left handed polarized beam respectively, was measured in the scattering of longitudinally polarized electrons  off a deuteron target. Since the electromagnetic parity conserving (PC) interaction is insensitive to the polarization of the beam, while the electroweak parity violating (PV) interaction changes sign upon flipping the spin of the incoming beam, taking the difference between $\sigma_R$ and $\sigma_L$ isolates the PV contribution to the cross section, which ultimately depends on the weak mixing angle. In the SLAC experiment,  the value of the weak mixing angle extracted from  $ A_{RL}$ indicated an amount of parity violation which was consistent with the SM. Beside establishing the SM as one of the most promising frameworks to describe the electroweak interaction, this experiment  also ruled out a number of possible models which were considered at the time as plausible. Subsequently, PV experiments involving scattering of electrons   off $^{12}$C~\cite{souder} and $^9$Be~\cite{heil} were performed to further test the SM at low energy. In the last few years, because of the constant improvement of experimental techniques, measurements of  $A_{RL}$ at the level of a few percent, are conceivable. Such high precision allows one to measure radiative corrections to the SM and look for deviations from the SM, opening a window on   possible  physics beyond the SM, and also to  have better insights into  the hadronic structure. A quantity that is being extensively studied to find deviations from the SM, is the weak mixing angle $\theta_W$, defined by the relation 
\bga
\sin^2\theta_W=\frac{g^{'2}}{g^{'2}+g^2}\,,
\ea 
and where $g'$ and $g$ are  $SU_L(2)\times U_Y(1)$ guage couplings. An important feature of $\sin^2\theta_W$, is that it value "runs" (i.e. changes) with energy. They way it changes with energy is described by the  renormalization group equation, which is a first order  differential equation, therefore, by measuring the angle at a given energy, one could predict its value at any energy. The standard model (SM) of electro-weak interactions makes a precise prediction on the running of the electro-weak angle, and measuring it in  high precision experiments in different kinematical regions, would be a powerful way to test the SM and to look for  new physics.

In this respect, a series of high precision low energy PV experiments have been recently performed, approved or proposed. Among them, it is worth to mention the  M{\o}ller experiment E-158~\cite{moller} performed at SLAC, where the weak mixing angle is extracted by measuring $A_{RL}$ in electron-electron scattering with an accuracy of $\delta(\sin^2\theta_W)\simeq .0007$, 
and the Qweak experiment~\cite{qweak} approved at the Jefferson Laboratory (JLAB) facility, which will measure the weak mixing angle by measuring the $A_{RL}$ on a proton target, aiming for a precision of $\delta(\sin^2\theta_W)\simeq 0.0005$. In both the experiments  the average $Q^2$ is 0.03 GeV$^2$. As pointed out in Ref.~\cite{mjrm1}, the sensitivity of an experiment to new physics depends on the kinematical region and on the target that is used. The experiments in Refs.~\cite{moller,qweak}, in tandem with the proposed experiments of Refs.~\cite{bosted,bosted1}, will form a  complementary program which will  explore a broad range of possibilities to extend the SM.   In Refs.~\cite{bosted,bosted1} they propose a measurement of $A_{RL}$ on a deuteron target in the deep inelastic region, with a momentum transfer squared of the order of 3 GeV$^2$ in Ref.\cite{bosted1}, and between 16 and 28 GeV$^2$ in Ref.\cite{bosted}, and extract the weak mixing angle from the asymmetry, with a precision of 0.5\% on
$\sin^2\theta_W$. In particular, the latter  experiment would explore a kinematical region far from the $Z$ pole, but at a sufficient high $Q^2$ so that, presumably, one should minimize the uncertainty due to the higher twist (HT) effects and also be able to test new physics. Following Ref.~\cite{mjrm1}, one could describe the effects of new physics, by introducing a four fermion contact interaction, whose Lagrangian is 
\bga\label{e300}
\mathcal{L}_{NEW}=\frac{4\pi k^2}{\Lambda^2}\sum_{q,i,j}h_{ij}^q \bar{e}_i\gamma_\mu e_i \bar{q}_j\gamma^{\mu}q_j\, ,
\ea 
where $\Lambda$ is the characteristic mass  that sets the new physics energy scale, $k^2$ is the coupling of the new interaction, and $h_{ij}$ are the helicity-dependent couplings ($i\,,j$ denote the helicity of the fermions).
Such a form of  interaction is suitable to describe different scenarios of new physics which could be tested by measuring the deep inelastic scattering (DIS) $A_{RL}$ on deuteron.  
For instance, super string inspired theories and some super symmetric theories admit the existence of an extra gauge boson, and the proposed experiment might set a lower bound on its mass.  One could also test compositeness  of fermions,  which could be described  by a contact interaction similar to the one in Eq. (\ref{e300}), and set limits on the mass scale $\Lambda$ at which the compositeness becomes manifest. Finally, Eq. (\ref{e300}) could also be applied to describe leptoquarks formation, {\it i.e.} the formation of an intermediate bound state of a lepton and a quark, in the process $e+q\rightarrow LQ \rightarrow e+q$ where $LQ$ is the leptoquark state. In this case the parameter $\Lambda$ represents the mass of the leptoquark state, and the proposed experiment could be used to set  a lower limit on it, given some reasonable assumptions on the scale of the couplings.  \\
A PV deep inelastic scattering (DIS) experiment might be proposed in the near future, whose goal is to measure HT effects~\cite{xion}. In this experiment, which might eventually run at JLAB, a PV asymmetry is measured using a deuteron target at a $Q^2$ of nearly 3 GeV$^2$. My study on  HT effect might provide a guidance in developing the details of the experiment, in particular at what kinematics and for which values of the Bjorken variable $x_B$, the HT effects could be prominent.\\

 Because of the high precision of these experiments, it is absolutely necessary to know how the hadronic effects might impact the interpretation of $A_{RL}$. In this study, which was mostly conducted along with Michael Ramsey-Musolf and Shi-Lin Zhu,  I consider a few of such effects which might be relevant to the interpretation of the different experiments.\\
\section{Complete Treatment of the Parity  Violating Asymmetry at Leading Twist}
In the second Chapter,  I consider an extensive study of the possible leading twist hadronic corrections which affect  $A_{RL}$ in a deep inelastic scattering (DIS) process. The main motivation for studying such corrections is that in Refs.~\cite{bosted,bosted1}, it has been proposed to use  the  deep inelastic polarized cross section   to extract $\sin^2\theta_W$ with an accuracy of 0.5$\%$. The corresponding precision required  to measure $A_{RL}$ is therefore of the order of 1$\%$, which is extremely challenging for this kind of experiment. Consequently, one has to make sure that all the possible theoretical uncertainties are below 1$\%$. In the approximate case in which  sea quarks are neglected, the mass of the target $M$ is negligible compared to the momentum transfer  $Q$, and the perturbative QCD (Quantum Chromo Dynamics) corrections are not included,   $A_{RL}$ for an isoscalar target like the deuteron assumes a particularly simple form~\cite{cahn}
\bga \label{ee1}
A_{RL}&=& -\frac{G_F q^2}{2\sqrt{2}\pi \alpha}\frac{9}{10}\Big[(1-\frac{20}{9} \sin^2\theta_W)+(1-4 \sin^2\theta_W)\lp\frac{1-(1-y)^2}{1+(1-y)^2}\rp\Big]\nn \\
\ea
where $G_F=1.16639(1)\times 10^{-5}$ Gev$^{-2}$ is the Fermi constant, $\alpha=\frac{1}{137}$ is the fine-structure constant, $q$ is the momentum transfer and $y=1-\frac{E'}{E}$ with $E$ ($E'$) the energy of the incoming (outgoing) electron. It's worth  noticing that, under the previous assumptions, the asymmetry does not depend on the structure of the target, but a  more complete treatment, which includes the hadronic corrections, is necessary if one wants to reach a precision of 1$\%$ in the interpretation of the measurement of $A_{RL}$. 
\subsection{Sea Quark, Perturbative QCD and Target Mass Contributions to $A_{RL}$ }
\begin{figure}
\epsfxsize=11.0cm
\centerline{\epsffile{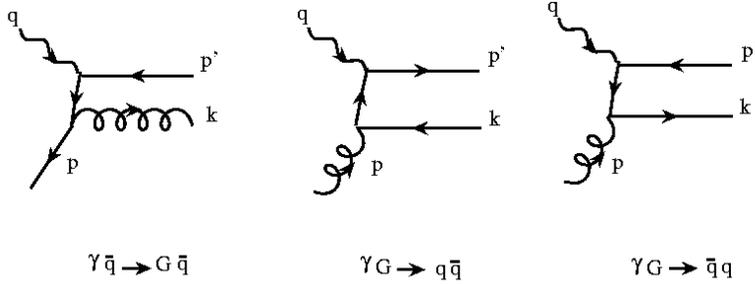}}
%\vspace{1cm}
\caption{
Order $\alpha_S$ contribution to $e\, P \rightarrow e' \, X $ DIS  process (only the hard part is shown) arising from antiquarks and gluons 
}
\label{f:lla}
\end{figure}

In Chapter \ref{ch:twist2}  I  present a detailed computation of $A_{RL}$ for a deuteron target, including the sea quark (see for instance Ref.~\cite{mjrm}), the perturbative QCD (PQCD)~\cite{alt} and the target mass (TM)~\cite{ellis,georgi} corrections. Eq. (\ref{ee1}) has been obtained by using  the quark parton model, where the cross section of the process of the lepton scattering off the target (deuteron, in this case), is viewed as an incoherent sum of cross sections of the lepton scattering off free pointlike particles (the partons). If one restricts the process to only valence quarks,  Eq. (\ref{ee1}) is recovered. The sea quark corrections, consist in including all the possible quarks which could participate to the process, compatibly with the kinematical constrains.
The PQCD corrections are corrections of first order in the strong coupling constant $\alpha_S$ and which cannot be reabsorbed in a redefinition of the parton  distribution functions (PDFs). Such terms arise from corrections to the processes depicted in Fig. \ref{f:lla} which go beyond the leading logarithmic approximation, and they are renormalization scheme dependent (see Sec. \ref{sec:asy}). Since these correction are expressed in terms of integrals of the leading twist PDFs, the are known once the latter are. In the PQCD correction,  also the gluon PDF contributes. 
  %%%%%%%%%%%%%%%%%%%%%%%%%%% Fig lla %%%%%%%%%%%%%%%%%%%%%%%%%%%%%
%%%%%%%%%%%%%%%%%%%%%%%%%%%%%%%%%%%%%%%%%%%%%%%%%%%%%%%
Finally the TM corrections are contributions which  are suppressed by a factor of $M^2/Q^2$, where $M$ is the target mass and $Q$ the momentum transfer. When deriving the naive parton model, one assumes that the nucleon is made up of  free partons that carry a faction $x$ (a<$x$<1) of the total its momentum $P_\mu$. If one considers also the possibility for the partons to interact among each other through the exchange of gluons, their momentum acquires also a  component which is  transverse  the nucleon momentum. Because of this transverse component of the parton momentum, the PDF obtained  in the naive parton model receive corrections which are proportional to $M^2/Q^2$~\cite{ellis,predazzi}. In DIS processes, where the momentum transfer  $Q$ is normally much larger that the nucleon mass $M$, those corrections are small, but they might become important at the kinematics proposed in~\cite{bosted,xion}, where $Q^2=2.8$ GeV$^2$. Also in this case they can be computed if the PDFs are known, but the gluon PDF does not contribute.

In  Sec. \ref{sec:dis} I  give an overview of the  DIS experiments to better understand the role they played in the last few decades in understanding the hadronic structure and how they lead to the formulation of the quantum chromo dynamics (QCD) gauge theory.
In Sec. \ref{sec:kinematics} I  introduce most of the nomenclature involved in DIS, from the kinematical variables, to  the hadronic and leptonic tensor, in  which terms one can express the cross section. Finally, in Sec. \ref{sec:asy}, I consider $A_{RL}$ in the case of deuteron target, including the corrections which were neglected in Eq. (\ref{ee1}). As mentioned before, all the corrections considered above can be expressed in terms of the PDFs.  These quantities, at the moment, cannot be computed from first principle (although there are some attempts to obtain them from lattice calculation~\cite{thomas}), and one has to rely on model independent extractions. Two groups, in particular, the CTEQ~\cite{cteq} and the MRST~\cite{mrst}, have, in the last years, continuously updated their PDFs extracted from the experiments. By using the most recent sets of PDFs provided by the two groups, I  compute the leading twist hadronic corrections to $A_{RL}$, and take the difference between the two predictions as a theoretical uncertainty arising form the PDFs. Although the two sets quite agree in describing the valence distribution, the sea quark and the gluon distributions might  substantially differ. These differences might represent a source of uncertainty in the interpretation of the measurement of $A_{RL}$, especially because of     the gluon distribution, which is the least known, and appears in the PQCD corrections.
 
\section{Higher Twist Corrections to $A_{RL}$} \label{HT_corr}
Chapters \ref{ch:twist} and  \ref{ch:two} are dedicated to the definition and estimate of the twist four correlation functions.
Similarly to the TM corrections,   the twist four corrections  appear in the DIS cross section and asymmetry suppressed by a factor of $\frac{\Lambda^2}{Q^2}$ compared to the twist two contributions, where $\Lambda$ is the QCD scale of the order of few hundreds MeV. Because of the small value of the momentum transfer which will be eventually used in the experiments proposed in~\cite{bosted,xion}, these corrections might give a substantial contribution to   $A_{RL}$.  The main difference between TM and HT corrections is that   the former can be computed in terms of PDFs, which are known from experiments .  As far as  the HT contributions are concerned, there exists only some model independent extractions from experimental data~\cite{sidorov,alekhin} which are not extremely precise. The problem is that it is very hard to disentangle and distinguish  between the contributions arising from the HT and those due to the higher order PQCD corrections, since at small momentum transfer $Q$, the logarithmic behavior of the PQCD corrections resembles the power behavior of the HT. It is therefore very important, when considering high precision experiments in the DIS region, to have an estimate of the HT.

 Already, more than twenty years ago, 
Politzer~\cite{Politzer}   pointed out the necessity of a complete theoretical description of  HT~\cite{gross} in DIS processes. In his pioneering work he gave for the first time 
a formal definition of the twist four operator of the nucleon using the operator product expansion (OPE)~\cite{wilson}. His  work was then extended by   Jaffe 
and   Soldate~\cite{jaffe11,Jaffe} and soon after   Ellis {\it et al.}~\cite{ellis} gave a different but  equivalent description which did not require the use of OPE. More than twenty years have  passed since those authors built the theoretical framework 
to compute the HT, but not many attempts have been done to do so~\cite{ellis1,signal,guo,casto,casto1}. The main problem is that in order to compute the HT one must solve the QCD 
equations which are currently intractable.  Most likely, years from now, people   will be able to compute the HT effects on the lattice but,  at this time,  even the computation of the   twist two PDFs presents some problem, and only a small number of moments of the PDFs have been computed~\cite{thomas}. Until this time will come, one has to rely on phenomenological models to estimate such effects. 

In this  part of the study I use the MIT bag model (MITBM) to describe the nucleon wave 
function and estimate the twist four contributions  to the different structure functions (SFs). 
The MITBM is a phenomenological model first proposed by the authors of Ref.~\cite{jaffe}  to describe how the quarks are confined inside the nucleon. 
Since no free quarks have been experimentally observed,   they must be tightly confined inside the hadrons. The main assumption of the 
model is that the quarks in the hadron reside in a region (the volume of the bag) of true vacuum (or perturbative vacuum), in which they behave as free particles. On the surface of the bag acts the pressure due to the  QCD vacuum 
(a continuous creation and annihilation of quark-antiquark pairs and  gluons) which keeps the quarks confined in the hadron. Such a model has been extensively used to estimate different physical quantities (hadron masses, magnetic moments, charge radii) and the results are   within twenty to forty percent of accuracy~\cite{greiner}. Keeping this in mind, I would like to emphasize that the main goal of this part of the study is not to give an exact quantitative description of the HT, but instead a   semi-quantitative one, in which the main concern is to obtain an order of magnitude estimate of such effects and see if they might represent a problematic source of uncertainty in the extraction of the weak mixing angle or in looking for hints of physics beyond the SM.  

In Sec. \ref{sec:twofermion_cf} I  present the contributions to the nucleon SFs of the two fermion correlation functions up to twist four (see Fig. \ref{fa1}). In this part of the study I estimate the two fermion correlations functions by computing their moments in the MIT bag model (MITBM) as it was originally suggested by the authors of Ref.~\cite{jaffe}, and then reconstruct them by means of the inverse Mellin transformation (IMT)~\cite{muta}. Indeed, given a function $f(t)$ sufficiently well behaved, one could compute the Mellin transformation
\bga
\phi (z)=\int_0^\infty dt\,t^{z-1} f(t)\,. 
\ea
The original funtion $f(t)$ can be obtained by applying the IMT to $\phi(z)$, 
\bga
f(t)=\frac{1}{2\pi i} \int_{c-i\infty}^{c+i\infty} dz\,t^{-z}\phi(z)\,, 
\ea
where $c$ is a positive constant. In this case I can compute the   moments of the PDFs (for the two fermion correlation functions) or of the SFs (for the four fermion correlation functions) from the relation~\cite{ellis}
\bga
M(N) = \int _0^1 dx\, x^{N-1} F(x)\,, 
\ea
where in this case $N$ is a positive integer and $F(x)$ is either a PDF or a SF. Unfortunately, by using the MITBM for the four fermion correlations (Fig. \ref{fa2}) I am only  able to compute a very limited (three to be exact) of moments, and the process of inversion provides a non unique solution (see Sec \ref{sec:fourfermion_tw4cf} for details). Theoretically, one would need an infinite number of moments, but it turned out that, for the case of the two fermion correlation functions, five moments were sufficient to provide a stable solution. 

 To have an idea of the order of accuracy of the  model, I also computed the twist two PDFs. These PDFs  are normally twice as big as the one extracted from data. As pointed out in Ref.~\cite{thomas1}, theses PDFs are obtained at the bag energy scale (which is of the order of few hundred MeV), which becomes a parameter of the model, that is fixed by evolving the PDFs at an higher energy scale by using the DGLAP evolution equations, and comparing them with the one extracted from experiments at that energy scale (see Sec. \ref{sec:twofermion}).  
After  evolving the PDFs obtained in the MITBM, I  have a quantitative agreement in magnitude, but only a qualitative agreement in shape, with the ones extracted from experiments (see Sec. \ref{sec:twofermion_cf}). However, since my intent is to provide  a semiquantitative estimate of the HT contributions, I consider the accuracy of the model adequate for my purpose. 

In Sec. \ref{sec:mmitbm} I   use a variation of the MITBM ( modified MITBM or MMITBM henceforth) suggested by the authors of Ref.~\cite{thomas1}.
In this model the authors overcome one of the major problem of the MITBM, namely the lack of translational invariance of the system, by application of a Peierls-Yoccoz projection~\cite{Peierls}. This model has the advantage of reproducing the twist two PDFs quite well, but at the expense of a larger number of parameters which need to be fitted. 
Moreover, in this model, it is extremely difficult to compute numerically the four fermion correlation functions, which I was able to estimate only in the MITBM. A common problem to both methods is the absence, at the moment, of the evolution equations (the equivalent of the DGLAP equations) for the twist four operators. In this study I compute the HT corrections to the SFs  at the bag model scale, and compare the results with some model independent extractions from experiments~\cite{sidorov,alekhin}. It is noticed that, if one trusts these extractions, the magnitude of the HT effects obtained with  my models is too large. This could be an indication that also for the twist four case, evolution might play an important role. To have an idea on how the twist four contribution might change upon evolution,  I make the ansatz of using the DGLAP evolution equations to evolve the two fermion twist four correlation functions. This procedure is quite arbitrary and I do not have any physical justification for applying  it, nevertheless I noticed that, after evolving the two fermion twist four correlation functions, the corresponding results were in reasonable agreement with some model independent extractions (see Sec. \ref{sec:tw4mmitbm}). I only applied evolution to the two fermion correlation functions, since the software I used~\cite{kumano}, requires that I introduce the distribution for each parton. For the four fermion correlation functions the partonic concept is lost (see the form of these contributions to the SFs in Sec. \ref{sec:fourfermion_tw4cf}), and the software cannot be used. It turned out that these contributions were small even if not evolved (see Figs. \ref{f215} and \ref{f216}, where the relative corrections to $A_{RL}$ for a deuteron target, arising from the four fermion correlation function are plotted).

In Sec. \ref{sec:final_arl}, I  present the results for $A_{RL}$, including all the corrections discussed in     Chapter \ref{ch:twist2} and the HT contributions.

\section{Bremsstrahlung Contribution to the Resonance Electroproduction Cross Section and $A_{RL}$}
In Chapter \ref{ch:e158} I  compute the parity violating asymmetry for the bremsstrahlung contribution to the proton electro-excitation, i.e.  the process    $e^-\, P\rightarrow e^-\, N^*$ (where $P$ is a proton, $e^-$  is an electron and  $N^*$ is a proton resonance). It was speculated that such a process could have been one of the major backgrounds in the experiment E-158 performed at SLAC~\cite{slac}.

 Since the M{\o}ller scattering consists in electron-electron scattering, one, normally, does not have to worry about hadronic backgrounds. But in order to reach the desired luminosity, atomic electrons in hydrogen atoms were used as a target in the experiment. Therefore, because of the presence of the protons,  a number of hadronic  backgrounds needs to be taken into account. Most of these backgrounds were estimated in the proposal~\cite{slac}, but the bremsstrahlung emission  was thought to be negligible, and therefore not considered. From preliminary experimental results, it was observed that the inelastic background constitutes 40$\%$ of the total asymmetry measured in the M{\o}ller ring, which implies that the prescribed accuracy of $8\%$ could only be reached if all the inelastic  backgrounds are known to a 20$\%$ level or better. 

The reason why I decided to compute the bremsstrahlung contribution to the proton electro-excitation can be understood after knowing a bit more about the detector used in the E-158 experiment. This detector might be thought as a series of concentric rings. The innermost is the M{\o}ller ring (which is actually made up of three rings), where the M{\o}ller events are detected. The outermost ring is the EP ring where most of the elastic and quasi-elastic events are restricted. Electrons which undergo inelastic scattering, might end up in both detectors. 
Very roughly, one might assume that, for 50 GeV electron beam, the events with energy between 12.5 and 25 GeV  end up in the M{\o}ller ring, while those with energy in the range 25 to 50 GeV are detected in the EP ring. The electron resulting from the proton electro-excitation, are normally confined in the EP ring. But if the electron undergoes a bremsstrahlung process, emitting a photon either before or after the scattering off the proton, it might lose enough energy so that it  ends up in the M{\o}ller ring. 

The main problem is that in order to compute the cross section and asymmetry for such a process, one needs to know the electro-magnetic and electro-weak form factors for the different resonances   
 for a large range of $Q^2$. Since we have a good knowledge of the electro-magnetic form factors just for a limited number of resonances, and the electro-weak ones are poorly known for all of them, I would like to investigate how the uncertainty on the resonance form factors  impacts the error on the total asymmetry measured in the M{\o}ller ring. To this end I  consider a simple model in which only two resonances contribute to the electro-excitation process, in particular I consider   the $P_{33}$,  which is a spin $\frac{3}{2}$ isospin $\frac{3}{2}$ resonance, and  the $D_{13}$ which is a 
spin $\frac{3}{2}$ isospin $\frac{1}{2}$ resonance, and by letting the form factors vary in a sensible range of value I  investigate the dependence of the asymmetry on their uncertainty. 

Chapter \ref{ch:e158} is organized as following. In Sec. \ref{sec:e-158} I  describe the E-158 experiment in some detail, while in Sec. \ref{sec:brem} I  present the calculation for the bremsstrahlung contribution to the  
$e^-\, P\rightarrow e^-\, N^*$ process for the $D_{13}$ and the $P_{33}$ resonances to the total asymmetry. In Sec. \ref{sec:infrared} I  outline the calculation of the loops corrections to the proton electro-excitation, which  needs to be added to the bremsstrahlung cross section to obtain a finite result.  Finally, in Secs. \ref{sec:results} and \ref{sec:conclusion} I  present the numerical results and the conclusions.

\section{ Electroweak Radiative Corrections to PV Electroexcitation of the $\Delta$}
In Chapter \ref{ch:delta} I  present a work done in  collaboration with Ramsey-Musolf, Zhu, Holstein and Maekawa~\cite{zhuu}. In this study, I contributed  in the computation of the chiral corrections to the axial $N\rightarrow \Delta$ electro-excitation  amplitude, where $N$ is a nucleon and $\Delta$ is the delta  resonance, which could help  the interpretation of the measurement of the axial $N \rightarrow \Delta$ matrix  element planned at Jefferson Laboratory~\cite{proposal}. The goal of the prospective  experiment, is the extraction of such matrix element to a 25$\%$ level of accuracy  by measuring the parity violating asymmetry $A_{RL}$ in the range of $Q^2$ between 0.1 and 0.6 GeV$^2$.
Here, I considered those corrections to $A_{RL}$ which are of order $\mathcal{O}(\alpha)$ compared to the leading term,  which is of order $\mathcal{O}(G_F)$, which arises from the $Z_0$ exchange. At first glance, one might think that, in a 25$\%$ determination of the matrix elements, the $\mathcal{O}(\alpha)$ corrections could be neglected. On the other hand, has it has been shown in Refs.~\cite{zhu,mike,mh}  similar corrections to the electron-proton PV amplitude, were  unexpectedly large and theoretically uncertain. If the same thing would happen for the $N\rightarrow \Delta$ matrix elements, the extraction of the axial form factors from $A_{RL}$ would be considerably more complicated than it was originally thought in the proposal.   

\begin{figure}
\epsfxsize=11.0cm
\centerline{\epsffile{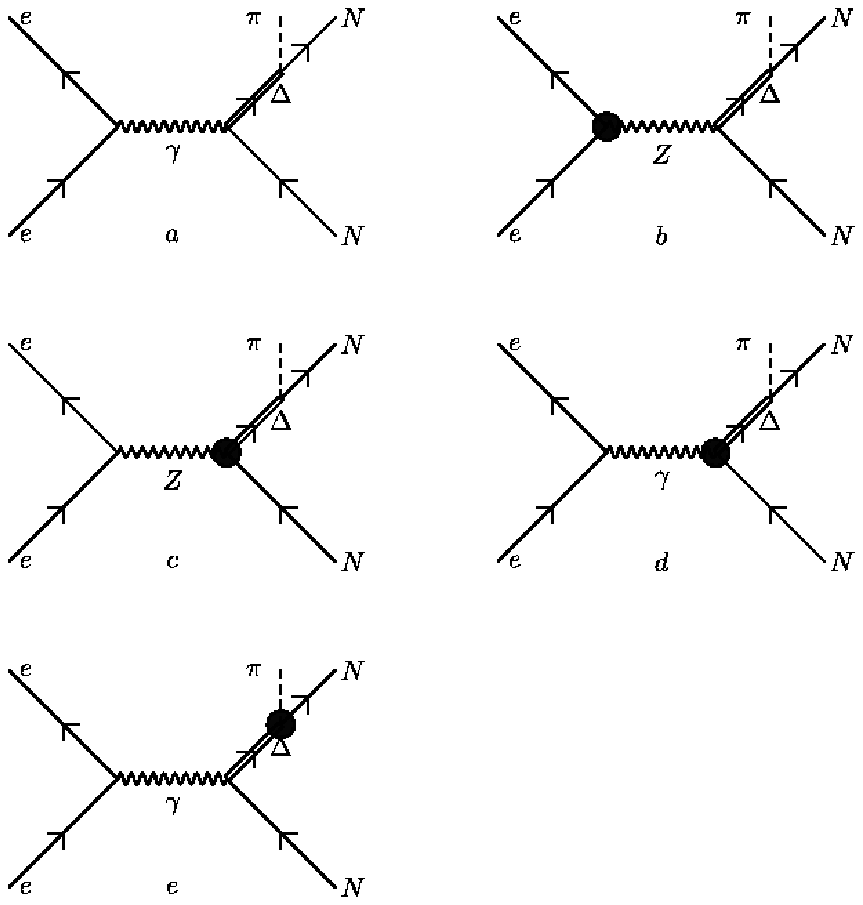}}
%\vspace{1cm}
\caption{
Feynman diagrams describing resonant pion electroproduction.
The dark circle indicates a parity violating coupling.
Fig. \ref{Fig.1a}d gives transition anapole and Siegert's term contributions.
Fig. \ref{Fig.1a}e leads to the PV d-wave $\pi N \Delta$ contribution.
}
\label{Fig.1a}
\end{figure}

The PV amplitude for the process $\vec{e}\, P\rightarrow e \Delta$ is generated by the diagrams in Figs.\ref{Fig.1a}b-e. The tree-level SM amplitude is
%%%%%%%%%%%%%%%%%%%%%%%%%%% Fig 1 %%%%%%%%%%%%%%%%%%%%%%%%%%%%%
%%%%%%%%%%%%%%%%%%%%%%%%%%%%%%%%%%%%%%%%%%%%%%%%%%%%%%%
\bga
iM^{PV}&=&iM_{AV}^{PV}+iM_{VA}^{PV} \,\, ,
\ea 
where
\bga
iM_{AV}^{PV}&=& i\frac{G_F}{2\sqrt{2}}l^{\lambda 5}\la \Delta |J_\lambda|N\ra\,\, ,
\ea
arise from the diagram in Fig.\ref{Fig.1a}b and
\bga
iM_{VA}^{PV}&=& i\frac{G_F}{2\sqrt{2}}l^{\lambda }\la \Delta |J_{\lambda 5}|N\ra\,\, ,
\ea
from the diagram in Fig.\ref{Fig.1a}c. The quantities $J_\lambda\,\,(J_{\lambda 5})$ and $l_\lambda\,\,(l_{\lambda 5})$ denote the vector (axial vector) weak neutral currents of the quarks and the electron, respectively~\cite{mjrm}. The diagram in Fig.\ref{Fig.1a}d contains two contributions, the so called Siegert term, and the anapole term whose amplitudes are
\bga
iM^{PV}_{Siegert}&=&-i\frac{(4 \pi \alpha)d_\Delta}{Q^2 \Lambda_\chi}\bar{e}\gamma_\mu e
\bar{\Delta}_\nu\big[(M-M_\Delta)g^{\mu\nu}-q^\nu \gamma^\mu\big]N\,\, ,
\ea
\bga
iM^{PV}_{Anapole}&=&-i\frac{(4 \pi \alpha)a_\Delta}{  \Lambda_\chi^2}\bar{e}\gamma_\mu e
\bar{\Delta}_\mu N \,\, , 
\ea
where $a_\Delta$ (called anapole) and $d_\Delta$ are low energy constants, which consist in a calculable part, and the counter term part which
 needs to be fixed from experiments (see Sec. \ref{sec:electroweakrc} for more details), $M$ and $M_\Delta$ are the masses of the nucleon and the $\Delta$ respectively, $N$ is the nucleon spinor, $\Delta_\mu$ is the $\Delta$ spinor, and $\Lambda_\chi \sim 1$ Gev is the chiral symmetry breaking scale. The amplitude arising from Fig.\ref{Fig.1a}e is the d-wave amplitude that also contributes to the radiative corrections. Its form is much more lengthy than the previous two, but can be found in Ref.~\cite{zhuu}. One thing worth noticing, is that, because of the factor $\frac{1}{Q^2}$ in the Siegert PV amplitude, this term is highly enhanced in the low $Q^2$ region accessed by the Jefferson Lab experiment, compared to the other amplitudes. This peculiarity of the Siegert amplitude, might actually be used to extract the counter term part of the low energy constant $d_\Delta$, since, as it has been shown in Sec. \ref{sec:electroweakrc}, for reasonable values  of  $d_\Delta$, the Siegert term dominates  in the low $Q^2$ region compared to the other contributions to $A_{RL}$.

Chapter \ref{ch:delta} is organized as following. After a general introduction in Sec. \ref{sec1} about the motivation and the general background of the study, I introduce, in Sec. \ref{sec2}, the definition of the various kinematical variables and the different quantities that  appear in the following  sections. Sec. \ref{sec:electroweakrc} is devoted to presenting the results of the one loop and $1/M_N$ chiral corrections to the low energy constants $a_\Delta$ and $d_\Delta$, and in Sec.  \ref{sec9}  to the counter term parts of  $a_\Delta$ and $d_\Delta$ is estimated, and the corrections due to the Siegert, anapole and d-wave contributions to $A_{RL}$, are computed. Finally, in Sec. \ref{sec10} I summarize the results of the study.
%page 2

%\input{ch1_1.tex}
\chapter{Complete Treatment of the Parity  Violating Asymmetry at Leading Twist}\label{ch:twist2}
In this chapter I  present the calculation for the parity violating (PV) asymmetry $A_{RL}$ at leading twist. Recently, some  experiments have been proposed in which a PV deep inelastic scattering (DIS) asymmetry, using deuteron target, will be measured to  a very high degree of accuracy~\cite{bosted,bosted1} to look for possible extension of the standard model. The intent is to extract $\sin^2\theta_W$ at a $0.5\% $, which translates into a $1 \%$ accuracy in measuring $A_{RL}^{DIS}$. A problem that needs to be considered is if our knowledge of the nucleon structure is good enough to allow such a high accuracy to be achieved. 
In the approximate case in which  sea quarks are neglected, the mass of the target $M$ is negligible compared to the momentum transfer, and  perturbative quantum chromo dynamics (PQCD) corrections are not included,   $A_{RL}^{DIS}$ for an isoscalar target, like the deuteron, assumes a particularly simple form~\cite{cahn}
\bga  \label{e202}
A_{RL}&=& \frac{G_F q^2}{2\sqrt{2}\pi \alpha}\frac{9}{10}\Big[(1-\frac{20}{9} \sin^2\theta_W)+(1-4 \sin^2\theta_W)\lp\frac{1-(1-y)^2}{1+(1-y)^2}\rp\Big]\nn \\
\ea
As  can be seen, under the previous assumptions, the asymmetry does not depend on the structure of the target. However a more complete treatment, which includes the hadronic corrections, is necessary if one wants to reach a precision of 1$\%$ in the interpretation of the measurement of $A_{RL}^{DIS}$. 

 In Sec. \ref{sec:dis} I  present an  overview of the DIS experiments since their advent in the late 1960's, to provide a better understanding of the fundamental role they played in expanding our knowledge of the nuclear and nucleon structure. I  also briefly  introduce the quark parton model (QPM). In Sec. \ref{sec:asy} I  present a complete calculation of $A_{RL}$ at leading twist, in which sea quarks, target mass (TM), and PQCD corrections are included. To do so I use two sets of PDFs parameterizations provided the CTEQ~\cite{cteq} and the MRST~\cite{mrst} groups. The idea is to use the difference in the prediction for $A_{RL}$ obtained by using the two parameterizations as an estimate of the  theoretical error due to the hadronic structure.
\section{Deep Inelastic Scattering and the Quark Parton Model} \label{sec:dis}
\begin{figure}
\centerline{\psfig{figure=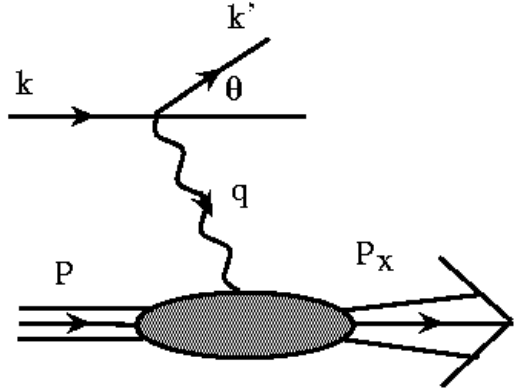}}
\caption{\label{f61}Feynman diagram of a   DIS process, where   an electron scatters off a nucleon target.}
\end{figure}

The scattering of electrons off atomic nuclei  has  proved to be one of the most effective ways to probe the  nucleon structure. The process normally involves an electron of four-momentum $k=(E,\mathbf{k})$ scattering off a nucleus target at rest, and measuring the energy and the angle of the scattered electron, whose four-momentum now is $k'=(E',\mathbf{k}')$ (see Fig. \ref{f61}). 
In this process  a virtual photon with four momentum $q=k-k'$ is transferred to the target. Since the wave length of the virtual photon is proportional to the inverse of the momentum transfer, the higher the momentum  the higher is the resolving power that can be achieved.
By increasing the $Q^2$, different layers of nuclear matter can be revealed, starting from the nucleus, going to the nucleons, and reaching the quarks. 
 For instance, one could consider  the scattering of an electron off a proton with mass $M$. From energy conservation, the Bjorken variable~\cite{bj} $x_B\equiv \frac{Q^2}{2M\nu}$, where $P$ is the proton momentum and $\nu=E-E'$, is equal to one for elastic scattering, and the   form factors (FFs) and cross section is proportional to a delta function $\delta(x_B-1)$ due to the kinematical constraint. The FFs could also have a dynamical $Q^2$ dependence due to the finite size of the target, which implies a different response depending on how deep inside we are probing the nuclear structure. This $Q^2$ dependence for proton elastic FFs is known to be approximately of type
\bga \label{equ1}
f(Q)\varpropto \lp \frac{1}{1+\frac{Q^2}{\Lambda_{proton}^2}}\rp^2\,,
\ea 
where $\Lambda^2_{proton}\simeq .7$ GeV$^2$. In  elastic scattering for $Q^2 \ll \Lambda_{proton}^2$,  $f(Q)\simeq 1$ and the proton is seen as  point like, and the internal structure is not resolved (no $Q^2$ dependence). On the other hand, for $Q^2\gg \Lambda_{proton}$, the FFs  decrease with $Q^2$, showing   a $Q^2$ dependence, and  therefore revealing the proton structure. Similarly, one can imagine to describe the $Q^2$ dependence  of the cross section of  electrons scattering off a nucleus, using a similar dependence as in Eq. (\ref{equ1}) (which is not necessarily true, but it is useful to describe the different layers of matter) introducing a new scale $\Lambda_{nucleus}$.  If one considers the scattering of electrons off a nucleus of $A$ atoms and mass $M_A$, when $Q^2\ll \Lambda_{nucleus}$ the electrons  scatter off the nucleus coherently, seeing the target as a whole. By increasing $Q^2$, the internal structure, {\it i.e } protons and neutrons, is revealed, and the most probable process is an incoherent elastic scattering of the electrons  off the constituent  nucleons. Considering the behavior of the nucleus  cross section as a function of the variable $x_A=\frac{Q^2}{2M_A(E-E')}$, it is peaked in the kinematical region where the incoherent elastic scattering of the nucleons  happens.  The elastic scattering of the nucleus, implies the kinematical constraint $x_B\simeq 1$, and therefore  one has $x_A\simeq \frac{M}{M_A}=\frac{1}{A}$, where $A$ is the number of constituents. 

In the late 1960 the first DIS experiments on proton and deuterium targets were performed at the Stanford Linear Accelerator  Center (SLAC). Since the $Q^2$ in these kinds of experiments is high, one can probe the internal structure on the target. It was surprising to notice that the electrons scattered at large angles, thirty time more copiously than expected~\cite{pano}, revealing   the composite structure of the proton  which appeared to be made of hard point like particles, named partons. Also in this case the cross section seemed to be dominated by the incoherent elastic scattering of the electrons on the constituents of the target, rather than on the target as a whole. The cross section showed a peak for a value of the Bjorken variable $x_B\simeq \frac{1}{3}$ suggesting that the number of partons forming the proton was   three. 

Another important feature which these experiments showed was that the nucleon structure functions (SFs), normally function of two kinematical variables chosen to be $Q^2$ and $x_B$,  seemed to depend only on $x_B$, and very mildly on $Q^2$. This phenomenon, called scaling, was predicted by Bjorken, who noticed that the SFs would exhibit this property in the limit $Q^2$ and $\nu \rightarrow \infty$   but keeping the ratio $Q^2/\nu$ constant~\cite{bj}. Moreover it was observed that the longitudinal cross section (the cross section of a longitudinally polarized virtual photon scattering off the nucleon) was practically zero, leading to the conclusion that the partons should be spin one-half particles. In order to explain the scaling behavior, Feynman~\cite{fey} gave an intuitive picture in which the proton is regarded as a collection of free particles off which the virtual photon scatters elastically, and the total cross section is the incoherent sum of the individual cross section weighted by the probability $q(x_B)$ (the parton distribution function) of a  parton of type $q$ to carry a fraction of the proton momentum between $x_B$ and $x_B+dx_B$.

 It was a quite remarkable property that those partons would behave as free particles, even though nobody was ever able to observe one outside the hadrons, implying that they would be tightly bind inside them. This property lead to the conclusion that the interaction among the partons  should exhibit the property of asymptotic freedom, i.e. the fact that the coupling constant should get smaller  as the momentum transfer used to probe the nucleon increases. Immediately the search for  a gauge theory  which satisfied this property began, until 't Hoff~\cite{hoff}, Gross and Wilczek~\cite{gross1} and Politzer~\cite{politzer} realized that non-Abelian theories possessed this feature. It became then evident that  Quantum Chromo Dynamics (QCD) would be the most promising candidate to describe the strong interaction.

Since then many DIS experiments have been performed, allowing one to better investigate the structure of the nucleon and, ultimately, the standard model (SM). In the late 1970's a parity violating DIS experiment was  performed at SLAC~\cite{prescot} to acquire a better knowledge of the PV neutral current. In the recent years the technological progress allows to measure SM parameters to a precision that was unimaginable just a decade ago. Recently proposed PV DIS experiments aim to measure the weak mixing angle to a very high degree of accuracy, allowing to test possible scenarios  of  physics  beyond the SM.         
\subsection{DIS Kinematics and Variables} \label{sec:kinematics}
In this section I  introduce the basic concepts and formalism of DIS. Most of the material can in found in Ref.~\cite{muta}. 

Consider a process where  a lepton   beam scatters inelastically on a nucleon target. In the deep inelastic regime, the momentum transfer is so high that the target is broken apart, and a multitude  of hadrons are produced.  We are interested in inclusive experiments, namely, those in which only the outgoing electrons are observed, while the produced hadrons are ignored~\cite{fey1}. The scattering amplitude for the process is depicted in Fig. \ref{f61} where, for simplicity, the lepton is taken to be an electron and the target a proton. From Fig. \ref{f61} the following kinematical quantities in the laboratory frame could be defined:
\bga
\theta &=& \cos^{-1}(\frac{\mathbf{k}\cdot\mathbf{k}'}{|\mathbf{k}|| \mathbf{k}'|})=\,\,\,\, \textrm{scattering angle}\\
k&=&(E,0,0,E)\,\,\,\, \textrm{incoming electron four momentum}\\
k'&=&(E',E'\sin\theta,0, E' \cos\theta) \,\,\,\, \textrm{outgoing electron four momentum} \\
P&=&(M,0,0,0)\,\,\,\, \textrm{proton four momentum}\\
q&=&k-k' \,\,\,\, \textrm{momentum transfer}\\
P_X&=&P+q \,\,\,\, \textrm{final hadronic system four-momentum}\\
Q^2&=&-q^2\simeq 4 E E' \sin^2\frac{\theta}{2}\,\,\,\,\, (me\ll E)\\
W^2&=&(P+q)^2=M^2+2M(E-E')-Q^2 \,\,\,\, \textrm{where } W \textrm{ is the invariant mass}\nn\\
&&\qquad\qquad\qquad\qquad\qquad\qquad\qquad\qquad\,\,\,\textrm{of the recoiling nucleus}\\
\nu&=& \frac{P\cdot q}{M}=(\textrm{   in lab frame   })=E-E'\\
 x_B&=&\frac{Q^2}{2M\nu} \, \in \, [0,1] \,\,\,\, \textrm{Bjorken variable} \\
\ea  
where the range of $x_B$ can be easily shown by noticing that for inelastic processes $W^2> M^2$, from which the constraint $x_B<1$ follows immediately.  In the previous relations the electrons are  considered massless, an approximation which is well justified for DIS processes in which  $Q^2\gtrsim 2$ GeV$^2$. 

Considering the Lagrangian describing the electron-proton electromagnetic interaction 
\bga
\mathcal{L}_{In}=\big[-e \bar{\psi}(x) \gamma_\mu \psi(x)+e J_\mu(x)\big]A^{\mu}(x)\, ,
\ea   
where $\bar{\psi}(x) \gamma_\mu \psi(x)$ is the electron current, $J_\mu(x)$ is the proton current and $A_\mu(x)$ the photon field, and applying the Feynman rules to the process in Fig. \ref{f61}, one finds, in the first order approximation, the scattering amplitude to be
\bga
\mathcal{M}=\frac{e^2}{Q^2}  \bar{u}(k',m')\gamma_\mu u(k,m)\la X| J_\mu(0)| PS \ra\,,
\ea
where $m$ and $m'$ are the spin components of the incoming and outgoing electrons respectively, $|PS\ra$ is the proton state with spin $S$ and $|X\ra$ is the hadronic final state. The differential inclusive unpolarized cross section is given by~\cite{muta}
\bga \label{e195}
E'\frac{d\sigma}{d k'}&=&\frac{1}{32(2\pi)^3 k\cdot P}\sum_{m',m,S} \sum_X (2\pi)^4 \delta^4(P_X+k'-k-P)|\mathcal{M}|^2\nn\\
&=& \frac{1}{k\cdot P} \lp \frac{\alpha}{q^2}\rp ^2 l_{\mu\nu}W^{\mu\nu}
\ea 
where $\alpha=\frac{e^2}{4\pi}$ is the fine structure constant, $l_{\mu\nu}$ is the leptonic tensor 
\bga
l_{\mu\nu}=\frac{1}{4}\sum_{m,m'}\Big(\bar{u}_{m'}(k')\gamma_\mu u_m(k)\Big)^*\Big( \bar{u}_{m'}(k')\gamma_\nu u_m(k)\Big)\, ,\nn\\
\ea
and $W_{\mu\nu}$ is the hadronic tensor

and
\bga
W_{\mu\nu}=\frac{1}{2\pi} \sum_X (2\pi)^4\delta^4(P_X-P-q) \frac{1}{2}\sum_S \la P S| J_\mu(0)| X\ra \la   X| J_\nu(0)| P S\ra \, .\nn \\
\ea
From the relation 
\bga
\sum_m u(k,m)\bar{u}(k,m)=k \sla +m_e
\ea
where $m_e$ is the electron mass, one finds for the leptonic tensor
\bga \label{e194}
l_{\mu\nu}&=& \frac{1}{4}\textrm{Tr}[(k\sla+m_e)\gamma_\mu (k \sla ' +m_e)\gamma_\nu] \nn \\
&=& k_{\mu}' k_\nu+k_\mu k_\nu' +\frac{q^2}{2} g_{\mu\nu}\, .
\ea 
For the hadronic tensor one has
\bga \label{e192}
W_{\mu\nu}&=&\frac{1}{2\pi} \sum_X 
\int d^4 x e^{i (q +P -P_X)\cdot x} \frac{1}{2}\sum_S \la P S| J_\mu(0)| X\ra \la   X| J_\nu(0)| P S\ra\nn \\
&=& \frac{1}{2\pi}  
\int d^4 x e^{i q\cdot x}  \la P | J_\mu(x)  J_\nu(0)| P \ra\, ,
\ea
where in the last step I have used translational  invariance which implies 
\bga
\la P  | e^{i P\cdot x}J_\mu(0) e^{-i P_X\cdot x} |X\ra=\la P  | e^{i \hat{\mathbf{P}}\cdot x}J_\mu(0) e^{-i \hat{\mathbf{P}}\cdot x} |X\ra=\la P  | J_\mu(x)  |X\ra \nn \\
\ea
and, since the states $X$s are not observed, I have also used the completeness  relation $ \sum_X | X\ra \la X|=  1$.
Moreover the  average over the target spin is understood, so that for an operator $\hat{O}$ I have
\bga
\la P|\hat{O}|P\ra \equiv \frac{1}{2}\sum_S \la P S| \hat{O}| PS\ra\, .
\ea
Even though the hadronic tensor cannot be directly computed - since the proton wave function is not known - one can still use Lorentz,  parity, time-reversal invariance and current conservation to constraint its form. From  current conservation, the hadronic tensor has to satisfy $q_\mu W^{\mu\nu}=q_\nu W^{\mu\nu}=0$ (which follows from the continuity equation $\partial_\mu J^\mu =0$). Moreover, as  can be seen from Eq. (\ref{e192}), the hadronic tensor has to be a second rank  tensor function of the momentum transfer and the target momentum; therefore, it has to be built  from the metric tensor $g_{\mu\nu}$ and   products of $q_\mu$ and $P_\mu$. 
The most general form for $W_{\mu\nu}$ satisfying the previous constraints is
\bga \label{e193}
W_{\mu\nu}&=&M W_1(x_B,Q^2)\lp -g_{\mu\nu}+\frac{q_\mu q_\nu}{q^2}\rp\nn \\
&+& \frac{W_2(x_B,Q^2)}{M} \lp P_\mu -\frac{P\cdot q}{q^2}q_\mu\rp\lp P_\nu- \frac{P\cdot q}{q^2} q_\nu \rp \, , \nn \\
\ea   
where the quantities $W_1$ and $W_2$ are the structure functions (SFs), which contain the information about the nucleon structure. This functions cannot yet be computed from first principles and they need to be measured from experiments. 
Substituting Eqs. (\ref{e194}) and (\ref{e193}) in (\ref{e195}) the unpolarized cross section becomes
\bga
\frac{d\sigma}{d \Omega dE'}=\frac{2 \alpha^2 E^{'2}}{M Q^4}\Big[ 2M W_1(x_B,Q^2)\sin^2\frac{\theta}{2}+  M W_2(x_B,Q^2) \cos^2\frac{\theta}{2}\Big] \, .
\ea 

It can be shown that, in the case in which the electrons scatter on a point-like object, the SFs assume a particular simple form. If, for instance, one considers electron-muon scattering the result is
\bga\label{ea30}
W_{1el}^{e\mu}&=&\frac{Q^2}{4 m_\mu^2 \nu}\delta\lp 1-\frac{Q^2}{2m_\mu \nu}\rp \, , \\
\nu W_{2el}^{e\mu}&=& \delta \lp 1-\frac{Q^2}{2 m_\mu \nu} \rp \, .
\ea
Analogously, if the electrons scatters off  a parton of charge $e_{q}$ and mass $m_q$, one would have  
\bga
W_{1el}^{e q}&=&e_q^2\frac{Q^2}{4 m_q^2 \nu}\delta\lp 1-\frac{Q^2}{2m_q \nu}\rp \, , \\
\nu W_{2el}^{e q}&=& e_q^2\delta \lp 1-\frac{Q^2}{2 m_q \nu} \rp \, .
\ea
Following the parton picture suggested by Feynman, one could write the     SFs as an incoherent sum over the elastic contributions from the scattering of the electrons off the constituent partons,  weighted by the probability distribution $q(x)$  for a parton to carry a fraction of the nucleon momentum between $x$ and $x+dx$, with $0\leq x\leq 1$. By using Eq. (\ref{ea30}) and the relation $m_q=x M$ one finds
\bga \label{e196}
W_1^P(x_B)&=&\sum_q \int_0^1 dx'q(x')e_q^2 \frac{Q^2}{4(x')^2 M^2\nu }\delta \lp 1-\frac{Q^2}{2 M x' \nu}\rp\nn\\
&=&\sum_q e_q^2 \int_0^1 dx'q(x')\frac{x'x_B}{2 x^{'2} M} \delta (x_B-x')\nn \\
&=& \frac{1}{2 M} \sum_q e_q^2 q(x_B)\, .
\ea 
Similarly for $W_2$ one gets
\bga \label{e197}
\nu W_2(x_B)= x_B\sum_q e_q^2 q(x_B)\,.
\ea
By introducing two new dimensionless SFs 
\bga
F_1^P(x_B)=M W_1^P(x_B)\,,
\ea
and
\bga
F_2^P(x_B)=\nu W_2^P(x_B)\,
\ea
one sees how the parton naturally explains the phenomenon of scaling, since the new SFs,  in this approximation, depend  only  on the Bjorken variable $x_B$ and not on $Q^2$.  Another important consequence that could be inferred from the previous equation is the relation~\cite{callan}
\bga
F_2(x_B)=2x_B F_1(x_B)\, ,
\ea
a very important result known as the Callan-Gross relation.  
\section{Parity Violating Asymmetry in Deuteron}
\label{sec:asy}
Parity violating experiments have played a crucial role in testing the SM of electro-weak interactions since the late 1970's~\cite{prescot}. In the recent years, thanks to a phenomenal technological progress, DIS PV asymmetries can be measured to a level of $1\%$, providing opportunity, in principle,  either to put more stringent constraints on the SM parameters, or eventually to look for possible  physics beyond the SM. In this section I  compute $A_{RL}$ considering all the possible leading twist corrections such as sea quark, TM and PQCD corrections, and check if their uncertainties  might vitiate the interpretation of the measurement to a $1\%$ level. To estimate the uncertainty  due these hadronic corrections I  adopt two different sets of PDFs parameterizations~\cite{cteq,mrst}. The difference between the two results is taken as a measure of the  uncertainty.

\begin{figure}
\centerline{\psfig{figure=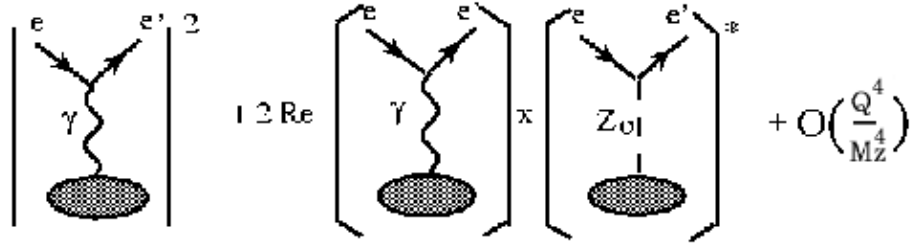}}
\caption{\label{f63} Polarized electron-nucleon cross section to first order in the Fermi constant $G_F$. }
\end{figure}

In order to isolate the PV part in a scattering process of an electron off a nucleon, one defines the polarized asymmetry
\bga \label{ee197}
A_{RL}\equiv \frac{\sigma_R-\sigma_L}{\sigma_R+\sigma_L}
\ea     
where $\sigma_{R(L)}$ stands for the cross section of a right handed (left handed) polarized electron beam scattering on a nucleon target. The cross section for such a process, up to  order  $\frac{Q^4}{M_Z^4}$ where $M_Z=91.1882(22)$ GeV is the mass of the neutral $Z_0$ boson, is represented schematically in Fig. (\ref{f63}). Since the electromagnetic part is parity conserving, the cross section is the same for right and left handed electrons. On the other hand, the interference term changes sign upon flipping the helicity of the electron; therefore,  the asymmetry  isolates the PV part. 
In analogy with what has been shown before for the electromagnetic case (see Eqs. (\ref{e194}) and (\ref{e193})), one could define a set of tensors for each of the processes in Fig. \ref{f63} in the case of polarized electrons. \\
The polarized electromagnetic tensor is~\cite{anselmino}
\bga 
l_{\mu\nu}^{\gamma}&=&\frac{1}{4}\sum_{m'}\Big(\bar{u}(k',m')\gamma_\mu u(k,\lambda)\Big)^*\Big( \bar{u}(k',m')\gamma_\nu u(k,\lambda)\Big) \nn\\
&=& \frac{1}{2}( k_{\mu}' k_\nu+k_\mu k_\nu' +\frac{q^2}{2} g_{\mu\nu}-i\lambda\epsilon_{\mu\nu\alpha\beta} k^{\alpha} k^{' \beta})
\ea
where $\lambda=\pm 1$ is the helicity of the initial electron. In deriving the previous relation I have used the relation
\bga
 u(k,\lambda)\bar{u}(k,\lambda)=(k \sla +m_e)\frac{1+\gamma_5 s\sla}{2}\,,
\ea
 In the case of relativistic electrons the spin is related to the momentum by the relation $s_\mu=\lambda\frac{k_\mu}{m_e}$. \\
The electron interacts weakly with the $Z_0$ boson through  a current 
\bga
j_\mu(x)=\bar{\psi} (x)\gamma_\mu (g_V^e+g_A^e\gamma_5)\psi (x) \, ,
\ea  
where $g_V^e=(-1+4 \sin^2\theta_W)$ and $g_A^e=1$ are the vector and axial coupling of the electron to the $Z_0$ respectively. In this case the leptonic tensor for the interference term becomes
\bga
l_{\mu\nu}^{\gamma Z}=(g_V^e+\lambda g_A^e) l_{\mu\nu}^{\gamma}\,.
\ea 
One may also want an interference hadronic tensor, given by the product of the hadronic electromagnetic current with the neutral electroweak one. 
\bga
W_{\mu\nu}^{\gamma Z}&=&\frac{1}{2\pi}  
\int d^4 x e^{i q\cdot x}  \la P | J_\mu^{\gamma}(x)  J_\nu^{\gamma Z}(0)| P \ra =\tilde{F}_1(x_B,Q^2)\lp -g_{\mu\nu} +\frac{q_\mu q_\nu}{q^2}\rp\nn \\
&+& \frac{\tilde{F}_2(x_B,Q^2)}{P\cdot q} \lp P_\mu -\frac{P\cdot q}{q^2}q_\mu\rp\lp P_\nu- \frac{P\cdot q}{q^2} q_\nu \rp
\nn \\
&+&\tilde{F}_3(x_B,Q^2)\frac{i\epsilon_{\mu\nu\alpha \beta}}{2 P\cdot q} P^{\alpha}q^{\beta}\,,
\ea 
where the $\tilde{F}_i(x_B,Q^2)$ are the electro-weak interference SFs, analogous of $F_{1,2}$ defined above. Note that the SF $\tilde{F_3}$, multiplies the  pseudo-tensor $\epsilon_{\mu\nu\alpha \beta}$.
The polarized cross section, neglecting terms of  order $\frac{Q^4}{M_Z^4}$, is~\cite{predazzi} 
\bga \label{e198}
\frac{d^2 \sigma_{L,R}}{d \Omega d E'}&=&
\frac{4 \pi \alpha^2 s}{Q^4} \bigg\{ \big[ x y^2 F_1(x,Q^2)+(1-y -\frac{x y M}{2 E} )F_2(x,Q^2)\big] +\nn \\
&-&\frac{Q^2}{M_Z^2}\frac{g_V^e\pm g_A^e}{8 \sin^2\theta_W \cos^2\theta_W}\big[ x y^2 \tilde{F}_1(x,Q^2)+\nn \\
&+&(1-y -\frac{x y M}{2 E} )\tilde{F}_2(x,Q^2)\pm (y-\frac{y^2}{2})x \tilde{F}_3(x,Q^2)\big]\bigg\}
\ea
where $L,R$ refers to a left handed or right handed electron (corresponding to the plus and minus sign respectively),   $s=(P+k)^2$ and $y= \frac{P\cdot q}{P\cdot k}$. Using  Eqs. (\ref{ee197}) and (\ref{e198}) the asymmetry can be written as
\bga \label{ea49}
A_{RL}=A_{RL}^0\frac{W^{(PV)}}{W^{(EM)}}
\ea
where
\bga
A_{LR}^0=-\frac{G_F Q^2}{2\pi \alpha \sqrt{2}}
\ea
and $G_F=1.6639(1)\times 10^{-5}$ GeV$^{-2}$ is the Fermi constant measured in muon decay, defined as
\bga 
 G_{F}=\frac{\pi\alpha}{\sqrt{2} M_Z^2 \sin^2\theta_W \cos^2\theta_W} \, .
\ea
The two quantities $W^{(PV)}$ and $W^{(EM)}$ are respectively defined as 
\bga 
W^{(PV)}&=&\frac{1}{4} \bigg\{2g_A^e\left[\tilde{F}_1x_B y^2+\tilde{F}_2\left((1-y)-\frac{x_B My}{2E}\rp\right]+    \nn \\
&+& g_V^e\tilde{F}_3x_B (1-(1-y)^2) \bigg\}
\ea 
\bga
W^{(EM)}=\left\{ F_1 y^2 x_B+ F_2\left[(1-y)-\frac{x_B My}{2E}\right]\right\}\, .
\ea 
In order to write the SFs in terms of the quark PDFs, it is convenient to express the hadronic electromagnetic and electroweak neutral  currents as sums over the different quark fields~\cite{mjrm} 
\bga
&&J_\mu^{EM}=\sum_{q}e_q \bar{q}\gamma_\mu q \nonumber \\ && J_\mu^{NC}=
\sum_{q}g_V^q \bar{q}\gamma_\mu q +\sum_{q}g_A^q \bar{q}\gamma_\mu\gamma_5 q  
\ea
where the sum runs over the $u,\, d ,\, s $ and $c$ quark, $e_q$ is the electric charge of a quark of type $q$, and $g_V^q$ and $g_A^q$ are the vector and axial-vector couplings of a quark $q$ to the neutral boson $Z_0$. In the SM one has 
\bga
g_V^u&=&g_V^c= 1-\frac{8}{3}\sin^2\theta_W\,\,\,\,\,\,\,\,\,\, , \,\,\,\,\,\,\,\,\,\,g_V^d=g_V^s=-1+\frac{4}{3}\sin^2\theta_W\nn \\
g_A^u&=&g_A^c= -1  \,\,\,\,\,\,\,\,\,\,\,\,\,\,\,\,\,\,\,\,\,\,\,\,\,\, \,\,\,\,\,\,\,\,\,\,,\,\,\,\,\,\,\,\,\,\, g_A^d=g_V^s=1 \, .
\ea  
Following Ref.~\cite{mjrm}, one could write the different hadronic currents in terms of their isospin content
\bga
J_\mu^{EM\,  T=0} =\frac{1}{\sqrt{3}} \hat{V}^{(c)}_\mu\,\,\,\, , \,\,\,\, J_\mu^{EM \, T=1}=\hat{V}_\mu^{(3)}
\ea 
where 
\bga
\hat{V}_\mu^{(c)}&=&\frac{1}{2\sqrt{3}} (\bar{u}\gamma_\mu u+\bar{d}\gamma_\mu d-2\bar{s}\gamma_\mu s +4\bar{c}\gamma_\mu c)\nn \\
\hat{V}_\mu^{(3)}&=&\frac{1}{2}(\bar{u}\gamma_\mu u -\bar{d} \gamma_\mu d)
\ea
so that
\bga
J_\mu^{EM}=J_\mu^{EM\, T=0}+J_\mu^{EM\, T=1} \, .
\ea
The superscript $T$ indicates the isospin, i.e. isoscalar for $T=0$ and isovector for $T=1$. The neutral current could then be decomposed as
\bga
J_\mu^{NC }&=& \sqrt{3} \xi_V^{T=0} J_\mu ^{EM\, T=0}+\xi_V^{ T=1}J_\mu ^{EM \, T=1}+\xi^{s}_V \bar{s}\gamma_\mu s +\xi^{c}_V \bar{c}\gamma_\mu c+   \xi_A^{T=0} \hat{A}_\mu ^{(8)}\nn \\ 
&+&\xi_A^{ T=1}\hat{A}_\mu ^{(3)}+\xi^{s}_A \bar{s}\gamma_\mu\gamma_5 s +\xi_A^{c} \bar{c}\gamma_\mu\gamma_5 c  
\ea
where the $\hat{A}_\mu^{(i)}$ can be obtained from the definitions of the $\hat{V}_\mu^{(i)}$ by replacing $\gamma_\mu \rightarrow  \gamma_\mu \gamma_5$ and the couplings are defined as 
\bga
\xi_V^{T=1}=g_V^u-g_V^d &\xi_A^{T=1}=g_A^u-g_A^d \\ \nn 
\xi_V^{T=0}=\sqrt{3}(g_V^u+g_V^d)&\xi_A^{T=0}=\sqrt{3}(g_A^u+g_A^d)\\ \nn
\xi_V^{s}=g_V^u+g_V^d+g_V^s& \xi_A^{s}=g_A^u+g_A^d+g_A^s\\ \nn
\xi_V^{c}=g_V^c-2g_V^u-2g_V^d& \xi_A^{c}=g_A^c-2g_A^u-2g_A^d \, .\\ \nn
\ea 
Notice that, in this, as opposed as in Ref.~\cite{mjrm}, I also considered the presence of charm quarks, since in the proposed experiment of Ref.~\cite{bosted1}, the $Q^2$ is much larger than the charm quark threshold production.  

In order to compute the SFs one has to consider the following matrix elements
\bga \label{e200}
\la P | J_\mu(x) J_\nu(0) | P \ra \nn \,. 
\ea
If now  we consider an isoscalar target as it has been proposed in~\cite{bosted,bosted1}, one would need to retain only those product of currents which transform as an isoscalar. Therefore, for the product of two elecromagnetic currents, which is needed to compute the electromagnetic  SFs, I have\footnote{Remember that the product of two isoscalars is still an isoscalar, the product of an isoscalar with an isovector is an isovector while  the product of two isovectors gives raise to an isoscalar, an isovector and an isotensor of rank two}
\bga   
J_{\mu}^{EM}(z)J_\nu^{EM}(0) \!\!\!&\simeq&\!\!\!J_{\mu}^{EM\, ,T=0}(z)J_\nu^{EM\, ,T=0}(0)+
J_{\mu}^{EM\, ,T=1}(z)J_\nu^{EM\, ,T=1}(0) \nn \\
\ea
where the symbol $\simeq$ indicates that this is not an identity, but I only retained those terms which give a nonzero contribution once they are sandwiched between isoscalar states. In a similar way, the only products needed for the interference SFs are   
\bga   
J_{\mu}^{EM}(z)J_\nu^{NC}(0) &\simeq& \xi_V^{T=1}J_{\mu}^{EM\, ,T=1}(z)J_{\nu}^{EM\, T=1}(0)+\nn \\  
&+&\sqrt{3}\xi_V^{T=0}J_{\mu}^{EM\, ,T=0}(z)J_{\nu}^{EM\, ,T=0}(0)+\nn \\
&+& \xi_V^sJ_{\mu}^{EM\, ,T=0}(z)\bar{s}(0)\gamma_\nu s(0)  + \nn \\
&+& \xi_V^{c}J_{\mu}^{EM\, ,T=0}(z)\bar{c}(0)\gamma_\nu c(0)  + \nn \\
&+&\xi_A^{T=1}J_{\mu}^{EM\, ,T=1}(z)A_{\nu}^{(3)}(0)+\nn \\
&+&\xi_A^{T=0}J_{\mu}^{EM\, ,T=0}(z)A_{\nu}^{(c)}(0)+\nn \\
&+&\xi_A^{s}J_{\mu}^{EM\, ,T=0}(z)\bar{s}(0)\gamma_\nu\gamma_5 s(0)  \nn \\
&+& \xi_A^{c}J_{\mu}^{EM\, ,T=0}(z)\bar{c}(0)\gamma_\nu\gamma_5 c(0)  \nn \\
\ea
The number of terms in the previous two relations could be further simplified by limiting ourselves to the case of leading twist contributions. In this approximation the product of two quark currents of different flavor can be neglected, since they give rise to twist four contributions (see Sec. \ref{sec:fourfermion_tw4cf}). The contributions to the SFs from a product of a unit charge currents such as  $\bar{q}(x) \gamma_\mu  q(x) \bar{q}'(0) \gamma_\mu (1+\gamma_5) q'(0)$ have been computed by the authors of Refs.~\cite{ellis,georgi,ji}. In Refs.~\cite{ellis,georgi} also the TM corrections have been included. The result is
\bga \label{ee202}
f_1^{q_i\, TM}(x_B,Q)&=& \frac{x_B}{2 k \xi} q_i(\xi)  
+ \frac{x^2  }{k^2}\frac{M^2}{Q^2} \int^1_{\xi} \frac{d \eta}{\eta} q_i(\eta)    
\ea  
\bga \label{ee200}
f_2^{q_i, TM}(x_B,Q)&=& \frac{x_B^2}{k^3 \xi}    q_i(\xi) + 
  \frac{6x_B^3 }{k^4} \frac{M^2}{Q^2}\int^1_{\xi} \frac{d \eta}{\eta}  q_i(\eta) 
\ea 
\bga\label{e201}
f_3^{q_i\, TM}(x_B,Q)&=&\pm \lp \frac{x_B}{ k^2 \xi} q_i(\xi)  
+ \frac{2x_B^2  }{k^3}\frac{M^2}{Q^2} \int^1_{\xi} \frac{d \eta}{\eta} q_i(\eta)  \rp  
\ea
where in (\ref{e201}) the plus sign refers to quarks and the minus to antiquarks since quarks couple to the $Z_0$ trough a vector minus axial current ($V-A$) while antiquarks couple through a $V+A$ current~\cite{predazzi}. 
The small case letters have been used   to emphasize the fact that they refer to a single quark (antiquark) SF and TM refers to the fact that the target mass corrections have been included.  The $q_i(x)$ is the PDF for the $i$ quark,    and I have introduced  the quantity
\bge
k=\lp 1+ 4 x_B^2\frac{M^2}{Q^2}\rp^{1/2}
\ee
and the Nachtmann variable $\xi$ 
\bge
\xi=\frac{2 x_B}{1+k} \, ,
\ee
which reduces to the Bjorken variable in the limit of zero target mass.

In addition to the TM corrections, there are other leading twist corrections to the SFs which are of first order in the strong coupling,
which I  call PQCD corrections. Such contributions to the SFs have been computed  by the authors of Refs.~\cite{alt,herrod}.
 As opposed to the leading logarithmic corrections (LLC) to the SFs, which do not depend on the process and the particular SF one is considering, and therefore
could be absorbed in the redefinition of the PDFs, these corrections are process dependent and they also differ depending on 
the specific SF. Moreover they are also renormalization scheme dependent, an important point that I  discuss 
later.\\
The contributions to the SFs $ F_1$, $F_2$ and $F_3$ in the modified  minimal subtraction scheme ($\overline{MS}$) are~\cite{herrod}
\bga
f_1^{q_i\, PQCD}(x_B)&=&\frac{\alpha_S(Q)}{4\pi }\frac{x_B}{\xi k}\int_{\xi}^1 \frac{dy}{y}\Big[\frac{C_F}{2}(F_q(\frac{\xi}{y})-4\frac{\xi}{y})q_i(y) \nn \\
&+& T_R (F_G(\frac{\xi}{y})-4\frac{\xi}{y}(1-\frac{\xi}{y}))g(y)\Big]
\ea
\bga
f_2^{q_i\, PQCD}(x_B)=\frac{\alpha_S(Q)}{4\pi}\frac{x_B^2}{k^3 \xi}\int_{\xi}^1\frac{dy}{y}\Big[C_FF_q(\frac{\xi}{y})q_i(y)+2T_R F_G(\frac{\xi}{y})g(y)\Big]
\ea
\bga
f_3^{q_i\, PQCD}(x_B)=\frac{\alpha_S(Q)}{4\pi}\frac{C_F x_B}{k^2 \xi}\int_{\xi}^1 \frac{dy}{y}\Big[F_q(\frac{\xi}{y})-2-2\frac{\xi}{y}\Big]q_i(y)\,,
\ea
where $g(x_B)$ is the gluon PDF,
and where $F_q(x_B)$ and $F_G(x_B)$ are defined as
\bga
F_q(x_B)&=&-\frac{3}{2}\frac{1+x_B^2}{(1-x)_+}+\frac{1}{2}(9+5 x_B)-2 \frac{1+x_B^2}{1-x_B}\log x_B\nn \\
&+&2(1+x_B^2)
\lp \frac{\log(1-x_B)}{1-x_B}\rp_+ -\delta (1-x_B)(9+\frac{2}{3}\pi^2)
\ea
and
\bga
F_G(x_B)= (1-2x_B+2x_B^2)\log \lp\frac{1-x_B}{x_B}\rp-1+8x_B(1-x_B)\, ,
\ea
where $C_F=\frac{4}{3}$ and $T_F=\frac{1}{2}$ and $g(x_B)$ is the gluon PDF.  Notice that the function $F_G(x_B)$ appears multiplied by a factor $\frac{1}{2}$ compared to way it appears in~\cite{herrod}, the reason being that in their expressions for the SFs, the sum runs only over the quarks, while in my case it runs over the quarks and antiquarks.    Once again I have used small case letters to indicate that the contribution refers to a single quark with unit coupling, and the TM corrections, which affect  also the PQCD corrections, have been included~\cite{bread}.

Since the SFs are physical quantities extracted from experiments, they cannot be scheme dependent. On the other hand, since the quantities $f_j^{q_i\,PQCD}(x_B)$ depend on the renormalization scheme adopted, also the quantities $f_j^{q_i\,TM}(x_B)$, and therefore  PDFs $q_j(x_B)$, have to, so that the dependence cancels out in the SFs. There are two different approaches normally used to define the PDFs, which are~\cite{predazzi}:
\begin{enumerate}
\item One specifies the renormalization scheme used to compute  the functions $f_i^{q_i\,PQCD}(x_B)$ and then extract the PDFs  in these scheme, which now carry a label indicating the scheme used (such as $MS$, $\overline{MS}$). In this case, as long as we are consistent in using the same scheme, the PDFs obtained are universal and can be used for any kind of process besides DIS.
\item Another approach is to consider a SF that can be precisely measured, usually taken to be $F_2$, and absorb all the PQCD corrections for that SF in the PDFs definition, namely, to any order in QCD $F_2$ is defined as
\bga
F_2(x_B,Q^2)\equiv x_B\sum_q e_q^2 q(x_B,Q^2)\,.
\ea 
In this case the PDFs obtained are physical quantities and do not depend on the renormalization scheme utilized. On the other hand, they are not universal, and if we are going to compute a process different from DIS, they would need some compensating factor. This scheme is called DIS scheme and in this case the PDFs  carry the label DIS.
\end{enumerate}  
I now proceed in writing the two electromagnetic and the three interference  SFs in the case of a deuteron target. I  assume isospin symmetry, which implies that  the up (down) quark distribution (for both quark and antiquark) in the proton (neutron) is the same as the down (up) distribution in the neutron (proton), i.e. $u^P=d^N$ ($d^P=u^N$). I also assume that the remaining sea  PDFs are the same for proton and neutron,  and assume symmetric distribution for the quark and antiquark, i.e. $s=\bar{s}$ and $c=\bar{c}$.\footnote{Recently, in order to reconcile the discrepancy between the SM value of the weak angle and the value measured in the NuTeV experiment~\cite{nutev}, it has been suggested that the previous assumption, isospin symmetry and symmetric sea quark distribution, might  not hold~\cite{kretzer,thomas3}. I do not consider this possibility here}. This translates in the following relations
\bga
s^P=s^N=\bar{s}^P=\bar{s}^N\,\,\,\,\,\,\,,\,\,\,\,\,\,\, c^P=c^N=\bar{c}^P=\bar{c}^N
\ea
 Moreover one defines the PDFs in the deuteron as
\bga
q^D=\frac{q^P+q^N}{2}
\ea  
Since, because of isospin symmetry, all the PDFs can be expressed in terms of the ones in the proton, I  omit the suffix $P$ from now on. The result for the five SFs is
\bga  
F_1 (x_B,Q^2)&=&  \frac{5}{18} \Big( f_1^{u\, TM}(x_B) +f_1^{\bar{u}\, TM}(x_B) +f_1^{d\, TM}(x_B)+f_1^{\bar{d}\,TM}(x_B)\nn\\
&+&f_1^{u\, PQCD}(x_B) +f_1^{\bar{u}\, PQCD}(x_B) +f_1^{d\, PQCD}(x_B) +f_1^{\bar{d}\,PQCD}(x_B)\Big)\nn \\
&+&1/9\Big(f_1^{s\, TM}(x_B) +f_1^{\bar{s}\, TM}(x_B) +f_1^{s\, PQCD}(x_B) +f_1^{\bar{s}\,PQCD}(x_B)\Big)\nn \\
&+&4/9\Big(f_1^{c\, TM}(x_B) +f_1^{\bar{c}\, TM}(x_B) +f_1^{c\, PQCD}(x_B) +f_1^{\bar{c}\,PQCD}(x_B)\Big)  \nn \\
\ea 
\bga
F_2 (x_B,Q^2)&=&  
\frac{5}{18} \Big( f_2^{u\, TM}(x_B) +f_2^{\bar{u}\, TM}(x_B) +f_2^{d\, TM}(x_B)+f_2^{\bar{d}\,TM}(x_B)\nn \\
&+&f_2^{u\, PQCD}(x_B) +f_2^{\bar{u}\, PQCD}(x_B) +f_2^{d\, PQCD}(x_B) +f_2^{\bar{d}\,PQCD}(x_B)\Big)\nn \\
&+&1/9\Big(f_2^{s\, TM}(x_B) +f_2^{\bar{s}\, TM}(x_B) +f_2^{s\, PQCD}(x_B) +f_2^{\bar{s}\,PQCD}(x_B)\Big)\nn \\
&+&4/9\Big(f_2^{c\, TM}(x_B) +f_2^{\bar{c}\, TM}(x_B)+f_2^{c\, PQCD}(x_B) +f_2^{\bar{c}\,PQCD}(x_B) \Big)  \nn \\
\ea 
\bga
\tilde{F}_1(x_B,Q^2)&=&\!  \frac{1}{4}\left(\xi_V^{T=1}\!+\!\frac{\xi_V^{T=0}}{3\sqrt{3}}\right)\Big(f_1^{u\, TM}(x_B) \!+\!f_1^{\bar{u}\, TM}(x_B) \!+\!f_1^{d\, TM}(x_B) \!+\!f_1^{\bar{d}\,TM}(x_B)\nn \\
&+&f_1^{u\, PQCD}(x_B) +f_1^{\bar{u}\, PQCD}(x_B) +f_1^{d\, PQCD}(x_B) +f_1^{\bar{d}\,PQCD}(x_B) \Big) \nn \\
&-& 1/3g_V^s\Big(f_1^{s\, TM}(x_B) +f_1^{\bar{s}\, TM}(x_B)+f_1^{s\, PQCD}(x_B) +f_1^{\bar{s}\,PQCD}(x_B) \Big) \nn \\
&+& 2/3g_V^c\Big(f_1^{c\, TM}(x_B) +f_1^{\bar{c}\, TM}(x_B) +f_1^{c\, PQCD}(x_B) +f_1^{\bar{c}\,PQCD}(x_B)\Big)  \nn \\
\ea
\bga
\tilde{F}_2(x_B,Q^2)&=&    \frac{1}{4}\left(\xi_V^{T=1}\!+\!\frac{\xi_V^{T=0}}{3\sqrt{3}}\right)\Big( f_2^{u\, TM}(x_B) \!+\!f_2^{\bar{u}\, TM}(x_B) \!+\!f_2^{d\, TM}(x_B)\! +\!f_2^{\bar{d}\,TM}(x_B)\nn \\
&+&f_2^{u\, PQCD}(x_B) +f_2^{\bar{u}\, PQCD}(x_B) +f_2^{d\, PQCD}(x_B) +f_2^{\bar{d}\,PQCD}(x_B)\Big)\nn   \\
&-&1/3g_V^s\Big(f_2^{s\, TM}(x_B) +f_2^{\bar{s}\, TM}(x_B)+f_2^{s\, PQCD}(x_B) +f_2^{\bar{s}\,PQCD}(x_B)\Big)  \nn \\
&+& 2/3g_V^c\Big(f_2^{c\, TM}(x_B) +f_2^{\bar{c}\, TM}(x_B) +f_2^{c\, PQCD}(x_B) +f_2^{\bar{c}\,PQCD}(x_B) \Big)  \nn \\
\ea 
\bga 
\tilde{F}_3(x_B,Q^2)&=& \frac{1}{4}\left(\xi_A^{T=1}\!+\!\frac{\xi_A^{T=0}}{3\sqrt{3}}\right)\Big(f_3^{u\, TM}(x_B)\! -\!f_3^{\bar{u}\, TM}(x_B)\! +\!f_3^{d\, TM}(x_B) \!-\!f_3^{\bar{d}\,TM}(x_B) \nn \\
&+&f_3^{u\, PQCD}(x_B) -f_3^{\bar{u}\, PQCD}(x_B) +f_3^{d\, PQCD}(x_B) -f_3^{\bar{d}\,PQCD}(x_B)\Big)    \nn \\
\ea
Let us now introduce some new quantities which allows us to rewrite $A_{RL}$ in a form more similar to the one in Refs.~\cite{bosted,bosted1}
\bga
R(x_B,Q^2)&=&\frac{F_2(x_B,Q^2)}{2 x_B F_1(x_B,Q^2)}-1\\
\tilde{R}(x_B,Q^2)&=&  \frac{\tilde{F}_2(x_B,Q^2)}{2 x_B \tilde{F}_1(x_B,Q^2)}-1\,,
\ea
which are both equal zero in the naive parton model because of the Callan-Gross relation~\cite{callan}, 
\bga
\mathcal{Q}_i(x_B,Q^2)&=& \frac{1}{x_B}\Big[f_2^{q_i\, TM}(x_B,Q^2)+f_2^{q_i\, PQCD}(x_B,Q^2)\Big]\\
\mathcal{Q}_{jv}(x_B,Q^2)&=&f_3^{q_j\, TM}(x_B,Q^2)-f_3^{\bar{q}_j\, TM}+f_3^{q_j\, PQCD}(x_B,Q^2)-f_3^{\bar{q}_j\, PQCD}(x_B,Q^2)\nn\\\,,
\ea
with $q_i=u,\bar{u},d,\bar{d},s,\bar{s},c,\bar{c}$ , $q_j=u,d$,  $\mathcal{Q}_i=\mathcal{U},\bar{\mathcal{U}},\mathcal{D},\bar{\mathcal{D}},\mathcal{S},\bar{\mathcal{S}},\mathcal{C},\bar{\mathcal{C}}$ and $\mathcal{Q}_j=\mathcal{U}_v,\mathcal{D}_v$. 
\bga
R_c(x_B,Q^2)&=& 2 \frac{\mathcal{C}(x_B,Q^2)+\bar{\mathcal{C}}(x_B,Q^2)}{\mathcal{U}(x_B,Q^2)+\bar{\mathcal{U}}(x_B,Q^2)+\mathcal{D}(x_B,Q^2)+\bar{\mathcal{D}}(x_B,Q^2)}\nn \\
R_s(x_B,Q^2)&=& 2 \frac{\mathcal{S}(x_B,Q^2)+\bar{\mathcal{S}}(x_B,Q^2)}{\mathcal{U}(x_B,Q^2)+\bar{\mathcal{U}}(x_B,Q^2)+\mathcal{D}(x_B,Q^2)+\bar{\mathcal{D}}(x_B,Q^2)} \nn \\
R_v(x_B,Q^2)&=&  \frac{\mathcal{U}_v(x_B,Q^2) +\mathcal{D}_v(x_B,Q^2) }{\mathcal{U}(x_B,Q^2)+\bar{\mathcal{U}}(x_B,Q^2)+\mathcal{D}(x_B,Q^2)+\bar{\mathcal{D}}(x_B,Q^2)}\nn \\
\ea 
\bga
C_{1,u}&=&-\frac{1}{2}g_A^eg_V^u\nn \\
C_{1,d}&=&-\frac{1}{2}g_A^eg_V^d\nn \\
C_{2,u}&=&-\frac{1}{2}g_V^eg_A^u\nn \\
C_{2,d}&=&-\frac{1}{2}g_V^e g_A^d
\ea
\bga \label{e204}
Y&=& \frac{1-(1-y)^2}{1+(1-y)^2-y^2\frac{R(x_B,Q^2)}{1+R(x_B,Q^2)}-\frac{y x_B M}{E}}\\ \label{e204a}
\tilde{Y}&=& \frac{1+(1-y)^2-y^2\frac{\tilde{R}(x_B,Q^2)}{1+\tilde{R}(x_B,Q^2)}-\frac{y x_B M}{E}}{1+(1-y)^2-y^2\frac{R(x_B,Q^2)}{1+R(x_B,Q^2)}-\frac{y x_B M}{E}}
\ea
After some tedious algebra it can be shown that 
\bga \label{e205}
A_{RL}&=&\frac{3 G_\mu Q^2}{2 \pi \alpha \sqrt{2}}\Bigg\{ \frac{\bigg(2 C_{1u} [1+R_c(x_B,Q)] -C_{1d}[ 1+R_s(x_B,Q)]\bigg)\tilde{Y} }{5+R_s(x_B,Q)+4R_c(x_B,Q)}\nn \\
&+& \frac{R_v(x_B,Q)(2C_{2u}-C_{2d})Y }{5+R_s(x_B,Q)+4R_c(x_B,Q)}\Bigg\}
\ea
It is worth to notice that one could  obtain Eq. (\ref{e202}) by keeping only the valence quark distributions and neglecting the TM, PQCD and the sea quark corrections. Under this assumption I have $R^s=R^c=R=\tilde{R}=0 $, $R_v=\tilde{Y}=1$ and Eq. (\ref{e202}) follows right away from (\ref{e205}). 
\section{Numerical Results}\label{sec:finalarllt}
In this section I  compute $A_{RL}$ in the case of a deuteron target, at the kinematics proposed in Refs.~\cite{bosted,bosted1}. In Ref.~\cite{bosted1}, the goal is a 1.1$\%$ measurement of $A_{RL}$ from which $\sin^2\theta_W$ should be extracted with an accuracy of $0.4\%$. The kinematical region chosen is such that $x_B\ge 0.3$, to reduce the contributions from the sea quark and from the gluons, and an average $Q^2$ of 20 GeV$^2$ so that HT contributions should be negligible (see Chap. \ref{ch:twist}).  The scattering angle, in the rest frame of the deuteron, is $\theta=12^\circ$.  The experiment is performed at two different  beam energies, at 36 and 39 GeV. For simplicity I choose a beam energy of 37 Gev and compute the asymmetry for a range of outgoing electron energies between 9.8 and 17.2 GeV, as suggested in the proposal. Correspondingly, the range for $x_B$ is 0.31 to 0.75, and the range for $Q^2$ is 15.8 to 27.8 GeV$^2$. It is worth noting that my expression for the asymmetry differs slightly from the one in the proposal. First, in the proposal the factor $\tilde{Y}$ is not present. It turns out that, in the kinematical region of the experiment, this function is very close to one (to a 0.1$\%$ level or better), at least in the case in which HT are not included. Another difference is that in tEqs. (\ref{e204}) and (\ref{e204a}), I have a term $(y x_B M)/E$ which is absent in their expression. This term is usually not included in the analysis of DIS experiments, because of the small ratio $M/E$. 

\begin{figure}
\centerline{\psfig{figure=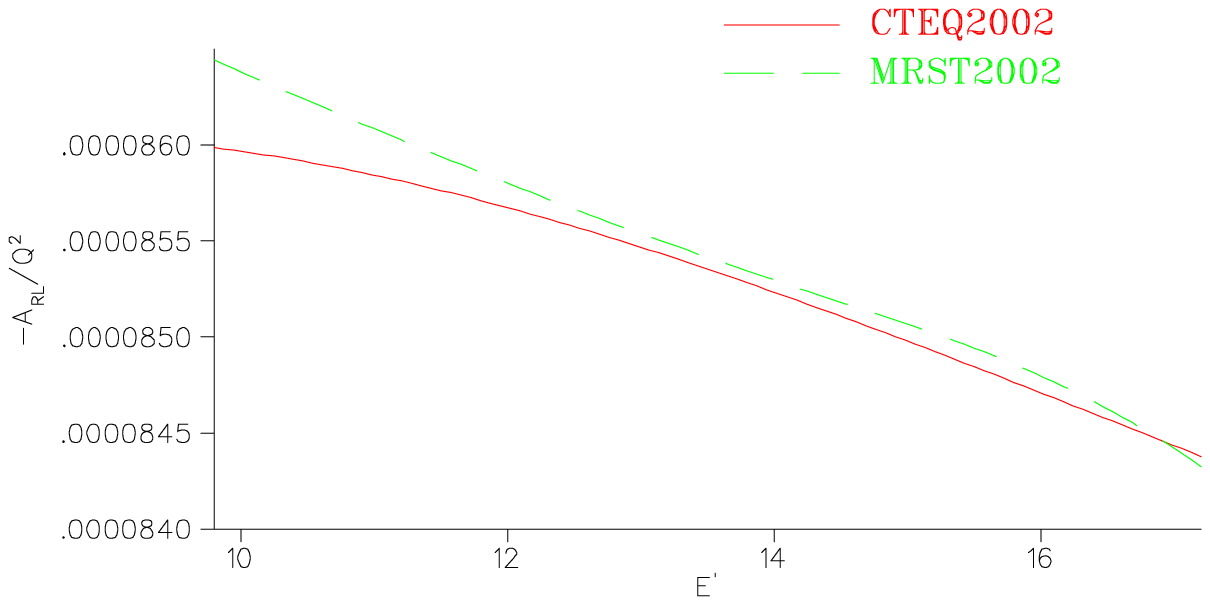}}
\caption{\label{f200} Predicted asymmetry at leading twist from the CTEQ (solid red line) and MRST (dashed green line) parameterizations for the kinematics proposed in Ref.~\protect\cite{bosted1}. }
\end{figure}

\begin{figure}
\centerline{\psfig{figure=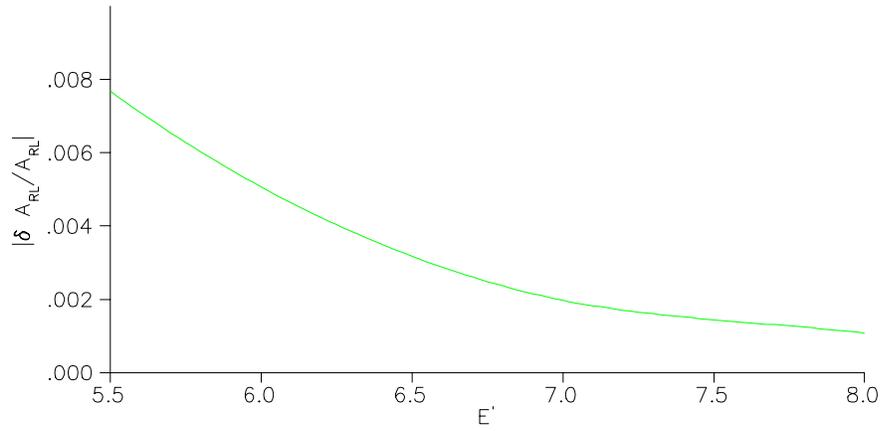}}
\caption{\label{f201}  Relative uncertainty in the asymmetry defined as in Eq. (\ref{eab1}) for the kinematics proposed in Ref.~\protect\cite{bosted1}. }
\end{figure}

Since the PDF cannot be computed from first principles, I need to resort to some available parameterization. I use two of the most recent, provided by the CTEQ~\cite{cteq} and the MRST~\cite{mrst} groups, and take the difference of the values obtained  for $A_{RL}$ as an indication of the theoretical uncertainty on the PDFs. In Fig. \ref{f200} I have plotted the two asymmetries computed by using the  two different parameterizations.
As it can be seen, the two predictions mostly differ in the low $x_B$ region    as shown in Fig. \ref{f201}, where I plotted the quantity
\bga\label{eab1}
 \frac{\delta A_{RL}}{A_{RL}}\equiv 2\left| \frac{A_{RL}^{CTEQ}-A_{RL}^{MRST}}{A_{RL}^{CTEQ}+A_{RL}^{MRST}}\right|,
\ea
(the factor of two come from taking the average of the two asymmetries), the uncertainty is, at the most, 0.6$\%$, which is below the aimed accuracy of the experiment. 

I proceed at the same way to compute the uncertainty on the asymmetry, for the kinematics proposed in Ref.~\cite{bosted}. In this proposal a single measurement at beam energy of 11 GeV, a scattering angle in the lab frame of 12.5$^\circ$, and a scattered electron energy of 5.5 GeV, is considered. I actually consider, similarly to the previous case, a larger kinematical region, which correspond to an outgoing electron energy ranging between 5.5 and 8 GeV. The corresponding range of the Bjorken variable $x_B$ is 0.28 to .72 and the range for $Q^2$ is 2.87 to 4.17 GeV$^2$. By doing so, I can have a better picture on the behavior of the twist two corrections over a broader kinematical range, allowing one to see where they can be minimized.

\begin{figure}
\centerline{\psfig{figure=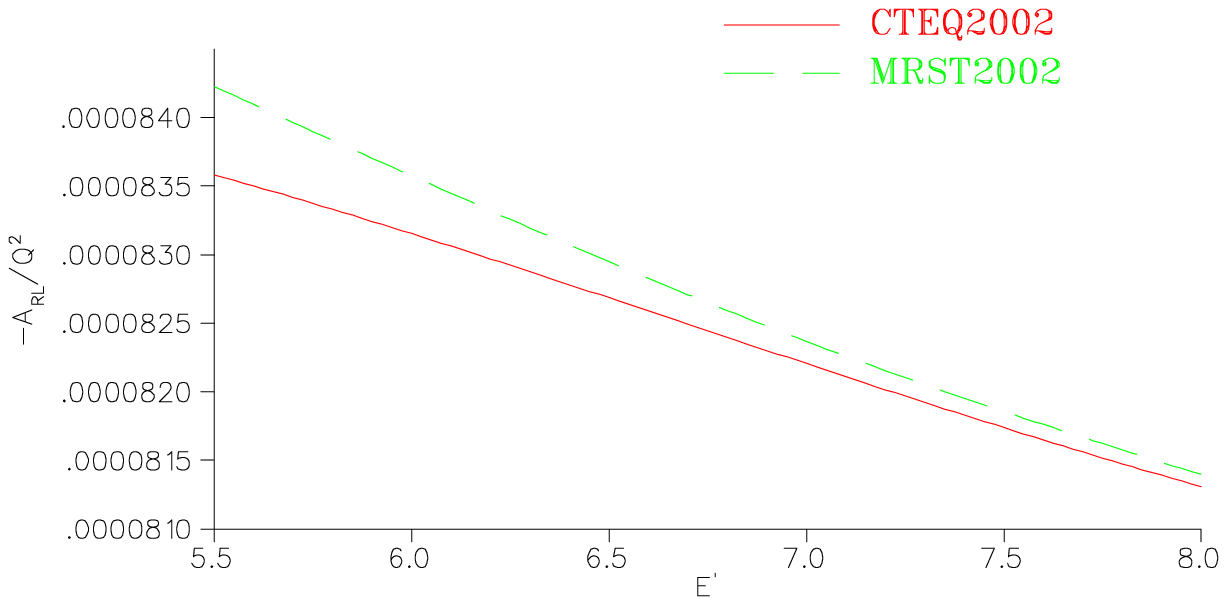}}
\caption{\label{f202} Predicted asymmetry at leading twist from the CTEQ (solid red line) and MRST (dashed green line) parameterizations for the kinematics proposed in Ref.~\protect\cite{bosted} . }
\end{figure}

\begin{figure}
\centerline{\psfig{figure=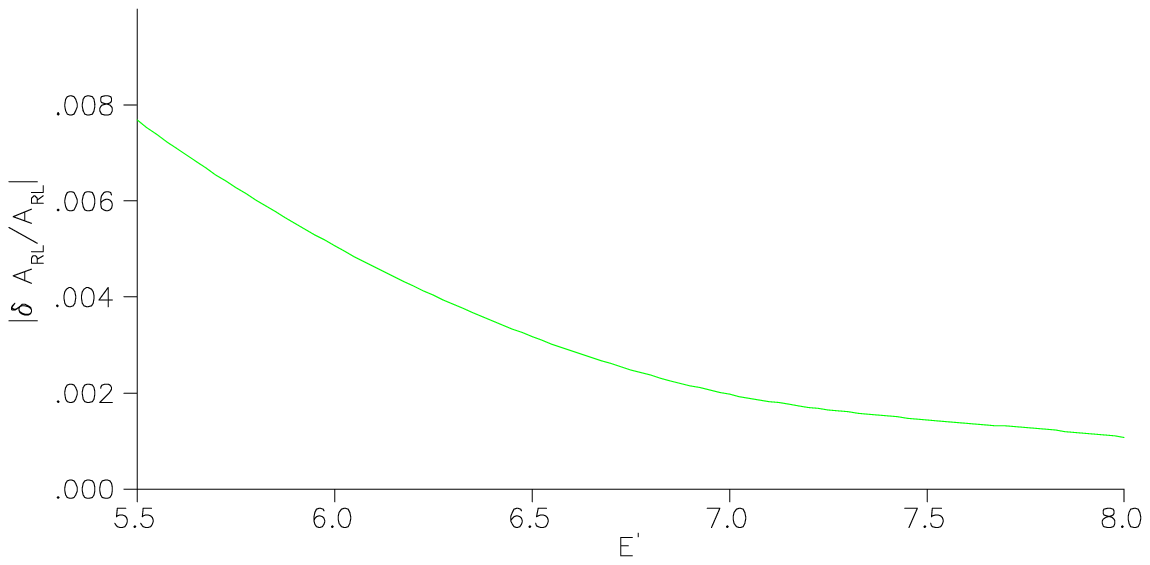}}
\caption{\label{f203}  Relative uncertainty in the asymmetry defined as in Eq. (\ref{eab1}) for the kinematics proposed in Ref.~\protect\cite{bosted}. }
\end{figure}

The asymmetry obtained from the two parameterizations is shown in Fig. \ref{f202}. Also in this case the largest discrepancy appears in the low $x_B$ region, and , as it can be seen from Fig. \ref{f203}, at an outgoing energy of 5.5 GeV, one has a maximum  value   of nearly 0.8$\%$, which is very close to the prescribed accuracy of the experiment of 1$\%$.It looks, therefore, that the kinematical point chosen for the experiment presents the highest uncertainty, and that performing the experiment at higher $E'$ would be more desirable if one's intent is to look for physics beyond the SM.  

%e205 f63 

%\input{ch_twist.tex}
\chapter{Introduction to Higher Twist}\label{ch:twist}
\section{Heuristic Definition of Higher Twist}
In this chapter I  introduce the concept of twist, and write down the set of operators that contributes to the hadronic tensor, up to twist four. Most of the material can be found in a beautiful paper by Ellis {\it et al.}~\cite{ellis}, where the reader is referred for more details. A somewhat  easier introduction may also be found in Ref.~\cite{ji}.\\

The original definition of twist of an operator is normally given in terms of its mass dimension and its spin. In order for an operator to have a definite spin, it has to be traceless, with definite symmetry property in the exchange of the Lorentz indices, and local (i.e. it must depend only on one space-time coordinate). If these properties are respected, then its spin is given by the number of Lorentz indices.\\
For instance, the totally symmetric, traceless, and local  operator defined as
\bga\label{ea0}
\theta_{\mu\nu}=\frac{1}{2}\Big[\bar{\psi}(0)\gamma_\mu \partial_\nu\psi(0) + \bar{\psi}(0)\gamma_\nu \partial_\mu\psi(0)\Big] -\frac{1}{4}g_{\mu\nu}\bar{\psi}(0)\partial \sla \psi(0)\,,
\ea
has spin equal two. The twist of an operator is then defined has $\tau=d-s$, where $d$ is the mass dimension of the operator, and $s$ is the spin~\cite{gross}. Since each field $\psi$ has mass dimension $3/2$ and the derivative has dimension one, one infers that the operator in Eq. (\ref{ea0}) has twist two. It is straightforward to see how a similar operator, but with an arbitrary number of derivatives, and appropriately symmetrized,  still has twist equal two. Indeed, each new derivative  counts for a new mass dimension, but also for a new Lorentz index, and therefore a new spin, keeping the value of the twist the same. It can be shown that an operator with twist $\tau$  enters a high energy physical process (such as cross section) scaled by a factor $\lp \frac{M}{Q}\rp^{\tau-2}$~\cite{jaffe_book}. 

As  just seen, this definition might require us to consider an infinite set of   different operators with all  the same twist, which is not very practical. Normally, a more convenient and less rigorous definition of twist is adopted (which is the one I  adopt henceforth in this study). In this case, the twist of an invariant matrix element of a bilocal  operator is defined as the order (plus two) in $\lp M/Q\rp$ at which it contributes in  DIS processes. 

This simpler definition allows one to identify the twist of a matrix element  by inspection. For instance consider the following matrix element, which appears in DIS processes 
\bga\label{ea1}
\la P| \bar\psi(0)\gamma_\mu \psi (\lambda n)|P\ra\,,
\ea 
where $|P\ra$ is the  state of a nucleon with momentum $P$ (I use the normalization condition $\la P|P'\ra =2E (2\pi)^3\delta^3(\vec{P}-\vec{P}')$, where $E$ is the nucleon energy), $\psi$ is a quark field, $\lambda$ is a light-cone coordinate, and $n$ is a light-cone vector defined as
\bga
n_\mu=\frac{1}{2  \mathbb{P}}(1,0,0,-1)\,,
\ea
with $\mathbb{P}$ being the boost parameter with dimension of  mass. In the limit $\mathbb{P}\rightarrow \infty$ one recovers the  infinite momentum frame, while $\mathbb{P}=M/2$  corresponds to the nucleon rest frame. Introducing  a second light-cone vector $p_\mu$ 
\bga
p_\mu=\mathbb{P}(1,0,0,1)\,,
\ea  
the nucleon momentum  becomes 
\bga\label{ea60}
P_\mu=p_\mu+1/2 M^2 n_\mu \,,
\ea 
where I  have chosen the system such that the nucleon moves along the $z$ axis  (notice that $P^2=M^2$ as required). Since the operator in Eq. (\ref{ea1}) can only be function of $\lambda$, $P_\mu$ and $n_\mu$, it can be decomposed as\label{pa3}
\bga \label{ea2}
\la P| \bar\psi^q(0)\gamma_\mu \psi^q (\lambda n)|P\ra=f_1(\lambda)p_\mu+f_2(\lambda) M^2 n_\mu\,,
\ea 
where
\bga \label{ea5}
f_1(\lambda)=\la P| \bar{\psi}^q(0)n\sla \psi^q (\lambda n)|P\ra\,,
\ea
and
\bga\label{ea6}
f_2(\lambda)=\frac{1}{M^2}\la P| \bar{\psi}^q(0)p\sla \psi^q (\lambda n)|P\ra\,,
\ea
where, in obtaining Eqs. (\ref{ea5}) and (\ref{ea6}) from Eq. (\ref{ea2}),  I used the fact that $n^2=p^2=0$ and $p\cdot n=1$.
Since the left hand side of Eq. (\ref{ea2}) has mass dimension one (three from the two quark fields and minus two from the two states  $|P>$ and $<P'|$),  one infers that the two functions $f_1(\lambda)$ and $f_2(\lambda)$ are dimensionless. At the end of the day, when computing physical processes involving the matrix element in Eq. (\ref{ea2}), the two functions $f_1$ and $f_2$  appear in a combination such that the term $f_2M^2$  acquires a multiplicative  factor $ 1/Q^2$, to make the $f_2$ contribution dimensionless (as $f_1$). Therefore one has that the function $f_2$  always appears suppressed by a factor $\lp M/Q\rp^2$ with respect to $f_1$ in high energy processes. One says that the matrix element $f_1$ has twist two, while the matrix element $f_2$ has twist four. It is worth  noticing that, as opposed to the previous more formal definition of twist for local operator, the matrix element of a bilocal operator does not have a definite twist. \\
After this heuristic introduction of the concept of twist, I am now going to give a more precise and  more technical definition of the twist two and four  matrix elements and their contribution to  the different structure functions (SFs). I closely follow the work done by Ellis {\it et al.} in Ref.~\cite{ellis}, where the second definition of twist is adopted. \\
\section{Formal Definition of Higher Twist}
\begin{figure}
\epsfxsize=9.0cm
\epsfysize=17.0cm
\centerline{\epsffile{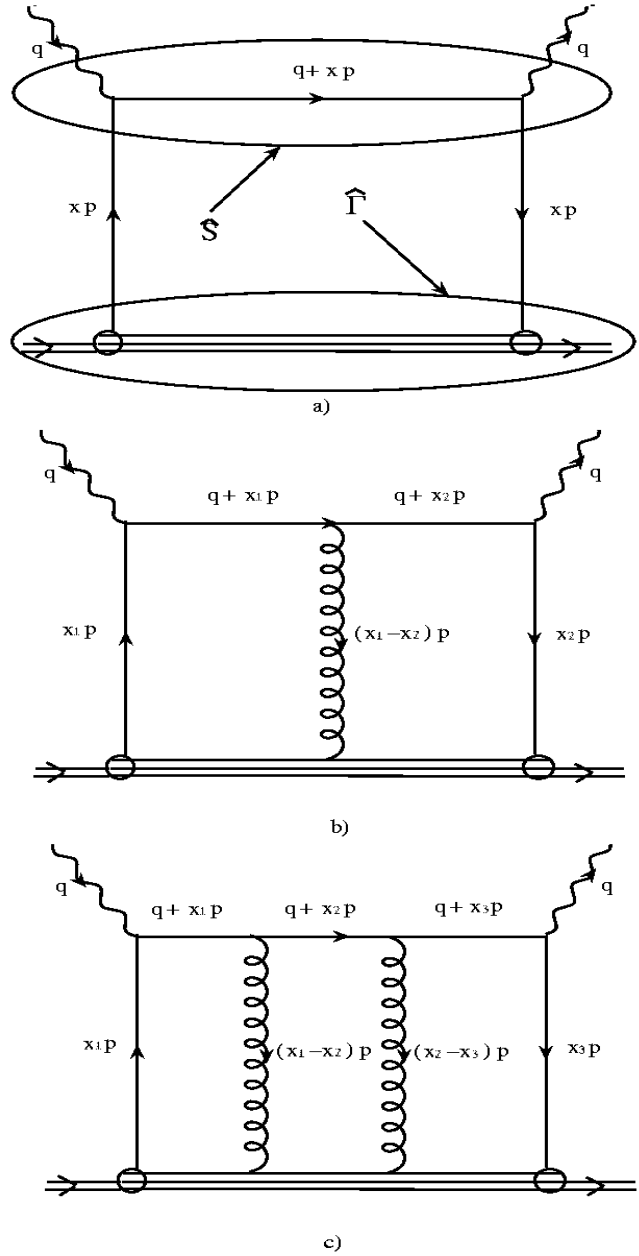}}
\caption{Feynman diagrams for DIS processes contributing up to twist four involving two fermions. In Fig. a) is shown the separation between the short distance part ($\hat{S}$) and the long distance part ($\hat{\Gamma}$). In a high energy process, the virtual photon strikes one of the quark inside the hadron. This virtual Compton scattering, indicated by $\hat{S}$, can be computed in perturbation theory, applying the well known Feynman rules. The bottom part of the graphs, the $\hat{\Gamma}$, requires the knowledge of the nucleon wave function to be computed, which at the moment is not available. In Fig. b) and c) the scattered quark exchanges one and two gluons with the remaining quarks inside the hadron.  
}
\label{fa1}
\end{figure}

\begin{figure}
\epsfxsize=9.0cm
\centerline{\epsffile{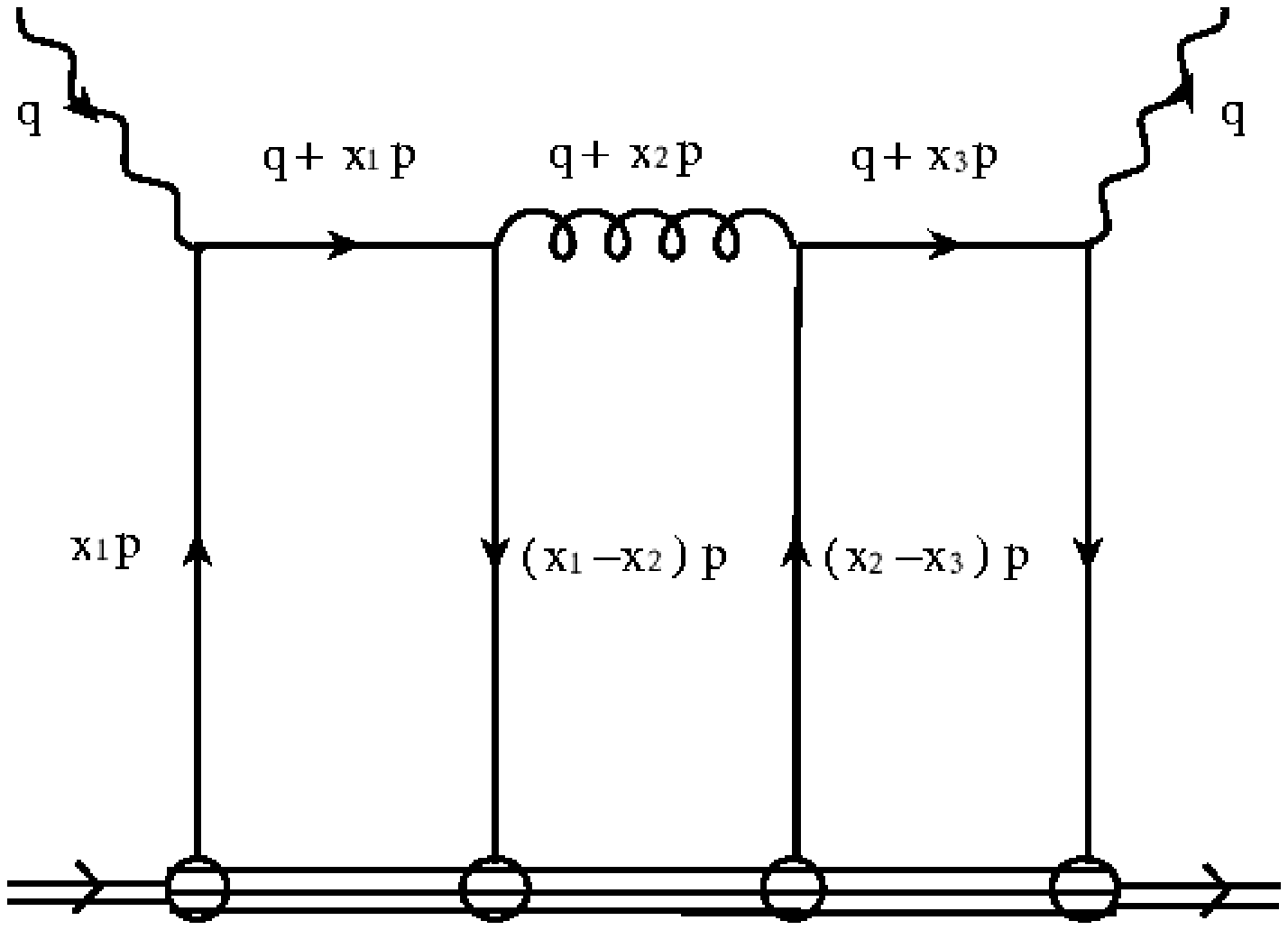}}
\caption{Feynman diagram for DIS process contributing  to twist four involving four fermions. There are other three similar diagrams obtained by crossing the first and/or the second pair of fermion lines.) 
}
\label{fa2}
\end{figure}

\begin{figure}
\epsfxsize=9.0cm
\centerline{\epsffile{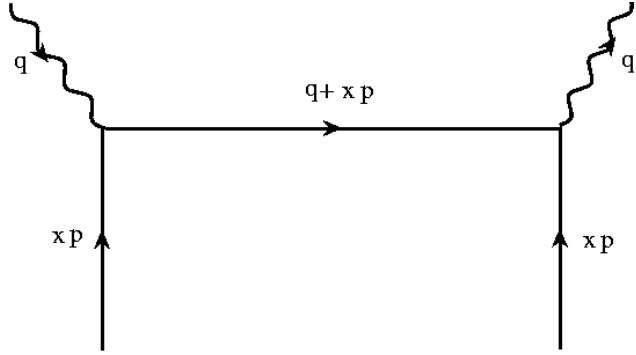}}
\caption{Feynman diagram corresponding to the forward Compton amplitude $\hat{S}(xp)$. This hard part of the process is computed using the normal Feynman rules.
}
\label{fa3}
\end{figure}

As  seen previously in Sec. \ref{sec:kinematics}, all the information on the hadronic structure is contained in the hadronic tensor. It can be shown that the hadronic tensor can be related to the forward virtual Compton amplitude $T_{\mu\nu}$ by means of the optical theorem~\cite{jaffe_book},
\bga
W_{\mu\nu}(x_B,Q^2)=\frac{1}{2i}\textrm{Im}\big[ T_{\mu\nu}(x_B,Q^2)\big]\,,
\ea
where Im stands for the imaginary part, $Q^2$ is related to the momentum transfer by $Q^2=-q^2$, and the forward virtual Compton amplitude is defined as
\bga
\frac{1}{8\pi} \int d^4 z e^{iqz} \la P| \textrm{T}\big[ J_\mu(0)J_\nu(z)\big]|P\ra\,,
\ea
where T is the time ordered product, and the hadronic current (I consider, for now, the couplings of the quarks to the  different bosons to be equal one), can be written in terms of the quark field $\psi^q(x)$ as $J_\mu(x)=\bar{\psi}^q(x)\gamma_\mu \psi^q(0)$ . It is assumed that in high energy processes, one can always factorize $T_{\mu\nu}$ in two parts, a short distance (or hard part) which can be computed by using perturbative QCD (PQCD) and depends on the hard momentum $q$, and a long distance part (or soft part), which cannot be computed perturbatively, and contains information on the hadronic structure (factorization theorem). \\
Omitting the Lorentz indices from now on, the forward amplitude, up to twist four, can be then written as
\bga\label{ea3}
\textrm{T}&=&[\hat{S}(k)\hat{\Gamma}(k)]+[\hat{S}_\mu(k_1,k_2)\hat{\Gamma}^\mu(k_1,k_2)]+[\hat{S}_{\mu\nu}(k_1,k_2,k_3)\hat{\Gamma}^{\mu\nu}(k_1,k_2,k_3)]\nn \\
&+&[\hat{S}(k_1,k_2,k_3)\hat{\Gamma}(k_1,k_2,k_3)] +\mathcal{O}\lp \frac{\Lambda^6}{Q^6}\rp\,,
\ea
where the four terms correspond to the processes depicted in Fig. \ref{fa1}a-c (two fermion correlation functions) and Fig. \ref{fa2} (four fermion correlation function), respectively. The restriction to diagrams involving no more that four fermion lines   is justified later on. 
%%%%%%%%%%%%%%%%%%%%%%%%%%%%%%%%%%%Figure  
%%%%%%%%%%%%%%%%%%%%%%%%%%%%%%%%%%%%%%%
In the previous equation, the square bracket indicates a trace over the color and spinor indices, the repeated $k$ and $k_i$ imply a four momentum integration in these variables, which links the upper part of the amplitudes (the $\hat{S}$'s) to the lower part (the $\hat{\Gamma }$'s).
%%%%%%%%%%%%%%%%%%%%%%%%%%%%%%%%%%%Figure  
%%%%%%%%%%%%%%%%%%%%%%%%%%%%%%%%%%%%%%%
The upper part of the graphs can be computed perturbatively, while the lower part is expressed in terms of matrix element of quark and gluon fields. Let us delineate, without going too much into details,  how such a factorization of the forward amplitude could be achieved. Considering the current matrix element in Eq. (\ref{ea3}), and keeping  the zero-th  order term in the perturbtive expansion one gets, 
\bga\label{ea7}
\la P| \textrm{T}\big[ J^{I}_\mu(0)J_\nu^I(z)\exp \lp-i\int H_I(t) dt\rp\big]|P\ra \simeq\la P| \textrm{T}\big[ J^{I}_\mu(0)J_\nu^I(z)\big]|P\ra\,,
\ea
where now the fields are in the interaction picture (indicated by the sub or super script $I$, which I  omit henceforth). 
The time ordered product of the four quark fields  gives, according to the Wick theorem (see, for instance Ref.~\cite{pesk}), a quark propagator from the contraction of two of the four fields, which is the upper part of the graph, and a nucleon matrix of the product of the two remaining fields, which represent the lower part, and that cannot be computed perturbatively. Broadly speaking, because of asymptotic freedom~\cite{hoff,gross1,politzer} (i.e. the fact that the strong coupling becomes small at high $Q^2$), the quarks inside the nucleon behave like free particles when probed by high energy virtual photons, and  perturbation theory can be applied to compute the upper part of the diagram. \\
After some algebra, it can then be shown that 
\bga
[\hat{S}(k)\hat{\Gamma}(k)]=\int dx S_\rho(x) \Gamma^\rho(x) \,,
\ea
where
\bga
S_\rho(x)&\equiv& [\gamma_\rho\hat{S}(x p)]\,,
\ea
where $\hat{S}(xp)$ is the amputated amplitude in Fig. \ref{fa3},
and
\bga
\Gamma_\rho(x)&\equiv&\frac{1}{2} \int \frac{d\lambda}{2\pi}\, e^{i\lambda x}  \la P|   \bar{\psi}^q(0) \gamma _\rho \psi^q (\lambda n )|P\ra\,.
\ea
One thing to notice is that now the short and long distance part of the process are linked by an integral over the light cone fraction $x$, instead  of being linked by the quark momentum $k$. The scattering amplitude of the quark $\hat{S}$ now  depends on $x p_\mu$, which means that the quark has  momentum parallel to $p_\mu$, that, in the massless limit (see Eq. (\ref{ea60})),  coincides with the nucleon momentum. 
This is an assumption  valid in a high energy process, where the interaction time between the virtual photon and the struck quark (which is proportional to the inverse of the energy of the virtual photon) is much smaller than the  time of interaction among the quarks (or partons) in the nucleon. Since the partons inside  the nucleon do not have time to interact, the system is frozen, and the photon "sees" one of the parton (the struck one) carrying a fraction $x$ of the  momentum of the nucleon, while the remaining $1-x$ fraction is carried by the other partons in the nucleon,  with no momentum in  the transverse direction (transverse to direction of motion of the nucleon), which would arise from the interaction among the partons.   \\
%%%%%%%%%%%%%%%%%%%%%%%%%%%%%%%%%%%Figure  
%%%%%%%%%%%%%%%%%%%%%%%%%%%%%%%%%%%%%%%

One can proceed similarly, considering higher terms in the expansion of the exponential in Eq. (\ref{ea7}), which are of the form
\bga
 \lp\int dt H_I(t)\rp^n \,, 
\ea
with $n$ integer, and where 
\bga
H_I(t)=\int d\vec{x}\,J_\mu(x)A^\mu(x) \,,
\ea
where $J_\mu$ is the quark current, and $A_\mu$ is the gluon field (color indices have been suppressed). 
 In the most general case of     a process involving 2F fermion lines and B gluon lines, the forward amplitude would then be written as
\bga
T=\sum_{F,B}\int d\{x\}S_{\rho_1\ldots\rho_F\mu_1\ldots\mu_B}(\{x\},x_B,Q^2) \omega^{\mu_1}_{\mu_1'}\ldots \omega^{\mu_B}_{\mu_B'}\Gamma^{\rho_1\ldots\rho_F\mu_1'\ldots\mu_B'}(\{x\},p,n,\Lambda)\,,\nn \\
\ea
where $S_{\rho_1\ldots\rho_F\mu_1\ldots\mu_B}$ represents the hard part of the process involving 2F fermions and B gluons, $\Gamma^{\rho_1\ldots\rho_F\mu_1\ldots\mu_B}$ is the soft part which consists in a nucleon matrix element with 2F quark fields and B gluon fields (actually B covariant derivatives), and $\omega_\mu^\nu=g_\mu^\nu-n_\mu p^\nu$ is a projector that removes the collinear component (i.e. the component along $p_\mu$) from an arbitrary vector. The quantity $\Lambda$ is the QCD energy scale, that sets the scale at which the long and short distance could be separated, and $\{x\}$ represent a set of light-cone fractions (fractions of the momentum   $p_\mu$ carried by the parton). One can express the soft part of the forward amplitude, by introducing  a set of dimensionless quantities $\Gamma_i(\{x\})$ such that
\bga
\Gamma^{\rho_1\ldots\rho_F\mu_1\ldots\mu_B}(\{x\},p,n,\Lambda)=\sum_i \Lambda^{\tau_i-2}  e_i^{\rho_1\ldots\rho_F\mu_1\ldots\mu_B}(n,p)\Gamma_i(\{x\})\,,
\ea
where $ e_i^{\rho_1\ldots\rho_F\mu_1\ldots\mu_B}(n,p)$ is a polarizer, which could be a function of $p_\mu$, $n_\mu$, $d_{\mu\nu}=g_{\mu\nu}-p_\mu n_\nu-n_\mu p_\nu$ and $\varepsilon_{\mu\nu}=\varepsilon_{\mu\nu\alpha\beta}p^\alpha n^\beta$, and $\tau_i$ is the twist of the amplitude $\Gamma_i$.

Before going any further, let us see the meaning of twist defined in this way.
By dimensional analysis, one can see that the forward amplitude T is dimensionless. This means that when multiplying the soft part by the hard part, which depends on $Q$, the terms $ \Gamma_i(\{x\})$ will be multiplied by a term  $(1/Q)^{\tau_i-2}$  to compensate the mass term $\Lambda^{\tau_i-2}$. Therefore one has that, in high energy processes, the higher is the twist, the more suppressed is the corresponding amplitude $ \Gamma_i(\{x\})$.\\
Let us now consider the minimum twist at which an amplitude might contribute.
Since, to an amplitude with 2F fermion lines and B gluon lines corresponds a nucleon  matrix element containing 2F quark fields and B gluon fields (or derivatives), the mass dimension of such matrix element is 3F+B-2, which comes from  the 2F quark fields, each of dimension 3/2, the B gluon fields or derivative, each of  dimension one, and the minus two comes from the dimension of the two states $<P|$ and $|P>$. Therefore, the minimum twist at which the amplitude $\Gamma^{\rho_1\ldots\rho_F\mu_1\ldots\mu_B}$ can contribute is
\bga
\tau_i^{min}=3F+B-\textrm{max}_{\{i\}}[\textrm{Dim}(e_i)]\,,
\ea
where Dim($e_i$) is the  dimension of the polarizer $e_i$. To compute the maximum dimension  of a polarizer, the key point to  notice is that all the gluon indices  $\mu_1\ldots \mu_B$ of $\Gamma^{\rho_1\ldots\rho_F\mu_1\ldots\mu_B}$ are projected onto   the $\omega$'s.  Because of the relation
\bga
\omega_{\mu}^{\nu} p^{\mu}= (g_{\mu}^{\nu}-n_\mu p^\nu)p^{\mu}=p^{\nu}-p^{\nu}= 0\,,
\ea
where I used the fact that $p\cdot n=1$, the polarizer cannot have any four vector $p_{\mu_i}$ in the gluon indices $\mu_1\ldots \mu_B$.  Since the available vectors and tensor to build up the polarizers are $p$, $n$, $d_{\mu\nu}$ and $\varepsilon_{\mu\nu}$, and since they  have dimension one, minus one and zero (for both tensors) respectively, the maximum dimension is achieved by considering F vectors $p_{\rho_i}$ in the fermion indices $\rho_1\ldots\rho_F$, and a combination of the two  tensors in the remaining  gluon indices. In the case  in which B is an odd number, it is not possible to use only tensors in the gluon indices, because they have an even number of indices, so one  needs to use a vector $n$. Keeping in mind this particular case, the maximum dimension is
\bga
\textrm{max}_{\{i\}}[\textrm{Dim}(e_i)]=F-\frac{1}{2}(1-(-1)^B)\,,
\ea   
which implies
\bga
\tau_{min}=2F+B+\frac{1}{2}(1-(-1)^B)\,.
\ea
This formula justifies the fact that, when considering  processes which contribute up to twist four, one has  to consider only the diagrams in Figs. \ref{fa1} and \ref{fa2}. The diagram in Fig. \ref{fa1}a has 2 fermion lines (F=1) and zero gluon lines (B=0), therefore the minimum twist of the corresponding matrix element is $\tau_{min}=2$. For the process in Fig. \ref{fa1}b, one has F=1, B=1 which implies $\tau_{min}=2 +1+ 1/2(1+1)=4$. Similarly, one can show that the remaining processes in Fig. \ref{fa1} and \ref{fa2} correspond also to matrix elements whose minimum twist is four.

 It can be shown that the complete set of matrix elements necessary to express the three independent SFs $F_1$, $F_L$ and $F_3$ up to twist four are \footnote{Notice that in Ref.~\cite{ellis} the longitudinal SF $F_L$ is defined as $F_L=F_2/x_B-2F_1$, in Ref.~\cite{ji} is defined as $F_L=F_2/2x_B-F_1$ while normally, in the literature, is defined as $F_L=F_2-2x_B F_1$~\cite{predazzi,bread,roberts}. I adopt the first definition. }
\begin{itemize}
\item two fermion twist two correlation function
\bga\label{ea10}
f_1^q(x_B)&=&\frac{1}{2}\int \frac{d\lambda}{2 \pi}e^{i\lambda x_B}\la P |\bar{\psi}^q(0)n \sla \psi^q (\lambda n) | P \ra\,.
\ea
\item two fermion twist four correlation functions
\bga \label{ea9}
\Lambda^2  T_1^q(x_B)&=& \frac{1}{2}\int \frac{d \, \lambda}{2 \pi}e^{i\lambda x_B}
\la P | \bar{\psi}^q(0) \gamma_\mu n \sla \gamma_\nu D_T^\mu(0) D_T^\nu(\lambda n)\psi^q (\lambda n)| P\ra\nn \\
\ea\bga 
\Lambda^2 T_2(x_2,x_1)&=& \frac{1}{2}\int \frac{d \, \lambda}{2 \pi}\frac{d \eta}{2 \pi}e^{i\lambda x_1}e^{i\eta(x_2-x_1)}\nn \\
&\times&\la P | \bar{\psi}^q(0) \gamma_\mu n \sla \gamma_\nu D_T^\mu(\eta n) D_T^\nu(\eta n)\psi^q (\lambda n)| P\ra \,,
\ea
where the  covariant derivative is defined as
\bga
D_\mu(z)=i\partial_\mu-g A_\mu(z)\,,
\ea 
and the transverse component of a vector $V$ is
\bga 
V_T^\mu=(g^{\mu\nu} -p^\mu n^\nu -n^\mu p^\nu)V_\nu\,.
\ea 
\item four fermion twist four correlation functions
 \bga
U_1(x,y,z)&=&\frac{g^2}{4\Lambda^2}\int \frac{d\lambda}{2\pi}\frac{d\mu}{2 \pi}
\frac{d\nu}{2\pi}e^{i\lambda x}e^{i\mu (y-x)}e^{i\nu(z-y)}\nn \\
&\times& \la P| \bar{\psi}^q(0)n \sla t^a \psi^q(\nu n) \bar{\psi}^{q'}(\mu n )n \sla t_a \psi^{q'} (\lambda n)|P\ra
\ea
 \bga
U_2(x,y,z)&=&\frac{g^2}{4\Lambda^2}\int \frac{d\lambda}{2\pi}\frac{d\mu}{2 \pi}\frac{d\nu}{2\pi}e^{i\lambda x}e^{i\mu (y-x)}e^{i\nu(z-y)}\nn \\
&\times& \la P| \bar{\psi}^q(0)n \sla \gamma_5 t^a \psi^q(\nu n) \bar{\psi}^{q'}(\mu n) n \sla t_a \gamma_5\psi^{q'} (\lambda n)|P\ra\,,
\ea
where $g$ is the strong coupling constant and $t_a=\frac{\lambda_a}{2}\,\,\,\, a=1\cdots 8$ are the $SU(3)$ color 
generators. 
\end{itemize}
The expressions of the SFs in terms of the different correlation functions can be found in Sec. \ref{sec_sftwofermion} for the two quark, and in Sec.  \ref{section_fourfermion} for the four quark correlation functions.\\

Before I end this chapter, let me mention that it can be shown, by using the equation of motion, that the operator in Eq. \ref{ea9} can be cast in the following form~\cite{ellis,ji} 
\bga \label{ea11}
\Lambda^2 T_1(x_B)=x_B^2  \int \frac{d\lambda}{2 \pi}e^{i\lambda x_B}\la P |\bar{\psi}^q(0)p \sla \psi^q (\lambda n) | P \ra\,,
\ea
 since I have thereby eliminated the gluon degree of freedom, this form is more suitable to be  computed in the MIT bag model. Taking the proton at rest, one has 
\bga
n_\mu=\frac{1}{M}(1,0,0,-1)\,\,\,\,\,\, \textrm{and} \,\,\,\,\,\, p_\mu=\frac{M}{2}(1,0,0,1)\,,
\ea
and the two matrix elements in Eqs. (\ref{ea10}) and (\ref{ea11}) can be expressed in terms of the "good" and "bad" components of the spinor $\psi$ defined as
\bga
\psi_+\equiv \frac{1}{2}(1+\gamma_0\gamma_3)\psi=\frac{1}{2}(1+\alpha_3)\psi
\ea
and 
\bga
\psi_-\equiv \frac{1}{2}(1-\gamma_0\gamma_3)\psi=\frac{1}{2}(1-\alpha_3)\psi\,,
\ea
respectively, so that, recalling that $n\sla=1/M(\gamma_0+\gamma_3)$ and $p\sla=M/2(\gamma_0-\gamma_3)$, I have
\bga \label{ea13} 
f_1^q(x_B)&=&\frac{1}{M}\int \frac{d\lambda}{2 \pi}e^{i\lambda x_B}\la P |\psi^{\dag q}_+(0) \psi^q_+ (\lambda n) | P \ra
\ea
and
\bga  \label{ea45}
\Lambda^2 T_1(x_B)= M x_B^2  \int \frac{d\lambda}{2 \pi}e^{i\lambda x_B}\la P |\psi^{\dag q}_-(0) \psi^q_- (\lambda n) | P \ra\equiv x_B^2 f_4^q(x_B)\,.
\ea
This form for the two correlation functions will be useful in Sec. \ref{sec:mmitbm} where they will be computed  using a modification of the MIT bag model.

\chapter{Estimate of the Twist Four Contribution in the MIT Bag Model}\label{ch:two}
When attempting to make precision tests of QCD inelastic processes, one has to face the problem of including the higher twist (HT) effects. In particular, the twist four effects contribute to DIS processes suppressed by a factor of $\lp \frac{\Lambda}{Q}\rp^2$, where $\Lambda$ is the QCD scale, of the order of few hundreds MeV, and $Q^2$ is (minus) the square of the four  momentum transfer. In the attempt of exploring all the possible windows in which to look for new physics, some DIS experiments have been suggested in which the momentum transfer is of the order of few GeV. The idea is to extract the weak mixing angle by measuring $A_{RL}$ to a one percent level of accuracy. Since the momentum transfer is not very large compared to the QCD scale, the HT effects might become significant, and eventually spoil the interpretation of the experiments, if not properly taken into account. In the past, not many   attempts to compute the HT effects have been done~\cite{ellis1,signal,guo,casto,casto1}, and I would like to provide my contribution on the matter. Besides, experimental data on the HT effects are almost inexistent, and a model estimate could be useful in planning future high precision DIS experiments.  In this chapter I  compute the twist four corrections to the nucleon SFs using two different variations of  the MIT bag model. In section \ref{sec:bag_model} I  briefly describe the MIT bag model as it was originally introduced in ref.~\cite{jaffe}. In section \ref{sec:twofermion_tw2cf} I  compute the moments of the twist two PDFs and reconstruct them by means of the inverse Mellin transform (IMT). In Secs. \ref{sec:twofermion_tw4cf} and \ref{sec:fourfermion_tw4cf} I  utilize the same approach used for the twist two PDFs to compute the two fermion and four fermion twist four corrections to the SFs. 
\section{The MIT Bag Model}
\label{sec:bag_model} 
In this section I  briefly describe the main features of the MITBM in its simplest form, as it was initially introduced by the authors of ref~\cite{jaffe}. The MITBM is a phenomenological model that describes how the quarks are confined inside the nucleon. 
Since no free quarks have been experimentally observed,   they must be tightly confined inside the hadrons. The main assumption of the 
model is that the quarks in the hadron reside in a region (the volume of the bag) of true vacuum (or perturbative vacuum), in which they behave as free particles. On the surface of the bag the  QCD vacuum (a continuous creation and annihilation of quark-antiquark pairs and  gluons)  exerts a pressure 
 which keeps the quarks confined in the hadron. The bag model can be described by introducing the lagrangian density~\cite{johnson}
\bga\label{e140}
\mathcal{L}_{Bag}=\lp\mathcal{L}_{QCD}-B\rp \theta(|\mathbf{x}|-R(\theta,\phi))
\ea
where $R(\theta,\phi)$ is the bag radius, $B$ is the bag constant, which physically represents the external pressure exerted by the QCD vacuum on the bag surface, $\theta(x)$ is the step function, and the QCD  lagrangian is defined as
\bga
\mathcal{L}_{QCD}=\sum_q \bar{q(x)}(i\partial \sla -m_q)q(x) \, .
\ea 
From now on I  assume massless quarks. Since the quarks are confined inside the hadrons, one must impose that  flux of quarks through the bag surface is zero. If one defines the quark current as
\bga
J_\mu^a(x)\equiv \Big(\bar{q}_r(x),\bar{q}_b(x),\bar{q}_g(x)\Big) \lambda^a \gamma_\mu 
\lp\!\!\!
 \begin{array}{c}
q_r(x) \\
q_b(x)\\
q_g(x)
\end{array}
 \!\!\!\rp
\ea 
where the indexes $r$, $b$ and $g$ are the color charges `red', `blue' and `green' and $\lambda^a\,\,\,\, a=1...8$ are the $SU(3)$ color matrices, the 
condition translates as
\bga \label{e141}
n^\mu J_\mu ^a|_{surface}=0 \,\,\,\,\,\, a=1\ldots 8
\ea
where $n_\mu=(0,\mathbf{n})$ is a unit vector normal to the bag surface. The equations of motion which follow from the lagrangian in Eq. (\ref{e140}) for massless quarks are
\bga \label{e142}
i\partial \sla q(x)=0
\ea
for $|\mathbf{x}| \le R$, and the boundary condition (\ref{e141}) becomes
\bga\label{e143}
i n_\mu \gamma^\mu q(x)=q(x)|_{surface}
\ea
One could also compute the outward pressure that the quarks inside the bag exert on its surface. Given the energy momentum tensor 
\bga
T_{\mu\nu}=\sum_q \frac{i}{2}\Big[ \bar{q}\gamma_\mu \overrightarrow{\partial} _\nu-\bar{q}\overleftarrow{\partial}_\nu \gamma_\mu q\Big]
\ea
the pressure is
\bga
P_D=n^\mu n^\nu T_{\mu\nu}=\frac{1}{2}n\cdot \nabla \sum_q \bar{q}q|_{surface}
\ea
Since this pressure has to be balanced by the exterior pressure due to the QCD vacuum, the previous relation also fixes the bag constant $B=P_D$.

Solving the equations of motion with the boundary condition (\ref{e143}) one gets, for a quark of type $q$ in the lowest energy state,
\bga \label{e145}
\psi_m(\mathbf{x},t)=N_m  {j_0(E|\mathbf{x}|)\chi_m \choose i \mathbf{\sigma }\cdot \hat{\mathbf{x}} j_1(E|\mathbf{x}|) \chi_m } \theta(R-|\mathbf{x}|)e^{-iEt}
\ea
where $E=\frac{\Omega}{R}$ and $\Omega\simeq 2.04$, $\hat{\mathbf{x}}$ is a unit vector in the $\mathbf{x}$ direction, $\mathbf{\sigma}$ is a vector whose components  are the Pauli spin matrices, $j_i(z)$ is the $i\!-\!th$ spherical Bessel function and
\bga
N_m^2=\frac{1}{4 \pi R^3}\frac{\Omega^4}{(\Omega^2-\sin^2\Omega)^2}\, .
\ea
is a normalization factor.

Since the quarks obey the Fermi-Dirac statistic, the total wave function (spatial-spin-flavor-color) has to be totally antisymmetric. Because of confinement, the color interaction is assumed to be restricted to the bag, so that from outside the hadron is seen as colorless. For a baryon, which is made up of the three quarks, the only colorless object that can be built is
\bga \label{e146}
\psi_{col}=\epsilon_{abc}q^a q^b q^c
\ea 
where $\epsilon_{abc}$ is the Levi-Civita antisymmetric tensor which makes the color wave function totally antisymmetric. As a consequence the spatial-spin-flavor part has to be combined in such a way that results totally symmetric. For a proton  with spin up the total wave function is~\cite{greiner}
\bga \label{e20}
\psi_P&=&\frac{1}{3 \sqrt{2}}\Big[ 2 u^\uparrow _1 u^\uparrow _2 d^\downarrow_3 +2 u^\uparrow _1
 d^\downarrow _2 u^\uparrow_3+2 d^\downarrow _1 u^\uparrow _2 u^\uparrow_3- u^\uparrow _1 u^\downarrow _2 d^\uparrow_3\nn \\
&-&u^\uparrow _1 d^\uparrow _2
 u^\downarrow_3- u^\downarrow _1 u^\uparrow _2 d^\uparrow_3- u^\downarrow _1 d^\uparrow _2 u^\uparrow_3 - d^\uparrow _1 u^\downarrow _2 
u^\uparrow_3- d^\uparrow _1 u^\uparrow _2u^\downarrow_3\Big]
\ea
Eq. (\ref{e20}) is the MITBM wave function, where $u$ and $d$ are the up and down quark fields and the arrow indicates the third component of the spin. The spatial part of the quark fields is given in Eq. (\ref{e145}) and for each combination $uud$ the color part is given by Eq. (\ref{e146}). This wave function has been extensively used to compute hadron masses, magnetic moments and charge radii and the results are within twenty to forty percent of accuracy~\cite{greiner}. Since my intent is just to estimate the HT contributions, such a precision is sufficient to reach my goal.
\section{Two Fermion Correlation Function}
\label{sec:twofermion_cf}
In this section I   consider the contributions of the two fermion twist two and twist four correlation functions to the different nucleon SFs. I  then  outline the method I use to compute the moments of the SFs, and the technique I use to reconstruct the SFs from them. I  first start with the twist two PDFs, so that I will be able to compare the results of my calculations with the available data, which will allow  to test my {\it modus operandi}, and then extend this approach to the twist four operators. 
\subsection{Contribution to the Structure Functions of the Two Fermion Twist Two and Twist  Four Operators}\label{sec_sftwofermion}
The contribution  to the deep inelastic scattering (DIS) SFs of the two quark operators
have been previously calculated~\cite{ellis,Jaffe,ji} up to twist four \footnote{Notice that in Ref.~\cite{ellis} the longitudinal SF $F_L$ is defined as $F_L=F_2/x_B-2F_1$, in Ref.~\cite{ji} is defined as $F_L=F_2/2x_B-F_1$ while normally, in the literature, is defined as $F_L=F_2-2x_B F_1$~\cite{predazzi,bread,roberts}. I adopt the first definition. }
\bga\label{e72}
F_L^{EM}(x_B)&=& 4\frac{\Lambda^2}{Q^2}\sum_q e_q^2T_1^q(\xi) 
\ea
\bga\label{e73}
F_1^{EM}(x_B,Q^2)&=& \frac{1}{2} \sum_q e_q^2\Big [f_1^q(\xi)\nn \\
&-&\frac{\Lambda^2}{Q^2}\xi
\int dx_1\,dx_2\frac{\delta(\xi-x_2)-\delta(\xi-x_1)}{x_2-x_1}T_2^q(x_2,x_1)\nn \\
\ea
\bga\label{e74}
F_L^{EMNC}(x_B,Q^2)&=& 4\frac{\Lambda^2}{Q^2}\sum_q e_qg_q^V T_1^q(\xi) 
\ea
\bga\label{e75}
F_1^{EMNC}(x_B,Q^2)&=& \frac{1}{2} \sum_q e_q g_q^V\Big [f_1^q(\xi)\nn \\
&-&\frac{\Lambda^2}{Q^2}\xi
\int dx_1\,dx_2\frac{\delta(\xi-x_2)-\delta(\xi-x_1)}{x_2-x_1}T_2^q(x_2,x_1)\nn \\
\ea
\bga \label{e76}
F_3^{EMNC}(x_B,Q^2)&=& -\sum_q(-1)^q e_q g_q^A\Big[ f_1^q(\xi)\nn \\
&-&\frac{\Lambda^2}{Q^2}\xi
\int dx_1\,dx_2\frac{\delta(\xi-x_2)-\delta(\xi-x_1)}{x_2-x_1}T_2^q(x_2,x_1)\Big]\, ,\nn \\
\ea
 where the $F_2(x_B)$ SF is related to $F_1(x_B)$ and $F_L(x_B)$ through the relation
\bga \label{e77}
F_2(x_B,Q^2)&=&x_B\big[ 2F_1(x_B)+F_L(x_B)\big]\, .
\ea
The superscript $EM$ indicates the electro-magnetic SFs,  while the $EMNC$ superscript labels the interference SFs, the ones arising from the interference of the electro-magnetic with the neutral current. The coupling $e_q$ is the charge of a quark of type $q$ while the couplings $g_q^V$ and $g_q^A$ are the vector and axial couplings respectively of a quark of type $q$ to the neutral boson $Z_0$. For the different couplings, I use the conventions in~\cite{mjrm}. 
I have introduced  the twist two PDF for a quark of type $q$ as (see Chap. \ref{ch:twist})
\bga \label{e71}
f_1^q(x_B)&=&\frac{1}{2}\int \frac{d\lambda}{2 \pi}e^{i\lambda x_B}\la P |\bar{\psi}^q(0)n \sla \psi^q (\lambda) | P \ra
\ea
and the two twist four correlation function for a quark of type $q$ as
\bga \label{e70}
\Lambda^2 T_1^q(x_B)&=& \frac{1}{2}\int \frac{d \, \lambda}{2 \pi}e^{i\lambda x_B}
\la P | \bar{\psi}^q(0) \gamma_\mu n \sla \gamma_\nu D_T^\mu(0) D_T^\nu(\lambda)\psi^q (\lambda)| P\ra\nn \\ 
\ea
\bga\label{e78}
\Lambda^2 T_2^q(x_2,x_1)&=& \frac{1}{2}\int \frac{d \, \lambda}{2 \pi}\frac{d \eta}{2 \pi}e^{i\lambda x_1}e^{i\eta(x_2-x_1)}\nn \\
&\times&\la P | \bar{\psi}^q(0) \gamma_\mu n \sla \gamma_\nu D_T^\mu(\eta) D_T^\nu(\eta)\psi^q (\lambda)| P\ra \, \nn \\  
\ea
where the transverse component of the covariant derivative is defined as 
\bga\label{e79}
D_T^\mu=(g^{\mu\nu} -p^\mu n^\nu -n^\mu p^\nu)D_\nu
\ea 
where $p_\mu$ and $n_\mu$ are the light cone vectors
\bga \label{e80}
p_\mu&=& \frac{M}{2}(1,0,0,1)
\ea
\bga \label{e81}
n_\mu&=& \frac{1}{M}(1,0,0,-1).
\ea
The variable $\xi$ is the Nachtmann variable defined as $\xi= \frac{Q^2}{2 p\cdot q}$, where $q$ is the momentum transfer and $Q^2=-q^2$. Moreover, $x_B=\frac{Q^2}{2 P\cdot q}$ is the Bjorken variable and $P$ is the nucleon momentum.

Using the equation of motion Eq.(\ref{e70}) can be rewritten in the following form~\cite{ellis,ji}
\bga \label{e82}
\Lambda^2 T_1^q(x_B)=x_B^2  \int \frac{d\lambda}{2 \pi}e^{i\lambda x_B}\la P |\bar{\psi}^q(0)p \sla \psi^q (\lambda) | P \ra\,.
\ea
\subsection{Moments of the Twist Two PDFs in the Bag Model}
\label{sec:twofermion_tw2cf}
In this section I would like to illustrate how the moments of the twist two PDFs could be computed it the bag model and then how the PDFs could be  reconstructed by using the IMT. Computing the twist two PDFs gives us also a way to test  the validity of my approach, by comparing my result with the available data.

Let up first consider the moments of the twist two PDF in Eq. (\ref{e71})~\cite{ellis}
\bga\label{e83}
M^q(N)&=&\int_0^\infty d x\, x^{N-1}f_1^q(x)\nn \\
&=& \int_0^\infty dx\, x^{N-1}\frac{1}{2} \int_{-\infty}^{\infty}\frac{d\lambda}{2\pi}e^{i\lambda x}\la P| \bar{\psi}^q(0)n \sla \psi^q(\lambda n)|P\ra \nn \\
\ea
By Taylor expanding the term
\bga
\psi^q(\lambda n)=\sum_{p=0}^{\infty}\frac{1}{p!}(n\cdot \partial)^p \psi^q(0)\,,
\ea
I have
\bga
M^q(N)&=&\frac{1}{2}\sum_{p=0}^\infty \frac{1}{p!}\la P|\bar{\psi}^q(0)n\sla (n\cdot\partial)^p \psi^q(0) |P\ra\nn \\
&\times& \int_0^\infty dx\, x^{N-1} \int_{-\infty}^{\infty}\frac{d \lambda}{2 \pi}e^{i\lambda x }\lambda^p \nn \\
&=&\frac{1}{2}\sum_{p=0}^\infty \frac{1}{p!} \la P|\bar{\psi}^q(0)n\sla (n\cdot\partial)^p\psi^q(0) |P\ra\nn \\
&\times& \int_0^\infty dx \,x^{N-1} (-i)^{p}\frac{\partial ^{p}}{\partial x^{p}}\delta(x)\nn \,,\\ 
\ea
where in the last step I used
\bga
\int_{-\infty}^{\infty}\frac{d \lambda}{2 \pi}e^{i\lambda x }\lambda^p=(-i)^{p}\frac{\partial ^{p}}{\partial x^{p}}\delta(x)\,.
\ea
Finally, by considering the relation
\bga
\int_0^\infty dx \,x^{n} \frac{\partial ^{p}}{\partial x^{p}}\delta(x)=(-1)^n n!\delta_{n,p} \nn \,,\\
\ea
I get
\bga
 M^q(N)=\frac{1}{2}i^{N-1} \la P|\bar{\psi}^q(0)n\sla (n\cdot\partial)^{N-1}\psi^q(0) |P\ra\, .
\ea

Because of the lack of translational invariance, in order to compute local operator matrix elements in the bag model, I need to switch from a continuum normalization for the state $|P\ra$ with normalization $\la P| P\ra =2M(2\pi)^3\delta^3(0)$ to a wave packet normalization for the state $|\hat{P}\ra\equiv [2M(2\pi)^3\delta^3(0)]^{-1/2} |P \ra $,with normalization $\la \hat{P}| \hat{P}\ra =1$ ~\cite{jaffe2,greiner}. Then, given a local operator of the type
\bga\label{e85}
\la P| O(0)|P\ra
\ea
it is replaced by
\bga\label{e86}
2 M \int d\mathbf{x}\la \hat{P}|O(\mathbf{x})| \hat{P}\ra\, . 
\ea
From Eqs.  (\ref{e81}) and (\ref{e83}) I get
\bga \label{e87}
M^q(N)&=&M  i^{N-1} \sum_m\int d\mathbf{x} \la \hat{P}|\bar{\psi}^q_m(\mathbf{x})n\sla (n\cdot\partial)^{N-1}\psi^q_m(\mathbf{x}) |\hat{P}\ra\nn \\
&=& \frac{i^{N-1}}{M^{N-1}}\sum_m\int d\mathbf{x} \la \hat{P}|\bar{\psi}^q_m(\mathbf{x},t)(\gamma_0+\gamma_3) \nn\\
&\times&(\partial_0+\partial_3)^{N-1}\psi^q_m(\mathbf{x},t) |\hat{P}\ra\nn\,. \\
\ea
By using the binomial formula $(a+b)^n=\sum_{k=0}^n {n \choose k} a^{n-k} b^k$ and substituting the expression for the bag model wave function in Eq.  (\ref{e145}), I have
\bga
M^q(N)
&=& \frac{i^{N-1}}{M^{N-1}}\sum_m\int d\mathbf{x} \la \hat{P}|\bar{\psi}^q_m(\mathbf{x},t)(\gamma_0+\gamma_3)\nn \\
&\times& \sum_{t=0}^{N-1}{N-1 \choose t}\partial_0^{N-1-t} \partial_3^t\psi_m^q(\mathbf{x},t) |\hat{P}\ra\nn \\
&=&\frac{   N_q N_m^2}{M^{N-1}} \lp \frac{\Omega}{R}\rp ^{N-4}
\sum_{t=0}^{N-1}i^{t}{N-1 \choose t} I_t\,,
\ea
where I have introduce the quantity
\bga \label{e88}
I_t&=&\int d \mathbf{y} \Big[ j_0(y)\partial^t_{y_3}j_0(y)+j_1(y) \frac{y_1^2+y_2^2}{y} \partial_{y_3}^t\lp \frac{j_1(y)}{y}\rp\nn \\
& +&j_1(y)\frac{y_3}{y}\partial_{y_3}^t\lp j_1(y)\frac{y_3}{y}\rp \Big]\nn \\
&+& i\int d\mathbf{y} \Big[ j_0(y) \partial_{y_3}^t \lp j_1(y) \frac{y_3}{y}\rp-j_1(y)\frac{y_3}{y}\partial^t_{y_3}j_0(y)\Big]
\ea
and  I have made a change of variable $\mathbf{y}= \frac{\Omega}{R} \mathbf{x}$ with $y=\sqrt{y_1^2+y_2^2+y_3^2}$, to make the integrals dimensionless. One could see, from symmetry consideration,  that the first integral in Eq. (\ref{e88}) contributes only for $t$ even, and the second integral contributes only for $t$ odd. In fact, the two spherical Bessel functions $j_0(y)$ and $j_1(y)$ are even functions of $\mathbf{y}$,  so that   the first integral is an even function of $\mathbf{y}$ for $t$ even (zero otherwise),  while the second integral the opposite is true.  The factor $N_q$ comes from the spin-flavor part of the wave function so that $N_u=2$ and $N_d=1$. From  the relation $\frac{M R}{\Omega}=4$~\cite{jaffe2}, Eq. (\ref{e87}) can be rewritten as
\bga \label{e100}
M^q(N)=N_q\lp\frac{1}{4}\rp^{N}\frac{\Omega}{\pi (\Omega^2-\sin^2\Omega)}\sum_{t=0}^{N-1}{N-1\choose t} (i)^t I_t
\ea
The first six moments of the twist two PDF for a $d$ quark are shown in Table \ref{t1}
\begin{table}
\caption{Moments for the $d$ quark twist two PDF obtained from the MITBM. }
\centerline{\begin{tabular}{|c|c|}
\multicolumn{2}{c}{}\\
\hline
\multicolumn{2}{|c|}{$M_N$ of $f_1^d(x_B)$  }\\
\hline
$M_1$& 1\\
$M_2$ &.167\\
$M_3$ & .042 \\
$M_4$ & .013\\
$M_5$ & .004\\
$M_6$ & .002 \\
\hline
\end{tabular}}
\label{t1}
\end{table}
\subsection{Moments Inversion of the Twist Two Correlation Functions}\label{sec:twofermion}
\begin{figure} 
\centerline{\psfig{figure=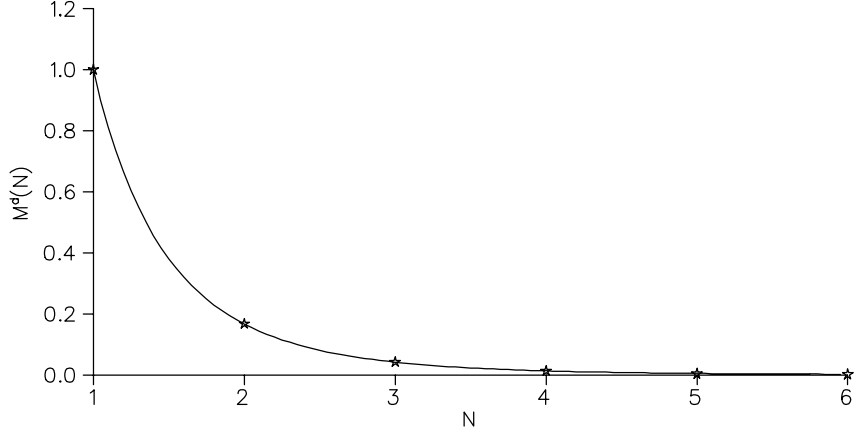}}
\caption{\label{f20} Moment function $M^d(N)$  obtained by fitting the first six moments in table \ref{t1}. The stars are the moments obtained from the MITBM.}
\end{figure}

\begin{figure}
\centerline{\psfig{figure=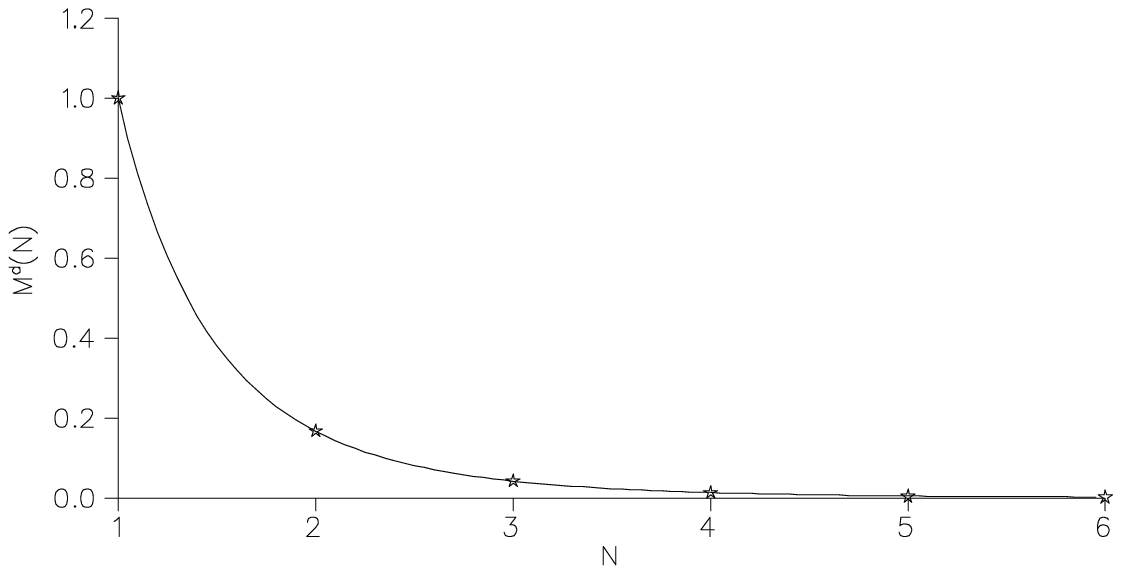}}
\caption{\label{f21} Moment function $M^d(N)$  obtained by fitting the first five moments in table \ref{t1}. The stars are the momonts obtained from the MITBM.}
\end{figure}

Now that I have computed the moments of the PDFs, I attempt to reconstruct them by using the IMT defined as~\cite{muta}
\bga
f_1^q(x_B)=\frac{1}{2\pi i}\int_{c-i\infty}^{c+i\infty} d N x_B^{-N} M^q(N)
\ea
where $N$ has been promoted to a continuum complex  variable.  I follow the method suggested in~\cite{thomas,gockeler} in which the moments are fitted to a function $M(N)$ for 
which the inverse Mellin transformation can be computed analytically. I use the following parametrization to fit the moments
\bga\label{e25a}
M(N)&=&a\,(\beta[1+c,b+N ] +d \,\beta[1+c,b+N+1]\nn \\
&+& e\,\beta[1+c,b+N+2])
\ea
where $\beta[a,b]$ is the beta function. 
The inverse transform for Eq. (\ref{e25a}) is
\bga\label{e101}
&&\frac{1}{2\pi i}\int_{c-i\infty}^{c+i\infty} d N x_B^{-N}\,a\Big(\beta[1+c,b+N ] \nn \\
&+&d\, \beta[1+c,b+N+1]+g\,\beta[1+c,b+N+2]\Big)\nn \\
&=&a\,x_B^b(1-x_B)^c(1+d\, x_B+e\, x_B^2)
\ea
which is the typical parametrization used in the  literature for the valence quark PDFs,  which are the ones one obtains from the bag model since no sea quark is included in it.
I fit the set of parameters ${a,b,c,d,e}$ using the first five and six moments. The results for the two sets of parameters are, 
\bga\label{e90}
a=7.582 \,\,\,\,\, b=.215 \,\,\,\,\, c=7.812\,\,\,\,\, d=7.532\,\,\,\,\, e=6.061
\ea
for the fit with six moments, and 
\bga\label{e91}
a=7.582 \,\,\,\,\, b=.212 \,\,\,\,\, c=7.778\,\,\,\,\, d=7.293\,\,\,\,\, e=6.155
\ea
for the fit with five moments. In Figs.\ref{f20} and \ref{f21} I have plotted the function $M^d(N)$ for the  parameterizations in Eqs. (\ref{e90}) and (\ref{e91}) respectively. The stars indicate the value of the moments obtained with the bag model. As it can be seen, there is not a relevant difference between the two parameterizations (when plotted on the same graph the two curves are indistinguishable). 

\begin{figure}
\centerline{\psfig{figure=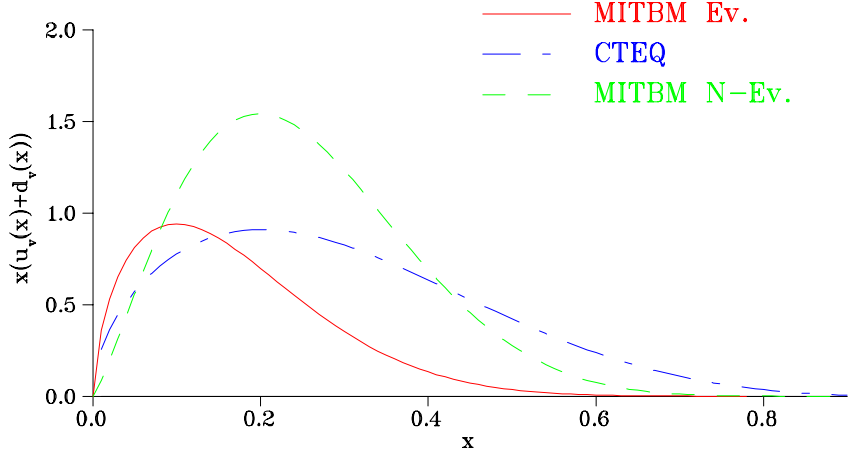}}
\caption{\label{f23} Comparison of the PDF combination $x_B[u_V(x_B)+d_V(x_B)]$ obtained from the bag model, evolved   and not evolved, with model independent extraction from CTEQ group. The green dashed line is the prediction obtained from the MITBM non evolved. The red solid line is the MITBM prediction evolved to a scale $Q^2=4$ GeV$^2$. The blue dot-dashed line represents the CTEQ parameterization at the scale $Q^2=4$ GeV$^2$.  } 
\end{figure}

One thing to notice is that the PDFs obtained  from the bag model are $Q^2$ independent. As suggested in ref.~\cite{thomas1} I consider this PDF as computed at the bag scale $\mu_{Bag}$ which has to be fixed. The idea is to evolve the PDF obtained from the bag model according to the Dokshitzer-Gribov-Lipatov-Altarelli-Parisi (DGLAP)~\cite{gribov,dokshitzer,altarelli} evolution equations, from an initial value of the scale $Q^2_{In}=\mu^2_{Bag}$, to a specific scale $Q^2$, and compare it with the PDF obtained from data at the scale $Q^2$. The bag scale $\mu_{Bag}$ is that value  
of $Q^2_{In}$ so that the evolved PDF obtained from the bag model, best resembles the PDF from data at the scale $Q^2$. In Fig.\ref{f23} I have plotted the PDF $x_B[u_V(x_B)+d_V(x_B)]$ obtained from the bag model  and evolved from the bag scale $\mu_{Bag}^2=.2$ GeV$^2$, to $Q^2=4$ GeV$^2$. I also plot the same PDF combination not evolved, together with the one obtained from data at $Q^2=4$ GeV$^2$ using the CTEQ parameterization~\cite{cteq}. The software I used to evolve the PDF has been kindly provided to us by  Miyama and Kumano. More details can be found in~\cite{kumano}. I choose $\mu_{Bag}$ so that the peaks of the two distributions have the same magnitude. As  can be seen, the evolved and the experimental distributions are not quite the same in this very simple model. On the other hand, my main goal is not to reproduce exactly the twist two PDFs, but instead to give a semi-quantitative estimate of the twist four correlation functions. In this respect, even if they are not perfect, the model provides results which are  within  the scope of my calculation. 
\subsection{Moments of the Two Fermion Twist Four Correlation Functions}
\label{sec:twofermion_tw4cf}
As  can be seen from Eqs. (\ref{e73}-\ref{e76}), there are two two-quark twist four correlation functions contributing to the different SFs, namely $T_1(x_B)$ and $T_2(x_1,x_2)$. I  use Eq. (\ref{e82}) since it  expresses the correlation function $T_1(x_B)$ in a form that makes straight forward the computation of the moments and the process of inversion. In this form I managed to eliminate the gluon degrees of freedom  in the covariant derivative, which I would not be able to include since they are not present in the MITBM.   

Defining the matrix element $f_4(x_B)$ as
\bga \label{e120}
f_4^q(x_B)\equiv\int \frac{d\lambda}{2 \pi}e^{i\lambda x_B}\la P |\bar{\psi}^q(0)p \sla \psi^q (\lambda) | P \ra\,,
\ea
one can    compute its  moments by following a   similar  derivation to that which leads to Eq. (\ref{e100}). The results is
\bga 
M^q_4(N)=N_q M^2\lp\frac{1}{4}\rp^{N}\frac{\Omega}{\pi (\Omega^2-\sin^2\Omega)}\sum_{t=0}^{N-1}{N-1\choose t} (i)^t J_t\,,
\ea
where $M^q_4(N)$ indicate the IMT of $f_4(x_B)$ and where 
\bga  
J_t&=&\int d \mathbf{y} \Big[ j_0(y)\partial^t_{y_3}j_0(y)+j_1(y) \frac{y_1^2+y_2^2}{y} \partial_{y_3}^t\lp \frac{j_1(y)}{y}\rp\nn \\
& -&j_1(y)\frac{y_3}{y}\partial_{y_3}^t\lp j_1(y)\frac{y_3}{y}\rp \Big]\nn \\
&+& i\int d\mathbf{y} \Big[ j_0(y) \partial_{y_3}^t \lp j_1(y) \frac{y_3}{y}\rp-j_1(y)\frac{y_3}{y}\partial^t_{y_3}j_0(y)\Big]\,.
\ea
 The correlation function $T_1^q(x_B)$ is then related to $f_4^q(x_B)$ through the relation (see Chap. \ref{ch:twist})
\bga \label{e121}
\Lambda^2\mathbf{ T}_1^q(x_B)= x_B^2f_4^q(x_B)
\ea
In   Table \ref{t2} are shown  the first six moments  of $f_4^d(x_B)$  
\begin{table}
\caption{Moments for the $d$ quark twist four PFD obtained from the MITBM. }
\centerline{\begin{tabular}{|c|c|}
\multicolumn{2}{c}{}\\
\hline
\multicolumn{2}{|c|}{$M_N$ of $f_4^d(x_B)$  }\\
\hline
$M_1$ & .880\\
$M_2$ & .293\\
$M_3$ & .110 \\
$M_4$ & .044\\
$M_5$ & .018\\
$M_6$ & .008 \\
\hline
\end{tabular}}
\label{t2}
\end{table}

I use the same approach I used for  the twist two PDF. I fit the moments to a function
of  type (\ref{e25a}) and reconstruct the correlation function $f^d_4(x_B)$ by using Eq. (\ref{e101}). I  report the result from the fit using only five moments and not the one with six moments, since either the moment function and the IMT, were indistinguishable in the two cases. 
\bga
a=25.778 \,\,\,\,\, b=2.154 \,\,\,\,\, c=7.445\,\,\,\,\, d=57.183\,\,\,\,\, e=13.848
\ea
\begin{figure} 
\centerline{\psfig{figure=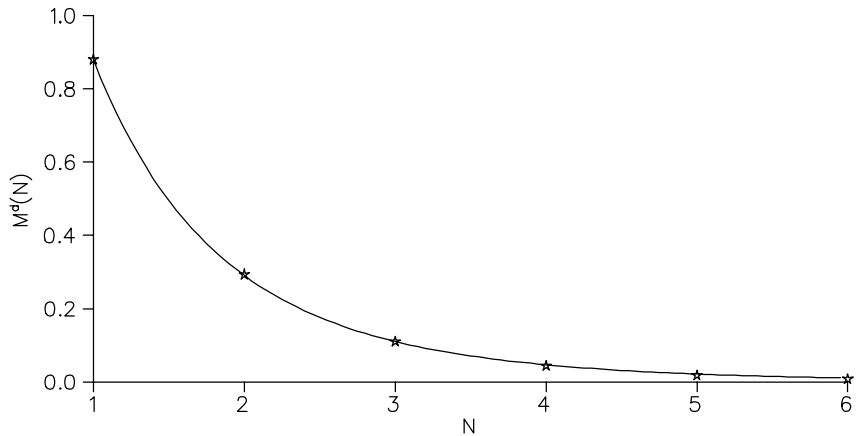}}
\caption{\label{f24} Moment function $M^d(N)$  relative to the twist four distribution $f_4^d(x_B)$ obtained by fitting the first six moments in table \ref{t2}. The stars are the moments obtained from the MITBM.}
\end{figure}
In Fig.\ref{f24} I have plotted the  moment function $M^d(N)$ for $f^d_4(x_B)$. In contrast with the twist two case,   a set of evolution equations, equivalent to the DGLAP equations for the twist two, is not available for the twist four case. As  will be  seen in Sec. \ref{sec:tw4mmitbm}, the HT contributions to the SFs obtained with the MITBM compute at the bag scale, are large compared with the model independent extractions of Refs.~\cite{sidorov,alekhin}. If one considers reliable these extractions, it might be  inferred that evolution could play an important role, reducing the magnitude of the HT effects at high $Q^2$. To get a sense on how the HT might change upon evolution, I make the ansatz of evolving them by using the DGLAP equation. It will seen  in Sec. \ref{sec:tw4mmitbm} that, although evolving the HT is just a guess, the magnitude I obtain is closer to the model independent extractions.  I  therefore  evolve the distribution $x_B\, f_4^d(x_B)$ from the bag scale $\mu_{Bag}$ I previously found for the twist two  case, to a scale $Q^2=4$ GeV$^2$. This last value is quite arbitrary, as long as it is greater than a couple of GeV$^2$. The reason is that  by using the DGLAP evolution equations, the evolved quantity shows  scaling property~\cite{bj,fey} and therefore it exhibits a smooth $Q^2$ dependence for $Q^2\gtrsim 2$ GeV$^2$. In Fig.\ref{f26} I show the distribution function $x_B\, f_4^d(x_B)$ in the two cases in which  it has evolved and in the case in which it as not. 
\begin{figure}
\centerline{\psfig{figure=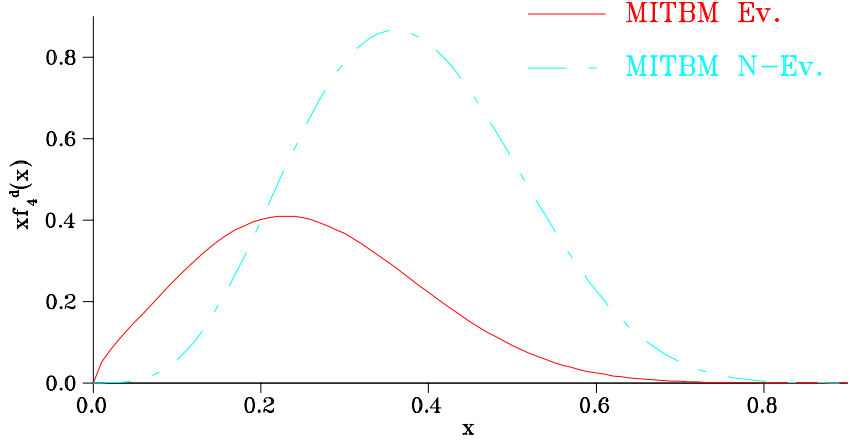}}
\caption{\label{f26} Twist four correlation $x_B f^d_4(x_B)$. The blue dot-dashed line refers to the correlation function computed at the bag scale. The solid red line shows the result of my ansatz, where the correlation function  has been  evolved using the DGLAP equation. }
\end{figure}
As opposite as to the matrix element $T_1^q(x)$, for which I was able to eliminate the gluon degree of freedom by using the equation of motion and compute its moments in the bag model, for   $T_2^q(x_1,x_2)$ such elimination is not possible,  and its moments cannot be computed directly in the bag model. However I can use an integral relation which relates $T_1^q(x_B)$ to $T_2^q(x_1,x_2)$~\cite{ellis,ellis1}
\bga\label{e122}
4\int d\,x T_1^q(x)=\int d\, x_1\,d \, x_2 T_2^q(x_1,x_2)\, . 
\ea
The authors of ref.~\cite{ellis1} suggested  expressing the correlation $T_1^q(x_B)$ in terms of the twist two PDF through the relation
\bga \label{e123}
\Lambda^2 T_1^q(x_B)=k^2 f_1^q(x_B)\, 
\ea 
where $k$ is a free parameter to be fitted, and then choose $T_2^q(x_1,x_2)$ in the form 
\bga \label{e124}
 T_2^q(x_1,x_2)=4 \delta(x_1-x_2) T_1^q(x_1)\, 
\ea
so that Eq. (\ref{e122}) is automatically satisfied. The presence of the $\delta(x_1-x_2)$ implies that the gluons exchanged in the process depicted in Fig. \ref{fa1}-c, are soft, i.e, they carry a small  fraction of the nucleon momentum. This simple form for $T_1^q(x_B)$ and $T_2(x_1,x_2)$ represents an educated guess which reproduces the $N$ dependence of the moments of the SFs as suggested by the authors in ref.~\cite{rujula}. I suggest a similar approach in which the matrix element $T_1^q(x)$ is computed in the MITBM using the IMT method, while to compute $T_2^q(x_1,x_2)$  I use the relation in Eq. (\ref{e124}). From Eq. (\ref{e124}) I get
\bga
&&\int dx_1\,dx_2\frac{\delta(x-x_2)-\delta(x-x_1)}{x_2-x_1} T_2^q(x_2,x_1)\nn \\
&=& - 4\frac{d}{dx}  T_1^q(x) 
\ea
which is useful to compute the SFs $F_1$ and $F_3$. One last thing to notice is that  the operator I compute in the bag model already includes the target mass corrections~\cite{ellis}\footnote{In Ref.~\cite{ellis} the matrix elements $T_1$ and $T_2$ including the TM are represented by bold-face letters}. To get the pure dynamic twist four contribution I need to subtract the target mass corrections. The term I need to subtract from  $\Lambda^2T_1(x_B)$ is
\bga
M^2 x_B^2\int_{x_B}^1 \frac{d\eta}{\eta} q(\eta)\,,
\ea
where $q(\eta)$ is the twist two PDF for a quark of type q.
I subtract the mass corrections at a scale $Q^2=4$ Gev$^2$. Since the mass corrections depend on the twist two PDF which exhibit scaling behavior (i.e. the depend on $x_B$ and very mildly on $Q^2$), the value of $Q^2$  I choose is not critical, as long as it is larger than few Gev$^2$. The plots of the two fermion twist four contributions to the different SFs can be found in Figs. (\ref{f53}-\ref{f58}) where I have displayed the results from the MITBM and modified MITBM (see Sec. \ref{sec:mmitbm}) compared with some model independent extractions of HT from data (see Sec. \ref{sec:tw4mmitbm}).   
\section{Four Fermion Twist Four Correlation Functions}
\label{sec:fourfermion_tw4cf}
In this section I  proceed as for  the case of the two fermion correlation function, and estimate the contributions to the SFs stemming from the four fermion operators by first computing the moments of such contributions and then reconstructing the SFs using the IMT.
\subsection{Contribution to the Structure Functions of the Four Fermion Twist Four Operators}
\label{section_fourfermion}
The contributions to the  DIS SFs 
of the four fermion twist four operators have been previously calculated by the authors of refs.
\cite{ellis,Jaffe,ji}   
\bga \label{e8}
F_1^{EM}(x_B)&=&\frac{ x_B}{2} \frac{\Lambda^2}{Q^2}\sum_{q,q'}\int dxdydz\Big[e_q e_{q'} U_1(x,y,z)\nn \\
& \times&\big( \Delta(x,y,z,x_B) +\Delta(y-x,y,y-z,x_B)\nn \\
&-&\Delta(y-x,y,z,x_B)-\Delta(x,y,y-z,x_B)\big)\nn \\
&+& e_q e_{q'}U_2(x,y,z)\nn \\&\times&\big( \Delta(x,y,z,x_B) +\Delta(y-x,y,y-z,x_B)\nn \\
&+&
\Delta(y-x,y,z,x_B)+\Delta(x,y,y-z,x_B)\big)\Big]
\ea
\bga\label{e9}
F_1^{EMNC}(x_B)&=& \frac{x_B}{2} \frac{\Lambda^2}{Q^2}\sum_{q,q'}\int dxdydz\Big[e_q g_{q'}^V U_1(x,y,z)\nn \\
& \times& \big( \Delta(x,y,z,x_B) +\Delta(y-x,y,y-z,x_B)\nn \\
&-&\Delta(y-x,y,z,x_B)-\Delta(x,y,y-z,x_B)\big)\nn \\
&+& e_q g^V_{q'}U_2(x,y,z)\nn \\
&\times&\big( \Delta(x,y,z) 
+\Delta(y-x,y,y-z,x_B)\nn \\
&+&\Delta(y-x,y,z,x_B)+\Delta(x,y,y-z)\big)\Big]
\ea
\bga\label{e10}
F_3^{EMNC}(x_B)&=& - x_B \frac{\Lambda^2}{Q^2}\sum_{q,q'}\int dxdydz
\Big[(e_q g_{q'}^A) U_1(x,y,z)\nn \\
& \times& \big( \Delta(x,y,z,x_B) -\Delta(y-x,y,y-z,x_B) \nn \\
&+& \Delta(x,y,y-z,x_B)-\Delta(y-x,y,z,x_B) \big)\nn \\
&+& e_q g^V_{q'}U_2(x,y,z)\nn \\
&\times&\big( \Delta(x,y,z) 
-\Delta(y-x,y,y-z,x_B)\nn \\
&+&\Delta(y-x,y,z,x_B)-\Delta(x,y,y-z)\big)\Big]
\ea
where $q$ and $q'$  are not necessarily quarks of the same flavor. For the four fermion correlation functions the longitudinal SF $F_L$ is zero, therefore I have the relation $F_2(x_B)=2x_BF_1(x_B)$ for both the $EM$ and the $EMNC$ contributions.
The functions $\Delta(x,y,z,x_B)$ are defined as
\bga \label{e1}
\Delta(x,y,z,x_B)&=&\frac{\delta(x-x_B)}{(y-x)(z-x)}+
\frac{\delta(y-x_B)}{(x-y)(z-y)}+\frac{\delta(z-x_B)}{(y-z)(x-z)}\nn \\
\ea
The four fermion operators $U_1$ and $U_2$ are 
\bga
U_1(x,y,z)&=&\frac{g^2}{4\Lambda^2}\int \frac{d\lambda}{2\pi}\frac{d\mu}{2 \pi}
\frac{d\nu}{2\pi}e^{i\lambda x}e^{i\mu (y-x)}e^{i\nu(z-y)}\nn \\
&\times& \la P| \bar{\psi}^q(0)n \sla t^a \psi^q(\nu n) \bar{\psi}^{q'}(\mu n )n \sla t_a \psi^{q'} (\lambda n)|P\ra
\ea
 \bga\label{ea80}
U_2(x,y,z)&=&\frac{g^2}{4\Lambda^2}\int \frac{d\lambda}{2\pi}\frac{d\mu}{2 \pi}\frac{d\nu}{2\pi}e^{i\lambda x}e^{i\mu (y-x)}e^{i\nu(z-y)}\nn \\
&\times& \la P| \bar{\psi}^q(0)n \sla \gamma_5 t^a \psi^q(\nu n) \bar{\psi}^{q'}(\mu n) n \sla t_a \gamma_5\psi^{q'} (\lambda n)|P\ra\,,
\ea
and
 \bge
n_\mu=\frac{1}{M}(1,0,0,-1)
\ee
where $g$ is the strong coupling constant and $t_a=\frac{\lambda_a}{2}\,\,\,\, a=1\cdots 8$ are the $SU(3)$ color 
generators. 

As for the case of the two quark correlation functions, I would like to compute the moments of the SFs 
\bga
M(N)&\equiv&\int _0^1dx_B\, x_B^{N-1} F_i(x_B)\,\,\,\,\,\, i=1,3
\ea 
and reconstruct them by using the inverse Mellin transform
\bga
F_i(x_B)=\frac{1}{2\pi i}\int_{c-i\infty}^{c+i\infty} dN\, x_B^{-N} M(N)\, .
\ea

To compute the moments, it is useful to consider following integral
\bga \label{e3}
&&\int dx\, dy\,dz\int dx_B  x_B^{N}\Delta(y-x,y-z,y,x_B) \nn \\
&\times&\int \frac{d\lambda}{2\pi}\frac{d\mu}{2 \pi}
\frac{d\nu}{2\pi}e^{i\lambda x}e^{i\mu (y-x)}e^{i\nu(z-y)} f_1(\nu n) f_2(\mu n) f_3(\lambda n)\nn\\
&=&
\int dx\, dy\,dz\,\sum_{j=1}^{N-1}(y-z)^{N-j-1}\sum_{k=0}^{j-1}(y-x)^{j-k-1} y^k\nn\\
&\times&\sum_{m,p,q=0}^{\infty}\frac{(n\cdot\partial)^m f_1(0)}{m!}\frac{(n\cdot\partial)^pf_2(0)}{p!}
\frac{(n\cdot\partial)^qf_3(0)}{q!}\nn \\
&\times&
 \int \frac{d\lambda}{2\pi}\frac{d\mu}{2 \pi}
\frac{d\nu}{2\pi}e^{i\lambda x}e^{i\mu (y-x)}e^{i\nu(z-y)} \nu^m\mu^p\lambda^q
\ea
where I used the following relation~\cite{ellis}
\bga \label{e2}
\int dx_B\, x_B^N \Delta(x,y,z,x_B)&=&\sum_{j=1}^{N-1}y^{N-j-1}\sum_{k=0}^{j-1}x^{j-k-1}z^k\,\, ,
\ea
and where $n\cdot \partial=n_\mu\partial^\mu$. From the fact that 
\bga
\int\frac{d\xi}{2\pi} e^{i\xi t} \xi^n=(-i)^n\frac{\partial^n}{\partial t^n}(\delta(t))
\ea 
I have for Eq. (\ref{e3})
\bga
&&\int dx\, dy\,dz\,\sum_{j=1}^{N-1}(y-z)^{N-j-1}\sum_{k=0}^{j-1}(y-x)^{j-k-1} y^k\nn \\
&\times&\sum_{m,p,q=0}^{\infty}(-i)^{m+p+q}\frac{(n\cdot\partial)^m f_1(0)}{m!}\frac{(n\cdot\partial)^pf_2(0)}{p!}
\frac{(n\cdot\partial)^qf_3(0)}{q!}\nn \\
&\times& \frac{\partial^m}{\partial z^m}(\delta(z-y))\frac{\partial^p}{\partial y^p}(\delta(y-x))\frac{\partial^q}{\partial x^q}(\delta(x))\nn\\
=&&\int dx\, dy\,dz\,\sum_{j=1}^{N-1}\sum_{t=0}^{N-j-1}{N-j-1\choose t}y^{N-j-1-t}(-z)^{t}\nn \\ 
&\times&\sum_{k=0}^{j-1} \sum_{s=0}^{j-k-1}{j-k-1\choose s}y^{j-k-1-s}(-x)^s y^k\nn \\
&\times&\sum_{m,p,q=0}^{\infty}(-i)^{m+p+q}\frac{(n\cdot\partial)^m f_1(0)}{m!}\frac{(n\cdot\partial)^pf_2(0)}{p!}
\frac{(n\cdot\partial)^qf_3(0)}{q!}\nn \\
&\times& \frac{\partial^m}{\partial z^m}(\delta(z-y))\frac{\partial^p}{\partial y^p}(\delta(y-x))\frac{\partial^q}{\partial x^q}(\delta(x))\nn\\
 \ea
By repeated use of the relation
\bga
\int dx\, x^n \frac{\partial^p}{\partial x^p}(\delta(x-y))&=&(-1)^n\frac{n!}{(n-p)!}y^{n-p} \,\,\,\left\{{ \textrm{for}\,\, p\leq n \atop \textrm{zero other wise }}\right.\nn \\
\ea
I get for Eq. (\ref{e3}) 
\bga \label{e4}
&&\int dx\, dy\,dz\int dx_B  x_B^{N}\Delta(y-x,y-z,y,x_B) \nn \\
&\times&\int \frac{d\lambda}{2\pi}\frac{d\mu}{2 \pi}
\frac{d\nu}{2\pi}e^{i\lambda x}e^{i\mu (y-x)}e^{i\nu(z-y)} f_1(\nu) f_2(\mu) f_3(\lambda)\nn\\
&=&\sum_{j=1}^{N-1}\sum_{k=0}^{j-1}  \sum_{s=0}^{j-k-1}\sum_{t=0}^{N-j-1}\sum_{m=0}^{s}\sum_{p=0}^{N-2-t-m}i^{N-2} (-1)^{s+t}
{N-j-1\choose t}\nn\\
 &\times&{j-k-1\choose s} {s \choose n} {N-2-t-m\choose p}\nn \\
&\times&(n\cdot\partial)^m f_1(0) (n\cdot\partial)^pf_2(0)(n\cdot\partial)^{N-2-p-m}f_3(0)\nn \\
\ea
Similar relations for the other three $\Delta$'s functions can be found in Appendix \ref{a3} 
and are used, along with Eq. (\ref{e4}), to calculate the moments of the SFs (\ref{e8}-\ref{e10}).

I now consider one of the possible contributions to the moment
of the  SFs, the remaining terms can be obtained  similarly. For instance consider the integral 
\bga
&&\int dx_B \,x_B^N \Delta(x,y,z,x_B) \int dxdydz U_2(x,y,z) \,.\nn \\
\ea
From Eq. (\ref{e5}) I have 
\bga\label{e16}
&&\int dx_B \,x_B^N \Delta(x,y,z,x_B) \int dxdydz U_2(x,y,z) \nn \\
&=& \frac{g^2}{4\Lambda^2}\sum_{j=1}^{N-1}\sum_{k=0}^{j-1} \sum_{m=0}^{j-k-1}\sum_{p=0}^{j-1-m}
i^{N-2}{j-1-m\choose p}{j-k-1\choose m}\nn\\
 &\times&\!\!\la P| \bar{q}_\sigma (0)n \sla \gamma_5 t^a(n\cdot\partial)^m   q_\sigma(0) (n\cdot\partial)^p \bar{q}_{\sigma '}' (0)n \sla\gamma_5 t_a
(n\cdot\partial)^{N-2-p-m}q_{\sigma '}'(0)|P\ra\nn \\
\ea

Consider first (omitting color matrices)
\bga \label{e18}
&& (n\cdot\partial)^p \bar{q}_{\sigma } (x)n \sla\gamma_5 
(n\cdot\partial)^qq_{\sigma }(x)\nn \\
&=&\frac{N_m^2}{M}\Big[ (n\cdot\partial)^p (j_0(E|\mathbf{x}|)e^{iEt} ) (n\cdot\partial)^q (j_0(E|\mathbf{x}|)e^{-iEt} )\chi_\sigma^\dag \sigma_3 \chi_\sigma\nn \\
&-&i  (n\cdot\partial)^q (j_0(E|\mathbf{x}|)e^{-iEt}\chi_\sigma^\dag)(n\cdot\partial)^p (j_1(E|\mathbf{x}|) \mathbf{\sigma}\cdot\hat{\mathbf{x}}e^{iEt} \chi_\sigma)\nn \\
&+&i  (n\cdot\partial)^p (j_0(E|\mathbf{x}|)
e^{iEt}\chi_\sigma^\dag)(n\cdot\partial)^q(j_1(E|\mathbf{x}|) \mathbf{\sigma}\cdot\hat{\mathbf{x}}e^{-iEt} \chi_\sigma) \nn \\
&+&\chi_\sigma^\dag(n\cdot\partial)^p (j_1(E|\mathbf{x}|)\mathbf{\sigma}\cdot \hat{x}e^{iEt})\sigma_3 (n\cdot\partial)^q (j_1(E|\mathbf{x}|) \mathbf{\sigma}\cdot\hat{x}e^{-iEt}
\chi_\sigma
\Big]\nn \\
&=&(-1)^{\sigma-1/2}\frac{N_m^2}{M}\Big[ (n\cdot\partial)^p (j_0(E|\mathbf{x}|)e^{iEt} ) (n\cdot\partial)^q (j_0(E|\mathbf{x}|)e^{-iEt} )\nn \\
&-&i  (n\cdot\partial)^q (j_0(E|\mathbf{x}|)e^{-iEt} )(n\cdot\partial)^p (j_1(E|\mathbf{x}|) \frac{z}{|\mathbf{x}|}e^{iEt}  )\nn \\
&+&i  (n\cdot\partial)^p (j_0(E|\mathbf{x}|)
e^{iEt} )(n\cdot\partial)^q (j_1(E|\mathbf{x}|)  \frac{z}{|\mathbf{x}|}e^{-iEt} ) \nn \\
&+&(n\cdot\partial)^p (j_1(E|\mathbf{x}|) \frac{z}{|\mathbf{x}|}e^{iEt}) (n\cdot\partial)^q (j_1(E|\mathbf{x}|) \frac{z}{|\mathbf{x}|}
e^{-iEt} )\nn \\
&-& (x^2+y^2)(n\cdot\partial)^p (\frac{j_1(E|\mathbf{x}|) }{|\mathbf{x}|}e^{iEt}) (n\cdot\partial)^q ( \frac{j_1(E|\mathbf{x}|)}{|\mathbf{x}|}
e^{-iEt} )
\Big]\nn \\
\ea
where I have made use of the fact that 
\bga
n \cdot \partial =\frac{1}{M}(\frac{\partial}{\partial t}+\frac{\partial}{\partial z})
\ea
only acts on the third spatial component.  Using Eq. (\ref{e18}) and setting $q=N-2-p-m$, Eq.  (\ref{e16}) becomes
\bga \label{e170}
&&\int dx_B \,x_B^N \Delta(x,y,z,x_B) \int dxdydz U_2(x,y,z) \nn \\
&=& \frac{g^2}{4\Lambda^2}k_{\sigma \sigma'} (-1)^{\sigma + \sigma'-1} \frac{N_m^2}{M} \sum_{j=1}^{N-1}\sum_{k=0}^{j-1} \sum_{m=0}^{j-k-1}\sum_{p=0}^{j-1-m}
i^{N-2}{j-1-m\choose p}{j-k-1\choose m}\nn\\
 &\times&\int d\mathbf{x}\Big[  (j_0(E|\mathbf{x}|)e^{iEt} ) (n\cdot\partial)^m (j_0(E|\mathbf{x}|)e^{-iEt} )\nn \\
&-&i  (n\cdot\partial)^m (j_0(E|\mathbf{x}|)e^{-iEt} ) (j_1(E|\mathbf{x}|) \frac{z}{|\mathbf{x}|}e^{iEt}  )\nn \\
&+&i   (j_0(E|\mathbf{x}|)
e^{iEt} )(n\cdot\partial)^m (j_1(E|\mathbf{x}|)  \frac{z}{|\mathbf{x}|}e^{-iEt} ) \nn \\
&+& (j_1(E|\mathbf{x}|) \frac{z}{|\mathbf{x}|}e^{iEt}) (n\cdot\partial)^m (j_1(E|\mathbf{x}|) \frac{z}{|\mathbf{x}|}
e^{-iEt} )\nn \\
&-& (x^2+y^2) (\frac{j_1(E|\mathbf{x}|) }{|\mathbf{x}|}e^{iEt}) (n\cdot\partial)^m ( \frac{j_1(E|\mathbf{x}|)}{|\mathbf{x}|}
e^{-iEt} )\Big]\nn \\
&\times&\Big[ (n\cdot\partial)^p (j_0(E|\mathbf{x}|)e^{iEt} ) (n\cdot\partial)^q (j_0(E|\mathbf{x}|)e^{-iEt} )\nn \\
&-&i  (n\cdot\partial)^q (j_0(E|\mathbf{x}|)e^{-iEt} )(n\cdot\partial)^p (j_1(E|\mathbf{x}|) \frac{z}{|\mathbf{x}|}e^{iEt}  )\nn \\
&+&i  (n\cdot\partial)^p (j_0(E|\mathbf{x}|)
e^{iEt} )(n\cdot\partial)^q (j_1(E|\mathbf{x}|)  \frac{z}{|\mathbf{x}|}e^{-iEt} ) \nn \\
&+&(n\cdot\partial)^p (j_1(E|\mathbf{x}|) \frac{z}{|\mathbf{x}|}e^{iEt}) (n\cdot\partial)^q (j_1(E|\mathbf{x}|) \frac{z}{|\mathbf{x}|}
e^{-iEt} )\nn \\
&-& (x^2+y^2)(n\cdot\partial)^p (\frac{j_1(E|\mathbf{x}|) }{|\mathbf{x}|}e^{iEt}) (n\cdot\partial)^q ( \frac{j_1(E|\mathbf{x}|)}{|\mathbf{x}|}
e^{-iEt} )
\Big]
 \nn \\
\ea
where $k_{\sigma \sigma '} $ is a factor arising from the spin-flavor-color part of the wave function. See appendix \ref{a3}
for the results arising from the $U_1$ opeator.

As an example let us consider the contribution to the  second moment of $F_1^{EM}(x_B)$ due  to the operator $U_2$ for a deuteron target. The reason I choose such a kind of target is that recent proposed DIS experiments might eventually use it~\cite{bosted,bosted1}. From Eq. (\ref{e2}), for $N=2$,
 I have 
\bga
\int dx_B\, x_B\, \Delta(x,y,z,x_B)=1 
\ea
for each $\Delta$, therefore from Eq. (\ref{e8}) I get (only $U_2$ contribution)
\bga\label{ea81}
M(N=2)&=& \frac{g^2  }{8Q^2}\frac{N_m^4}{M E^3 }\sum_{q,q'}\sum_{\sigma,\sigma'}e_qe_{q'}(-1)^{\sigma+\sigma'-1}k_{\sigma\sigma'}\nn\\
&\times&\int d\mathbf{y}\Big[ j_0^2(|\mathbf{y}|)+\frac{(y_1^2+y_2^2-y_3^2)}{|\mathbf{y}|^2} j_1^2(|\mathbf{y}|)\Big]\,,
\ea 
where I made a change of variable $\mathbf{y}=E\mathbf{x}$ to make the integral dimensionless, and $y=\sqrt{y_1^2+y_2^2+y_3^2}$
(the extra factor of $2M$ is due to the different normalization when I go from continuum normalization to wave packet normalization. See Sec. \ref{sec:twofermion_tw2cf}). 
Since
\bga
N_m^2&=& \frac{1}{4 \pi R^3}\frac{\Omega^4}{\Omega^2-\sin^2\Omega}\, ,
\ea
and $\alpha_{S}=\frac{g^2}{4 \pi}$ 
and $\frac{R M}{\Omega}=4$~\cite{jaffe2}, using the spin-flavor  factors from Eq. \ref{e35} and the color factors in tables \ref{t36}-\ref{t38}, I have
\bga
M(N=2)&=&2\lp-\frac{500}{81}-\frac{320}{81}\rp \frac{\alpha_S  }{4^6 \pi Q^2}\frac{\Omega^2 M^2}{(\Omega^2-\sin^2\Omega)^2 } \nn \\
&\times& \int d\mathbf{y}\Big[ j_0^2(|\mathbf{y}|)+\frac{(y_1^2+y_2^2-y_3^2)}{|\mathbf{y}|^2} j_1^2(|\mathbf{y}|)\Big]\,.
\ea 
The factor $\frac{1}{2}\lp-\frac{500}{81} -\frac{272}{81}\rp$ comes from the charge-spin-flavor-color term in Eq. (\ref{ea81}), which, recalling the form of the operator $U_2$ in Eq. (\ref{ea80}), is equal to
\bga\label{ea83}
&&\sum_{q,q'}\sum_{\sigma,\sigma'}e_qe_{q'}(-1)^{\sigma+\sigma'-1}k_{\sigma\sigma'}\nn \\
&=&\sum_{q,q'}\sum_{\sigma,\sigma'}(-1)^{\sigma+\sigma' -1}e_q e_{q'}\,_{SFC}\la P| \psi^{\dag \,q}_{\sigma}\lambda_a\psi^q_\sigma \psi^{\dag \, q'}_{\sigma '} \lambda^a \psi^{q'}_{\sigma '}| P\ra_{SFC}\,,
\ea
where $|P\ra_{SFC}$ stands for the spin-flavor-color part of the wave function. The first term $-500/81$ comes from the proton contribution, while the second term $-272/81$ comes from the neutron contribution, and the factor $1/2$ is due to the definition of the deuteron SFs as average of the two.  Let us compute explicitly, as an example,  the proton contribution. The flavor and color parts for a proton wave function   can be found in  appendix \ref{a2}. Since I am working in the bag model, the sum over the quark flavors is restricted to  $u$ and  $d$, hence giving the following choices for $q$ and $q'$:
\bga
 q=q'=u \,\,\,; \,\,\,q=q'=d\,\,\,;\,\,\, q=u\,,\,q'=d\,\,\,;\,\,\, q=d\,,\, q'=u\,. 
\ea
Obviously, the last two cases give the same contribution. The color contributions are easily computed. From tables \ref{t36}, \ref{t37} and \ref{t38},  I have that the color factor, which is given by the sum of the all entrees in the second column,  is -8/3 for  $q=q'=u$, 4/3 for $q=q'=d$ and -16/3 for $q=u\,,\,q'=d$ and $q=d\,,\, q'=u$ (remember that the repeated color indices in the SU(3) matrices  $\lambda_a$ implies a sum over such indices). For the flavor part, whose contribution for the different pairs are shown in Eq. 
(\ref{e35}), one needs to be a little careful because of the factor $(-1)^{\sigma+\sigma'-1}$, where $\sigma$ and $\sigma '$ are the third component of the spin of the first and second pair, respectively. If the two pairs have same spin component, then the factor give one, if they have opposite spin component, the factor is minus one. For $q=q'=u$ the flavor factor is 8/3, for $q=q'=d$ the result is 1 and for $q=u\,,\,q'=d$ and $q=d\,,\, q'=u$ I have
-4/3. Finally, including  the product of the quark charges, the factor in Eq. (\ref{ea83}) is
\bga
\frac{4}{9}\lp -\frac{8}{3} \rp \frac{8}{3}-2 \frac{2}{9}\lp -\frac{16}{3}\rp\lp-\frac{4}{3}\rp+\frac{1}{9}\frac{4}{3}1=-\frac{500}{81}\,.
\ea
The contribution from the neutron wave function can be similarly computed, except that now, in the flavor and color factors, I need to exchange $u$ with $d$.\\

 The value of the strong coupling to be used requires some attention. The authors of ref.~\cite{jaffe11} take it to be
$\sqrt{\alpha_S(\mu_{Bag}^2) \alpha_S(Q^2)}$ where $\alpha_S(\mu_{Bag}^2)$ is the value of the coupling constant at the bag scale, 
while $\alpha_S(Q^2)$ is the value of the strong  coupling at the $Q^2$ I intend to evaluate the HT contributions. 
I take   the strong coupling constant to be equal one, therefore my estimate provides an upper limit. Notice that according to the value of the bag constant I found in Sec. \ref{sec:twofermion_tw2cf}, $\mu^2_{Bag}\simeq .2$ GeV$^2$, for a $Q^2\simeq 1$ GeV$^2$ and for a value of $\Lambda_{QCD}=.25$ GeV (value I used  for the evolution software provided by~\cite{kumano}), the coupling constant would be of the order of $\alpha_S\simeq .9$, decreasing to $\alpha_S\simeq .7$ for a $Q^2=4$ GeV$^2$.    
 Taking  a value of $\alpha_S=1$, performing the integral and taking $M=.938$ GeV and 
$Q=1$ Gev, I get the result in the table \ref{ta2}.   The complete set of results for the moments of the different SFs are summarized  in   Tables \ref{ta1}- \ref{t6}. All the results are for $Q=1$ Gev. 
It must be noticed that,  because of the crossing symmetry in the virtual forward Compton amplitude, one knows that for $F_1$
only even moments contribute while for $F_3$ only odd moments contribute~\cite{yndu} (which is true in general, not only for the four fermion twist four SFs). Moreover, since Eq. (\ref{e2}) is zero for $N=1$, we
 have the exact result that the first
moment of  $F_3$ is zero. 

\begin{table}
\label{ta1}
\caption{Moment function for the SF $F_1^{EM}$ obtained from the MITBM. Contribution arising from the $U_1$ operator. }
\centerline{\begin{tabular}{|c|c|}
\multicolumn{2}{c}{}\\
\hline
\multicolumn{2}{|c|}{$M_N$ of $F_1^{EM}(x_B)$ from $U_1$}\\
\hline
$M_2$& 0\\
$M_4$ &$7.77 \times 10^{-5}$\\
$M_6$ & $2.81\times 10^{-5}$\\
\hline
\end{tabular}}
\end{table} 
\begin{table}
\label{ta2}
\caption{Moment function for the SF $F_1^{EM}$ obtained from the MITBM. Contribution arising from the $U_2$ operator. }
\centerline{\begin{tabular}{|c|c|}
\multicolumn{2}{c}{}\\
\hline
\multicolumn{2}{|c|}{$M_N$ of $F_1^{EM}(x_B)$ from $U_2$}\\
\hline
$M_2$&$ -35.85\times 10^{-4}$\\
$M_4$ &$-3.97 \times 10^{-4}$\\
$M_6$ & $-0.05 \times 10^{-4}$\\
\hline
\end{tabular}  }
\end{table}
\begin{table}
\label{t3}
\caption{Moment function for the SF $F_1^{EMNC}$ obtained from the MITBM. Contribution arising from the $U_1$ operator. }
\centerline{\begin{tabular}{|c|c|}
\multicolumn{2}{c}{}\\
\hline
\multicolumn{2}{|c|}{$M_N$ of $F_1^{EMNC}(x_B)$ from $U_1$}\\
\hline
$M_2$& 0\\
$M_4$ &$1.97 \times 10^{-4}$\\
$M_6$ & $0.71 \times 10^{-4}$\\
\hline
\end{tabular} }
\end{table}
\begin{table}
\label{t4}
\caption{Moment function for the SF $F_1^{EMNC}$ obtained from the MITBM. Contribution arising from the $U_2$ operator. }
\centerline{\begin{tabular}{|c|c|}
\multicolumn{2}{c}{}\\
\hline
\multicolumn{2}{|c|}{$M_N$ of $F_1^{EMNC}(x_B)$ from $U_2$}\\
\hline
$M_2$&$ -36.15 \times 10^{-4}$\\
$M_4$ &$-4.00 \times 10^{-4}$\\
$M_6$ & $-.05 \times 10^{-4}$\\
\hline
\end{tabular}  }
 \end{table}
\begin{table}
\label{t5}
\caption{Moment function for the SF $F_3^{EMNC}$ obtained from the MITBM. Contribution arising from the $U_1$ operator.}
\centerline{\begin{tabular}{|c|c|}
\multicolumn{2}{c}{}\\
\hline
\multicolumn{2}{|c|}{$M_N$ of $F_3^{EMNC}(x_B)$ from $U_1$}\\
\hline
$M_1$& 0\\
$M_3$ &$-31.72 \times 10^{-4}$\\
$M_5$ &  $-6.30\times 10^{-4}$\\
\hline
\end{tabular}}
\end{table}
\begin{table}
\label{t6}
\caption{Moment function for the SF $F_3^{EMNC}$ obtained from the MITBM. Contribution arising from the $U_2$ operator. }
\centerline{\begin{tabular}{|c|c|}
\multicolumn{2}{c}{}\\
\hline
\multicolumn{2}{|c|}{$M_N$ of $F_3^{EMNC}(x_B)$ from $U_2$}\\
\hline
$M_1$& 0\\
$M_3$ &$18.97 \times 10^{-4}$\\
$M_5$ & $3.21 \times 10^{-4}$\\
\hline
\end{tabular} }
\end{table}
\subsection{Moment Inversion of the Four Fermion Twist Four Correlation Functions }
To compute the contributions of the four fermion twist four matrix elements to the SFs,  I use the same approach previously used for the  case of the two fermion correlation functions. As before I attempt to reconstruct SFs by means of the IMT
\bga
F_i(x_B)=\frac{1}{2\pi i}\int_{c-i\infty}^{c+i\infty} d N x_B^{-N} M(N)
\ea
The problem now is to numerically invert the previous relation by using the only 
three moments that I obtained. Because of the small number of moments available, a complete and reliable numerical reconstruction 
is not  possible. As before I fit the moments to a function of type 
\bga\label{e25}
M(N)&=&a\,(\pm\beta[1+c,b+N ] +d \,\beta[1+c,b+N+1]\nn \\
&+& e\,\beta[1+c,b+N+2])
\ea
The plus sign is used to fit those moments for which the lowest one   is non zero and the minus 
sign for those for which the lowest moment is zero. 
In this case, since the number of parameters is larger than the number of points used for the fit, the solution is not unique. Indeed, it happens that totally
different sets of parameters fit the moments equally well. 
I also consider an alternative type of function to fit the moments to test how much the inversion processes  depends on the function chosen. The function is 
\bga\label{e26}
M(N)=\sum_{i=j}^{j+2}(\frac{a_i}{N^i})\,,
\ea
where $j$ is any integer.
The corresponding Mellin Transform is
\bga
\frac{1}{2\pi i}\int_{c-i\infty}^{c+i\infty} d N x_B^{-N}\sum_{i=j}^{j+2}(\frac{a_i}{N^i})
=\sum_{i=j}^{j+2}a_i \frac{(-\log{x_B})^{i-1}}{(i-1)!} \,.
\ea
Because of the presence of the term $(\log x_B)^{i-1}$ in the IMT, I   restrict the calculation  to the case   $j=1$ and $j=2$, to avoid unnatural large values of the quantities $x_B F_i(x_B)$ in the limit in which $x_B$ goes to zero.

I  present  the results for the fits for two of the possible contributions to the SFs due to the operator $U_2$ 
(the rest of the results for the other contributions of $U_1$ and $U_2$ to the SFs  are presented in the appendix).
As an example I   consider the contributions to the moments of $F_1^{EM} (x_B)$ coming from $U_2$ operator. 
In this case, for a parametrization of the kind of Eq. (\ref{e25}), I have found 
two sets of parameters $\{a,b,c,d,e\}$ which fit the moments with the same degree of accuracy. The first set is
\bga \label{e27}
&&\!\!\!\!\!\!\!\!\!\!\!\!a=-302.360\times 10^{-4}\,\,\,\,\, b=1.545\,\,\,\,\,c=9.362\,\,\,\,\,d=265.331\,\,\,\,\,e=3.006\nn\\
\ea 
The second set is
\bga\label{e28}
&&\!\!\!\!\!\!\!\!\!\!\!\!a=-77.404\times 10^{-4}\,\,\,\,\, b=-.112\,\,\,\,\,c=7.477\,\,\,\,\,d=5.447\,\,\,\,\, e=179.207\nn\\
\ea 
 The set of parameters relative to the parametrization of the type (\ref{e26}) are, for $j=1$
\bga \label{e29}
a_1=8.709 \times 10^{-4}\,\,\,\,\, a_2= -48.282\times 10^{-4}\,\,\,\,\, a_3 =-36.746\times 10^{-4}&&\nn\\
\ea 
and for $j=2$
\bga \label{e30}
a_2=59.569 \times 10^{-4}\,\,\,\,\,a_3= -433.076\times 10^{-4}\,\,\,\,\, a_4 =431.403\times 10^{-4}&&\nn\\
\ea 
\begin{figure}   
\centerline{\psfig{figure=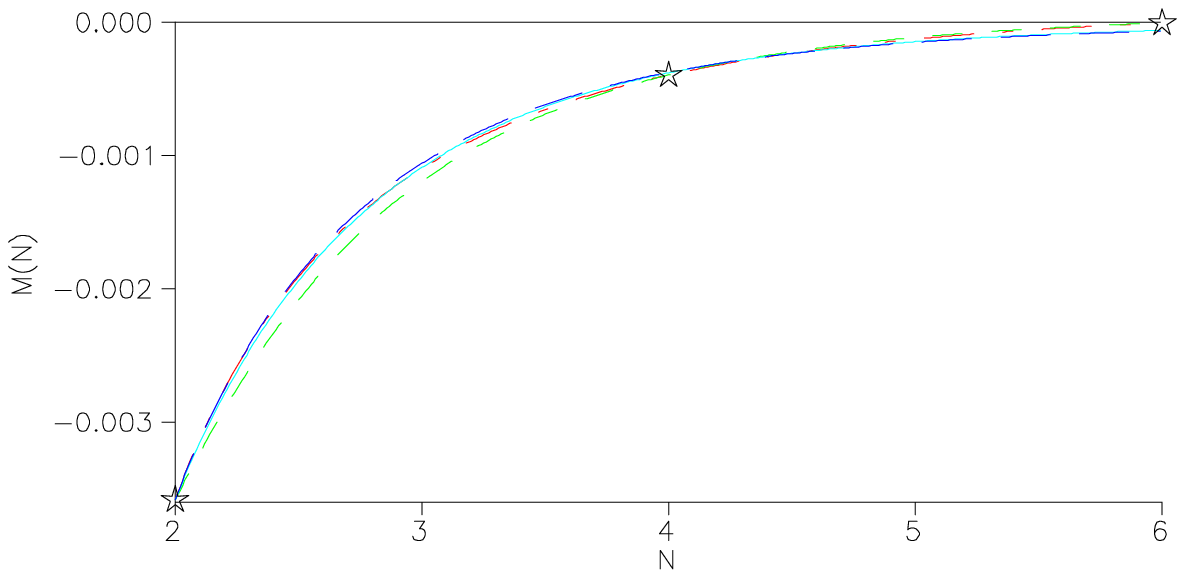}}
\caption{\label{f7}  Comparison of the different fits for the moment function $M(N)$ relative to the $F_1^{EM}(x_B)$ SF ($U_2 $ contribution). The solid light blue line 
is relative to the set of parameters in Eq. (\ref{e27}), 
the long dashed blue line is relative to the set of parameters in Eq. (\ref{e28}), the dot-dashed red  line is relative to the set of parameters in Eq. (\ref{e29}),
and the short dashed green line is relative to the set of parameters in Eq. (\ref{e30}). The three moments obtained from the bag model are represented by stars.}
\end{figure}
\begin{figure}  
\centerline{\psfig{figure=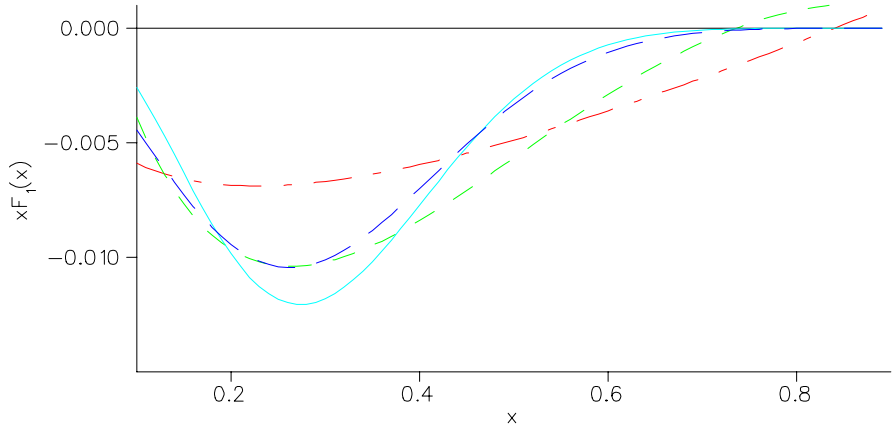}}
\caption{\label{f8} Reconstruction of the SF $x_BF_1^{EM}(x_B)$ through the use of the inverse Mellin transform technique ($U_2$ contribution). 
The type of line used for each 
parametrization is the same as in Fig.\ref{f7}.}
\end{figure}
 In Fig. \ref{f7} I have plotted the different moment functions $M(N)$ relative to the different set of parameters (Eqs. (\ref{e27}-\ref{e30})).  
The stars represent the points  obtained from the bag model. It can be seen how all the different parameterizations fit the data
quite well and, in this case, the corresponding  curves are very similar in shape and magnitude.  On the other hand, after computing
the inverse Mellin transform, those different parameterizations give rise to different $x_B$ dependence for the SFs as one can see
from Fig.\ref{f8}. Nevertheless,
although each parametrization corresponds to a different shape, the overall magnitude is quite similar. I would like to emphasize that I am not interested in
an exact calculation of the $x_B$ dependence, since it is already known, from section \ref{sec:twofermion}, that the simple bag model does not reproduce such a dependence for the twist two PDFs. What we
  would like to investigate, instead, is if the magnitude of the contribution to the SFs of these four fermion  twist four corrections  is indeed negligible compared to the twist two contribution (or other twist four 
contribution stemming from two fermion correlation functions). All the contributions from the four fermion correlation functions have not been evolved, since, as will be seen in Sec. \ref{sec:final_arl}, their contribution is already small even if they are not evolved (note that evolution decreases the magnitude of the function evolved).

\begin{figure}
\centerline{\psfig{figure=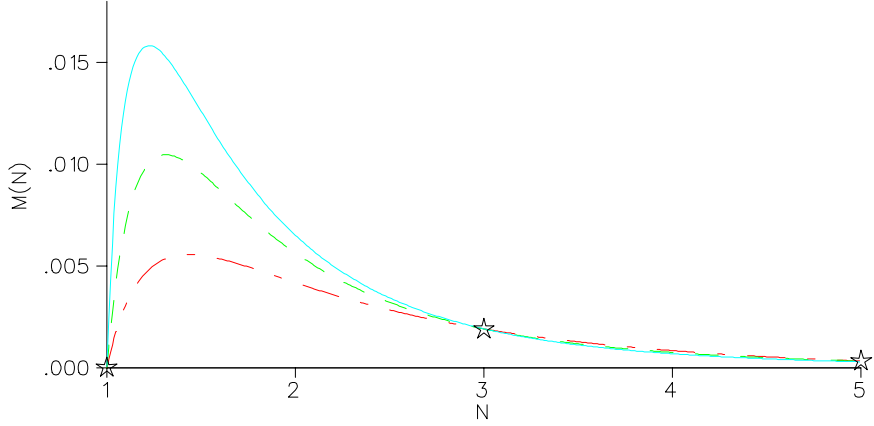}}
\caption{\label{f9} Comparison of the different fits for the moment function $M(N)$ relative to the $F_3^{EMNC}(x_B)$ SF ($U_2$ contribution).
 The solid blue line is relative
 to the set of parameters in Eq. (\ref{e30}), 
the dashed green line is relative to the set of parameters in Eq. (\ref{e31}) and  the dot-dashed red line is relative to the set of parameters in Eq. (\ref{e32}). 
The three moments 
obtained from the bag model are represented by stars. }
\end{figure}

As a second example I consider 
the contribution to the SF $F_3$ coming from the operator $U_2$. In this case I was only able to find one set of parameters relative to the 
parametrization (\ref{e25}) (that does not mean that there is only one nor that the one I found is the best one). 
The set of parameters  for the different parameterizations are
\bga\label{e31}
&&\!\!\!\!\!\!\!\!\!\!\!\!a=41.369\times 10^{-4}\,\,\,\,\, b=-.821\,\,\,\,\,c=5.514\,\,\,\,\,d=30.754\,\,\,\,\, e=43.527\nn\\
\ea 
\bga \label{e32}
a_1=-26.773 \times 10^{-4}\,\,\,\,\, a_2= 195.114 \times 10^{-4}\,\,\,\,\, a_3 =-168.341\times 10^{-4}&&\nn\\
\ea 
\bga \label{e33}
a_2=-45.854 \times 10^{-4}\,\,\,\,\,a_3= 447.451\times 10^{-4}\,\,\,\,\, a_4 =-401.604\times 10^{-4}&&\nn\\
\ea 
The different moments functions
are plotted in Fig. \ref{f9}. In this case, although they all fit the points obtained from the bag model calculation, they have quite different shape (at least in the
range of values between $N=1$ and $N=3$). However,
also in this case, the magnitude of the inverse transform is quite similar (see Fig. \ref{f10}).  The parameterizations for the remaining terms of the SFs are
 in the appendix \ref{a1}.
\begin{figure}
\centerline{\psfig{figure=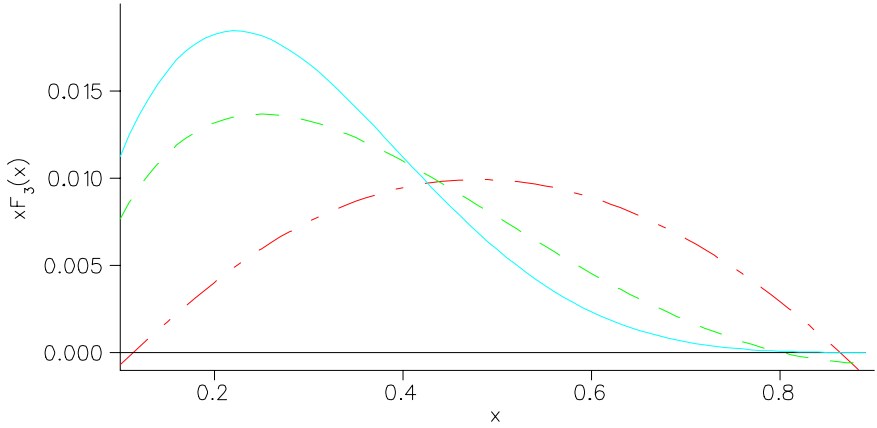}}
\caption{\label{f10} Reconstruction of the SF $x_BF_3^{EMNC}(x_B)$ through the use of the inverse Mellin transform technique ($U_2$ contribution). 
The type of line used for each 
parametrization is the same as in Fig.\ref{f9}.}
\end{figure}

\section{Two Fermion Twist Two and Four Correlation Functions in the Modified MIT Bag Model }
\label{sec:mmitbm}
In this section I try to use a modified version of the bag model (MMITBM) suggested by the authors of Ref.~\cite{thomas1}, to compute the twist two and twist four contributions to the SFs. In this case the correlation functions (\ref{e71}) and (\ref{e70}) are directly computed, without calculating the moments and then obtaining the SFs through an IMT. The same method applied to the four fermion twist four correlation functions, involves the numerical evaluation of multidimensional integrals, which I was not able to compute in a reasonable amount of time.
\subsection{Modified MIT Bag Model}
In the MIT bag model, since the quarks are confined in the bag,  translational invariance is lost. The authors of~\cite{thomas1} suggest  using the Peierls-Yoccoz~\cite{Peierls} projection, to ensure that the proton wave function is a momentum eigenstate, thereby making the system translational invariant. They define the action of the quark field $\psi(\mathbf{x})$ on a $n-$quark state  as
\bga\label{e163}
\psi(\mathbf{x})|\mathbf{x}_1\mathbf{x}_2\ldots \mathbf{x}_n\ra =\delta(\mathbf{x}-\mathbf{x}_1)|\mathbf{x}_2\ldots \mathbf{x}_n\ra+\,\,\textrm{permutations}\,.
\ea
A translational invariant n-quark state with momentum $p_n$ is introduced in coordinate space as
\bga\label{ea40}
\la \mathbf{x}_1\ldots\mathbf{x}_n|\mathbf{p}_n\ra\equiv \frac{1}{\phi_n(\mathbf{p_n})} \int d\mathbf{R} e^{i\mathbf{p_n}\cdot\mathbf{R}}\psi(\mathbf{x}_1-\mathbf{R})\ldots \psi(\mathbf{x}_n-\mathbf{R})\,,
\ea
where the normalization factor $\phi(\mathbf{p}_n)$ is given by the normalization condition
\bga
\la \mathbf{p}_n| \mathbf{p'}_n\ra =2M (2\pi)^3\delta(\mathbf{p}_n-\mathbf{p'}_n)\,,
\ea
where $M$ is the target mass, from which one finds
\bga\label{ea41}
|\phi_n(\mathbf{p}_n)|^2=\frac{1}{2M}\int d\mathbf{x}e^{-i\mathbf{p_n}\cdot \mathbf{x}}\Big[\int d \mathbf{y} \psi^\dag (\mathbf{y}-\mathbf{x})\psi(\mathbf{y})\Big]^n\,.
\ea
Recalling the form of the twist two PDF in Eq. (\ref{ea13})
\bga 
f_1^q(x_B)&=&\frac{1}{M}\int \frac{d\lambda}{2 \pi}e^{i\lambda x_B}\la P |\psi^{\dag q}_+(0) \psi^q_+ (\lambda n) | P \ra\,,
\ea
 and inserting a complete set of states 
\bga
1=\frac{1}{2M}\int\frac{d\mathbf{p}_n}{(2\pi)^3} |p_n \ra \la p_n| 
\ea 
I get
\bga\label{e150}
f_1^d(x_B)&=&\frac{1}{2M^2}\sum_{m}\la \alpha |P_{d,m}^n| \alpha \ra\int \frac{d\lambda}{2\pi}e^{i\lambda x_B}\int\frac{d\mathbf{p}_n}{(2\pi)^3} \la p| \psi^{\dag q}_+(0)|p_n \ra\la p_n| \psi_+(\lambda n ) |p\ra\nn \\
&=& \frac{1}{2M^2}\sum_{m}\la \alpha |P_{d,m}^n| \alpha \ra\int\frac{d\mathbf{p}_n}{(2\pi)^3} \delta(x_B-\frac{(p_n^+-p^+)}{M})|\la p_n| \psi_+(0) |p\ra|^2\,,\nn \\
\ea
where the plus component of a four-vector $v_\mu$ is defined as $v^+=v_0+v_3$, $|\alpha\ra$ is the spin-flavor part of the nucleon wave function,  and  $\la \alpha | P_{q,m}^n| \alpha \ra$ is the projector operator onto  flavor $q$ and spin $m$ for a $n$ quark intermediate state. In the last step in Eq. (\ref{e150}) I have used translational invariance, such that
\bga
\la p_n | \psi_+(\lambda n) |p\ra =\la p_n| e^{-i \lambda \hat{P}\cdot n} \psi_+(0) e^{i \lambda \hat{P}\cdot n}| p\ra=e^{-i\frac{(p_n^+-p^+)}{M}}\la p_n|  \psi_+(0) | p\ra\,,
\ea 
where $\hat{P}$ is the momentum operator, and where I used the relation $v\cdot n=\frac{v_+}{M}$.
\begin{figure} 
\centerline{\psfig{figure=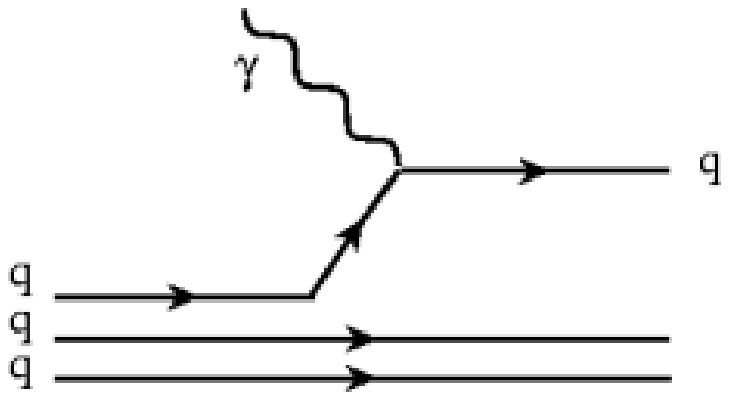}}
\caption{\label{f36}  Two quark intermidiate state process contributing to the  quark twist two  PDF.}
\end{figure}
\begin{figure} 
\centerline{\psfig{figure=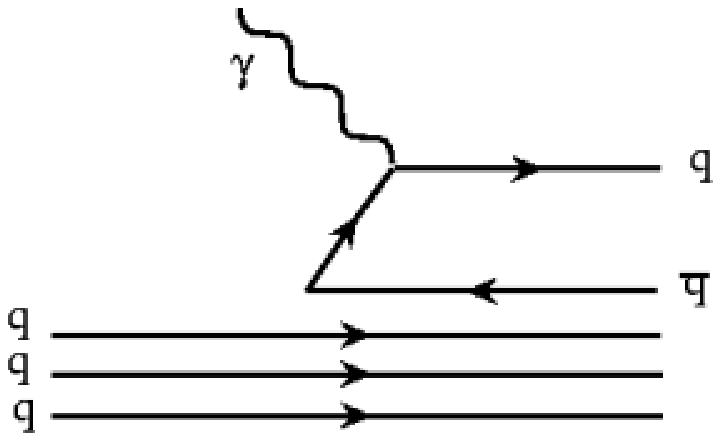}}
\caption{\label{f37}  Four quark intermidiate state process contributing to the quark twist two PDF.}
\end{figure}
\begin{figure}
\centerline{\psfig{figure=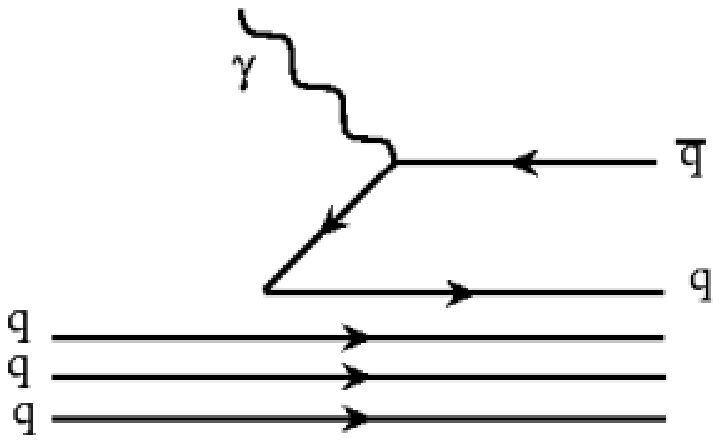}}
\caption{\label{f38}  Two quark intermidiate state process contributing to the antiquark twist two  PDF.}
\end{figure}
Since the quark field operator $\psi_+(0)$ can annihilate a quark in the position $\mathbf{0}$, but also add an antiquark in the same position, there are two possible intermediate states $| p_n\ra$, one with two quarks  and one with  three quarks and an antiquark   (see Figs. \ref{f36} and \ref{f37}). The PDF for an antiquark can be obtained from Eq. (\ref{ea13}) by considering the  operator $\psi_+^\dag$ instead of $\psi_+$~\cite{thomas1}, which yields
\bga
\bar{f}_1^{\bar{d}}(x_B)=\frac{1}{2M^2}\sum_{m}\la \alpha |P_{\bar{d},m}^n| \alpha \ra\int\frac{d\mathbf{p}_n}{(2\pi)^3} \delta(x_B-\frac{(p_n^+-p^+)}{M})|\la p_n| \psi^\dag_+(0) |p\ra|^2\,.
\ea
 Notice that, in this case, the operator $\psi^\dag_+(0)$ either creates a quark or annihilates an antiquark. Since in the MIT proton  wave function there are no antiquarks  to be annihilated, the only contribution comes from adding a quark to the proton, giving rise to the four quark intermediate state represented in Fig. \ref{f38}. Inserting a complete set of coordinate states in Eq. (\ref{e150}) one gets
\bga\label{e160}
f_1^d(x_B)&=& \sum_{m}\la \alpha |P_{d,m}^2| \alpha \ra\int \frac{d\mathbf{p}_2}{2 M (2\pi)^3} \delta(M(x_B-1)-p_2^+)\nn\\ 
&\times& \Big[|\int d\mathbf{x}_1d\mathbf{x}_2 d \mathbf{x}_3\la p_2 | \psi^d_+(0)| \mathbf{x}_1\mathbf{x}_2 \mathbf{x}_3 \ra \la \mathbf{x}_1\mathbf{x}_2 \mathbf{x}_3|p\ra |^2\Big]\nn \\
&+& \sum_{ m}\la \alpha |P_{d,m}^4| \alpha \ra\int \frac{d\mathbf{p}_4}{2 M (2\pi)^3} \delta(M(x_B-1)-p_4^+)\nn \\
&\times& \Big[| \int d\mathbf{x}_1d\mathbf{x}_2 d \mathbf{x}_3\la p_4 |\psi_+^d(0)| \mathbf{x}_1\mathbf{x}_2 \mathbf{x}_3 \ra \la \mathbf{x}_1\mathbf{x}_2 \mathbf{x}_3|p\ra |^2\Big] 
\ea
Let us consider part of the  second term in the previous equation.  Working in the target rest frame so that $|p=0\ra$, I get
\bga
&&\int d\mathbf{x}_1d\mathbf{x}_2d \mathbf{x}_3  \la p_4 | \psi^d_+(0)| \mathbf{x}_1\mathbf{x}_2 \mathbf{x}_3 \ra \la \mathbf{x}_1\mathbf{x}_2 \mathbf{x}_3|p=0\ra  \nn \\
&=&  \int d\mathbf{x}_1d\mathbf{x}_2 d \mathbf{x}_3 \la p_4 |  \mathbf{x}_1\mathbf{x}_2 \mathbf{x}_3 \mathbf{0}_{\bar{d}}\ra \la \mathbf{x}_1\mathbf{x}_2 \mathbf{x}_3|p=0\ra \nn \\
&=& \int d \mathbf{x}_1 d \mathbf{x}_2 d \mathbf{x}_3 \frac{1}{\phi_3(0)\phi_4^*(p_4) } \nn \\
&\times& \int d\mathbf{R} e^{-i \mathbf{p_4}\cdot \mathbf{R}}
\psi^\dag (\mathbf{x}_1\!-\!\mathbf{R})\psi^\dag(\mathbf{x}_2\!-\!\mathbf{R})\psi^\dag(\mathbf{x}_3\!-\!\mathbf{R})\psi^{(\bar{d})}_+(-\mathbf{R})\nn \\
&\times&\int d\mathbf{R}' 
\psi (\mathbf{x}_1\!-\! \mathbf{R}')\psi(\mathbf{x}_2\!-\!\mathbf{R}')\psi(\mathbf{x}_3\!-\!\mathbf{R'})\,,\nn\\
\ea
where I used Eq. (\ref{ea40}) for a three  quark state with momentum $\mathbf{p}=\mathbf{0}$ and a four quark state with momentum $\mathbf{p}_4$. By making the change of variables $\mathbf{R}\rightarrow -\mathbf{R}$ and $\mathbf{x}_i +\mathbf{R}\rightarrow \mathbf{x}_i$ ($i=1,2,3$), I have
\bga
&&\int d\mathbf{x}_1d\mathbf{x}_2d \mathbf{x}_3  \la p_4 | \psi^d_+(0)| \mathbf{x}_1\mathbf{x}_2 \mathbf{x}_3 \ra \la \mathbf{x}_1\mathbf{x}_2 \mathbf{x}_3|p=0\ra  \nn \\
&=&\int d \mathbf{x}_1 d \mathbf{x}_2 d \mathbf{x}_3 \frac{1}{\phi_3(0)\phi_4^*(p_4) } 
\int d\mathbf{R} e^{i \mathbf{p_4}\cdot \mathbf{R}}
\psi^\dag (\mathbf{x}_1 )\psi^\dag(\mathbf{x}_2 )\psi^\dag(\mathbf{x}_3 )\psi^{(\bar{d})}_+(\mathbf{R})\nn \\
&\times& \int d\mathbf{R}' 
\psi  (\mathbf{x}_1\!-\!\mathbf{R}'\!-\!\mathbf{R})\psi (\mathbf{x}_2\!-\!\mathbf{R}'\!-\!\mathbf{R})\psi (\mathbf{x}_3\!-\!\mathbf{R'\!-\!\mathbf{R}})\nn\\,.
\ea
By shifting the variable $\mathbf{R}'+\mathbf{R}\rightarrow \mathbf{R}'$ and using Eq. (\ref{ea41}) I arrive to the result
\bga \label{ea44}
&&\int d\mathbf{x}_1d\mathbf{x}_2d \mathbf{x}_3  \la p_4 | \psi^d_+(0)| \mathbf{x}_1\mathbf{x}_2 \mathbf{x}_3 \ra \la \mathbf{x}_1\mathbf{x}_2 \mathbf{x}_3|p=0\ra  \nn \\ 
&=& 2M\frac{\phi_3^*(0)}{\phi_4^*(p_4)} \tilde{\psi}_+^{(\bar{d})}(\mathbf{p}_4)  \,,
\ea
where I have defined the quantity
\bga
\tilde{\psi}_+^{(\bar{d})}(\mathbf{p}_4) =\int d\mathbf{R} e^{i\mathbf{p}_4\cdot \mathbf{R}}\psi_+^{(\bar{d})}(\mathbf{R})\,.
\ea
The spinor $\tilde{\psi}_+^{\bar{d}}(\mathbf{p}_4)$ is obtained by considering the antiquark bag wave function
\bga  
\psi_{m}^{\bar{q}}(\mathbf{x},t)=N_m  {- i \mathbf{\sigma }\cdot \hat{x} j_1(E|\mathbf{x}|) \chi_m \choose j_0(E|\mathbf{x}|)\chi_m  } \theta(R-|\mathbf{x}|)e^{-iEt}
\ea
A similar relation can be found for the first term in Eq. (\ref{e160}), so that I have
\bga\label{e161}
f_1^d(x_B)&=&2M\sum_m \la \alpha |P_{d,m}^2| \alpha \ra \int\frac{d\mathbf{p_2}}{(2\pi)^3}\delta(M(1-x_B)-p_2^+)\left| \frac{\phi_2(p_2)}{\phi_3(0)}\right|^2 |\tilde{\psi}_+^{(d)}(\mathbf{p}_2)|^2\nn \\
&+&2M\sum_m \la \alpha |P_{d,m}^4| \alpha \ra \int\frac{d\mathbf{p_4}}{(2\pi)^3}\delta(M(1-x_B)-p_4^+)\left| \frac{\phi_3(0)}{\phi_4(p_4)}\right|^2 |\tilde{\psi}_+^{(\bar{d})}(\mathbf{p}_4)|^2\nn \\
\ea
Similarly one can compute the antiquark distribution, obtaining
\bga\label{e162}
f_1^{\bar{d}}(x_B)&=&2 M\sum_m \la \alpha |P_{\bar{d},m}^4| \alpha \ra \int\frac{d\mathbf{p_4}}{(2\pi)^3}\delta(M(1-x_B)-p_4^+)\left| \frac{\phi_3(0)}{\phi_4(p_4)}\right|^2 |\tilde{\psi}_+^{\dag( d)}(\mathbf{p}_4)|^2\nn\\
\ea
Using the bag model wave function (\ref{e20}), one finds the following relations~\cite{thomas1,stratmann}
\bga
|\phi_n(p_n)|^2=\frac{4\pi \Omega^n}{[2(\Omega^2-\sin^2\Omega)]^n}\lp\frac{2R}{\Omega}\rp^3 \int_0^{\Omega} dv\, v^{2-n}T^n(v)j_0(\frac{2uv}{\Omega}) 
\ea
where $u=|\mathbf{p}_n |R$ and the function $T(v)$ is
\bga
T(v)=\lp \Omega\!-\!\frac{\sin^2\Omega}{\Omega}\!-\!v\rp\sin 2v\!-\!\lp \frac{1}{2}\!+\!\frac{\sin 2\Omega}{2\Omega}\rp\cos 2v\! +\!\frac{1}{2}\!+\!\frac{\sin 2\Omega}{2\Omega}\!-\!\frac{\sin^2\Omega}{\Omega^2}v^2\, .
\nn \\
\ea 
Moreover, one has
\bga\label{e168}
&&\tilde{\psi}_+^{(d)}(\mathbf{p}_n)=\tilde{\psi}_+^{\dag(d)}(\mathbf{p}_n)=\tilde{\psi}_+^{(\bar{d})}(-\mathbf{p}_n)\nn \\
&=& \frac{\pi R}{2}\frac{\Omega^2}{(\Omega^2-\sin^2\Omega)}\Big[s_1^2(u)+2\frac{p_n^z}{|\mathbf{p}_n|}s_1(u)s_2(u)+s_2^2(u)\Big]
\ea
where $s_1(U)$ and $s_2(u)$ are defined as
\bga
s_1(u)=\frac{1}{u}\Big[\frac{\sin(u-\Omega)}{u-\Omega} -\frac{\sin(u+\Omega)}{u+\Omega}\Big]
\ea
and
\bga
s_2(u)=2j_0(\Omega)j_1(u)-\frac{u}{\Omega}s_1(u)\, .
\ea
In Eqs. (\ref{e161}) and (\ref{e162}) one can perform the angular integration 
\bga \label{e165}
\int d\mathbf{p}_n\delta (M(1-x_B)-p_n^+)=2\pi \int _{|M^2(1-x_B)^2-M_n^2|/(2M(1-x_B))}^{\infty} d|\mathbf{p}_n|\,|\mathbf{ p}_n| 
\ea
where the orthogonal component $\mathbf{p}_{\perp}$ is understood to be
\bga
\mathbf{p}_{\perp}^2=2M(1-x_B)\sqrt{M_n^2+\mathbf{p}_n^2}-M^2(1-x_B)^2-M_n^2
\ea
and $M_n$ is the mass of the $n$ quark intermediate state. 

One thing to notice is the difference in some of my relations with the ones in Ref.~\cite{thomas1}. For the four fermion intermediate states the ratio in the normalization factors is reversed, {\it i.e.} they have $|\phi_4/\phi_3|$, while I have $|\phi_3/\phi_4|$, but the whole calculation that leads to Eq. (\ref{ea44}) with the normalization factor reversed is not shown.  In the rest of the calculation I  adopt Eq. (\ref{ea44}).  As observed in Ref.~\cite{thomas1}, the Peierls-Yoccoz projection is a non relativistic approximation, which is valid only for $|\mathbf{p}_n| \lesssim M_n$. As a consequence, the integrand in Eq. (\ref{e165}) has an upper limit cut off equal to $M_n$. 

\begin{figure} 
\centerline{\psfig{figure=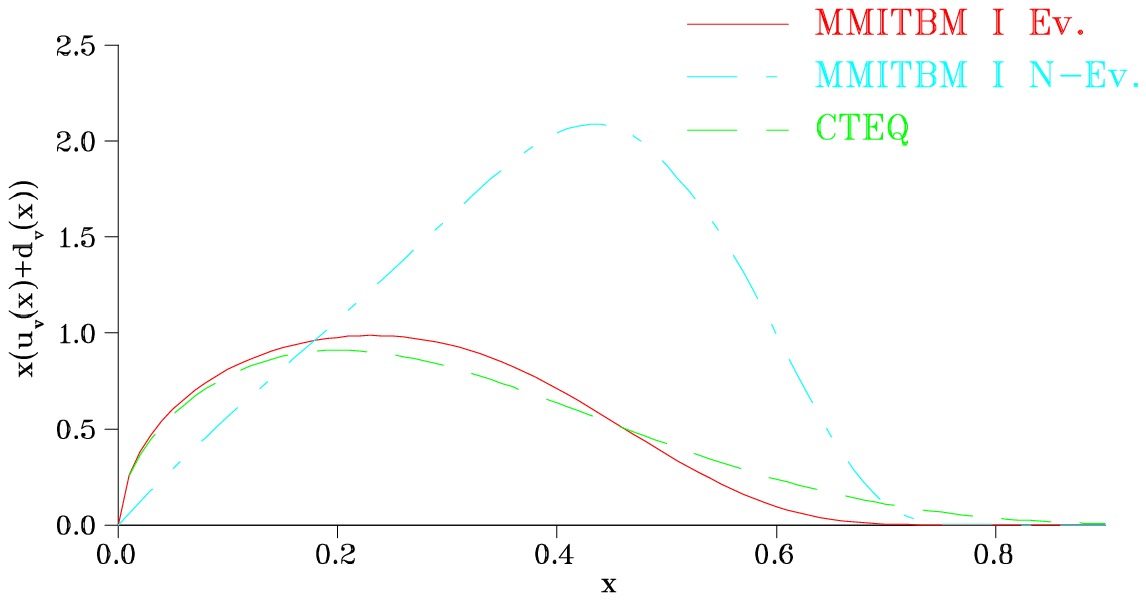}}
\caption{\label{f50} Comparison of the PDF combination  $x_B[u_v(x_B)+d_v(x_B)]$ obtained from the MMITBM with the CTEQ parameterization (dashed green line), before (dot-dashed  blue line) and after evolution (solid red line). The parameters are $R=1 $ fm, $M_2=700$ MeV, $M_4= 1111.6$ MeV and $\mu_{Bag}= .45 $ MeV}
\end{figure}

\begin{figure} 
\centerline{\psfig{figure=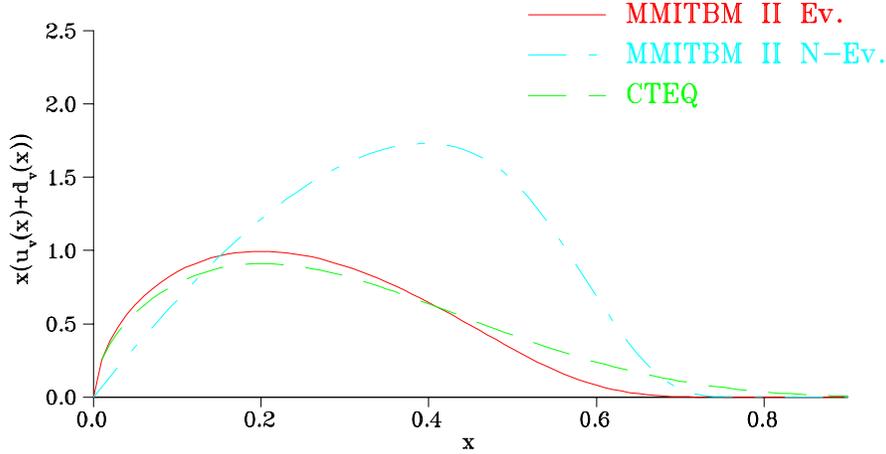}}
\caption{\label{f51} Comparison of the PDF combination  $x_B[u_v(x_B)+d_v(x_B)]$ obtained from the MMITBM with the CTEQ parameterization (dashed green line), before (dot-dashed blue line) and after evolution (solid red line). The parameters are $R=.9 $ fm, $M_2=750 $ MeV, $M_4=1032.9
$  MeV and $\mu_{Bag}= .48$ MeV}
\end{figure}

The valence distributions $q_v$ have to satisfy the conditions
\bga\label{e166}
\int dx_B {u_v(x_B)\choose d_v(x_B)}={2\choose 1}\,,
\ea  
where 
\bga
q_v(x_B)=f_1^q(x_B)-f_1^{\bar{q}}(x_B)\, .
\ea
In Ref.~\cite{thomas1} the normalization condition cannot be satisfied. The contribution stemming from the four fermion intermediate state   counts only for a nearly 2\% of the normalization, while the contribution from the two fermion intermediate counts, at the most, for 82\%. 
In order to saturate the condition in Eq. (\ref{e166}) they added a term
 {\it ad hoc} of the form $(1-x_B)^7$.  With my normalization, I am  always able,  by varying the parameters, namely the radius $R$, the two masses $M_2$ and $M_4$ of the two and four quark intermediate state, and the bag scale $\mu_{Bag}$ (see Sec. \ref{sec:twofermion} for more details on how to fix this last parameter) in reasonable range of values, to fulfill  the normalization condition, and obtaining a valence  distribution that resembles the one extracted from data (see Figs. \ref{f50} and \ref{f51}).

One of the problems of this model is that it yields a $u$ quark valence distribution such that $u_v=2d_v$, which is known not to be the case. The model could be improved to accommodate the different $x_B$ dependence in the two distributions by    assuming that the two quark intermediate state could form either a spin singlet or a spin triplet state with different masses~\cite{close}.  I am not trying to reproduce perfectly the twist two PDFs, but instead to estimate the twist four contributions; therefore, for my purpose, I  consider the simplified model.  In Fig. \ref{f50}  and \ref{f51} I have plotted the distribution function $x_B[u_v(x_B)+d_v(x_B)]$ obtained from the MMITBM before and after evolution, and compared it with the CTEQ parametrization for two  different sets of parameters. In Fig. (\ref{f50}) the parameters are: 
\bga \label{e180}
R=1  \,\textrm{fm}\,\,\,\, M_2=700\, \textrm{ MeV}\,\,\,\, M_4= 1111.6 \,\textrm{MeV}\,\,\,\, \mu_{Bag}= .45 \,\textrm{ MeV}\,.  
\ea
In Fig. (\ref{f51}) the parameters are:
\bga \label{e181}
R=0.9 \, \textrm{fm}\,\,\,\, M_2=750\, \textrm{ MeV}\,\,\,\, M_4= 1032.9\, \textrm{MeV}\,\,\,\, \mu_{Bag}= .48 \,\textrm{ MeV}\,.  
\ea
Both sets of parameters are such that the normalization condition (\ref{e166}) is satisfied and, as it can be seen from the plots, they both describe the data quite well.
\subsection{Twist Four Contributions to the Structure Functions in the Modified MIT Bag Model}
\label{sec:tw4mmitbm}
\begin{figure} 
\centerline{\psfig{figure=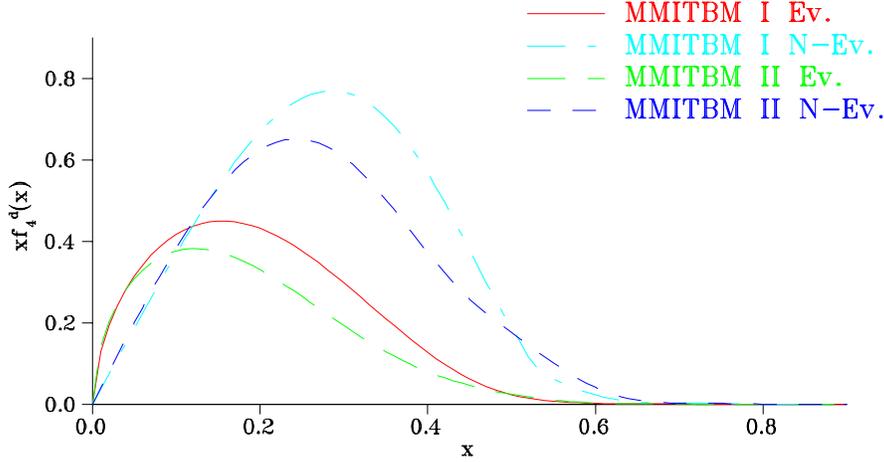}}
\caption{\label{f52} Correlation function $x_B f_4^d(x_B)$ obtained from the MMITBM using the two set of parmaters in Eqs. (\ref{e180}) and (\ref{e181}) before and after evolution. }
\end{figure}

Having fixed the parameters of my model, I can now proceed in estimating the twist four contribution to the SFs. The approach I follow has been described in Sec. \ref{sec:twofermion_tw4cf}. I  compute the twist four matrix element
\bga
f_4^q(x_B)=M  \int \frac{d\lambda}{2 \pi}e^{i\lambda x_B}\la P |\psi^{\dag q}_-(0) \psi^q_- (\lambda n) | P \ra\,,
\ea
which is related to $T_1(x_B)$ by Eq. (\ref{ea45}). 
As before,  since the matrix element $T_2(x_1,x_2)$ cannot be directly computed, I  estimate it by making the ansatz 
\bga 
 T_2^q(x_1,x_2)=4 \delta(x_1-x_2) T_1^q(x_1)\,,
\ea
and  I  evolve the twist four function $x_B f_4^q(x_B)$ using the DGLAP equation  (refer to Sec. \ref{sec:twofermion_tw4cf} for details). Following the same steps that lead to Eq. (e161), and using the relation
\bga
\psi^{(q)}_+(\mathbf{p}_n)=\psi^{(q)}_-(-\mathbf{p}_n)\,,
\ea
one has
\bga 
f_4^d(x_B)&=&2M^3\sum_m \la \alpha |P_{d,m}^2| \alpha \ra \int\frac{d\mathbf{p_2}}{(2\pi)^3}\delta(M(1-x_B)-p_2^+)\left| \frac{\phi_2(p_2)}{\phi_3(0)}\right|^2 |\tilde{\psi}_-^{(d)}(\mathbf{p}_2)|^2\nn \\
&+&2M^3\sum_m \la \alpha |P_{d,m}^4| \alpha \ra \int\frac{d\mathbf{p_4}}{(2\pi)^3}\delta(M(1-x_B)-p_4^+)\left| \frac{\phi_3(0)}{\phi_4(p_4)}\right|^2 |\tilde{\psi}_-^{(\bar{d})}(\mathbf{p}_4)|^2\nn \\
\ea
and
\bga 
f_4^{\bar{d}}(x_B)&=& 2M^3\sum_m \la \alpha |P_{\bar{d},m}^4| \alpha \ra \int\frac{d\mathbf{p_4}}{(2\pi)^3}\delta(M(1-x_B)-p_4^+)\left| \frac{\phi_3(0)}{\phi_4(p_4)}\right|^2 |\tilde{\psi}_-^{\dag( d)}(\mathbf{p}_4)|^2\, .\nn\\
\ea
In Fig. (\ref{f52}) I have plotted the valence distribution function $x_Bd_v^{HT}(x_B)$ where 
\bga
q_v^{HT}(x_B) = f_4^q(x_B)-f_4^{\bar{q}}(x_B)
\ea
before and after evolution for the two different sets of parameters $\{R,M_2,M_4,\mu_{Bag}\}$. As it can be seen, shape and order of magnitude are similar to the ones of the twist two correlation functions, suggesting  that the HT might play an important role in processes involving a low  momentum transfer, where the scaling factor $\frac{M^2}{Q^2}$ is of the order of unity.

\begin{figure} 
\centerline{\psfig{figure=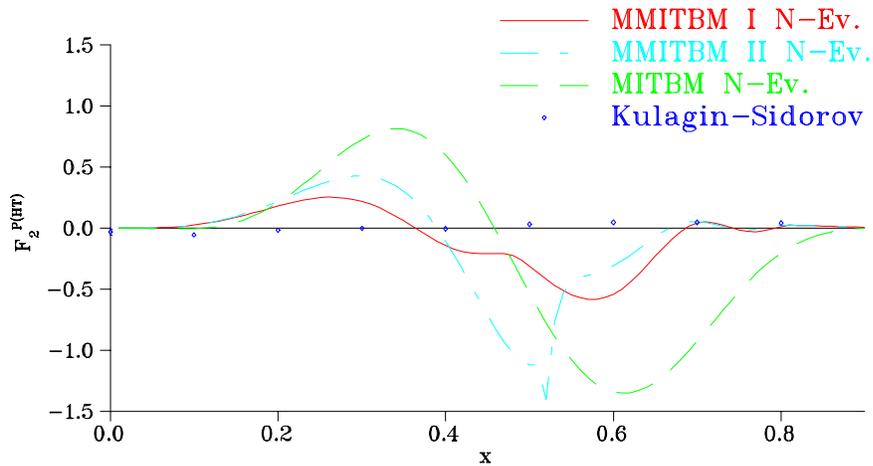}}
\caption{\label{f53} Twist four contribution to the $F_2$ SF obtained from the MITBM (dashed green line) and the MMITBM (solid red line and dot-dashed blue line) with no evolution, compared to the model indepenent extraction of ~\protect\cite{alekhin} .}
\end{figure}

\begin{figure} 
\centerline{\psfig{figure=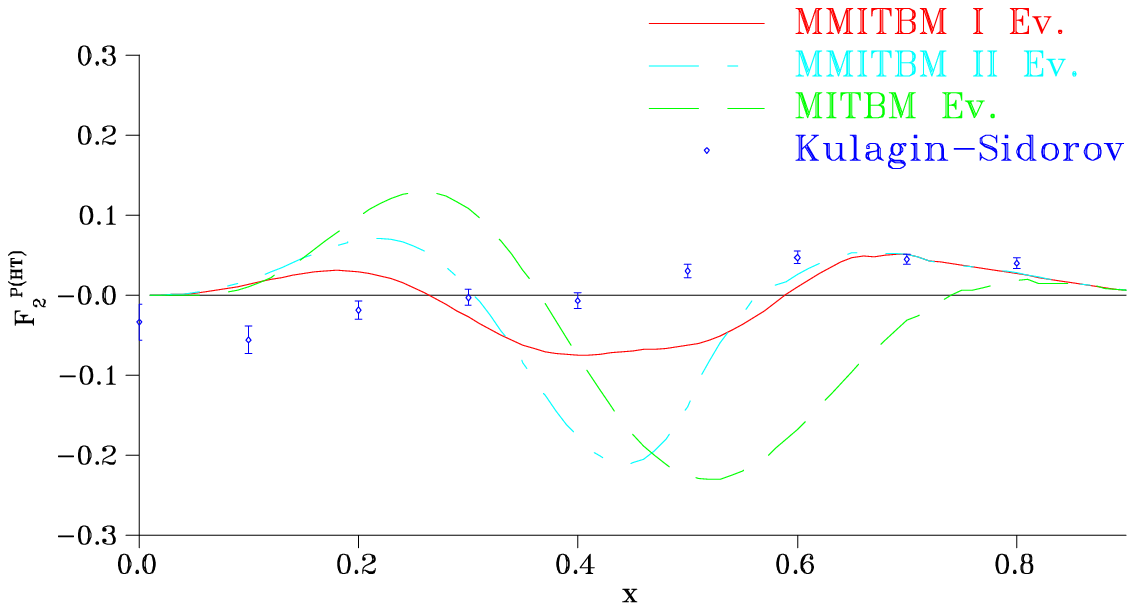}}
\caption{\label{f54} Twist four contribution to the $F_2$ SF obtained from the MITBM (dashed green line) and the MMITBM  (solid red line and dot-dashed blue line) with  evolution, compared to the model indepenent extraction of ~\protect\cite{alekhin}.  }
\end{figure}

\begin{figure} 
\centerline{\psfig{figure=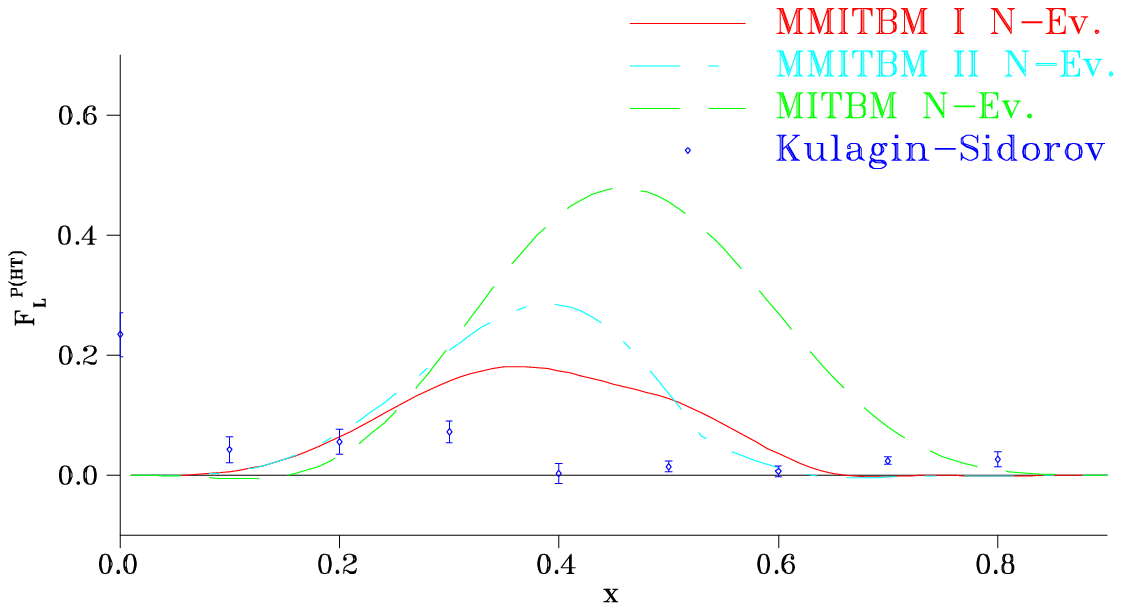}}
\caption{\label{f55} Twist four contribution to the $F_L$ SF obtained from the MITBM  (dashed green line) and the MMITBM  (solid red line and dot-dashed blue line) with no evolution, compared to the model indepenent extraction of ~\protect\cite{alekhin}.  }
\end{figure}

\begin{figure} 
\centerline{\psfig{figure=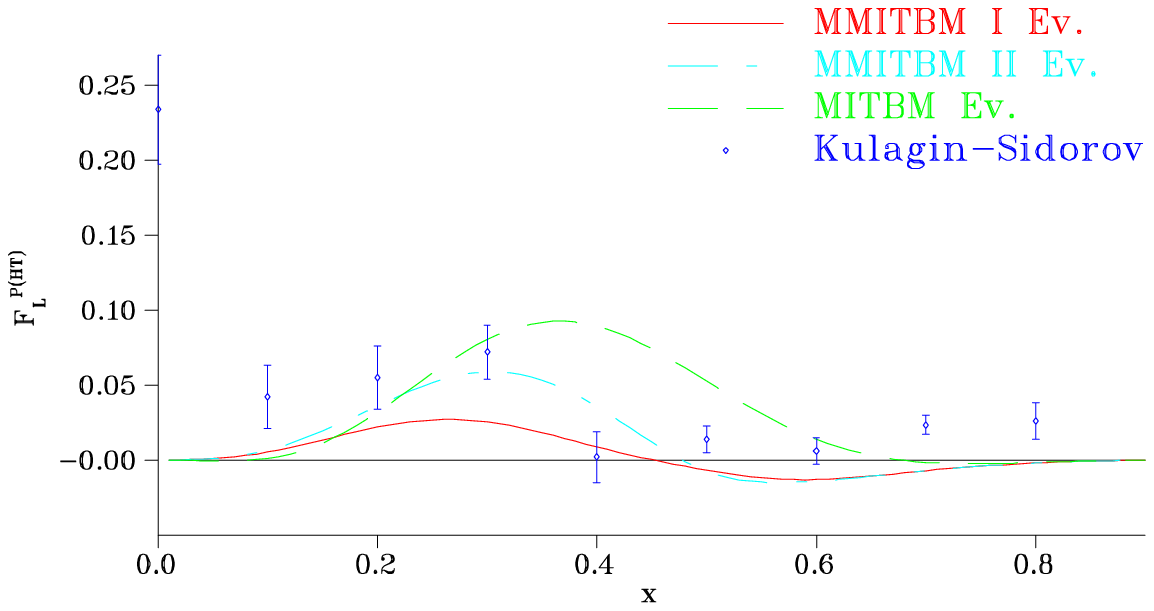}}
\caption{\label{f56} Twist four contribution to the $F_L$ SF obtained from the MITBM  (dashed green line) and the MMITBM  (solid red line and dot-dashed blue line) with  evolution, compared to the model indepenent extraction of ~\protect\cite{alekhin}.  }
\end{figure}

\begin{figure} 
\centerline{\psfig{figure=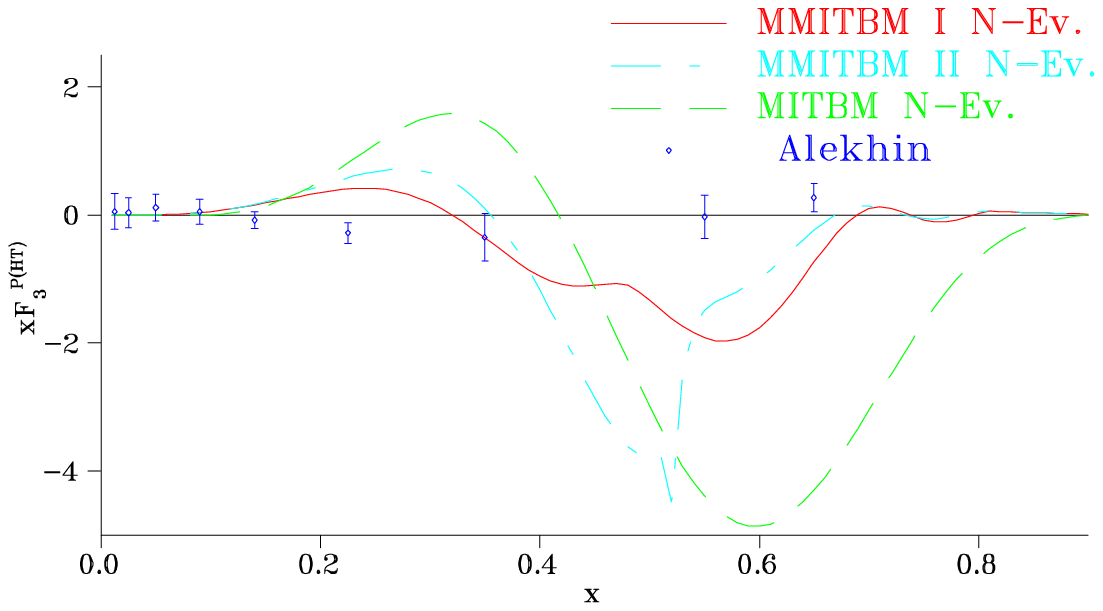}}
\caption{\label{f57} Twist four contribution to the $x_BF_3$ SF obtained from the MITBM  (dashed green line) and the MMITBM  (solid red line and dot-dashed blue line) with no evolution, compared to the model indepenent extraction of \protect \cite{sidorov}.  }
\end{figure}

\begin{figure} 
\centerline{\psfig{figure=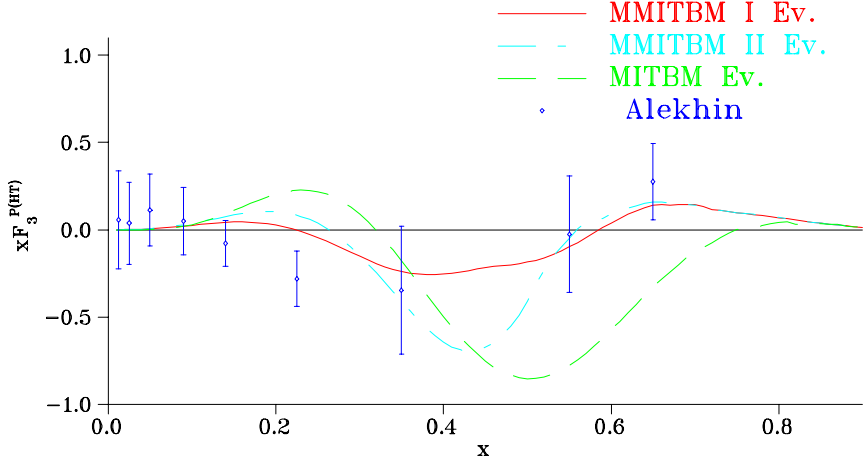}}
\caption{\label{f58} Twist four contribution to the $x_B F_3$ SF obtained from the MITBM  (dashed green line) and the MMITBM (solid red line and dot-dashed blue line)  with  evolution, compared to the model indepenent extraction of \protect \cite{sidorov}.  }
\end{figure}

I now proceed as in Sec. \ref{sec:twofermion_tw4cf} to obtain the twist four corrections to the different SFs. In Figs. \ref{f53}-\ref{f58} I plot the HT contribution to SFs $F_2$, $F_L$ and $x_B F_3$ obtained with the two models I used, namely the MITBM and the MMITBM, with some model independent extraction  from data. In Figs. (\ref{f53}-\ref{f55}) the SFs are computed by using the  non evolved  correlation function $x_B f_4^d(x_B)$, while, in Figs. (\ref{f56}-\ref{f58}) it has been evolved. I want to emphasize the fact that the HT have never  been directly measured. Normally, what is done, is to introduce a new term in the fitting function for the SFs of the type
\bga
F_i(x_B,Q^2)=F_i^{LTMC}(x_B,Q^2)+\frac{H_i^{HT}(x_B)}{Q^2}
\ea
where $F_i^{LTMC}$ is the leading twist (LT) SF including the target mass corrections, and $H_i^{HT}(x_B)$ represents the HT corrections. It is not easy to judge how reliable these extractions could be. First of all the function $H^{HT}(x_B)$ is taken to be $Q^2$ independent, assuming that the scaling as $\frac{1}{Q^2}$ dominates with respect to possible $\log Q^2$ dependence arising from PQCD corrections. Moreover it is not always straightforward to discern between the $\frac{1}{Q^2}$ dependence of the twist four and the $\log Q^2$  dependence arising from the higher order PQCD corrections to the LT, meaning that sometimes, what is thought to be HT contribution, could partially be  accounted for, once  higher order PQCD corrections are taken into account~\cite{sidorov}. Nonetheless, this is the only attempt done to extract HT from experiment, and the only way I have to  compare my results with some model independent extractions. In the next section I  compute the HT corrections to the PV asymmetry for a deuteron target,   using the results obtained in this section.
%%%%%%%%%%%%%%%%%%%%%%%%%% figures
%%%%%%%%%%%%%%%%%%%%%%%%%%%%%%%%%%%%%%%%%%%%%%%%%%%%%%%%%%5
\newpage\newpage\newpage
\section{Higher Twist Corrections to $A_{RL}$ for Deuteron Target }\label{sec:final_arl}
Similarly to what has been done in Sec. \ref{sec:finalarllt}, in this section I  consider the corrections to the LT asymmetry due the the HT contributions for the kinematics proposed in refs.~\cite{bosted1,bosted}. For simplicity, I computed the LT asymmetry using only the CTEQ parameterization. First, I consider the HT contributions from the  two quark twist four correlation functions. In all of the following figures, what is plotted is the quantity
\bga\label{eab2}
\frac{\delta A_{RL}}{A_{RL}}\equiv 2\left| \frac{A_{RL}^{LT}-A_{RL}^{HT}}{A_{RL}^{LT}+A_{RL}^{HT}}\right|\, ,
\ea
where $A_{RL}^{LT}$ is the LT asymmetry, whereas the $A_{RL}^{HT}$ corresponds to the asymmetry in which HT have been included, and the factor of two comes from averaging the two asymmetries. 
%%%%%%%%%%%%%%%%%%%%%%%%%%%%%%% figure
\begin{figure} 
\centerline{\psfig{figure=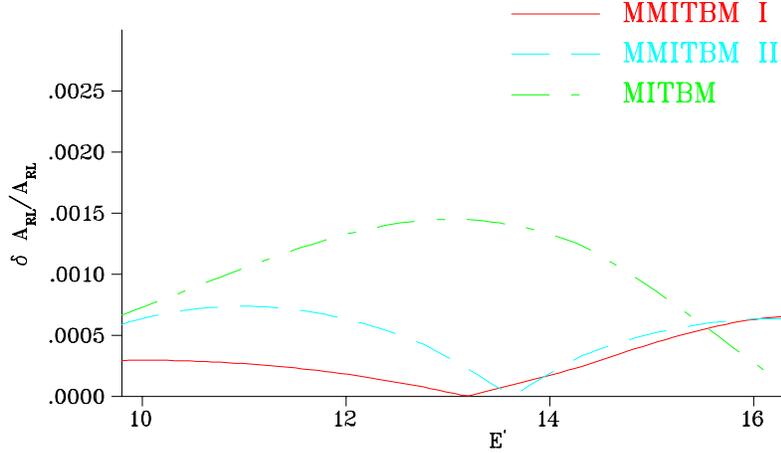}}
\caption{\label{f204} Relative uncertainty in the asymmetry defined in Eq. (\ref{eab2}) due to the HT two fermion contributions for the kinematics proposed in Ref.~\protect \cite{bosted1}. The red solid line is relative to the HT contributions obtained in the MMITBM using the set 1 of parameters, the blue dashed line using the set 2 of parameters, and the green dot-dashed line is obtained from the MITBM.  }
\end{figure}
%%%%%%%%%%%%%%%%%%%%%%%%%%%%%%%%%
I start by  considering the kinematical region proposed in Ref. ~\cite{bosted1}, which consists in a incoming beam energy $E=37$ GeV, a scattering angle of twelve degrees, and an energy of the outgoing electrons ranging between 9.8 to 17.2 GeV. In Fig. \ref{f204} I have plotted the relative uncertainty  in the asymmetry for the three cases in which  the HT were estimated by using the MITBM and the MMITBM (I considered the contributions relative to both sets of parameters, the first set  (set 1 henceforth)   $R=1 $ fm, $M_2=700$ MeV, $M_4= 1111.6$ MeV and  $\mu_{Bag}= 0.45 $ MeV and the second set (set 2) $R=0.9 $ fm, $M_2=750 $ MeV, $M_4=1032.9
$  MeV and $\mu_{Bag}= 0.48$ MeV ) and evolved according to  the DGLAP equations. It can be seen that for all three cases the relative magnitude of the HT effects is, at the most 0.15$\%$. In Fig. \ref{f205}, I have instead plotted the relative asymmetry obtained with the two MMITBMs in the case in which the SFs have not been evolved. In this case the maximum value of the uncertainty is a bit more than 0.6$\%$.
%%%%%%%%%%%%%%%%%%%%%%%%%%%% figure
\begin{figure} 
\centerline{\psfig{figure=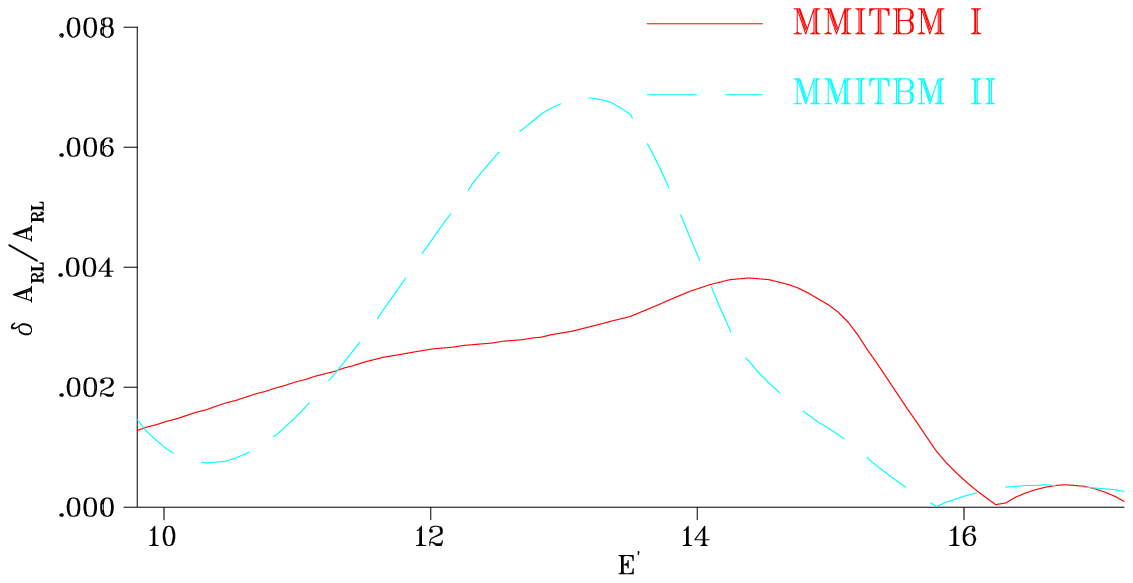}}
\caption{\label{f205} Relative uncertainty in the asymmetry due to the HT two fermion contributions for the kinematics proposed in Ref.~\protect \cite{bosted1}. The red solid line is relative to the HT contributions obtained in the MMITBM using the set 1 of parameters, the blue dashed line using the set 2 of parameters.  }
\end{figure}
%%%%%%%%%%%%%%%%%%%%%%%%%%%%%%%%%%%%%%%5

\begin{figure} 
\centerline{\psfig{figure=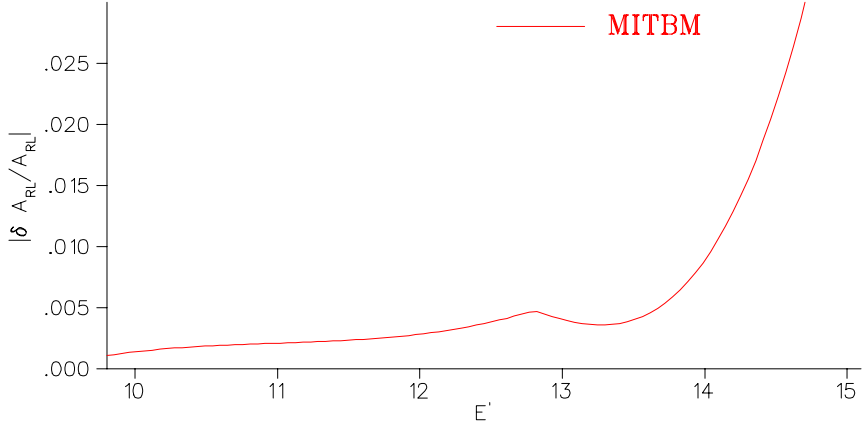}}
\caption{\label{f206} Relative uncertainty in the asymmetry due to the HT two fermion contributions computed with the MITBM for the kinematics proposed in Ref.~\protect \cite{bosted1} .}
\end{figure}

\begin{figure} 
\centerline{\psfig{figure=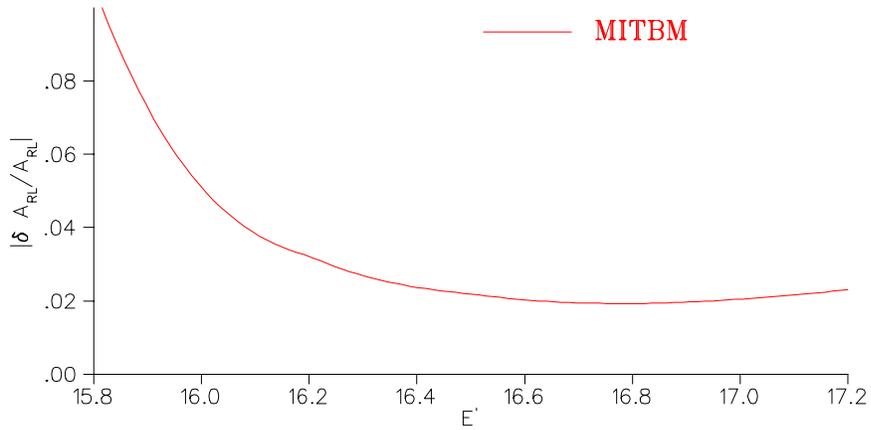}}
\caption{\label{f207} Relative uncertainty in the asymmetry defined in Eq. (\ref{eab2}) due to the HT two fermion contributions for the kinematics proposed in Ref.~\protect \cite{bosted}. The red solid line is relative to the HT contributions obtained in the MMITBM using the set 1 of parameters, the blue dashed line using the set 2 of parameters, and the green dot-dashed line is obtained from the MITBM.   }
\end{figure}

Next I consider the  relative asymmetry obtained with the non evolved MITBM. I split the energy range in two parts, which are considered in Figs. \ref{f206} and \ref{f207}. As it can be seen, the  corrected asymmetry  becomes unexpectedly large  around $E'=15$ GeV. This situation is actually unphysical, since this singular behavior is due to the fact that the electromagnetic cross section, which appears in the denominator on the asymmetry, becomes negative around $E'=15.5$.

The main cause for the cross section to become negative is the possibility for the HT contribution to the different SFs to be negative, in conjunction with the relatively large size of the HT contribution  arising from my model, when evolution is not applied. This fact might suggest that my estimate of the HT contribution are too large when they are not evolved, and, therefore, that evolution could be a determinant ingredient in the study of HT effects. Another possibility is that twist higher than four could become important and, if added, they might prevent the cross section from becoming negative.  
%%%%%%%%%%%%%%% figure
%%%%%%%%%%%%%%%%%%%% 

%%%%%%%%%%%%%%%%%%%%%%%%%%%% figure

%%%%%%%%%%%%%%%%%%%%%%%%%%%%%%%%%%%%%%%5

I now consider the corrections in $A_{RL}$ due to the two fermion twist four contributions at the   kinematics suggested in Ref.~\cite{bosted}. As before, even though the proposed experiment consists in only one measurement for a   beam energy of $E=11$ GeV, a scattering angle of twelve and a half degree, and an outgoing electron energy of $5.5$ GeV, I actually  consider a wider range of outgoing electron energies, so that one could see at which kinematics the theoretical uncertainties due to the HT effects, could be minimized. The range is from 5.5  to 8.0 GeV.

In this case, the enhancement in the asymmetry due to the smallness of the EM cross section, is present in all three cases in which the HT contribution have not been evolved, since the $Q^2$ is much smaller than in the previous case. 

\begin{figure} 
\centerline{\psfig{figure=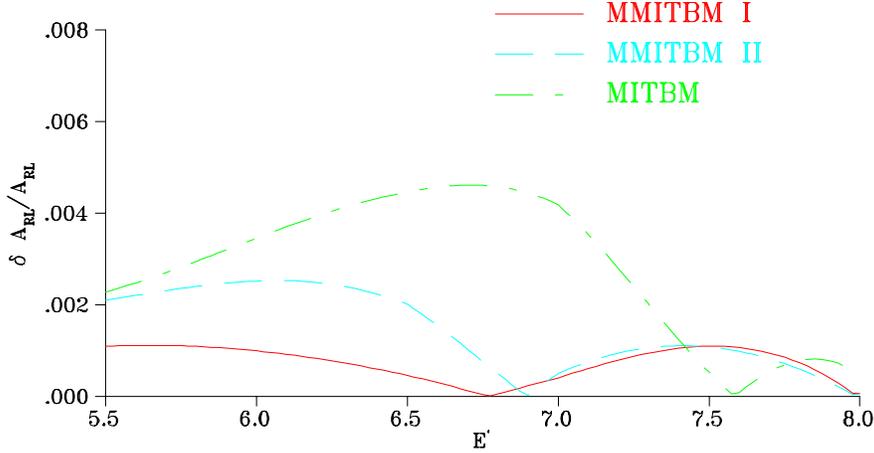}}
\caption{\label{f208} Relative uncertainty in the asymmetry due to the HT two fermion contributions for the kinematics proposed in Ref.~\protect \cite{bosted}.  }
\end{figure}

\begin{figure} 
\centerline{\psfig{figure=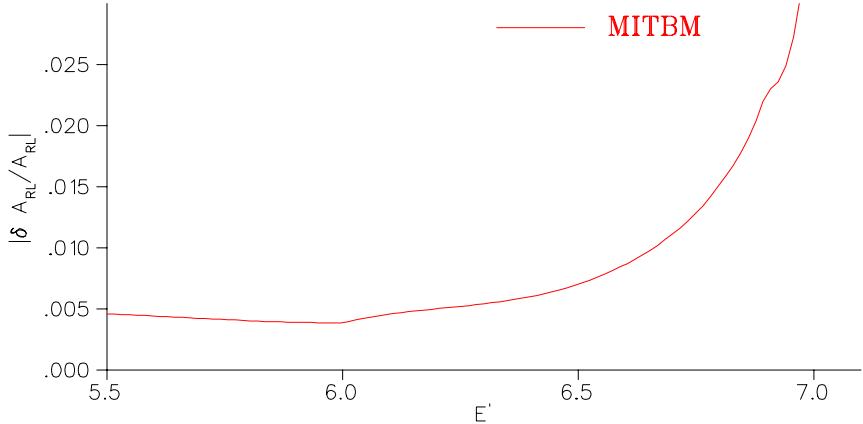}}
\caption{\label{f209} Relative uncertainty in the asymmetry due to the HT two fermion contributions computed with the MITBM for the kinematics proposed in Ref.~\protect \cite{bosted} .}
\end{figure}

\begin{figure} 
\centerline{\psfig{figure=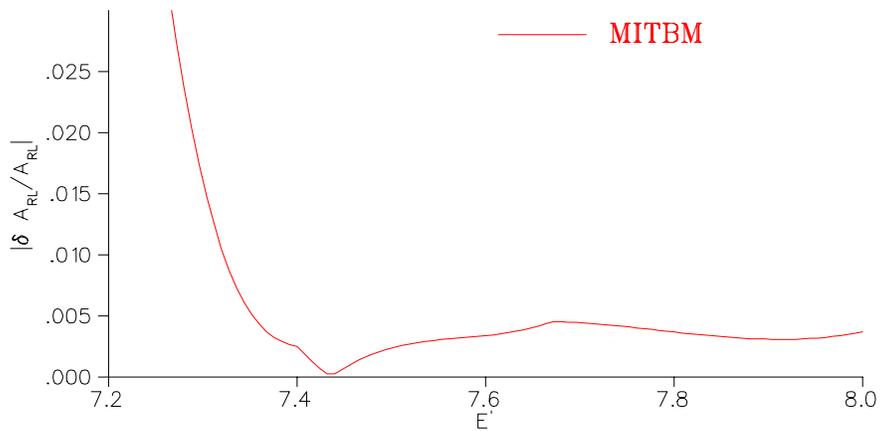}}
\caption{\label{f210} Relative uncertainty in the asymmetry due to the HT contributions computed with the MITBM for the kinematics proposed in Ref.~\protect\cite{bosted}.   }
\end{figure}

\begin{figure} 
\centerline{\psfig{figure=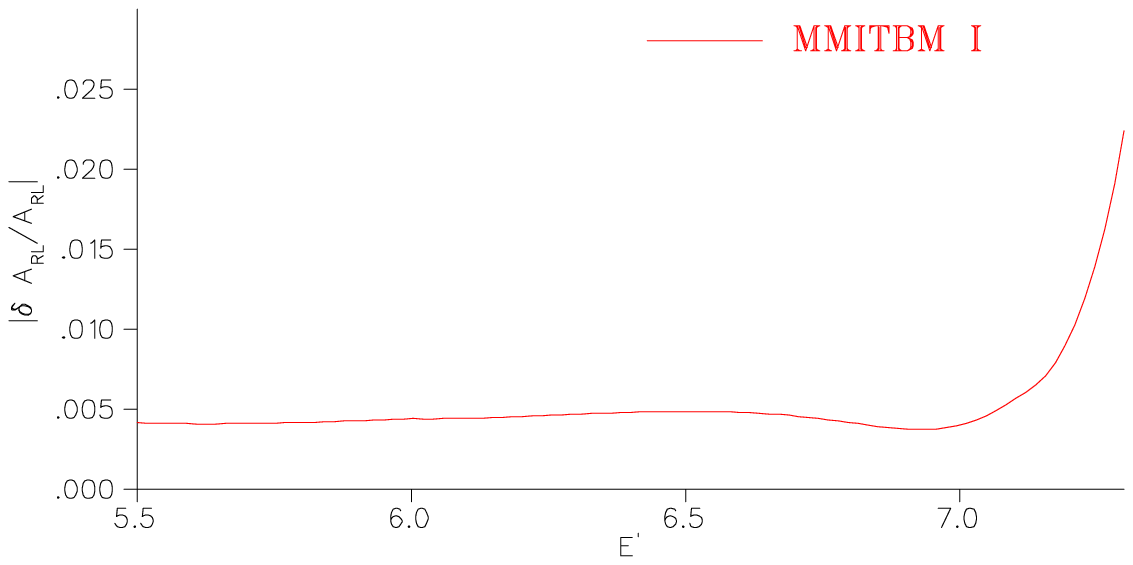}}
\caption{\label{f211} Relative uncertainty in the asymmetry due to the HT two fermion contributions computed with the MMITBM set 1 for the kinematics proposed in Ref.~\protect \cite{bosted} .}
\end{figure}

\begin{figure} 
\centerline{\psfig{figure=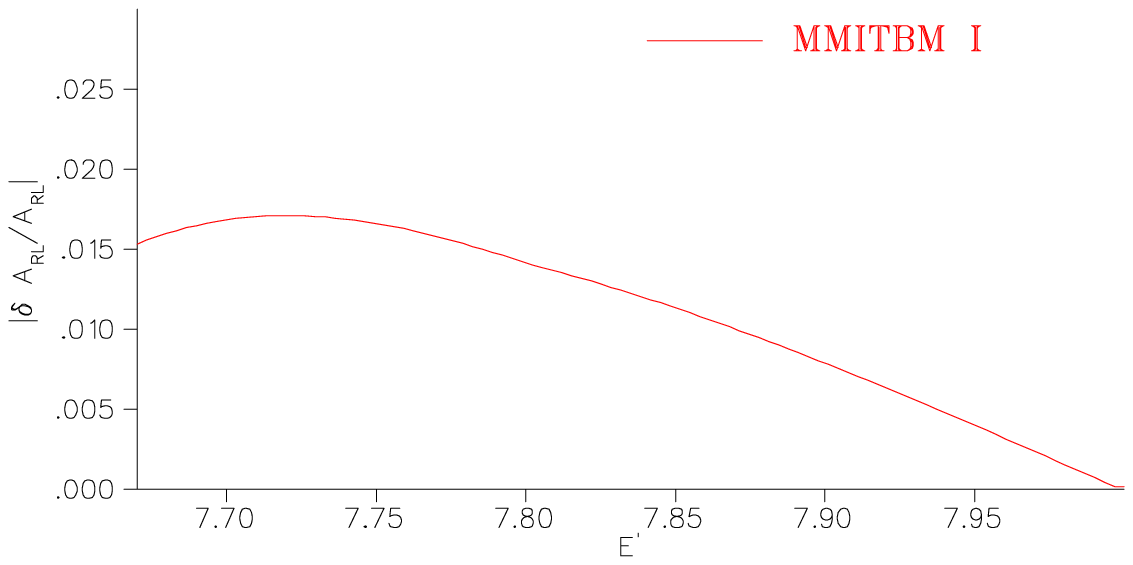}}
\caption{\label{f212} Relative uncertainty in the asymmetry due to the HT two fermion contributions computed with the MMITBM set 1 for the kinematics proposed in Ref.~\protect\cite{bosted}.   }
\end{figure}

\begin{figure} 
\centerline{\psfig{figure=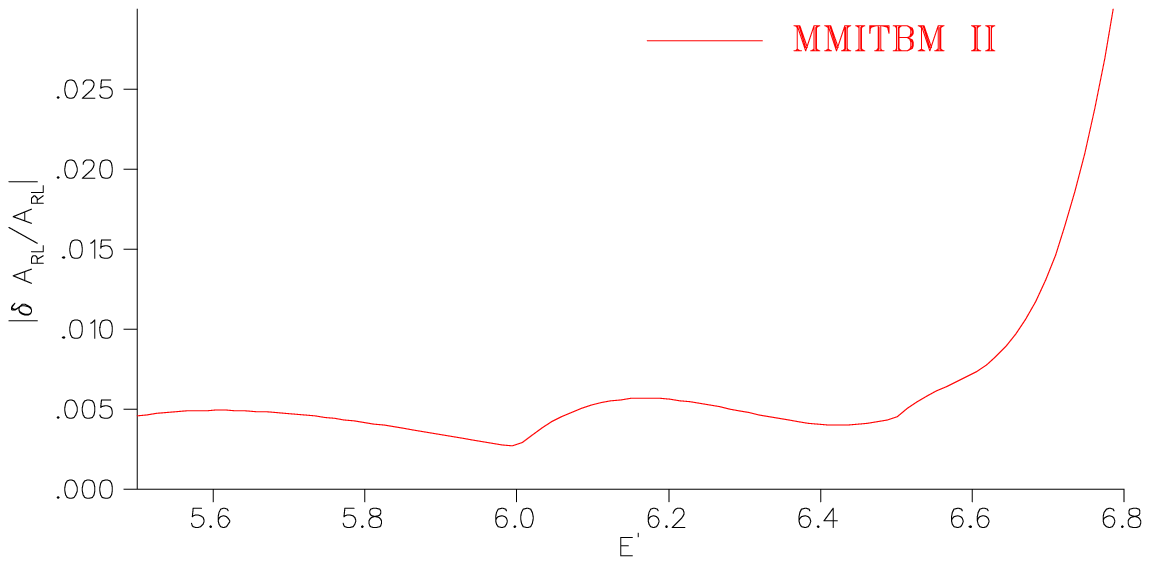}}
\caption{\label{f213} Relative uncertainty in the asymmetry due to the HT two fermion contributions computed with the MMITBM  set  2 for the kinematics proposed in Ref.~\protect \cite{bosted} .}
\end{figure}

\begin{figure} 
\centerline{\psfig{figure=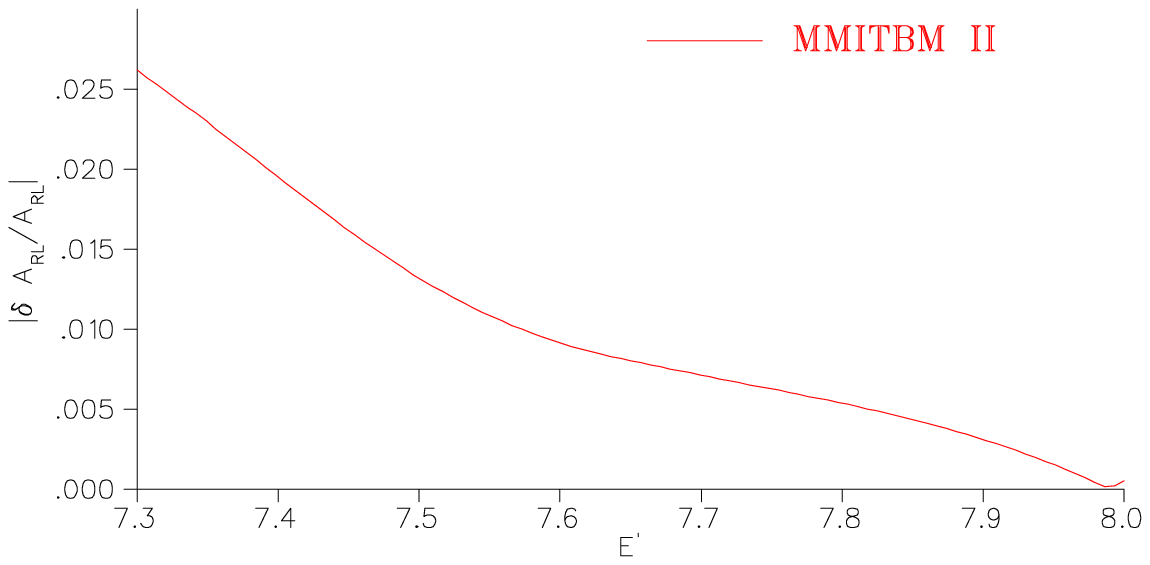}}
\caption{\label{f214} Relative uncertainty in the asymmetry due to the HT two fermion contributions computed with the MMITBM set 2 for the kinematics proposed in Ref.~\protect\cite{bosted}.   }
\end{figure}

In Fig. \ref{f208} I have plotted the relative asymmetry for the same three cases as in Fig. \ref{f204}, except that now the kinematics is the one in Ref.~\cite{bosted}. 
Also in this case the relative uncertainty is smaller than the prescribed accuracy, being always less than 0.5\% . 
In Figs. \ref{f209}-\ref{f214} I have plotted the relative asymmetry in  the case in which the HT have not been evolved. As can be seen in all three models there exists a point in which the EM cross section vanishes, causing the asymmetry to diverge.

  As mentioned before the reason for the asymmetry to be divergent is that the EM cross section can become zero, if the HT could be large in size and negative. Of course the  cross section cannot be negative, nevertheless, it might happen that it becomes extremely small, causing the asymmetry to be large. Let us consider the  HT contributions obtained  with the evolved MITBM,  which, as  can be  seen from Figs.  \ref{f56} and \ref{f58}, it  seems to reproduce the same order of magnitude of the HT effects  found in  Ref.~\cite{alekhin,sidorov}. I tried to multiplying all the HT corrections in $A_{RL}^{HT}$  by a factor of two, and I notice that the difference between the LT and the HT asymmetry could be easily larger than  1.5\%, 
reaching a difference of 20\%
if the HT SF are scaled by a factor of 2.8. In this case the electromagnetic cross section does not become negative, but simply small enough so that  the HT asymmetry becomes large. I emphasize that all I have done, is to take my model estimate of the HT contributions, which seems to agree with some model independent extraction, and scale it by a factor between 2 and 2.8.

Notice that such enhancement in $A_{RL}$, cannot happen in  other models such as the one used in Ref.~\cite{casto,casto1}, where  the HT contributions are estimated by computing the first moment HT correction to the different SFs, and where the $x_B$ dependence is assumed to be of the same type of the twist two SFs.  For instance, the correction $F^{HT}_{2\, corr.}$ to the $F_2$ SF would be given by 
\bga
F^{HT}_{2\, corr.}(x_B)=\left[\int_0^1 dx F_2^{HT}(x)\right]x_B(u_v(x_B)+d_v(x_B))\, ,
\ea  
where the first moment of $F_2$,
\bga 
\int_0^1 dx F_2^{HT}(x)\, ,
\ea
is computed in the MITBM. For such a form of the HT, since all the SFs in the asymmetry formula in Eq. (\ref{ea49}), including the HT corrections, will have the same $x_B$ dependence,  such a  dependence will cancel out in the asymmetry, yielding a $x_B$ independent shift in the asymmetry.  

  It seems therefore clear that, in order to perform high precision  PV DIS experiments, it is absolutely necessary to have a very good knowledge, not only in the magnitude, but also in the sign and $x_B$ dependence of the HT, which is the main cause for the electromagnetic cross section to become small and give rise to an enhancement in $A_{RL}$.

%%%%%%%%%%%%%%%%%%%%%%%%%%%% figure
%%%%%%%%%%%%%%%%%%%%%%%%%%%%%%%%%%%%%%%5

\begin{figure} 
\centerline{\psfig{figure=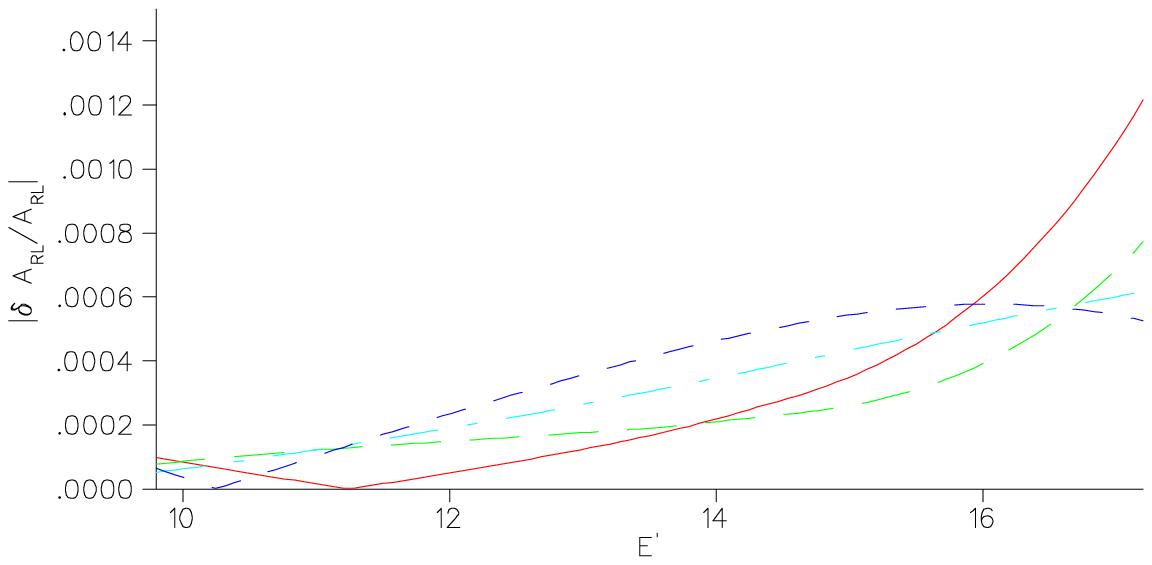}}
\caption{\label{f215} Relative uncertainty in the asymmetry due to the HT four fermion contributions computed with the MITBM  for the kinematics proposed in Ref.~\protect \cite{bosted1} .}
\end{figure}

\begin{figure} 
\centerline{\psfig{figure=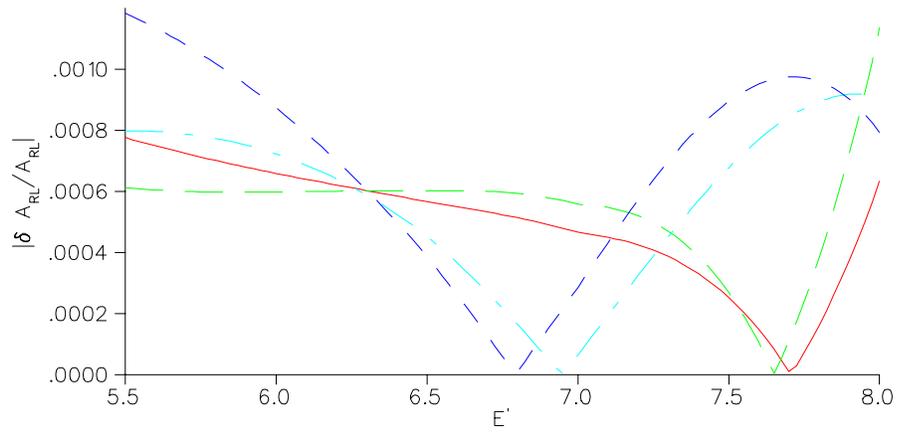}}
\caption{\label{f216} Relative uncertainty in the asymmetry due to the HT four fermion contributions computed with the  for the kinematics proposed in Ref.~\protect\cite{bosted}.   }
\end{figure}

I next consider the relative uncertainty arising from the four fermion twist four correlation functions. The results for the kinematics of Ref.~\cite{bosted1} are presented in Fig. \ref{f215}. As  can be seen, for all the different set of parameters I use to compute the IMT, the relative uncertainty is of the order of 0.1\%. 
This result could be considered as an upper limit, since evolution, which decreases the magnitude of the SFs, has not been applied. Moreover I have taken the strong coupling to be equal one, while in reality it should be less than this value, and decreasing with increasing $Q^2$.  In Fig. \ref{f216} I have plotted the relative uncertainty  for the four fermion twist four correlation functions for the kinematics proposed in  Ref.~\cite{bosted1}, and also in this case the uncertainty is, at the most, of the order of 0.1\%. 

According to my model it seems that the normal belief that the four fermion  HT correlation functions should be negligible is confirmed. 
 %%%%%%%%%%%%%%%%%%%%%%%%%%%% figure
%%%%%%%%%%%%%%%%%%%%%%%%%%%%%%%%%%%%%%%5

%\input{ch3.tex}
\chapter{Bremsstrahlung Contribution to the Resonance Electroproduction Cross Section and $A_{RL}$}\label{ch:e158}
In this chapter I  present a complete calculation of the bremsstrahlung contribution to the cross section and the PV asymmetry $A_{RL}$ in the process $e^-\, P\rightarrow N^* \, e^-$, where $N^*$ represents a resonance. This process could be a major source of background in the  experiment E-158~\cite{slac}, recently completed at the Stanford Linear Accelerator Center (SLAC). In this experiment the PV asymmetry $A_{RL}$ in the M{\o}ller scattering process $e^-\, e^- \rightarrow e^-\, e^-$ is measured, and the  weak minxing angle $\sin^2\theta_W$ is extracted. Since the electron beam scatters on atomic electrons in a hydrogen target, there are many backgrounds that need to be considered due to the scattering  of the electrons on the protons. Most of the backgrounds have been estimated in the proposal~\cite{slac}, but from preliminary results it was noticed that the inelastic background accounted for $40\%$ of the total asymmetry measured. The goal of the experiment was to measure the M{\o}ller asymmetry to a $8\%$ level, which requires a knowledge of the inelastic background  to a $20\%$ of accuracy or better. I  study one of the possible inelastic background, namely the bremsstrahlung contribution to the $A_{RL}$ in the $e^-\, P\rightarrow N^* \, e^-$ process. In order to do so, I  resort to a simple model in which only two resonances (the $\Delta(1232)$ and the $N(1520)$) contribute. I  investigate if the theoretical uncertainties on the resonances form factors, might produce  an uncertainty on the asymmetry that is of the order of 8\%.

 The  chapter is divided as follow: in Sec. \ref{sec:e-158} I  describe the experiment E-158, while in Sec. \ref{sec:brem} I  present the calculation for the polarized cross section for the bremsstrahlung contribution to the proton electro-excitation. In Sec. \ref{sec:infrared} I  focus on the computation of the one loop corrections to the process $e^-\, P\rightarrow e^-\, N^*$, which are necessary to make the bremsstrahlung process infrared finite. Finally, in Sec. \ref{sec:results} I  present the numerical results of the calculation and in Sec. \ref{sec:conclusion} the conclusions.
\section{The E-158 Experiment.}\label{sec:e-158}
The E-158 experiment~\cite{slac} performed at SLAC,  is part of an extensive program of high precision experiments, with the objective of searching for physics beyond the SM. Experiments such as the LEP I and II at CERN, and the SLD at SLAC, have measured neutral current observables with an extremely high precision at the $Z_0$ resonance energy region~\cite{delphi}. The E-158 experiment  tried to provide an  high precision measurement of an electroweak observable, such as $A_{RL}$, away from the $Z_0$ pole, in order to explore a different  kinematical region in which to look for physics beyond the SM.

\begin{figure}   
\centerline{\psfig{figure=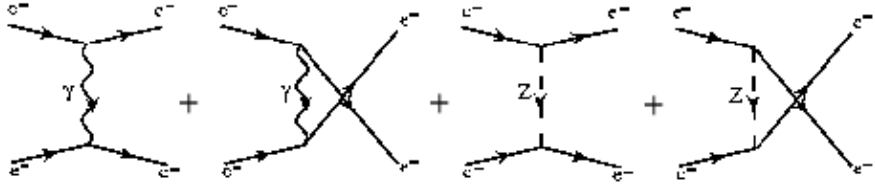}}
\caption{\label{f65} Neutral current amplitudes contributing to the leading order to the M{\o}ller cross section and $A_{RL}$.}
\end{figure}

\begin{figure}   
\centerline{\psfig{figure=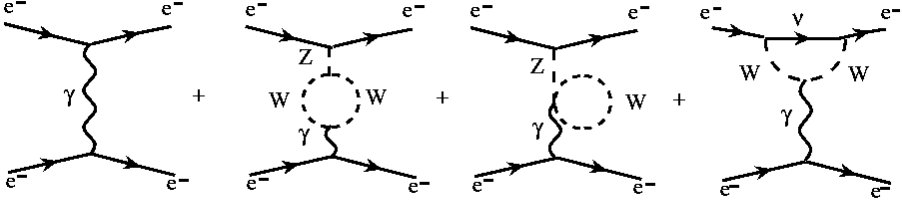}}
\caption{\label{f66} Dominant contributions of the  electroweak radiative correction to the polarized asymmetry $A_{RL}$.}
\end{figure}

At tree level, the processes contributing to $A_{RL}$ are depicted  in Fig. \ref{f65}, and the result is~\cite{mar,slac}
\bga
A_{RL}&=& -m E \frac{G_F}{\sqrt{2} \pi \alpha}\frac{16 \sin^2\Theta}{(3+ \cos^2 \Theta)^2}g_{ee}
\ea
where $G_F=1.16639(1)\times 10^{-5}$ GeV$^{-2}$, $m$ and $E$ are the electron mass and incoming energy, respectively, $\Theta$ is the scattering angle in the center of mass frame (CMF), and $g_{ee}$ is the pseudo-scalar weak neutral current coupling, that at tree level reads
\bga
g_{ee}\equiv -\frac{1}{4}g_V^eg_A^e=\frac{1}{4} -\sin^2\theta_W\, .
\ea 
Radiative corrections reduce the magnitude of the tree level value of $g_{ee}$ up to $40\%$~\cite{mar}. Some of the processes contributing to the radiative corrections are shown in Fig. \ref{f66}. 
In the experiment, electrons with scattering angle in the CMF such that $-0.5\le \cos \Theta\le 0$ are detected. For an electron beam of $50$ GeV this translates in the detection of electrons with energy between $12.5$ and $25$ GeV. All the events are collected in a cylindrical  detector, which is made up of concentric rings. The M{\o}ller detector, in which most of the M{\o}ller events are detected, consists in three rings: the IN, the MID and the OUT ring. Concentric to the M{\o}ller detector there is the $DEAD$ ring, whose only goal is to separate the M{\o}ller  ring from the outermost  EP ring. In this last ring,  most of the elastic and inelastic events,  from the scattering of the electrons of the protons, are collected. 

Because  the electrons in the target  are atomic electrons in hydrogen atoms, the presence of the protons  is source of a number of backgrounds which need to be considered. Most of them were estimated in the experiment proposal, but preliminary results from the experiment showed that the inelastic background was as big as  $40\%$ of the total asymmetry. Since the stated goal of the experiment was the measurement of $A_{RL}$ to a $8\%$ level of accuracy, the inelastic backgrounds need to be known to a $20\%$ level or better. It was speculated that the bremsstrahlung process, in which a hard photon is emitted after the scattering with the protons in the hydrogen atoms,  could   be a major source of the inelastic backgrounds. I  divide this background in two major contributions: {\it (i)} the bremsstrahlung process in which the electron emits an hard real photon either before or after the scattering with the proton which gets excited  into one of its resonances (Fig. \ref{f67}), and {\it (ii)} the  bremsstrahlung process in which the scattering with the proton is deeply inelastic, breaking up the target in a multitude of particles. In both cases the main concern is that, because of the fact that the electron emits an hard photon, it could lose sufficient energy to end up in the M{\o}ller  instead of in the EP detector as expected. Even though the deep inelastic and the electro-excitation cross sections are much smaller than the M{\o}ller one, the asymmetry for these two processes is one order of magnitude or more larger than the M{\o}ller one, potentially making the total asymmetries  comparable. 
Another problem is that only for few resonances the electromagnetic form factors are well known, and even for these, the corresponding  the PV FFs are usually very uncertain for most of them. These two facts, the possible large value of the resonance asymmetry and the uncertainty in the resonance form factors, suggest that a complete numerical computation of the bremsstrahlung contributions to the proton electro-excitation asymmetry be performed.  However, a complete calculation is unrealistic.
In this model I  simplify the situation by dividing the events in two kinematical regions. I assume that all the events with energy between 30 and 50 GeV are detected in the EP ring, while the events with energy between 12.5 to 30 GeV  end up in the M{\o}ller ring. To see how the uncertainty on the form factors might impact the measurement of $A_{RL}$, I  compute the  contribution of  two resonances to the bremsstrahlung asymmetry  in the EP ring (in this energy range the M{\o}ller contribution is negligible), by assuming some sensible value of the form factors. For these conditions I  also  compute the total asymmetry in the M{\o}ller ring, i.e. the asymmetry given by the sum of the  M{\o}ller and    bremsstrahlung process for the two resonances. This  is be my reference value, which is used for comparisons. I   then  allow the form factors to change in a reasonable range, in such a way that $A_{RL}$ in the EP remains the same, and then compute the new M{\o}ller asymmetry, and compare it with my reference value. In practice, the idea is to see how much  a measurement of the asymmetry in the EP might constrain the asymmetry in the M{\o}ller ring. The value of $A_{RL}$ in the EP ring I kept constant might be considered as my "measurement". Of course, there could be different values of the form factors that might  provide such a outcome. The issue I am trying to address here is if these different values, all compatible with the potential measurement in the EP, will result in a large uncertainty in $A_{RL}$ extracted in the M{\o}ller ring.  
\section{Cross Section and $A_{RL}$ for Proton Electro-Excitation.}\label{sec:brem}
In this section I  focus on the computation of the bremsstrahlung contribution in the proton electro-excitation to the cross section  and the polarized asymmetry. I  consider a simple model in which only two  resonances contribute, the isovector $\Delta(1232)$ and the isoscalar $N(1520)$. The electromagnetic and the neutral current amplitudes, contributing to the process I am considering, are depicted in Figs. \ref{f67} and \ref{f67a}. Let us consider the case in which the incoming electrons move along the $z$ axis and scatter in  the $(x,z)$ plane. The following kinematical quantities may be defined:
\begin{figure}   
\centerline{\psfig{figure=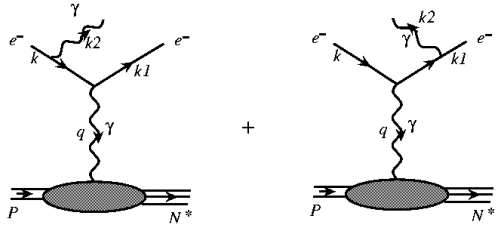}}
\caption{\label{f67} Electromagnetic current amplitude for bremsstrahlung process.}
\end{figure}
\begin{figure}  
\centerline{\psfig{figure=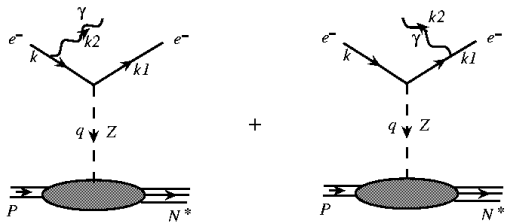}}
\caption{\label{f67a}Neutral current amplitude for bremsstrahlung process.}
\end{figure}
\bga
k&=& (E,0,0,\sqrt{E^2-m^2}) \, \,\textrm{incoming electron four momentum}\nn \\
s&=& \pm \frac{1}{m}(\sqrt{E^2-m^2},0,0,E) \,\, \textrm{incoming electron spin four vector}\nn \\
k_{1}&=& (E',\sqrt{E'^2-m^2} \sin\theta,0,\sqrt{E'^2-m^2}\cos\theta)\,\, \textrm{outgoing 
electron}\nn \\
&& \textrm{ four momentum}\nn \\
k_{2}&=& k_{0}(1,\sin\alpha \cos\phi,\sin\alpha \sin\phi,\cos\alpha)\,\,\textrm{real photon four momentum}
\nn \\
P&=&(M,0,0,0) \,\, \textrm{proton four momentum}\nn \\
P_{N^{*}}&=& (P_{N^{*}0},P_{N^{*}1},P_{N^{*}2},P_{N^{*}3}) \,\,\textrm{resonance four momentum}\nn \\
q&=&k-k_{1}-k_{2} \,\, \textrm{four momentum  transferred}\nn \\
Q^2&=&-q^2\, . 
\ea    
With such definition the amplitude in Fig. \ref{f67} can be computed, and the result is
\bga
M^{em}&=&\frac{e^3}{q^2} \la P_{\sigma_{1}}|J_{\mu}^{em}(0)|N^{*}_{\sigma_{2}}\ra \nn \\
&\times& \bar{u}_{s_2}(k_1)\Big[\epsilon^i_{\alpha} \frac{(\gamma^\alpha k\sla _2 +2 k_1^\alpha)}{2 k_1\cdot k_2}\gamma^\mu -  \gamma^\mu\frac{(\gamma^\alpha k\sla_2 +2 k^\alpha)}{2 k\cdot k_2}\epsilon^i_{\alpha} \Big]u_{s_1}(k)\, ,
\ea
where $e$ is the electron charge and $\epsilon_\alpha^i(k_2)$ is the photon polarization vector and $i$ labels th, $s_1$ and $s_2$ are the spins of the incoming and outgoing electrons, respectively,  polarization state of the outgoing photon, $\sigma_1$ is the proton spin and $\sigma_2$ is the resonance spin. The differential cross section then becomes
\bga\label{e207}
\frac{d\sigma}{d\Omega}^{brem}&=& \frac{1}{16 M E }\sum_{s_{1}s_{2}}\sum_{\sigma_{1}\sigma_{2}}\nn \\
&&\int\frac{d\! E' E'^{2}}{2E'(2 \pi)^{3}}\int\frac{d^{3} k_{2} }{2k_{0}(2 \pi)^{3}}
\int\frac{d^{3} P_{N^{*}} }{2P_{0 N^{*}}(2 \pi)^{3}}\nn \\
&&(2 \pi)^{4}\delta^4\!\Big(\! P_{N^{*}}\!-\!(P\!+\!k\!-\!k_{1}\!-\!k_{2})\Big)|M^{em}|^2=\nn \\
&=&\frac{2 \alpha^3}{ (4 \pi)^2  M E }\int d\! E' \frac{E'}{Q^4}\int\frac{d^{3} k_{2} }{k_{0}}\delta\!\lp M_{N^*}^2\!-\!(P\!+\!k\!-\!k_{1}\!-\!k_{2})^2\rp W_{\mu \nu }^{em} L_{em}^{\mu\nu} \nn \\
\ea
where I have introduced the hadronic and leptonic tensors $W_{\mu \nu }^{em} $ and $L^{\mu\nu}_{em}$ respectively defined as
\bga  \label{e211}
\delta\lp M_{N^*}^2-(P+k-k_{1}-k_{2})^2\rp W_{\mu\nu}^{em}=&&\nn \\ 
\frac{1}{4}\int\frac{d^{3} P_{N^{*}} }{2P_{0 N^{*}}(2 \pi)^{3}}\sum_{\sigma_{1}\sigma_{2}} (2 \pi)^{3}\delta^4\lp P_{N^{*}}-(P+k-k_{1}-k_{2})\rp&&\nn \\
\la P_{\sigma_{1}}|J_{\mu}^{em}(0)|N^{*}_{\sigma_{2}} \ra  \la N^{*}_{\sigma_{2}}|J_{\nu}^{em}(0)| P_{\sigma_{1}} \ra \nn \\ 
\ea
and
\bga
L_{\mu\nu}^{em}&=&\sum_{s_1 s_2}\sum_i  \bar{u}_{s_2}(k_1)\Big[\epsilon^i_{\alpha} \frac{(\gamma^\alpha k_2\slab +2 k_1^\alpha)}{2 k_1\cdot k_2}
\gamma^\mu -  \gamma^\mu\frac{(\gamma^\alpha k_2\slab -2 k^\alpha)}{2 k\cdot k_2}\epsilon^i_{\alpha}\Big] u_{s_1}(k)\nn \\
&\times&\bar{u}_{s_1}(k)\Big[\gamma^\nu\frac{(\gamma^\beta k_2\slab +2 k_1^\beta)}{2 k_1\cdot k_2}\epsilon^{i*}_{\beta }-  \epsilon^{i*}_{\beta} \frac{(\gamma^\beta k_2\slab -2 k^\beta)}{2 k\cdot k_2}\gamma^\nu  \Big]u_{s_2}(k_1)\\
&=& -\textrm{Tr}\Big\{ \Big[\frac{(\gamma^\alpha k_2\sla +2 k_1^\alpha)}{2 k_1\cdot k_2}\gamma^\mu -  \gamma^\mu\frac{(\gamma^\alpha k_2\slab -2 k^\alpha)}{2 k\cdot k_2}\Big](k\sla +m) \Big[ \gamma^\nu\frac{(\gamma^\alpha k_2\slab +2 k_1^\alpha)}{2 k_1\cdot k_2}\nn \\
& -&    \frac{(\gamma^\alpha k_2\slab -2 k^\alpha)}{2 k\cdot k_2}\gamma^\nu \Big](k_1\slab+m)\Big\}\,,
\ea
where I have used the relations
\bga
&&\sum_s u_s(p)\bar{u}_s(p)=p\sla+m \label{e216}\\
&& \sum_i \epsilon_\alpha^{i*}(k)\epsilon_\beta^i(k)= -g_{\alpha\beta} +\ldots\,,
\ea
where in the last relation  the ellipsis indicates terms which do not contribute to the $L_{\mu\nu}$. The leptonic tensor, although tedious and lengthy,  can  be computed. The result is presented in appendix \ref{app:tensors}. 

Using the delta function to eliminate the integration over $E'$, Eq. (\ref{e207}) may be written as
\bga \label{e208}
 \frac{d\sigma }{d\Omega}^{brem}&=&\frac{2 \alpha^3}{ (4 \pi)^2  M E }\int\frac{d^{3} k_{2} }{k_{0}}\frac{ W_{\mu \nu }^{em} L_{em}^{\mu\nu}}{\left| \frac{df}{dE'}\right|}\frac{E'}{Q^4} \nn \\
\ea
where I have defined
\bga
\frac{df}{dE'}&=& 2(M+E(1-\cos\theta ) -k_{0}(1-\sin\theta\cos\phi\sin\alpha-\cos\alpha\cos\theta))\, ,\nn \\
\ea
and where $E' $ is now equal to
\bga
E' &=&\frac{-M_{N^{*}}^2+M^2+2m^2+2ME-2Mk_{0}-2Ek_{0}(1-\cos\alpha)}{2(M+E(1-\cos\theta ) -k_{0}(1-\sin\theta\cos\phi\sin\alpha-\cos\alpha\cos\theta))}\,. \nn \\
\ea
On general grounds, invoking Lorentz,  space-inversion  and time-reversal invariance, and current conservation, the hadronic tensor can be casted in the following form:
\bga \label{e212}
W_{\mu\nu}^{em}&=&\lp -g_{\mu\nu}+\frac{q_{\mu}q_{\nu}}{q^2}\rp W_1^{em} \nn \\
&+&\lp P_{\mu} -\frac{P\cdot q}{q^2} q_{\mu}\rp\lp P_{\nu} -\frac{P\cdot q}{q^2} q_{\nu}\rp\frac{W_2^{em}}{M^2}\,,\nn \\
\ea  
and the structure functions (SFs) $W_1^{em}$ and $W_2^{em}$ contain information on the structure of the resonance. Let us now express the SFs in terms of the hadronic current form factors. 
For the two spin $3/2$ resonances,   I take the hadronic  current matrix element to be of the following form \cite{smith,nath}
\bga  \label{e209}
&&\la N^*_{\sigma_2}|J_{\mu}^{em}(0)| P_{\sigma_1} \ra  =\nn \\
&&\bar{u}^{\lambda}_{\sigma_2}(P_{N^*})\bigg[ \Big( \frac{C_3^{\gamma}}{M}\gamma^{\nu}+\frac{C_4^{\gamma}}{M^2}P_{N^*}^{\nu} +\frac{C_5^{\gamma}}{M^2}P^{\nu}\Big)(g_{\lambda \mu}g_{\rho\nu}-g_{\lambda \rho}g_{\mu\nu})q^{\rho}\gamma_5 \bigg]u_{\sigma_1}(P)\,. \nn \\
\ea
For the $\Delta(1232)$  resonance the following assumptions describe the data correctly \cite{dufner}:
\bge  \label{e210}
C_5^{\gamma}=0, \,\,\,\,\,\,\,\, C_4^{\gamma}=-\frac{M}{M_{N^*}}C_3^{\gamma}\,.
\ee
From Eqs. (\ref{e211}) and (\ref{e209}), and from the relation~\cite{liu}
\bga
\sum_{\sigma_2} u^{\lambda}_{\sigma_2}(P_{N^*})\bar{u}^{\rho}_{\sigma_2}(P_{N^*})&=&\Big[g^{\lambda \sigma}-\frac{2}{3} \frac{ P_{N^*}^\lambda P_{N^*}^\rho}{ M_{N^*}^2}-\gamma^{\lambda}\gamma^\rho +\frac{1}{3 M_{N^*}}(\gamma_\lambda P_{N^*}^\rho-\gamma_\rho P_{N^*}^\lambda  )\Big]\nn \\
&\times&\frac{\gamma \cdot P_{N*}+M_{N^*}}{2 M_{N^*}}
\ea
it is straight forward to compute the  hadronic tensor  and,  after performing some Dirac algebra and equating the result to equation
(\ref{e212}), I get, for the structure functions $W_1^{em}$ and $W_2^{em}$, the following results:
\bge \label{e213}
W_1^{em}= \frac{c M^2_{N^*}}{6M^2}a^2 C_3^{\gamma}\,\!^2
\ee
\bge \label{e214}
W_2^{em}=\frac{2 Q^2 M^2_{N^*}}{3}a C_3^{\gamma}\,\!^2\,,
\ee
and I have introduced the following kinematical factor
\bge
a=(M+M_{N^*})^2+Q^2
\ee
and
\bge
c=(M-M_{N^*})^2+Q^2 \,.
\ee
In order to simply the problem, I  also assume that the relations in Eq. (\ref{e210}) are true also for the $N(1520)$ resonance. 

Next, I want to compute the polarized asymmetry defined as
\bga\label{e215}
A_{RL}=\frac{
\lp\frac{\delta\sigma_
{R}}{\delta \Omega}
\rp^{brem}
-\lp\frac{\delta\sigma_
{L}}{\delta \Omega}
\rp^{brem}}{\lp\frac{\delta\sigma_
{R}}{\delta \Omega}\rp^{brem}+\lp\frac{\delta\sigma_
{L}}{\delta \Omega}\rp^{brem}}\, ,
\ea
where the subscripts $R,L$ refer to the helicity of the incoming electrons. Notice that, when considering  different scattering processes, the total asymmetry is defined as
\bga
A_{RL}^{TOT}=\sum_k f_k A_{RL}^k\,,
\ea
where 
\bga
f_k=\frac{\sigma_k}{\sum_k \sigma_k}\,,
\ea
and $A_{RL}^{k}$ and $\sigma_k$ are the asymmetry and the cross section for each single process, respectively. In this case I consider three processes: the M{\o}ller scattering, and the  bremsstrahlung processes of the two resonances. 

  Since the electromagnetic contribution to the cross section is insensitive to the polarization of the beam, taking the  difference in the asymmetry isolates the PV part of the process, which changes sign upon flipping the helicity of the electrons. In this case the amplitude of the neutral current process, in the limit $M_Z^2 \gg Q^2$,  is
\bga
M^{nc}&=& \frac{e^3}{16 \sin^2\theta_W \cos^2\theta_W M_Z^2} \la P_{\sigma_{1}}|J_{\mu}^{nc}(0)|N^{*}_{\sigma_{2}}\ra \nn \\
&\times& \bar{u}_{s_2}(k_1)\Big[\epsilon^i_{\alpha} \frac{(\gamma^\alpha k_2\slab +2 k_1^\alpha)}{2 k_1\cdot k_2}\gamma^\mu(g_V^e+g_A^e\gamma_5)\nn \\
& -&  \gamma^\mu(g_V^e+g_A^e\gamma_5)\frac{(\gamma^\alpha k_2\slab -2 k^\alpha)}{2 k\cdot k_2}\epsilon^i_{\alpha} \Big]u_{s_1}(k)\,,\nn \\
\ea  
and the numerator of Eq. \ref{e215} becomes 
\bga \label{e221}
\frac{d\sigma_{RL}}{d\Omega}^{brem}&\equiv&\frac{\delta\sigma_
{R}}{\delta \Omega}^{brem}-\frac{\delta\sigma_
{L}}{\delta \Omega}^{brem} = \frac{1}{16 M E }\sum_{s_{2}}\sum_{\sigma_{1}\sigma_{2}}\nn \\
&&\int\frac{d\! E' E'^{2}}{2E'(2 \pi)^{3}}\int\frac{d^{3} k_{2} }{2k_{0}(2 \pi)^{3}}
\int\frac{d^{3} P_{N^{*}} }{2P_{0 N^{*}}(2 \pi)^{3}}\nn \\
&&(2 \pi)^{4}\delta^4\!\Big(\! P_{N^{*}}\!-\!(P\!+\!k\!-\!k_{1}\!-\!k_{2})\Big)(M^{em}M^{nc*} +M^{nc}M^{em*})\nn \\
&=&\frac{2 G_F \alpha^2}{\sqrt{2} (4 \pi)^2  M E } \int\frac{d^{3} k_{2} }{k_{0}} W_{\mu \nu }^{nc} L_{nc}^{\mu\nu} \frac{E'}{Q^2}\, ,\nn \\
\ea
where $G_F=1.6639(1)\times 10^{-5}$ GeV$^{-2}$ is the Fermi constant.

In complete analogy with what I have done before, I have defined the interference leptonic tensor as
\bga
L_{\mu\nu}^{nc}&=&\sum_{ s_2}\sum_i  \bar{u}_{s_2}(k_1)\Big[\epsilon^i_{\alpha} \frac{(\gamma^\alpha k_2\slab +2 k_1^\alpha)}{2 k_1\cdot k_2}
\gamma^\mu(g_V^e+g_A^e\gamma_5) \nn \\
&-&  \gamma^\mu(g_V^e+g_A^e \gamma_5)\frac{(\gamma^\alpha k_2\slab -2 k^\alpha)}{2 k\cdot k_2}\epsilon^i_{\alpha}\Big] u_{s_1}(k)\nn \\
&\times&\bar{u}_{s_1}(k)\Big[\gamma^\nu\frac{(\gamma^\beta k_2\slab +2 k_1^\beta)}{2 k_1\cdot k_2}\epsilon^{i*}_{\beta }-  \epsilon^{i*}_{\beta} \frac{(\gamma^\beta k_2\slab -2 k^\beta)}{2 k\cdot k_2}\gamma^\nu  \Big]u_{s_2}(k_1)\nn\\
&=& -\textrm{Tr}\Big\{ \Big[\frac{(\gamma^\alpha k_2\sla +2 k_1^\alpha)}{2 k_1\cdot k_2}\gamma^\mu(g_V^e+g_A^e \gamma_5) -  \gamma^\mu(g_V^e+g_A^e \gamma_5)\frac{(\gamma^\alpha k_2\slab +2 k_1^\alpha)}{2 k_1\cdot k_2}\Big]\nn \\
&\times& (k\sla +m)\frac{1+\gamma_5 s_1 \slab}{2} \Big[ \gamma^\nu\frac{(\gamma^\alpha k_2\slab +2 k_1^\alpha)}{2 k_1\cdot k_2} 
-    \frac{(\gamma^\alpha k_2\slab +2 k_1^\alpha)}{2 k_1\cdot k_2}\gamma^\nu \Big](k_1\slab+m)\Big\}\,.\nn\\
\ea
Notice that in this case there is no sum over the incoming electron spin, for this reason Eq. (\ref{e216}) now reads
\bga
u_s(p)\bar{u}_s(p)=(p \sla +m)\frac{1+\gamma_5 s \sla}{2}\,.
\ea
The full expression of the interference leptonic tensor can also be found in appendix \ref{app:tensors}. The hadronic tensor for the interference process is
\bga
\delta\lp P_{N^{*}}^2-(P+k-k_{1}-k_{2})^2\rp W_{\mu\nu}^{nc}=&&\nn \\ 
\frac{1}{4}\int\frac{d^{3} P_{N^{*}} }{2P_{0 N^{*}}(2 \pi)^{3}}\sum_{\sigma_{1}\sigma_{2}} (2 \pi)^{3}\delta^4\lp P_{N^{*}}-(P+k-k_{1}-k_{2})\rp&&\nn \\
\la P_{\sigma_{1}}|J_{\mu}^{em}(0)|N^{*}_{\sigma_{2}} \ra  \la N^{*}_{\sigma_{2}}|J_{\nu}^{nc}(0)| P_{\sigma_{1}} \ra \, ,\nn \\ 
\ea
and similarly with what has been done before for  the electromagnetic case, its most general form compatible Lorentz and time-reversal invariance, and current conservation, is
 \bga 
W_{\mu\nu}^{nc}&=&\lp -g_{\mu\nu}+\frac{q_{\mu}q_{\nu}}{q^2}\rp W_1^{nc} \nn \\
&&\lp P_{\mu} -\frac{P\cdot q}{q^2} q_{\mu}\rp\lp P_{\nu} -\frac{P\cdot q}{q^2} q_{\nu}\rp\frac{W_2^{nc}}{M^2}\nn \\
&-&i\epsilon _{\mu\nu\alpha\beta} P^{\alpha}q^{\beta} \frac{W_3^{nc}}{M^2} \,.
\ea
If I now assume a form for the  matrix element of the hadronic neutral current as follows~\cite{nath}
\bga  
&&\la P_{N^{*}}|J_{\mu}^{nc}(0)| P \ra  =\nn \\
&&\bar{u}^{\lambda}(P_{N^*})\bigg[ \lp  \frac{C_{3V}^{Z}}{M}\gamma^{\nu} +\frac{C_{4V}^Z}{ M^2} P ^{\nu}_{N^*}\rp (g_{\lambda \mu}g_{\rho\nu}-g_{\lambda \rho}g_{\mu\nu})q^{\rho}\gamma_5 \bigg]u(P) \nn \\
&&+\bar{u}^{\lambda}(P_{N^*})\bigg[  \frac{C_{4A}^{Z}}{M^2}P_{N^*}^{\nu} (g_{\lambda \mu}g_{\rho\nu}-g_{\lambda \rho}g_{\mu\nu})q^{\rho}+C_{5A}^{Z} g_{\lambda \mu} \bigg]u(P)
\ea
the hadronic form factors could be computed, and what I obtain is
\bge  
W_1^{nc}= \frac{c M^2_{N^*}}{6M^2}a^2 C_{3V}^{Z}C_3^\gamma
\ee
\bge \label{eq13}
W_2^{nc}=\frac{2 Q^2 M^2_{N^*}}{3}a C_{3V}^{Z}C_3^\gamma
\ee
\bge 
W_3^{nc}=\frac{2a }{3 M M_{N^*}}C_{3}^\gamma (\frac{1}{2} b C_{4A}^Z -M^2 C_{5A}^Z)\,,
\ee
where also in this case, for simplicity, I made the   approximation $C_{4V}^Z=-\frac{M}{M_{N^*}}C_{3V}^Z$ as for the electromagnetic case, and $b$ is defined as
\bga
b=(M+M_{N^*})(M-M_{N^*})+Q^2\,.
\ea
 Following Ref.~\cite{mjrm}, the hadronic currents can be decomposed in terms of their isospin content (see also Sec. \ref{sec:asy})
\bga
J_{\mu}^{em}(0)=J_{\mu}^{ T=1}(0) +J_{\mu}^{T=0}(0)\,,
\ea
and 
\bga
J_{\mu}^{nc}(0) = \xi_V^{T=1}J_{\mu}^{T=1}(0)+\sqrt{3} \xi _V^{T=0}J_{\mu}^{T=0}(0)+A_\mu(0)\,,
\ea
where 
\bga
\xi_V^{T=1}=2(1-2\sin^2\theta_W)&\,\,\,\,\,\,,\,\,\,\,\,\,& \xi_V^{T=0}=-\frac{4}{\sqrt{3}}\sin^2\theta_W\,,
\ea
and the superscript $T=1$ and $T=0$ stands for isovector and isoscalar, respectively. The vector $A_\mu(0)$ represents the axial part of the hadronic neutral current, which I am  not interested in decomposing.
Because of the fact that the $\Delta(1232)$ resonance is an isospin $\frac{3}{2}$ particle, while the proton has isospin $\frac{1}{2}$, the isoscalar component of the currents does not contribute to the matrix elements
$\la P_{N^{*}}|J_{\mu}^{em,nc}(0)| P \ra $. As a consequence, I have that the interference vector form factor is related to the electromagnetic one by the relation
\bga\label{e230}
C^{Z}_{3V}=\xi_V^{T=1}C^{\gamma}_3\,.
\ea 
In the case of the $N(1520)$, since it is an isospin $\frac{1}{2}$ particle, both components of the hadronic current contribute, so that I have
\bga\label{ee231}
C^{\gamma}_{3}=C^{\gamma T=1}_3 + C^{\gamma T=0}_3
\ea
\bga\label{e232}
C^{Z}_{3V}=\xi_V^{T=1}C^{\gamma T=1}_3 +\sqrt{3}\xi_V^{T=0} C^{\gamma T=0}_3\,.
\ea 
In order to compute the bremsstrahlung cross section and asymmetry, according to Eqs. (\ref{e208}) and (\ref{e221}), I would need now to compute the two tensor contractions $L^{em}_{\mu\nu}W_{em}^{\mu\nu}$ and $L_{\mu\nu}^{nc}W_{nc}^{\mu\nu}$. The results can be found in the appendix \ref{app:tensors}. 
\section{Infrared Cancellation}\label{sec:infrared}
\begin{figure}  
\centerline{\psfig{figure=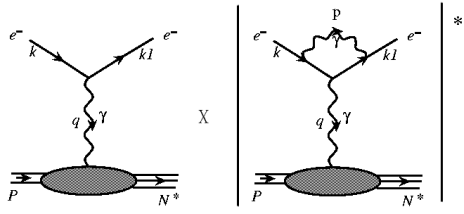}}
\caption{\label{f69}One loop correction to the EM cross section.}
\end{figure}

\begin{figure} 
\centerline{\psfig{figure=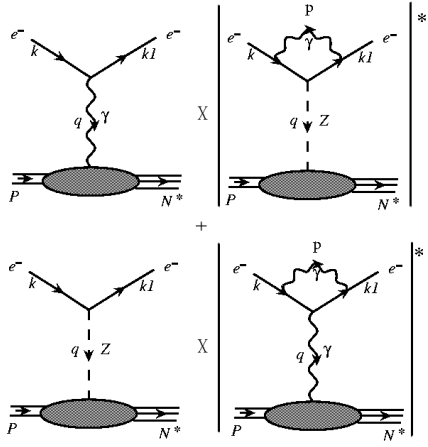}}
\caption{\label{f70}One loop correction to the polarized   cross section $\sigma_L-\sigma_R$. }
\end{figure}

In this section I   show how the bremsstrahlung cross section and asymmetry become infrared finite once I had the one loop vertex corrections shown in Figs. \ref{f69} and \ref{f70} to the electro-excitation process. 
As  will be shown in a moment, the integral over the photon energy in the bremsstrahlung cross section and $A_{RL}$,  diverges if the lower limit is taken to be zero.  In order to keep it finite, I  introduce  a lower limit cut off. Of course, since the cross section and $A_{RL}$ are physical quantities, they need to be finite and cut off independent. Adding the one loop contribution will indeed make them finite and cut off independent. The reason why  I compute to soft  bremsstrahlung contribution   is that this process, even though does not contribute to the asymmetry in the M{\o}ller ring, it still needs to be considered when computing  the total asymmetry in the EP ring. 

First of all let us consider the zero photon  energy limit in the square amplitudes $W_{\mu\nu}^{em}L_{em}^{\mu\nu}$ and $W_{\mu\nu}^{nc} L_{nc}^{\mu\nu}$. The lengthy relations in Eqs. (\ref{e218}) and (\ref{e219}) become
\bga
\lim_{k_0\rightarrow 0}W_{\mu\nu}^{em}L^{\mu\nu}_{em}&=&2\Big( -4W_1^{em}(m^2-Q^2)+W_2^{em}(4E^2-Q^2)\Big)\nn \\ &&\lp\frac{2 k\sd k_1}{(k_1\sd k_2) (k\sd k_2) }-\frac{m^2}{(k\sd k_2)^2}-\frac{m^2}{(k_1\sd k_2)^2}\rp 
\ea
and
\bga
\lim_{k_0\rightarrow 0}W_{\mu\nu}^{nc}L^{\mu\nu}_{nc}&=&\!2\Big( g_A^e[4W_1^{nc}Q^2+W_2^{nc}(4E^2-Q^2)] \nn \\ 
&& \!2g_V^e Q^2 \frac{E\!+\!E'}{M} W_3^{nc}\Big)\lp\frac{2 k\sd k_1}{(k_1\sd k_2) (k\sd k_2) }\!-\!\frac{m^2}{(k\sd k_2)^2}\!-\!\frac{m^2}{(k_1\sd k_2)^2}\rp \,.\nn \\ 
\ea
In order to obtain the cross section and $A_{RL}$, I need to perform the following integral
\bga
\int \frac{d^3 k_2}{k_0}\lp\frac{2 k\sd k_1}{(k_1\sd k_2) (k\sd k_2) }-\frac{m^2}{(k\sd k_2)^2}-\frac{m^2}{(k_1\sd k_2)^2}\rp =&&\nn \\
\int_\mu^{El} \frac{d k_0}{k_0}\int d\Omega\lp\frac{2 k\sd k_1}{(k_1\sd \hat{k}_2) (k\sd \hat{k}_2) }-\frac{m^2}{(k\sd \hat{k}_2)^2}-\frac{m^2}{(k_1\sd \hat{k}_2)^2}\rp \,,&&\nn \\
\ea
where $\hat{k}_2^{\nu}=k_2^\nu /k_0$, $\mu$ is the lower limit cut off, and $E_l$ is an upper limit cut off for which the soft photon approximation is still valid. The integration over the solid angle of the last two terms   is straight forward. Taking the four-vector $k^\mu$ (or $k_1^\mu$) along the $z$ axis, I have 
\bga
\int d\Omega\frac{m^2}{(k_1\sd \hat{k}_2)^2}&=&\int d\Omega\frac{m^2}{(k\sd \hat{k}_2)^2}= \int d\Omega \frac{m^2}{(E-|\mathbf{k}|\cos\theta)^2 }\nn \\
&=& 2\pi\int_{-1}^{1}dy \frac{m^2}{(E-|\mathbf{k}|y)^2}=\frac{2\pi m^2}{|\mathbf{k}|}\lp \frac{1}{E-|\mathbf{k}|}
-\frac{1}{E+\mathbf{|k|}}\rp\nn \\
&=& 4 \pi \frac{m^2}{k^2}=4 \pi\,,\nn \\
\ea 
while for the first term I have
\bga 
&&\int d\Omega\frac{2 k\sd k_1}{(k_1\sd \hat{k}_2) (k\sd \hat{k}_2) }=\int_0^1dx \int d\Omega\frac{2 k\sd k_1}{\Big(x k_1\sd \hat{k}_2+(1-x) k\sd \hat{k}_2 \Big)^2}=\nn\\ 
&&=\int_0^1dx \int d\Omega\frac{2 k\sd k_1}{\Big( \hat{k}_2 \sd (-x q 
+ k) \Big)^2}=4\pi \int_0^1 dx \frac{2 k\sd  k_1 }{(-xq+k)^2  }\nn \\
&&=4\pi \int_0^1 dx \frac{2m^2+Q^2 }{(Q^2x(1-x)+m^2)  }\,,
\ea
where I introduced the Feynman relation
\bge
\frac{1}{AB}=\int_0^1 \frac{dx}{(Ax+B(1-x))^2}\,.
\ee
Putting all together and performing the integral over the photon energy, I get for Eqs. (\ref{e208}) and (\ref{e221}) in the soft photon approximation 
\bga \label{e229}
\lp\frac{d\sigma }{d\Omega}\rp^{brem}&=& \frac{\alpha^3 E' \Big( W_2^{em}(4E^2-Q^2)-4W_1^{em}(m^2-Q^2)\Big)} {2\pi M E [M+E(1-\cos\theta)]Q^4}\log\frac{E_l}{\mu}\times\nn \\
&\times&\lp \int_0^1\frac{dx(Q^2+2m^2)}{(Q^2x(1-x)+m^2)}-2\rp\,,
\ea
\bga \label{e229a}
\lp\frac{\delta\sigma_
{LR}}{\delta \Omega}\rp^{brem}  &=& \frac{ \alpha^2 E' G_F } {8  \sqrt{2} \pi^2  M E [M+E(1-\cos\theta)]  Q^2}\log\frac{E_l}{\mu}\nn \\ 
&\times&\Big( g_A^e[W_2^{nc}(4E^2\!-\!Q^2)\!+\!4W_1^{nc}Q^2]
\!+\!2 g_V^e Q^2 \frac{(E+E')}{M}W_3^{nc}\Big)\nn \\
&\times&\lp \int_0^1\frac{dx(Q^2+2m^2)}{(Q^2x(1-x)+m^2)}-2\rp \,,
\nn \\
\ea
where I did not consider finite terms in the limit in which $\mu$ goes to zero\footnote{In the actual calculation, the sum of the bremsstrahlung and the loop contribution has been performed numerically, so the finite terms left out in Eqs. (\ref{e229}) and (\ref{e229a}) are included}. 
As  can be seen, in such a limit,  both  expressions are logarithmically divergent. In order to obtain a finite expression I need to add the one loop corrections. This contribution to   the cross section is lengthy and cumbersome to obtain, so I delineate only the main steps, referring the reader to standard books. My derivation follow closely the one that is found in \cite{pesk}. 
Considering  the process represented in Fig. \ref{f69}, the cross section is
\bga\label{e222}
\lp\frac{d\sigma }{d\Omega}\rp^{loop}&=&\frac{\alpha^3 2 \pi E'}{M E[M+E(1-\cos\theta)]Q^4 }\tilde{L}_{\mu\nu}^{em} W^{\mu\nu}_{em}\,,
\ea 
where I have introduced a new leptonic tensor relative to the vertex correction 
\bga
\tilde{L}_{\mu\nu}\!&=&\!\int\! \frac{d^4p}{(2\pi)^4}\frac{\textrm{Tr}\Big[\gamma_\alpha(p\sla\!+\!q\sla\!+\!m)\gamma_\nu(p\sla\!+\!m)\gamma^\alpha(k\sla\!+\!m)\gamma_\mu(k\sla\!+\!q\sla\!+\!m)\Big]}{(p^2-m^2)((p+q)^2-m^2)((k-p)^2-\mu^2)}
\,,\nn \\
\ea
while the hadronic tensor is the same as for the bremsstrahlung case.
Notice that I have introduced a fictitious mass $\mu$ in the photon propagator to make the integral finite and which, at the end of the day,   cancels the cut off dependence in the bremsstrahlung cross section. Using standard procedures in performing the integral  and keeping only terms which are singular in the limit $\mu$ goes to zero, equation (\ref{e222}) gives 
\bga \label{e223}
\frac{d\sigma }{d\Omega}^{loop}&=&\frac{\alpha^3 2 \pi E'}{(4\pi)^2M E[M+E(1-\cos\theta)]Q^4 }\nn \\
&&\int_0^1dz\int_0^{1-z}dy\Bigg[\frac{F(y,z,Q^2)}{\Big(m^2(1-z)^2+Q^2y(1-z-y)+\mu^2 z\Big)} 
\Bigg]\,,\nn \\
\ea
where I have defined the function $F(y,z,Q^2)$ as
\bga
F(y,z,Q^2)&=&16W_1^{em}\Big([(z-4)z+1]m^4+m^2Q^2[y^2+(z-1)y\nn\\
&+&z^2+z-1]-Q^4(y-1)(y+z)\Big)\nn \\
&-&4W_2^{em}\Big(4\{ m^2[(z-4)z+1]+Q^2(y-1)(y+z)\} E^2\nn \\
& -&Q^2 \{-m^2[z(z+2)-1]+Q^2(y-1)(y+z)\}\Big)\,.
\ea
The  integral in Eq. (\ref{e223}) in the limit $\mu$ goes to zero, becomes divergent in the region $z=1$ and $y=0$. Since I am concerned in the divergent pieces of the cross section,  I   take such a limit where possible. The function $F(y,z,Q^2)$ becomes  then
\bga
F(0,1,Q^2)&=&-4(2m^2+Q^2)\Big((4E^2-Q^2)W_2^{em}-4(m^2-Q^2)W_1^{em}\Big)\nn \\
\ea 
so that equation (\ref{e223}) yields
\bga \label{e224}
\frac{d\sigma }{d\Omega}^{loop}&=&-\frac{\alpha^3 \Big((4E^2-Q^2)W_2^{em}-4(m^2-Q^2)W_1^{em}\Big)  E'}{2\pi M E[M+E(1-\cos\theta)]Q^4 }\nn \\
&&\int_0^1dz\int_0^{1-z}dy\Bigg[\frac{(2m^2+Q^2)}{\Big(m^2(1-z)^2+Q^2y(1-z-y)+\mu^2z \Big)} 
\Bigg]\,.
\ea
Beside being infrared divergent, the one loop cross section is also ultraviolet divergent. Making it finite is part of the renormalization program. In this case I eliminate the divergent part by doing the following substitution in the integral in Eq. (\ref{e224}) (see \cite{pesk} chapters 6 and 7 for more details)
\bga
&&\int_0^1dz\int_0^{1-z}dy\Bigg[\frac{(2m^2+Q^2)}{\Big(m^2(1-z)^2+Q^2y(1-z-y)+\mu^2 z\Big)}\Bigg]\nn \\
&\rightarrow& \int_0^1dz\int_0^{1-z}dy\Bigg[\frac{(2m^2+Q^2)}{\Big(m^2(1-z)^2+Q^2y(1-z-y)+\mu^2 z \Big)}\nn \\
&-&\frac{2m^2 }{\Big(m^2(1-z)^2 +\mu^2 z\Big)}\Bigg]
\ea
where basically I have subtracted the same quantity calculated at $Q^2=0$ (on-shell renormalization scheme). Performing the change of variable $z=1-t$ and $y=t x$ the first integral of the previous relation becomes
\bga
\int_0^1dz\int_0^{1-z}dy\frac{(2m^2+Q^2)}{\Big(m^2(1-z)^2+Q^2y(1-z-y)+\mu^2z \Big)}&=&\nn\\
\int_0^1 dx \int_0^1 dt \frac{t}{t^2(m^2+Q^2x(1-x)) +\mu^2(1-t)}&=&\nn\\
\int_0^1dx\frac{m^2+Q^2}{2[m^2+Q^2x(1-x)]}\log\lp \frac{m^2+Q^2x(1-x)}{\mu^2}\rp\,.
\ea
The first integral is straight forward and the result is
\bga
\int_0^1dz\int_0^{1-z}dy\frac{2m^2 }{\Big(m^2(1-z)^2 +\mu^2 \Big)}=\log\frac{m^2}{\mu^2}\,,
\ea
so that the loop contribution yields
\bga \label{e225}
\frac{d\sigma }{d\Omega}^{loop}&=&-\frac{\alpha^3 \Big((4E^2-Q^2)W_2-4(m^2-Q^2)W_1\Big)  E'}{2\pi M E[M+E(1-\cos\theta)]Q^4 }\nn \\
&&\Bigg[\int_0^1dx\frac{(2m^2+Q^2)}{\Big(m^2 +Q^2x(1-x)  \Big)}\log\lp \frac{m^2+Q^2x(1-x)}{\mu^2}\rp-\log\frac{m^2}{\mu^2}
\Bigg] \,.\nn\\
\ea
Now adding equations (\ref{e229}) and (\ref{e225}) I get
\bga \label{e227}
\frac{d\sigma }{d\Omega}^{brem}+\frac{d\sigma }{d\Omega}^{loop}&=&\frac{\alpha^3 \Big((4E^2-Q^2)W_2^{em}-4(m^2-Q^2)W_1^{em}\Big)  E'}{2\pi M E[M+E(1-\cos\theta)]Q^4 }\nn \\
&&\Bigg[\int_0^1dx\frac{(2m^2+Q^2)}{\Big(m^2 +Q^2x(1-x)  \Big)}\log\lp \frac{E_l}{\sqrt{m^2+Q^2x(1-x)}}\rp\nn \\
&-&2 \log\frac{E_l}{m}
\Bigg]\,,
\ea
which is independent on the cut off $\mu$ as expected. At the same  way the one loop corrections to the  polarized cross section   for the process in Fig. \ref{f70} have  been calculated, leading to
\bga
\frac{\delta\sigma_
{LR}}{\delta \Omega}^{brem}&+&\frac{\delta\sigma_
{LR}}{\delta \Omega}^{loop} = \frac{ \alpha^2  E' G_F} {8 \sqrt{2}  \pi^2  M E [M+E(1-\cos\theta)] Q^2}\nn \\ 
&&\Big( g_A^e[W_2(4E^2-Q^2)+4W_1Q^2]+2 g_e Q^2 \frac{(E+E')}{M}W_3\Big)\nn \\
&&\Bigg[\int_0^1dx\frac{(2m^2+Q^2)}{\Big(m^2 +Q^2x(1-x)  \Big)}\log\lp \frac{E_l}{\sqrt{m^2+Q^2x(1-x)}}\rp\nn \\
&-&2 \log\frac{E_l}{m}
\Bigg]\,.
\ea
\section{Numerical Studies} \label{sec:results}
\begin{figure}  
\centerline{\psfig{figure=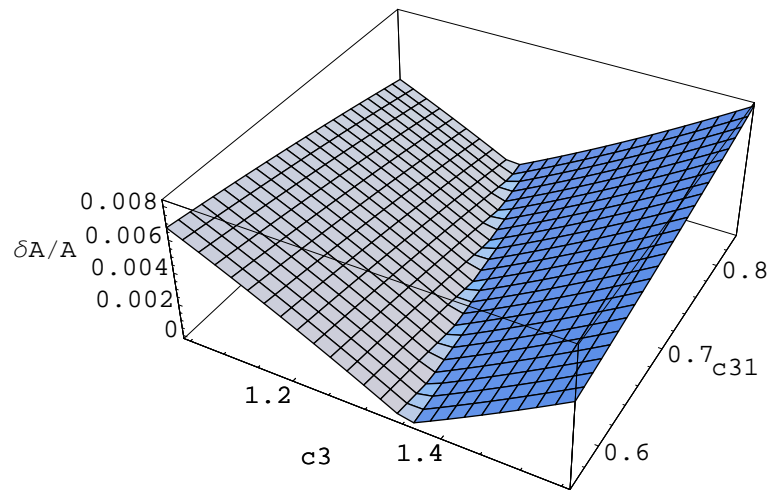}}
\caption{\label{f75}Relative uncertainty in the total asymmetry defined as in Eq. (\ref{eab3}) as a function of $c_3$ and $c_{31}$ with initial values $c_3=1.3$, $c_{31}=.7$ and $c_{30}=1/10c_{31}$. The parameters $c_3$ and $c_{31}$ have been changed up to 20\% of their initial value, while $c_{30}$ has been changed accordingly, in order to keep the total asymmetry in the EP ring constant. }
\end{figure}

In this section I  present the numerical results of my two resonance model. The idea, as I mentioned before, is to compute the asymmetry in the EP region,  which in my case is the range of energy of the ougoing electrons that goes from 30 to 50 GeV. All cross sections and $A_{RL}$'s are integrated over the angular resolution, which corresponds to a range of scattering angle in the lab frame between $4.5$ and $8.0$ mrd. In order to simplify the problem,  I  assume the same  $Q^2$ dependence in all the form factors. Specifically, I  assume a dipole dependence of the type
\bga\label{ea86}
C_i=\frac{c_i}{\lp 1+ Q^2/a  \rp}\,, 
\ea
where the parameter $a$ is taken to be the same for all the form factors. For the delta resonance electromagnetic form factor, I have
\bge
C_3^\gamma(Q^2)=\frac{c_{3}}{(1+Q^2/a)^2}
\ee
and, for the interference one, according to equation (\ref{e230}), I have 
\bge
C_{3V}^{Z}(Q^2)=\xi _V^{T=1}C_3^\gamma(Q^2)\,.
\ee
For the $N(1520)$ resonance I write
\bga
C_3^\gamma(Q^2)&=&\frac{(c_{31}+c_{30})}{(1+Q^2/a)^2}\equiv C_{3}^{\gamma\, T=1}(Q^2) + C_{3}^{\gamma\, T=0}(Q^2) \nn \\
\ea
and, according to equation (\ref{e232}) 
\bge
C_{3V}^{Z}(Q^2)=\xi _V^{T=1}C_{3}^{\gamma\, T=1}(Q^2)+\sqrt{3} \xi^{T=0}_V C_{3}^{\gamma\, T=0}(Q^2)\,.
\ee
In order to fix the initial parameters I have considered the cross sections $e^- \, P \rightarrow \pi ^+\, N$ and  $e^-\, P \rightarrow \pi^0 \, P$. The MAID group  provides a computational tool to compute the aforementioned  cross sections, allowing one to isolate the contribution arising from each single resonance (it can be found in their web site at http://www.kph.uni-mainz.de/MAID/)~\cite{drechsel}. I have then best fit the parameters to the cross sections, and what I get is 
\bga \label{e231}
a=.54 \quad c_3=1.3\quad c_{31}=.7 \quad c_{30}=\frac{c_{31}}{10}\,.
\ea
Another approximation I have made, is to consider a zero width resonance,  which considerably reduces the computational time. At this point I  consider the total asymmetry in the EP ring obtained for this initial set of values of the form factors, and consider it as a possible  result from the measurement. I  also  compute the total asymmetry in $A_{RL}$ in the M{\o}ller ring, and consider it as a reference value $A_{RL}^{M 0}$. What I want to see is if this measurement in the EP ring could actually  constrain  the  measurement in the M{\o}ller ring. The same value of the asymmetry in the EP ring could have been obtained also for some different value of the form factors. I  change the value of $c_{3}$ and $c_{31}$ and treat them as independent variables, and find the value of $c_{30}$   such  that  the asymmetry in the EP ring does not change. I   then  compute the total asymmetry in the M{\o}ller ring for these new values of the form factors and compare it to the reference value $A_{RL}^{M 0}$ to check if the variation in the form factor induced a large difference in $A_{RL}$ computed in the M{\o}ller ring. The contribution to $A_{RL}$ arising from the PV SF $W_3^{nc}$ are not explicitly included, since it turned out to be negligible (remember that this term always appears multiplied by the vector coupling constant $g_V^e\simeq .075$). The total asymmetry in the EP ring is 
\bga
A_{RL}^{EP}=\frac{1}{\sigma^{\Delta} +\sigma^{N}}\Big[(\sigma^{\Delta}_R-\sigma^{\Delta}_L)+(\sigma^N_R-\sigma^N_L) \Big] \,
\ea
in which I have neglected the contribution from the M{\o}ller process since it is much smaller than the bremsstrahlung.
Considering the values of the parameters in Eq. (\ref{e231}) I get the following initial values for the EP ring
\bge
\sigma^{\Delta}=3.18 \times 10^{-4} \textrm{GeV$^{-2}$}
\ee
\bge
\sigma^{\Delta}_L-\sigma^{\Delta}_R=2.61\times 10^{-9} \textrm{GeV$^{-2}$}
\ee
\bge
\sigma^{N}=9.38  \times10^{-5} \textrm{GeV$^{-2}$}
\ee
\bge
\sigma^{N}_L-\sigma^{N}_R=6.35\times 10^{-10} \textrm{GeV$^{-2}$}\,,
\ee
giving a total asymmetry of
\bge
A_{RL}^{EP}=-7.88 \times 10^{-6}\,.
\ee
In the M{\o}ller ring I consider both processes, the bremsstrahlung and the M{\o}ller, so that the total asymmetry can be written as
\bga
A_{RL}^{MTOT}&=&\frac{1}{\sigma^M \!+\!\sigma^{\Delta}\!+\!\sigma^{N}}\Big((\sigma_R^{M}\!-\!\sigma_L^{M})\!+\! (\sigma_R^{\Delta}\!-\!\sigma_L^{\Delta})\!+\!(\sigma_R^{N}\!-\!\sigma_L^{N})\Big)\,,\nn \\
\ea
where the M{\o}ller differential cross section and polarized cross sections   are \cite{slac}
\bge
\frac{d\sigma}{d\Omega}^M=\frac{\alpha^2}{2 m E}\frac{(3+\cos^2\Theta)^2}{\sin^4\Theta}
\ee
and
\bge
\frac{d\sigma^M_L}{d\Omega}-\frac{d\sigma^M_R}{d\Omega}=8\frac{G_F\alpha}{\sqrt{2}\pi }\frac{g_{ee}}{\sin^2\Theta}\,,
\ee
and where I used for the coupling $g_{ee}$ its tree level value corrected by a factor of 0.4 due to radiative corrections~\cite{mar}.
The total cross sections for the M{\o}ller process integrated over the angular resolution are
\bge
\sigma^{M}=4.18 \times 10^{-2}\textrm{Gev$^{-2}$}
\ee
\bge
\sigma^{M}_L-\sigma^M_R=6.23 \times10^{-9} \textrm{Gev$^{-2}$}\,,
\ee
while, for the two resonances I get
\bge
\sigma^{\Delta}=2.2 \times 10^{-5} \textrm{GeV$^{-2}$}
\ee
\bge
\sigma^{\Delta}_L-\sigma^{\Delta}_R=9.77\times 10^{-11} \textrm{GeV$^{-2}$}
\ee
\bge
\sigma^{N}=6.72  \times 10^{-5} \textrm{GeV$^{-2}$}
\ee
\bge
\sigma^{N}_L-\sigma^{N}_R=2.44 \times 10^{-11} \textrm{GeV$^{-2}$}\,
\ee
so that my reference value for the total asymmetry is
\bga
A_{RL}^{M0}=-1.52\times 10^{-7}\,.
\ea
In Fig. \ref{f75} I have plotted the quantity 
\bga \label{eab3}
\frac{\delta A}{A}\equiv 2\frac{A_{RL}^{MTOT}-A_{RL}^{M 0}}{A_{RL}^{MTOT}+A_{RL}^{M 0}}\,,
\ea   
where $A_{RL}^{MTOT}$ is the value of the total asymmetry for which I allow
 the parameters $c_3$ and $c_{31}$ to change up to $20 \%$ of their initial  value.  Notice that the factor of two in the definition of the relative error, comes from taking the average of two asymmetries. It seems quite clear that in this simple model, even a very poor knowledge of the resonance form factors (which is not the case for the ones I used) does not impact severely the extraction of the total asymmetry, since allowing a very "generous" mistake on the form factors of $20 \%$, will only reflect in a less than $1\%$ error on the asymmetry. Nevertheless, including less known resonances to the model,  might result in a theoretical uncertainty of the order of the experimental error.
%e231 f75
\section{Comments and Conclusions} \label{sec:conclusion}
In this chapter I have investigated the contribution of the bremsstrahlung radiation to the proton electro-excitation process, by using a simple two resonance model. I have studied how the possible uncertainties on the resonances form factors might be  source of error in the measurement of $A_{RL}$ in the E-158 experiment. As it turned out, even knowing the  form factors to 20$\%$ level,  only affects $A_{RL}$ with an error of the order of less then one percent, which is much smaller than the prescribed precision of the experiment, which is 8$\%$. I tried to see what would be the error on the form factors that  produces a 8\% 
uncertainty in the asymmetry.  What I found is that the error should be of the order of 150\%. Even though such uncertainty might seem large for two resonances such as the $Delta(1232)$ and the $N(1520)$, there are still many resonances that contribute to the bremsstrahlung process, whose FFs are not very well known.  A more complete study, including other resonances, seems necessary in order to completely exclude the possibility that the bremsstrahlung contribution to the proton electro-excitation might introduce an uncertainty of the order of the prescribed precision of the experiment. I also changed the $Q^2$ dependence by varying the $a$ coefficient in Eq. (\ref{ea86}) between 0.4 and 0.7, but the result did not change appreciably.  On the other hand, in my simple model, I have made some assumptions. I assumed that the  dipole dependence for the form factors was the same for both resonances. Moreover, for the isoscalar resonance I took the $Q^2$ dependence to be the same for the isovector and the isoscalar form factors, which is not necessarily true. It would be also interesting to be able to include the continuous DIS region in the calculation. The main contribution to the bremsstrahlung process arises from the kinematical region in which the photon is emitted parallel either to the incoming or the outgoing electron. In this region the $Q^2$ it is no more than $0.1$ GeV$^2$, so, in order to compute the DIS contribution, one would need to know the proton SFs at a very small $Q^2$, but this kind of knowledge is lacking at the moment.  In conclusion, it seems that   in this two resonance model, the bremsstrahlung contribution  to the proton electro-excitation should not vitiate the extraction on the M{\o}ller asymmetry at a $8\%$ level.
However, I also showed that a 150\% uncertainty in the resonances FFs, might produce an uncertainty in the asymmetry in the M{\o}ller ring as big as the prescribed experimental precision. This result suggests that a more detailed study be done, including a larger number of resonances contributing to the bremsstrahlung process.  

\chapter{Electroweak Radiative Corrections to the Parity-Violating Electroexcitation of the $\Delta$}\label{ch:delta}
In this chapter I am going to present the results of the work I have done in collaboration with Ramsey-Musolf, Zhu, Holstein and Maekawa~\cite{zhuu}. In this work, we analyze the degree to which parity violating electro-excitation  of the resonance $\Delta(1232)$ may be used to extract the weak neutral axial vector transition  form factors. My contribution consisted in double checking the computation of  the chiral loops and the $1/M_N$ corrections to the parity violating asymmetry arising from the anapole, the Siegert, and the d-wave parity violating lagrangian. In the next sections I will summarize the calculation and present the main results of this work. More details can be found in ref.~\cite{zhuu}.
\section{Introduction}\label{sec1}
The electroweak form factors associated with the excitation of the
$\Delta(1232)$ resonance are of considerable interest to hadron structure
physicists. In the large $N_c$ limit, the $(N,\Delta)$
form a degenerate multiplet under spin-flavor SU(4) symmetry \cite{dashen},
and one
expects the structure of the lowest-lying spin-$1/2$ and spin-$3/2$ $qqq$
states
to be closely related. The electroweak transition form factors may provide
important insights into this relationship and shed light on QCD-inspired
models of the lowest lying baryons. These
form factors describe $N\to\Delta$ matrix elements of the vector and axial
vector currents \cite{jones,adl,sch}:
\begin{equation}\label{46}
<\Delta^+ (p')| V^3_\mu |N> =\bar \Delta^{+\nu}(p')
\{[{C_3^V \over M}\gamma^\lambda +{C_4^V \over M^2}p'^\lambda
+{C_5^V \over M^2}p^\lambda](q_\lambda g_{\mu\nu}-q_\nu g_{\lambda\mu})
+C_6^V g_{\mu\nu}\} \gamma_5 u(p)
\end{equation}
\begin{equation}\label{47}
<\Delta^+ (p')| A^3_\mu |N> =\bar \Delta^{+\nu}(p')
\{[{C_3^A \over M}\gamma^\lambda +{C_4^A \over M^2}p'^\lambda]
(q_\lambda g_{\mu\nu}-q_\nu
g_{\lambda\mu}) +C_5^A g_{\mu\nu}+{C_6^A\over M^2} q_\mu q_\nu\}  u(p)
\end{equation}
where the baryon spinors are defined in the usual way.
The form factors $C_3^V$ and $C_5^A$are the $N\to\Delta$ analogues of the
nucleon's
electroweak form factors $F_1$ and $G_A$.
At present, there exist considerable
data on the vector current transition form factors $C_i^V$ $(i=3-6)$
obtained with
electromagnetic
probes. A comparison with theoretical predictions points to significant
disagreement (see ref. \cite{muk}
for a tabulation of theoretical predictions).
For example, lattice QCD calculations of the magnetic transition form factor
yield a value $\sim 30\%$ smaller than obtained from experiment \cite{Lei92},
and constituent quark
models based on spin-flavor SU(6) symmetry similarly underpredict the
data\cite{HHM95}.
One hopes that additional input, in tandem with theoretical progress, will
help
identify the origin of these discrepancies.

The situation involving the axial vector transition form factors $C_i^A$
$(i=3-6)$ is less
clear than in the vector case, since existing data -- obtained from charged
current
experiments -- have considerably larger uncertainties than for the vector
current
channel. While QCD-inspired models tend to underpredict the central value
for
the axial matrix elements by $\sim 30\%$ as they do for the vector form
factors,
additional and more precise experimental information
is needed in order to make the test of
theory significant.
To that end, an extraction of the
axial vector $N\to\Delta$ matrix element using parity-violating electron
scattering (PVES) is planned at the Jefferson Laboratory \cite{proposal}.
The goal of this measurement is to perform a $\simle 25\%$ determination for
$|q^2|$ in the range of $0.1-0.6$ (GeV$/c)^2$. If successful, this
experiment
would considerably sharpen the present state of experimental knowledge of
the axial vector transition amplitude.

Here we examine the interpretation of the prospective
measurement.
In a previous work \cite{muk}, the impact of non-resonant backgrounds was
studied and found not to present a serious impediment to the
extraction of the $C_i^A$. Here, we compute the
electroweak radiative corrections, which arise from
${\cal O}(\alpha G_F)$ contributions to the PV axial transition amplitude.
We correspondingly
characterize the relative importance of the corrections by discussing the
ratio
$\RAd$ of the higher-order to tree-level amplitudes. This ratio  is
nominally ${\cal O}(\alpha)$, so that one might naively justify
neglecting radiative corrections when interpreting a 25\%
determination of the axial term.
However, previous work on the axial vector radiative corrections $\RAp$ to
PV elastic electron-proton scattering suggests that the relative importance
of such corrections can be both unexpectedly large
${\it and}$ theoretically uncertain
\cite{zhu,mike,mh}.
Moreover, results obtained by the SAMPLE collaboration \cite{sample} suggest
that $\RAp$ may be substantially larger than given by the best theoretical
estimate\cite{zhu}.
The origin of this apparent enhancement is presently not understood.
Were similar uncertainties to occur for PV electroexcitation of the
$\Delta$, the task of
extracting the desired axial transition form factors from the PV asymmetry
would
become considerably more complicated than assumed in the original
incarnation of the
experimental proposal.

In studying the axial vector radiative corrections, it is important to
distinguish
two classes of contributions. The first involves electroweak radiative
corrections to the elementary $V(e)\times A(q)$ amplitudes, where $q$ is
any one of the
quarks in the hadron and $V$ ($A$) denotes a vector (axial vector) current.
These terms, referred to henceforth as \lq\lq
one-quark" radiative corrections, are calculable in the Standard Model. For
elastic
scattering from the proton, they contain
little theoretical uncertainty apart from the gentle variation with Higgs
mass,
long-distance QCD effects involving light-quark loops in the $Z-\gamma$
mixing
tensor, and SU(3)-breaking effects in octet axial vector matrix elements
$\bra{p} A_{\lambda}^{(3,8)}\ket{p}$. Such one-quark contributions to
$\RAp$ and $\RAd$ can be
large, due to the
absence in loop terms of the small $(1-4\sstw)$ factor appearing in the tree
level
$V(e)$ coupling and the presence of large logarithms of the type $\ln
(m_q/\mz)$.

The second class of radiative corrections, which we refer to as \lq\lq
many-quark" corrections, involve weak interactions among quarks in the
hadron. In refs. \cite{zhu,mike,mh}, the many-quark corrections were shown
to
generate considerable theoretical uncertainty in the PV, axial vector $ep$
amplitude. A particularly important subset of these effects are associated
with
the nucleon anapole moment (AM), which constitutes the leading-order, PV
$\gamma NN$ coupling. The result of the SAMPLE measurements,
which combine
PV elastic $ep$ and quasielastic $ed$ scattering to isolate the isovector,
axial vector
$ep$ amplitude, implies that the one-quark/Standard Model plus
many-quark/anapole contributions
significantly underpredict the observed value of $\RAp$.

In what follows, we compute the analogous radiative corrections $\RAd$ for
the axial
$N\to\Delta$ electroexcitation amplitude. In principle,
as in the elastic case, the one-quark
corrections are determined completely by the Standard Model,
although long-distance QCD effects -- which
are finessed for the $ep$ channel using SU(3) symmetry plus nucleon and
hyperon $\beta$-decay
data -- are not controlled in the same manner for the $N\to\Delta$
transition. We
make no attempt to estimate the size of such effects here.
Instead, we focus on the
many-quark contributions which, as in the elastic case, can be
systematically organized
using chiral perturbation theory ($\chi$PT). We compute these corrections
through ${\cal O}(p^3)$. We find:

\begin{itemize}

\item [(i)] As in the case of $\RAp$, the correction $\RAd$ is
both substantial and theoretically uncertain. Thus, a proper interpretation
of
the PVES $N\to\Delta$ measurement {\em must} take into account ${\cal O}(\alpha
G_F)$
effects.

\item [(ii)] In contrast to the elastic PV asymmetry, the $N\to\Delta$
asymmetry
does not vanish at $q^2=0$. This result follows from the presence of an
${\cal O}(\alpha
G_F)$ contribution -- having no analog in the elastic channel -- generated
by a new PV $\gamma N\Delta$ electric dipole coupling
$d_\Delta$. Specifically, we show below that
\begin{equation}
\label{eq:photon1}
\alrd(q^2=0) \approx -{2d_\Delta\over C_3^V}{M_N\over\lamchi}+\cdots
\end{equation}
where $\alrd(q^2)$ is the PV  asymmetry on the $\Delta$
resonance, $\lamchi=4\pi F_\pi\sim 1$ GeV is the scale of chiral symmetry
breaking, $C_3^V\sim 2$ is the dominant $N\to \Delta$ vector transition
form factor,
$d_\Delta$ is a low-energy constant whose scale is set by hadronic weak
interactions, and the $+\cdots$ denote non-resonant, higher order chiral,
and $1/M_N$ corrections.

\item [(iii)] The experimental observation of
surprisingly large SU(3)-violating contributions to
hyperon radiative decays suggests that the effect
of $d_\Delta$ could be significantly enhanced over
its \lq\lq natural" scale, yielding an $N\to\Delta$ asymmetry $\sim 10^{-6}$
or larger at the photon point\footnote{For a PV photoproduction asymmetry of
this
magnitude,
a measurement using polarized photons at Jefferson Lab would be an
interesting -- and
potentially feasible\cite{CJ01} -- possibility. An analysis of the
 real $\gamma$ asymmetry appears in a separate communication
\cite{Zhunew}.}.

\item [(iv)] The presence of the PV $d_\Delta$ coupling implies that
the $q^2$-dependence of the axial vector transition amplitude
entering PV electroexcitation of the $\Delta$ could differ
significantly from the $q^2$-dependence of the
corresponding amplitude  probed with neutral current
neutrino excitation of
the $\Delta$. As we demonstrate below, it may be possible to separate the
$d_\Delta$
contribution from other effects by exploiting the unique $q^2$-dependence
associated with this new term. We illustrate this possibility by considering
a
low-$|q^2|$, forward angle asymmetry measurement.

\item [(v)]  An experimental separation of the $d_\Delta$ contribution
from the remaining terms in the axial vector response would be of interest
from
at least two standpoints. First, it would provide a unique window  -- in the
$\Delta S=0$ sector -- on the dynamics underlying the poorly understood PV
$\Delta
S=1$ radiative and nonleptonic decays. Second, it would help to remove a
significant
source of theoretical uncertainty in the interpretation of the $N\to\Delta$
asymmetry, thereby allowing one to extract the $N\to\Delta$ axial vector
form factors
with less ambiguity.

\item [(vi)] A comparison of PV electroexcitation of the $\Delta$
with  more precise, prospective neutrino excitation measurements
would be particularly interesting, as inelastic neutrino scattering
is insensitive to
the large $\gamma$-exchange
effects arising at ${\cal O}(\alpha G_F)$ which contribute to PV
electron scattering \cite{mike,mh}.

\end{itemize}

\begin{figure}
\epsfxsize=11.0cm
\centerline{\epsffile{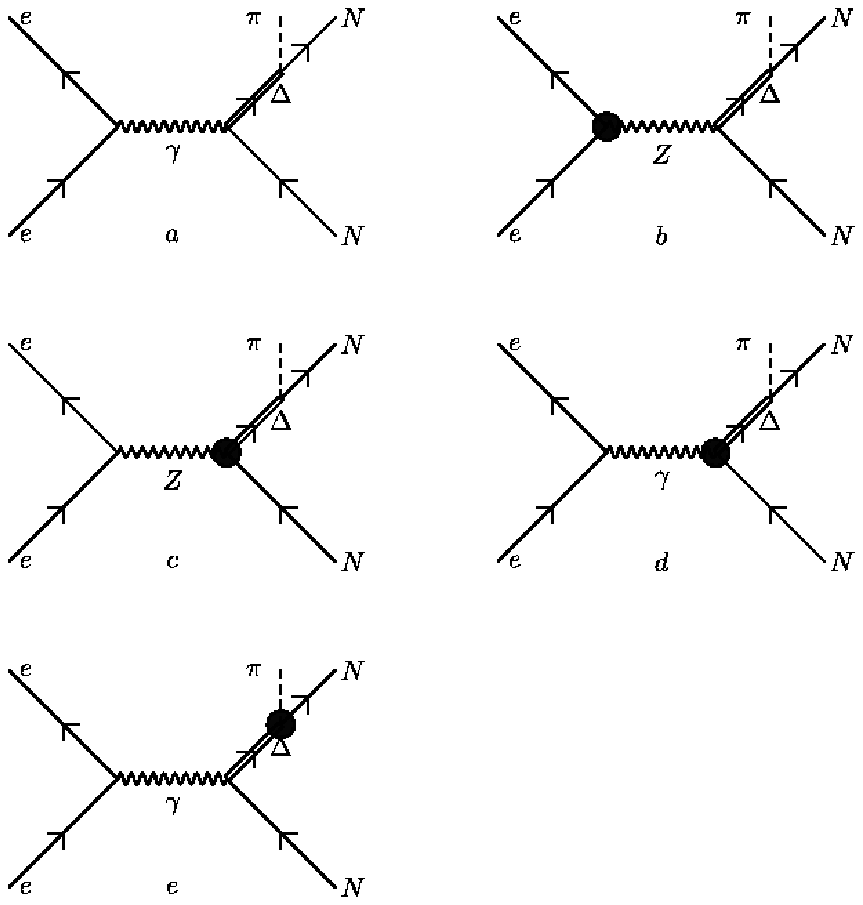}}
%\vspace{1cm}
\caption{
Feynman diagrams describing resonant pion electroproduction.
The dark circle indicates a parity violating coupling.
Fig. \ref{Fig.1}d gives transition anapole and Siegert's term contributions.
Fig. \ref{Fig.1}e leads to the PV d-wave $\pi N \Delta$ contribution.
}
\label{Fig.1}
\end{figure}

While the remainder of the paper is devoted to a detailed discussion of
these points, several aspects deserve further comment here.
First, the origin of the nonvanishing $\alrd(q^2=0)$ in 
Eq. (\ref{eq:photon1}) is readily
understood in terms of Siegert's theorem \cite{Siegert,friar}, familiar
in nonrelativistic nuclear physics. For electron scattering processes such
as shown in Fig. \ref{Fig.1}, the leading PV $\gamma$-hadron coupling (Fig.  \ref{Fig.1}d)
corresponds to
matrix elements of the transverse
electric multipole operator ${\hat T}^\sst{E}_{J=1\lambda}$, and according to
Siegert's Theorem,
matrix elements of this operator can be written in the form \footnote{We
adopt the \lq\lq extended" version of Siegert's theorem derived in ref.
\cite{friar}.}
%%%%%%%%%%%%%%%%%%%%%%%%%%% Fig 1 %%%%%%%%%%%%%%%%%%%%%%%%%%%%%
%%%%%%%%%%%%%%%%%%%%%%%%%%%%%%%%%%%%%%%%%%%%%%%%%%%%%%%
\begin{equation}
\label{eq:Siegert1}
\bra{f}{\hat T}^\sst{E}_{J=1\lambda}\ket{i} = -{\sqrt{2}\over
3}\omega\bra{f}
\int\ d^3x\ xY_{1\lambda}(\Omega){\hat\rho}(x)\ket{i} +{\cal O}(q^2)\ \ \ ,
\end{equation}
where the $\omega=E_f-E_i$. The leading component in
Eq. (\ref{eq:Siegert1}) is $q^2$-independent and proportional to $\omega$
times the
electric dipole matrix element. Up to overall numerical factors, this $E1$
matrix element is simply
$d_\Delta/\lamchi$. It does not contribute to PV elastic electron
scattering, for
which $\omega=0$.  The remaining terms of ${\cal O}(q^2)$ and higher contain
matrix
elements of the anapole operator \cite{hax89,mh}, which generally do not
vanish for
either elastic or inelastic scattering.  When
$\bra{f}{\hat T}^\sst{E}_{J=1\lambda}\ket{i}$ is inserted into the full
electron scattering amplitude, the $1/q^2$ from the photon propagator
cancels
the leading $q^2$ from the anapole term, yielding a $q^2$-independent
contact interaction. In contrast,
for inelastic processes such as electroexcitation of the $\Delta$,
$\omega=m_\Delta-\mn$ does not vanish, and the dipole matrix element
in Eq. (\ref{eq:Siegert1}) generates a contribution to the PV scattering
amplitude $M_\sst{PV}$ behaving as $1/q^2$ for low-$|q^2|$. Since the
parity-conserving (PC) amplitude $M_\sst{PC}$ -- whose interference with
$M_\sst{PV}$ gives rise to $\alrd$ -- also goes as $1/q^2$, the inelastic
asymmetry
does not vanish at the photon point. Henceforth, we refer to the dipole
contribution
to the asymmetry as $\alrds$, and the corresponding ${\cal O}(\alpha)$
correction
to the ${\cal O}(G_F)$ $Z^0$-exchange, axial vector neutral current
amplitude as
$\RAs$. We note that the importance of $\alrds$ --  relative to the anapole
and
$Z^0$-exchange contributions to the asymmetry -- increases as one approaches
the photon point, since the latter vanish for $q^2=0$.

It is straightforward to recast the foregoing discussion in a
covariant framework using effective chiral Lagrangians. The dipole term
in Eq.  (\ref{eq:Siegert1}) corresponds
to the operator \cite{zhu,CJ01b}
\begin{equation}
\label{eq:Siegert2}
{\cal L}^{\mbox{Siegert}} = i{e d_\Delta\over\lamchi}{\bar
\Delta^+}_\mu\gamma_\lambda p
F^{\mu\lambda}+{\mbox{H.c.}}
\end{equation}
while the transition anapole contribution arises from
the effective interaction
\begin{equation}
\label{eq:anapole1}
{\cal L}^{\mbox{anapole}}={e a_\Delta\over\lamchis}{\bar\Delta^+}_\mu
p{\partial}_\lambda
F^{\lambda\mu}+{\mbox{H.c.}}\ \ \ .
\end{equation}
\section{Electroexcitation: general features }
\label{sec2}
%%%%%%%%%%%%%%%%%%%%%%%%%%%%%%%%%%%%%%%%%%%%%%%%%%%%%%%%%%%%%%%%%%%%%%%%%%%%
%%%

\begin{figure}
\epsfxsize=12.0cm
\centerline{\epsffile{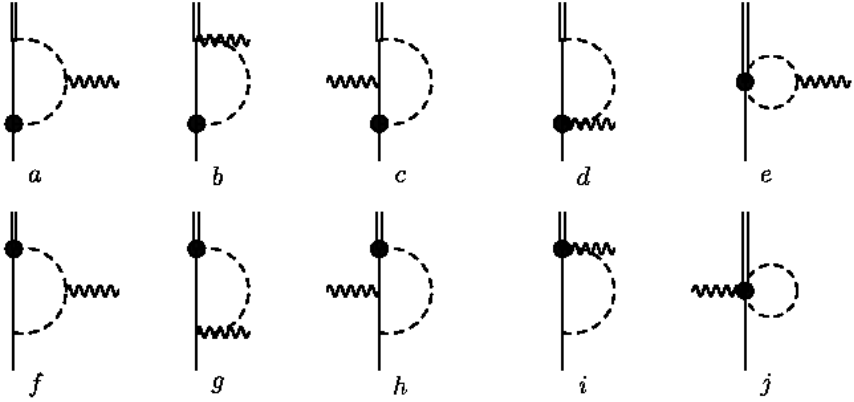}}
%\vspace{1cm}
\caption{
Meson-nucleon intermediate state contributions to the
$N\to\Delta$ transition anapole and Siegert couplings
$a_\Delta$ and $d_\Delta$, respectively. The shaded
circle denotes the PV vertex. The single solid,
double solid, dashed,  and curly lines correspond
to the $N$, $\Delta$, $\pi$, and $\gamma$,
respectively.
}
\label{Fig.2}
\end{figure}

\begin{figure}
\epsfxsize=10.0cm
\centerline{\epsffile{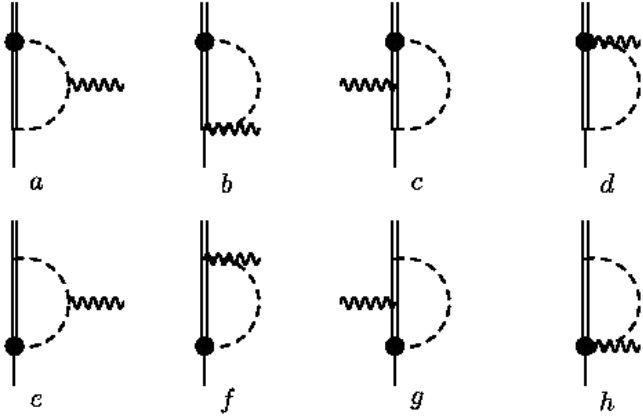}}
%\vspace{1cm}
\caption{
Same is Fig. \ref{Fig.2} but with $\Delta$-$\pi$ intermediate states.
}
\label{Fig.3}
\end{figure}

\begin{figure}
\epsfxsize=10.0cm
\centerline{\epsffile{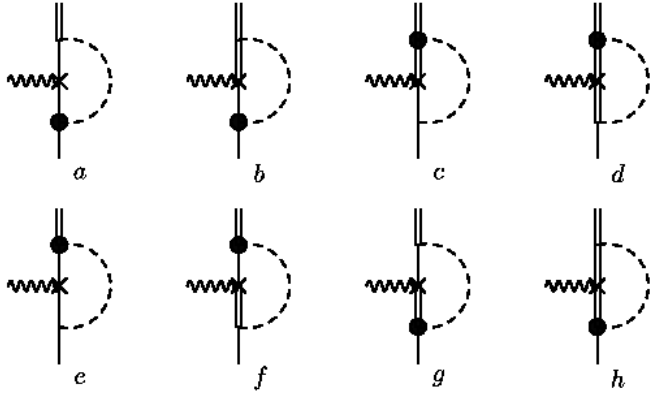}}
%\vspace{1cm}
\caption{
Same as Fig. \ref{Fig.2} but involving insertions of the
baryon magnetic moment operator,
denoted by the cross. }
\label{Fig.4}
\end{figure}

\begin{figure}
\epsfxsize=8.0cm
\centerline{\epsffile{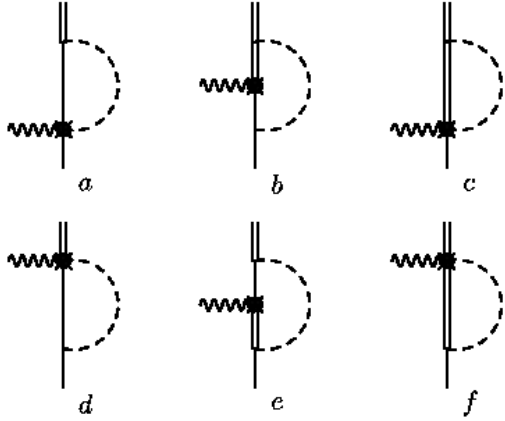}}
%\vspace{1cm}
\caption{
Same as Fig. 2 but with PV electromagnetic insertions,
denoted by the overlapping cross and shaded circle.
}
\label{Fig.5}
\end{figure}

The amplitudes relevant to PV electroexcitation of the $\Delta$ are shown in
Fig. \ref{Fig.1}. The asymmetry arises from the interference of the PC amplitude of
Fig. 1a with the PV amplitudes of Figs. \ref{Fig.1}b-e. In Fig. \ref{Fig.1}b-d, the shaded
circle
denotes an axial gauge boson (V)-fermion (f) coupling, while the remaining
V-f couplings are vector-like. In Fig. \ref{Fig.1}e, the shaded circle indicates the
PV
$N\Delta\pi$ d-wave vertex. All remaining $N\Delta\pi$ vertices in Fig. \ref{Fig.1}
involve
strong, PC couplings. In general, the interaction vertices of Fig. \ref{Fig.1} contain
loop effects as well as tree-level contributions. The loops relevant to the
PV interactions (up to the chiral order of our analysis) are shown in Figs.
\ref{Fig.2}-\ref{Fig.5}.
%%%%%%%%%%%%%%%%%%%%%%%%%%%%%%% Fig 2 %%%%%%%%%%%%%%%%
%%%%%%%%%%%%%%%%%%%%%%%%%%%%%%%%%%%%%%%%%%%%%%%%%%%%
The formalism for treating the contributions to $\alrd$ from Figs. \ref{Fig.1}a-c is
discussed in
detail in ref. \cite{muk}. Here, we review only those elements most germane
to
the discussion of electroweak radiative corrections. We also discuss
general features
of the new contributions from Figs. \ref{Fig.1}d and \ref{Fig.1}e not previously analyzed.

\medskip
\subsection{Kinematics and PV Asymmetry}
\medskip
We define the appropriate kinematic variables for the reaction
\begin{equation}
\label{eq16}e^{-}\left( k\right) +N\left( p\right) \rightarrow
e^{-\prime }(k^{\prime}) +\Delta \left( p_\Delta \right) \rightarrow
e^{-\prime }\left( k^{\prime }\right) +
        N^{\prime }\left( p^{\prime }\right) +\pi \left( p_\pi \right) ,
\end{equation}
In the laboratory frame one has
\begin{equation}
\label{eq17}s=\left( k+p\right) ^2,\quad q=p_\Delta -p=k-k^{\prime },\quad
p_\Delta =p^{\prime }+p_\pi ,
\end{equation}
where ${\bf p}=0$, and
\begin{equation}
\label{eq18}s=k^2+2k\cdot p+p^2=m^2+2M\epsilon+M^2,
\end{equation}
$\epsilon$ being the incoming electron energy, $m$ and $M=m_N$ being the
electron
and nucleon masses, respectively. One may relate the square of the four
momentum transfer
\begin{equation}
Q^2=|{\vec q}|^2-q_0^2
\end{equation}
to $s$ and the electron scattering angle $\theta$ as
\begin{equation}
\label{eq27}\sin^2\theta/2=\frac{M^2Q^2}{\left( s-M^2\right) \left(
s-M_\Delta ^2-Q^2\right) }.
\end{equation}
%%%%%%%%%%%%%%%% Fig 3 %%%%%%%%%%%%%%%%%%%%%%%%%
%%%%%%%%%%%%%%%%%%%%%%%%%%%%%%%%%%%%%%%%%%%%%%%%
The energy available in the nucleon-gauge boson ($\gamma$ or $Z^0$)
center of mass (CM) frame is $W\equiv\sqrt{p_\Delta^2}$ and
the energy of the gauge boson in the CM frame is
\begin{equation}
\label{kin1}
q_0\ =\ {{W^2-Q^2-M^2}\over{2W}}.
\end{equation}
%%%%%%%%%%%%%%%%%%%%%%%%%%% Figure 4
%%%%%%%%%%%%%%%%%%%%%%%%%%%%%%%%%%%%%%%%%%%%%
\medskip

\medskip
As shown in ref. \cite{muk}, one may distinguish three
separate dynamical contributions to the
PV asymmetry. Denoting these terms by
$\Delta^\pi_{(i)}$ ($i=1,\ldots , 3$), one has
\begin{equation}
\label{alr1}
A_{LR}={{N_+ - N_-}\over{N_++N_-}}=
\frac{-G_\mu}{\sqrt{2}}\frac{Q^2}{4\pi \alpha }\left[
\Delta^\pi_{(1)} + \Delta^\pi_{(2)}+\Delta^\pi_{(3)}\right] ,
\end{equation}
where $N_{+}$ ($N_{-}$) is the number of detected, scattered electrons
for an incident beam of positive (negative) helicity electrons, $\alpha$ is
the
electromagnetic fine structure constant, and
$G_\mu$ is the Fermi constant measured in $\mu$-decay. The
$\Delta^\pi_{(1,2)}$
contain the vector current response of the target, arising from the
interference
of the amplitudes in Figs. \ref{Fig.1}a,b, while the term $\Delta^\pi_{(3)}$contains the
axial vector response function,
generated by the interference of Figs. \ref{Fig.1}a and \ref{Fig.1}c-e.

The leading term, $\Delta^\pi_{(1)}$, is nominally independent
of the hadronic structure -- due to cancellations between the numerator and
denominator
of the asymmetry -- whereas $\Delta^\pi_{(2,3)}$ are sensitive to details of
the hadronic
transition amplitudes. Specifically, one has
\begin{equation}
\label{delpi1}\Delta^\pi_{(1)}=g_{A}^e{\xi_V^{T=1}} \ \ \ ,
\end{equation}
which includes the entire resonant hadronic vector current contribution to
the
asymmetry. Here, $g_{A}^e$ is the axial vector electron coupling to the
$Z^0$
and ${\xi_V^{T=1}}$ is the isovector hadron-$Z^0$ vector current
coupling \cite{mjrm,mus92a}:
\begin{equation}
g_A^e \xi_V^{T=1} = -2(C_{1u}-C_{1d})
\end{equation}
where the $C_{1q}$ are the standard $A(e)\times V(q)$ couplings in the
effective
four fermion
low-energy Lagrangian \cite{pdg}. At tree level,
$g_A^e\xi_V^{T=1}=2(1-2\sstw)\approx
1$. Vector current conservation and the approximate isospin symmetry of the
light
baryon spectrum protects $\Delta^\pi_{(1)}$ from receiving large and
theoretically
uncertain QCD corrections. In
principle, then, isolation
of $\Delta^\pi_{(1)}$ could provide a test of fundamental electroweak
couplings. As shown
in ref. \cite{muk}, however, theoretical uncertainties associated with the
non-resonant
background contribution $\Delta^\pi_{(2)}$ and axial vector contribution
$\Delta^\pi_{(3)}$
would likely render such a program unfeasible.
%%%%%%%%%%%%%%%%%%%%%%%%%%%%%% Fig 5
%%%%%%%%%%%%%%%%%%%%%%%%%%%%%%%%%%%%%%%%%%%%%%%%%%%%%%%%%%
The interest for the Jefferson Lab measurement\cite{proposal} lies in the
form factor
content of the axial vector
contribution $\Delta^\pi_{(3)}$.
For our purposes, it is useful to distinguish between
the various contributions to this response according to the amplitudes of
Fig. \ref{Fig.1}. From the interference of Figs. \ref{Fig.1}a and \ref{Fig.1}c
we obtain the axial vector neutral current response:
\begin{equation}
\label{eq:delta3}
\Delta^\pi_{(3)}({\mbox{NC}}) \approx g_V^e\xi_A^{T=1} F(Q^2,s)\ \ \ ,
\end{equation}
where
\begin{equation}
\label{eq:xia1}
g_V^e\xi_A^{T=1} = -2(C_{2u}-C_{2d})
\end{equation}
in the absence of target-dependent, QCD contributions to the one-quark
electroweak radiative corrections. The $C_{2q}$ are the $V(e)\times A(q)$
analogues of the $C_{1q}$ \cite{pdg}, while
the function $F(Q^2,s)$ gives the
dependence of
$\Delta^\pi_{(3)}({\mbox{NC}})$ on the axial couplings
$C_i^A$. Following ref. \cite{muk} we obtain
\begin{equation}
\label{60'}
 F(Q^2, s) =
\frac{C_5^A}{C_3^V}
\left[ 1+\frac{M_\Delta ^2-Q^2-M^2}{2 M^2}
\frac{C_4^A}{C_5^A} +{q_0+W-M\over 2M} {C_3^A\over C_5^A}\right] {\cal
P}\left(
Q^2,s\right) ,
\end{equation}
where
\begin{equation}
\label{61'}
{\cal P}\left(Q^2,s\right) =
\frac{MM_\Delta \left( \left(s-M^2\right) +
  \left( s-M_\Delta ^2\right) -Q^2\right) }
{\frac 1 2\left(Q^2+\left( M_\Delta +M\right) ^2\right)
   \left(Q^2+\left( M_\Delta -M\right) ^2\right)
 +\left( s-M^2\right) \left( s-M_\Delta ^2\right) -Q^2s}\ \ \ .
\end{equation}
In arriving at Eqs. (\ref{eq:delta3}-\ref{61'}) we have included only
resonant contributions
from the $\Delta$. Non-resonant background effects have been analyzed in
refs. \cite{muk,ham}. Note that $F(Q^2,s)$ is a frame-dependent quantity,
depending as it does on $q^0$. However, for simplicity of notation, we have
suppressed
the $q^0$-dependence in the list of the arguments.

The interference of Figs. \ref{Fig.1}a and \ref{Fig.1}d generates the transition anapole and
Siegert contributions associated
with the interactions of Eqs. (\ref{eq:Siegert2},\ref{eq:anapole1}):
\begin{equation}
\label{eq:xia2}
\Delta^\pi_{(3)}({\mbox{Siegert}})+\Delta^\pi_{(3)}({\mbox{anapole}})\ \ \ ,
\end{equation}
while the interference of Figs. \ref{Fig.1}a and \ref{Fig.1}e generates the response associated
with the
PV $N\Delta\pi$ d-wave interaction:
\begin{equation}
\Delta^\pi_{(3)}({\mbox{d-wave}})\ \ \ .
\end{equation}
From the total contribution
\begin{equation}
\Delta^\pi_{(3)}({\mbox{TOT}})=\Delta^\pi_{(3)}({\mbox{NC}})+\Delta^\pi_{(3)
}
({\mbox{Siegert}})+\Delta^\pi_{(3)}({\mbox{anapole}})+
\Delta^\pi_{(3)}({\mbox{d-wave}})
\end{equation}
we may define the overall ${\cal O}(\alpha)$ correction $\RAd$ to the ${\cal
O}(G_F)$
axial response via
\begin{equation}
\label{eq:radcor1}
\Delta^\pi_{(3)}({\mbox{TOT}})=2(1-4\sstwo)(1+\RAd) F(Q^2,s)\ \ \
\end{equation}
where $\theta_\sst{W}^0$ is the weak mixing angle at tree-level in the
Standard Model:
\begin{equation}
\label{eq:sstwdef}
\sstwo(1-\sstwo) = {\pi\alpha\over\sqrt{2} G_\mu\mzs} \ \ \ ,
\end{equation}
or
\begin{equation}
\sstwo = 0.21215\pm 0.00002\ \ \ .
\end{equation}
One may decompose the ${\cal O}(\alpha)$ effects described by $\RAd$
according to several sources:
\begin{equation}
\label{eq:xia3}
\RAd=\RAewk+\RAs+\RAa+\RAdw+\cdots \ \ \ ,
\end{equation}
where the $+\cdots$ indicate possible contributions from other
many-quark and QCD effects not included here.
The quantity $\RAewk$ denotes the one-quark radiative corrections,
\begin{equation}
\label{eq:raewkdef}
\RAewk={C_{2u}-C_{2d}\over C_{2u}^0-C_{2d}^0}-1
\end{equation}
with the superscript \lq\lq 0" denoting the tree-level values of the $C_{2q}$.
The correction $\RAewk$ denotes both the effects of ${\cal O}(\alpha)$
corrections
to the relation in Eq. (\ref{eq:sstwdef}) as well as the ${\cal O}(\alpha
G_F)$ contributions
to the neutral current $e$-$q$ amplitude. While the tree-level weak mixing
angle is
renormalization scheme-independent, both $\sstw$ and the correction
$\RAewk$ depend on
the choice of renormalization scheme. In what follows, we quote results for
both the
on-shell renormalization (OSR) and ${\overline{\mbox{MS}}}$ schemes. Note
that our
convention for the $R_\sst{A}^{(k)}$ differs from the convention adopted in
our
earlier work of ref. \cite{zhu}, where we normalized to the scheme-dependent
quantity $1-4\sstw$.

The remaining corrections are defined by
\begin{eqnarray}
\RAs&=&\Delta^\pi_{(3)}({\mbox{Siegert}})/\Delta^\pi_{(3)}({\mbox{NC}})^0\\
R_A^{\mbox{anapole}}&=&\Delta^\pi_{(3)}({\mbox{anapole}})/\Delta^\pi_{(3)}({
\mbox{NC}})^0\\
R_A^{\mbox{d-wave}}&=&\Delta^\pi_{(3)}({\mbox{d-wave}})/\Delta^\pi_{(3)}
({\mbox{NC}})^0\ \ \ ,
\end{eqnarray}
where the \lq\lq 0" denotes the value of the NC contribution
at tree-level.

\medskip
\section{ Electroweak radiative corrections}\label{sec:electroweakrc}

The parity violating amplitude for the process ${\vec e} p\to e \Delta$ is
generated by the diagrams in Fig. \ref{Fig.1}b-e. At tree-level in the Standard
Model,
one has
\begin{equation}
iM^\sst{PV} = iM^\sst{PV}_\sst{AV} + iM^\sst{PV}_\sst{VA}\ \ \ ,
\end{equation}
where
\begin{equation}\label{a}
iM^\sst{PV}_\sst{AV}= i{G_\mu \over 2\sqrt{2}} l^{\lambda 5} < \Delta
|J_\lambda
|N>
\end{equation}
from Fig. \ref{Fig.1}b and
\begin{eqnarray}\label{b}
iM^\sst{PV}_\sst{VA}&=& i{G_\mu \over 2\sqrt{2}} l^\lambda
< \Delta |J_{\lambda 5}|N>   \; .
\end{eqnarray}
from Fig. \ref{Fig.1}c.
Here, $J_\lambda$ ($J_{\lambda 5}$) and $ l_\lambda$ ($l_{\lambda 5}$)
denote the vector
(axial vector) weak neutral currents of the quarks and electron,
respectively \cite{mjrm}. Note that the vector leptonic weak neutral
current
contains the factor $g_V^e=(-1+4\sstw)\approx -0.1$, which
strongly suppresses the
leading order $Z^0$-exchange amplitude of Fig. \ref{Fig.1}c.

The interactions given in Eqs. (\ref{eq:Siegert2},\ref{eq:anapole1})
generate additional contributions to
$M^\sst{PV}_\sst{VA}$  when a photon is
exchanged between the nucleon and the electron (Fig. \ref{Fig.1}d). The
corresponding
amplitudes are
\begin{eqnarray}
\label{eq:mSiegert}
iM^\sst{PV}_{\mbox{Siegert}} & = & -i{(4\pi\alpha)d_\Delta\over Q^2\lamchi}
{\bar e}\gamma_\mu e{\bar\Delta_\nu}\left[(M-M_\Delta)
g^{\mu\nu}-q^\nu\gamma^\mu\right]N\\
\label{eq:manapole}
iM^\sst{PV}_{\mbox{anapole}} & =& i{(4\pi\alpha)a_\Delta\over\lamchis}
{\bar e}\gamma^\mu
e{\bar \Delta_\mu} N \  \ \ \ .
\end{eqnarray}
We note that, unlike $M^\sst{PV}_\sst{VA}$, the amplitudes in Eqs.
(\ref{eq:mSiegert})
and (\ref{eq:manapole})
contain no $(1-4\sstw)$
suppression. Consequently,
the relative importance of the PV $\gamma$-exchange many-quark amplitudes is
enhanced by $1/(1-4\sstw)\sim 10$ relative to the leading order neutral
current
amplitude.

The constants $d_\Delta$ and $a_{\Delta}$ contain contributions from loops
(L) generated by the PV 
heavy baryon chiral Lagrangians (see ref.~\cite{zhuu})  and from counterterms (CT) in
the tree-level effective Lagrangian of Eqs.
(\ref{eq:Siegert2},\ref{eq:anapole1}):
\begin{eqnarray}
d_{\Delta} &= & d_{\Delta}^\sst{L} + d_{\Delta}^\sst{CT}\\
a_{\Delta} &= & a_{\Delta}^\sst{L} + a_{\Delta}^\sst{CT}\ \ \ .
\end{eqnarray}
In HB$\chi$PT, only the parts of the loop amplitudes non-analytic in quark
masses $m_q$ can be unambigously
indentified with $d_\Delta^\sst{L}$ and $a_{\Delta}^\sst{L}$. Contributions
analytic in the $m_q$ have the same form as operators appearing the
effective
chiral Lagrangian, and since the latter carry {\em a priori} unknown
coefficients
which must be fit to experimental data, one has no way to distinguish their
effects from loop contributions analytic in $m_q$. Consequently, all
remaining analytic terms may be incorporated
into $d_\Delta^\sst{CT}$ and $a_{\Delta}^\sst{CT}$.  In principle,
$d_{\Delta}^\sst{CT}$ and $a_{\Delta}^\sst{CT}$ should be
determined from experiment. At present, however, we know of no independent
determination of these constants to use as input in predicting $\RAd$, so
we rely on model estimates for this purpose.

The structure arising from the PV d-wave amplitude (Fig. \ref{Fig.1}e) is considerably
more complex than those associated with Figs. \ref{Fig.1}b-d, and it will not be  presented here (see ref.~\cite{zhuu} for more details). We mention,however, that the amplitude of Fig. \ref{Fig.1}e --
like
its partners in Fig. \ref{Fig.1}d -- does not contain the $1-4\sstw$ suppression
factor
associated with the ${\cal O}(G_F)$ amplitude of Fig. \ref{Fig.1}c.

For future reference, it is useful to give expressions for the various
contributions to $\Delta^\pi_{(3)}$ as well as the corresponding
contributions
to $\RAd$ and the total asymmetry $\alrd$. For the response function, we
have
\begin{eqnarray}
\Delta^\pi_{(3)}({\mbox{Siegert}})& = & {8\sqrt{2}\pi\alpha\over G_\mu
Q^2}{d_\Delta
\over C_3^V}\left[{q_0+W-M_N\over 2\lamchi}\right] {\cal
P}\left(Q^2,s\right) \\
\Delta^\pi_{(3)}({\mbox{anapole}})& = & -{8\sqrt{2}\pi\alpha\over G_\mu
\lamchis}
{a_\Delta\over C_3^V} {\cal P}\left(Q^2,s\right)\\
\Delta^\pi_{(3)}({\mbox{d-wave}})& = & -{8\sqrt{2}\pi\alpha\over G_\mu
\lamchis} \left[{\lamchi\over M_\Delta+M_N}\right] {f_{N\Delta\pi}\over g_{\pi
N\Delta}} H(Q^2,s){\cal P}\left(Q^2,s\right)\ \ \ .\nn \\
\end{eqnarray}
The appearance of ${\cal P}\left(Q^2,s\right)$ results from the different
kinematic
dependences generated by the transverse PC and axial vector PV contributions
to the
electroexcitation asymmetry\cite{muk,mjrm}. The function $H(Q^2,s)$ is a
gently
varying function of $Q^2$,  and numerical calculation over the kinematics of  the Jefferson Lab maesurement shows that
\bga
|H(Q^2,s)| &<& 0.1\, .
\ea

The corresponding radiative corrections are
\begin{eqnarray}
\label{eq:rasscale}
\RAs & = & {8\sqrt{2}\pi\alpha\over
G_\mu\lamchis}{1\over 1-4\sstwo}{d_\Delta\over
2C^A_5}{\lamchis\over Q^2}{q_0+W-M\over 2\lamchi} f(Q^2,s)^{-1}\\
\label{eq:raascale}
\RAa & =&  -{8\sqrt{2}\pi\alpha\over
G_\mu\lamchis}{1\over 1-4\sstwo}{a_\Delta\over
2C^A_5} f(Q^2,s)^{-1}\\
\RAdw&=& - {4\sqrt{2}\pi\alpha\over
G_\mu\lamchis}{1\over 1-4\sstwo}{\Lambda_\chi\over m_\Delta +m_N}
{f_{N\Delta\pi}\over g_{\pi N\Delta}} {C_3^V\over C_5^A} H(Q^2, s)
 f(Q^2,s)^{-1}\ \ \ ,\nn \\
\end{eqnarray}
where
\begin{equation}
f(Q^2,s) = 1+\frac{M_\Delta ^2-Q^2-M^2}{2 M^2}
\frac{C_4^A}{C_5^A} +{q_0+W-M\over 2M} {C_3^A\over C_5^A}\sim 1 \ \ \  .
\end{equation}
In order to set the overall scale of $\RAs$, $\RAa$, and $\RAdw$, we follow
ref.
\cite{zhu} and
set $d_\Delta\sim a_\Delta\sim f_{N\Delta\pi}\sim g_\pi$,
where $g_\pi=3.8\times 10^{-8}$ is the
\lq\lq natural" scale for charged current hadronic PV effects
\cite{ddh,zhu2}.
Using $C_5^A\sim 1$, $C_3^V/C_5^A\sim 1.6$, $g_{\pi N\Delta}\sim 1$,
$f(Q^2,s)\sim 1$ and
$H(Q^2, s)\sim 0.1$, we obtain
\begin{eqnarray}
\RAs & \sim & 0.0041 \ (\lamchis/Q^2) \\
\RAa & \sim & -0.0041 \\
\RAdw & \sim & -0.0002
\ \ \ .
\end{eqnarray}
As we show below, $\RAa$ may be significantly enhanced over this general
scale.

Finally, the total contribution to the asymmetry from the various response
functions is given by
\begin{eqnarray}\nonumber
\alrd[\Delta^\pi_{(1)}] & = & {G_\mu Q^2\over
4\sqrt{2}\pi\alpha}2(C_{1u}-C_{1d})\\
&\approx& -9\times 10^{-5} [Q^2/({\mbox{GeV}}/c)^2] \\ \nonumber
\alrd[\Delta^\pi_{(3)}({\mbox{NC}})] & = & {G_\mu Q^2\over
4\sqrt{2}\pi\alpha}2(C_{2u}-C_{2d}) F(Q^2,s) \\
&\approx& -6.3\times10^{-6} F(Q^2,s) [Q^2/({\mbox{GeV}}/c)^2]\\ \nonumber
\alrd[\Delta_\pi^{(3)}({\mbox{Siegert}})] & = & -{2 d_\Delta\over C_3^V}
{\delta\over\lamchi} {\cal
P}(Q^2,s)\\ &\approx & -2\times 10^{-8}\ \left[ {d_\Delta/g_\pi\over
C_3^V}\right] {\cal P}(Q^2,s)\\ \nonumber
\alrd[\Delta_\pi^{(3)}({\mbox{anapole}})] & = & {2 a_\Delta\over C_3^V}
{Q^2\over\lamchis} {\cal P}(Q^2,s)\\
&\approx & 2.8\times 10^{-8} \left[ {a_\Delta/g_\pi\over C_3^V}\right]
{\cal P}(Q^2,s)
[Q^2/({\mbox{GeV}}/c)^2]\\ \nonumber
\alrd[\Delta^\pi_{(3)}({\mbox{d-wave}})]
&=& {f_{N\Delta\pi}\over
g_{\pi N\Delta}}
H(Q^2, s) {\cal P}(Q^2, s) {2Q^2\over \Lambda_\chi (m_\Delta +m_N)}\\
& \approx & 3.0\times 10^{-8}
[{f_{N\Delta\pi}/g_\pi\over g_{\pi N\Delta}}]
H(Q^2, s) {\cal P}(Q^2,s) [Q^2/({\mbox{GeV}}/c)^2]
\ \ \ .\nn
\end{eqnarray}

At this point, using the chiral Lagrangian describing the PV $\gamma N \Delta$ transition~\cite{zhuu}, one could compute the contributions to $a_\Delta$ and $d_\Delta$ generated by the loops of Figs.(\ref{Fig.2}-\ref{Fig.5}). Loop corrections to the PV $\pi N\Delta$ d-wave interaction contribute
at higher order than considered here, so we do not discuss them explicitly. The final result we get by summing up all the possible contributions from the different diagrams is   
\begin{eqnarray}
\label{eq:adtotal}\nonumber
a_\Delta^L(TOT) &=&-{\sqrt{3}\over 6\pi} g_{\pi N\Delta} h_{\pi}
\left[{\Lambda_\chi\over m_\pi } F_0^N -{1\over 24}{\Lambda_\chi\over m_N}
G_0 +{\Lambda_\chi\over m_N}F_1^N
+{\Lambda_\chi\over m_N}{\delta\over m_\pi}F_2^N\right] \\ \nonumber
&&+{\sqrt{3}\over 18\pi} g_{\pi N\Delta} h_{\Delta}
\left[{\Lambda_\chi\over m_\pi } F_0^\Delta -{11\over 24}{\Lambda_\chi\over
m_N}
G_0 -{\Lambda_\chi\over m_N}
({\delta^2\over m_\pi^2}-1)F_1^\Delta\right]\\  \nonumber
&&+{\sqrt{6}\over 36}g_{\pi N\Delta} (h_V^0+{4\over 3}
h_V^2) G_0\\ \nonumber
&&+{1\over 6}g_{\pi N\Delta} ({h_V^{\Delta^+
\Delta^0}\over
\sqrt{3}} +h_V^{\Delta^{++} \Delta^+})
G_0\\
&&-{1\over
6}(h_A^{p\Delta^+\pi^+\pi^-}-h_A^{p\Delta^+\pi^-\pi^+})
G_0
\end{eqnarray}

\begin{eqnarray}
\label{eq:ddtotal}\nonumber
d_\Delta^L(TOT)&=& -{\sqrt{3}\over 3\pi}h_\pi g_{\pi N\Delta}
\Bigl[{1\over 4} G_0
+{\delta\over m_\pi} F_3^N +{m_\pi\over m_N}F_4^N\\ \nonumber
&&+ {\pi\over 2} {m_\pi\over m_N}-{\delta\over 2m_N}G_0
 -{\delta^2\over m_Nm_\pi}F_5^N\Bigr]\\
&& -{\sqrt{3}\over 9\pi}h_\Delta g_{\pi N\Delta}
\left[{1\over 4} G_0-{\delta\over m_\pi} F_3^\Delta
- {\delta^2-m_\pi^2\over m_N m_\pi}F_4^\Delta
\right]\; .
\end{eqnarray}
As we mention previously, the constants $a_\Delta$ and $d_\Delta$ consist of two pieces, the loop contributions $a_\Delta^L$ and $d_\Delta^L$, and the counter term pieces $a_\Delta^{CT}$ and $d_\Delta^{CT}$, which are unknown {\it a priori} and should be determined from experiments. Also the low energy constant $f_{N\Delta \pi}$ is unknown and should be extracted from data. If one uses naive dimensional analysis (NDA), it can be shown that all three quantities, $a_\Delta^{CT}$, $d_\Delta^{CT}$  and $f_{N \Delta \pi}$, are ${\cal O} \lp g_\pi \rp$. Using vector meson dominance for $a_\Delta^{CT}$  and resonance saturation method for $d_\Delta^{CT}$ and $f_{N \Delta \pi}$, one could find a reasonable range for this quantities which might significantly deviate from the value obtained by using NDA~\cite{zhuu}. The results are summarized in table \ref{tab4}
\begin{table}
\begin{center}~
\begin{tabular}{|c||c|c|}\hline
\hbox{Coupling} & Best value & Reasonable range
\\\hline\hline
$|d_\Delta^{CT}(RES)|$  & $25 g_\pi$ & $0 \to 100 g_\pi$  \\
$a_\Delta^{CT}(VMD)$  &  $15 g_\pi$ & $(-15\to 70) g_\pi$  \\
$|f_{N\Delta\pi}|$ & $4 g_\pi$& $0 \to 16 g_\pi$\\
\hline
\end{tabular}
\end{center}
%\vspace{0.5cm}
\caption{\label{tab4}
Best values and reasonable ranges for $d_\Delta^{CT}$, $a_\Delta^{CT}$.}
\end{table}
\section{The scale of radiative corrections}
\label{sec9}
%\setcounter{equation}{0}
%%%%%%%%%%%%%%%%%%%%%%%%%%%%%%%%%%%%%%%%%%%%%%%%%%%%%%%%%%%%%%%%%%%%

In the absence of target-dependent QCD effects, the ${\cal O}(\alpha G_F)$
contributions to $\Delta_\pi^{(3)}$ are determined entirely by the one-quark
corrections $\RAewk$ as defined in Eq. (\ref{eq:raewkdef}). As noted above,
$\RAewk$ incorporates the effects of both the ${\cal O}(\alpha)$ corrections
to the definition of the weak mixing angle in Eq. (\ref{eq:sstwdef}) as well
as the ${\cal O}(\alpha G_F)$ contributions to the elementary $e$-$q$
neutral
current amplitudes. The precise value of $\RAewk$ is renormalization
scheme-dependent,
due to the truncation of the perturbation series at ${\cal O}(\alpha G_F)$.
In Table
\ref{tab55}, we give the values of $\sstw$, $-2(C_{2u}-C_{2d})$, and
$\RAewk$ in the
OSR and ${\overline{\mbox{MS}}}$ schemes. We note
that the impact of the ${\cal O}(\alpha)$ one-quark corrections to the
tree-level
amplitude
is already significant, decreasing its value by $\sim 50\%$. As noted in
Sec. \ref{sec1},
this sizable suppression results from the absence in various loops of the
$1-4\sstw$ factor appearing at tree-level, the appearance of large
logarithms
of the type $\ln m_q/\mz$, and the shift in $\sstw$ from its tree-level
value\footnote{At this order, the scheme-dependence
introduces a 10\% variation in the amplitude, owing to the omission of
higher-order (two-loop and beyond) effects.}.

In discussing the impact of many-quark corrections, it is useful to
consider a number of
perspectives. First, we compare the relative importance of the one- and
many-quark corrections
by studying the ratios $R_\sst{A}^{(i)}$. Using the results previously obtained,
we derive numerical expressions for these ratios in terms of the various
low-energy
constants. For the relevant input parameters we use
current amplitude $1-4\sstwo$, $g_A=1.267\pm0.004$ \cite{pdg},
$g_{\pi N \Delta} =1.05$ \cite{hhk}, $G_\mu =1.166 \times 10^{-5}$
GeV$^{-2}$,
$\delta =0.3$ GeV, $\mu =\Lambda_\chi =1.16$ GeV,
$f_\rho=5.26$, $f_\omega =17$ \cite{sakurai}, $g_\pi =3.8 \times 10^{-8}$,
$C_5^A=0.87$ and $C_3^V=1.39$ \cite{HHM95}. It is worthwhile mentioning that
$2C_5^A$ is normalized such that this factor becomes $g_A$ for
polarized $e p$ scattering. We find then
\begin{eqnarray}\nonumber
\RAa&=&0.01\times {1.74\over 2C_5^A}\times\{ -0.04 h_\pi -0.07 h_V
+0.006h_\Delta
-0.18 h_V^\Delta  \\
\label{num}
&&+0.17 h_A^{N\Delta\pi\pi}
+0.09 |h_{\Delta N\rho}^0 +h_{\Delta N\rho}^1-h_{\Delta N\rho}^{'1}|
+0.025|h_{\Delta N\omega}^1|\}\nn\\
&&\\
\RAs&=&0.01\times {1.74\over 2C_5^A}\times \Bigl[ 0.83 d^{CT}_\Delta-
0.09 h_\pi -0.03 h_\Delta \Bigr]\nn \\
&\times& {0.1\hbox{GeV}^2\over |q^2|} {q_0+W-M\over
0.6\hbox{GeV}} \\
\RAdw&=&0.00105\times f_{N\Delta\pi}\times (C_3^V/C_5^A) \times H(Q^2,s)
\end{eqnarray}
where
\begin{eqnarray}
h_V & = &h_V^0+{4\over 3}h_V^2 \\
h_V^\Delta & = & {h_V^{\Delta^+\Delta^0}\over \sqrt{3}}
+h_V^{\Delta^{++}\Delta^+}\\
h_A^{N\Delta\pi\pi} & = &
h_A^{p\Delta^+\pi^+\pi^-}-h_A^{p\Delta^+\pi^-\pi^+}
\end{eqnarray}
and where all PV couplings are in unit of $g_\pi$ and $|H(Q^2,s)|<0.1$.
\begin{table}
\begin{center}~
\begin{tabular}{|c||c|c|c|}\hline
\hbox{Scheme} & $\sstw$ & $-2(C_{2u}-C_{2d})$ & $\RAewk$
\\\hline\hline
Tree Level & $0.21215\pm 0.00002$ & $0.3028$& 0 \\
OSR  & $0.22288\pm 0.00034$ & 0.1404 & $-0.536 $ \\
${\overline{\mbox{MS}}}$  & $0.23117\pm 0.00016$ & 0.1246 & $-0.589$ \\
\hline
\end{tabular}
\end{center}
%\vspace{0.5cm}
\caption{\label{tab55}
Weak mixing angle and one-quark ${\cal O}(\alpha G_F)$ contributions to
isovector axial transition current. }
\end{table}
The expressions in Eqs. (\ref{num}) illustrate the sensitivity of the
radiative
corrections to the various PV hadronic couplings. As expected on general
grounds,
the overall size of the $R_\sst{A}^{(i)}$ is about one percent when the PV
couplings
assume their \lq\lq natural" scale (NDA). The relative importance of the
Siegert's
term correction, however, grows rapidly when $Q^2$ falls below $\sim 0.1$
(GeV$/c)^2$.
The hadron resonance models we have used to estimate the size of the counter terms, may yield significant
enhancements
of the $R_\sst{A}^{(i)}$ beyond the naive dimensional analysis scale. To obtain a range of values
for the
corrections, we list in Table \ref{tab5} the available theoretical estimates for the
PV
constants, including both the estimates given above as well as those
appearing in
refs. \cite{fcdh,ddh}. We observe that the couplings $h_A^i$, $h_V^i$,
$d_\Delta$ and
$h^i_{\Delta N\rho}$ are weighted heavily in the expressions of Eqs.
(\ref{num}).
At present, these couplings are unconstrained by
conventional analyses of hadronic PV and there exist no model estimates
for $h_A^i$ and $h_V^i$. Consequently, we allow the various combinations
of
these constants appearing in Eq. (\ref{num}) to range between $10 g_\pi$ and
$-10 g_\pi$,
using $g_\pi$ as a reasonable guess for their best values.

\begin{figure}
\epsfxsize=10.0cm
\centerline{\epsffile{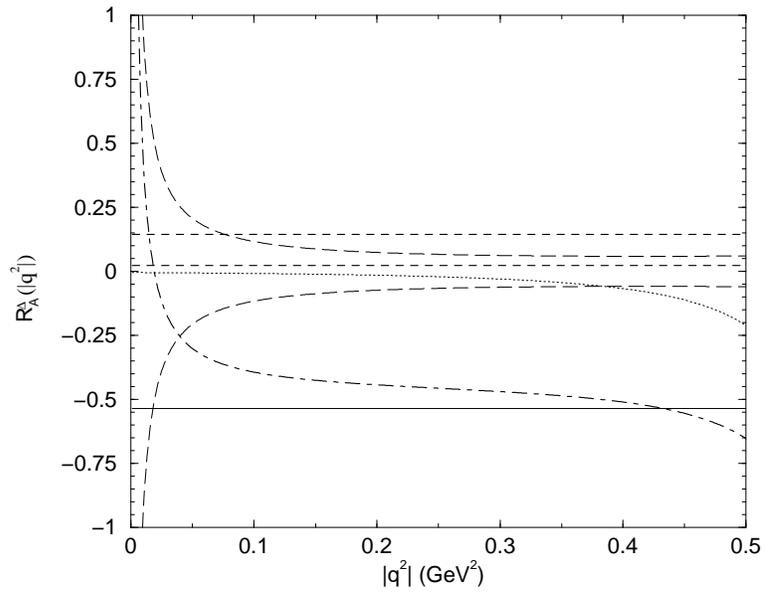}}
%\vspace{1cm}
\caption{
Contributions to the electroweak radiative correction
$R_A^\Delta$ at beam energy $0.424$ GeV. The short-dashed lines
show the upper and lower bounds of the reasonable range for anapole
contribution.
The solid line is the one-quark contribution. The  upper (lower) long-dashed
line
is the Siegert term with $d_\Delta= 25 g_\pi\ (-25 g_\pi)$.
The dotted line is the d-wave contribution.
}
\label{Fig.8}
\end{figure}

The resulting values for the $R_\sst{A}^{(i)}$ are shown in Table \ref{tab6}
and
Fig. \ref{Fig.8}.
%%%%%%%%%%%%%%%%%%%%%%%%%%%%%%%%% Figure 8
%%%%%%%%%%%%%%%%%%%%%%%%%%%%%%%%%%%%%%%%%%%%%%%%%%%%%%%%5
For the ratio $\RAs$, we quote results for two overall signs $(\pm)$ for
$d_\Delta$, since
at present the overall phase is uncertain. From both Table \ref{tab6} and
Fig. \ref{Fig.8}
we observe
that the importance of the many-quark corrections can be significant in
comparison to the
one-quark effects $\RAewk$. Moreover, the theoretical {\em uncertainty},
resulting from the
reasonable ranges for the PV parameters in Table \ref{tab5}, can be as large
as
$\RAewk$ itself. It is
conceivable that the total correction $\RAd$ could be as much as $\pm 1$
near the lower
end of the kinematic range for the Jefferson Lab $N\to\Delta$ measurement.
While this
result may seem surprising at first glance, one should keep in mind that the
${\cal O}(\alpha G_F)$ one-quark effects already yield a 50\% reduction in
the tree-level
axial amplitude, while the absence of the leading factor of
$Q^2$ in the Siegert
contribution
to $\alrd$ enhances the effect of the unknown constant $d_\Delta$ for
low momentum transfer. If
the Siegert operator is enhanced by the same mechanism proposed to account
for the
violation of Hara's~\cite{hara} theorem in $\Delta S=1$ hyperon radiative decays, then
the magnitude
of the effects shown in Table \ref{tab6} and Fig. \ref{Fig.8} is not unreasonable.
Conversely, should a
future measurement imply $\RAd\sim\RAewk$, then one may have reason to
question the
resonance saturation model for both $d_\Delta$ and the hyperon decays.

\begin{table}
\begin{center}~
\begin{tabular}{|c||c|c|c|}\hline
\hbox{Coupling constants} &\hbox{Source}  &\hbox{Best
values}
&
\hbox{Range}
\\\hline\hline
$h_\pi$  & \cite{fcdh} (\cite{des}) & 7 (7) & $0\to 17$ \\
$h_\Delta$  & \cite{fcdh} (\cite{des}) & -20 (-20) & $-51\to 0$ \\
$h^1_{\Delta N\omega}$  & \cite{fcdh} (\cite{des}) & 11 (10) & $5\to 17$ \\
$h^0_{\Delta N\rho}$  & \cite{fcdh} (\cite{des}) & 20 (30) & $-54\to 152$ \\
$h^1_{\Delta N\rho}$  & \cite{fcdh} (\cite{des}) & 20 (20)& $17\to 26$ \\
$h^{'1}_{\Delta N\rho}$  & \cite{fcdh} (\cite{des}) & 0 (0)& $-0.5\to 2$
\\
$h_V$& \cite{zhu} & 1 & $-10\to 10$ \\
$h_V^\Delta$ & \hbox{this work} & 1 & $-10\to 10$ \\
$h_A^{N\Delta\pi\pi}$ & \cite{zhu1} &  1 & $-10\to 10$ \\
\hline
\end{tabular}
\end{center}
%\vspace{0.5cm}
\caption{\label{tab5}
Range and the best values for the available PV coupling constants (in units
of $g_\pi$) from refs. \protect\cite{fcdh,des,zhu,zhu1} and this work.}
\end{table}

\begin{figure}
\epsfxsize=10.0cm
\centerline{\epsffile{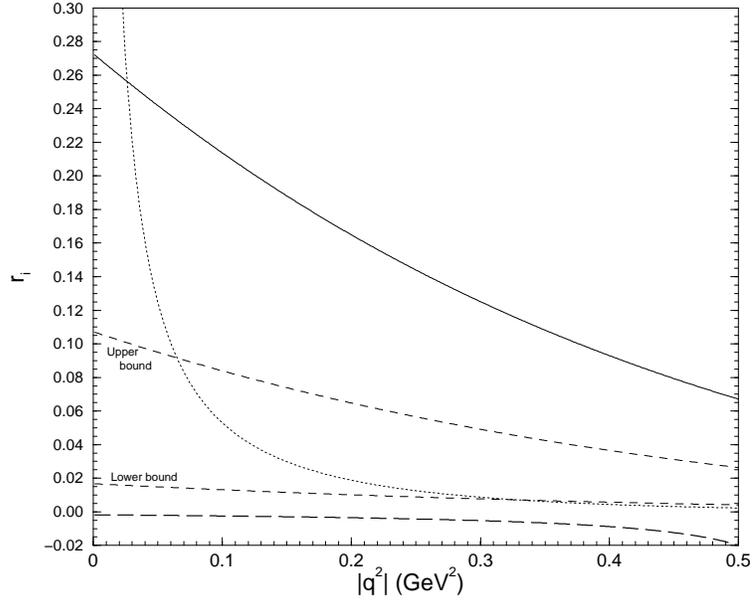}}
\vspace{1cm}
\caption{
Ratio of asymmetry components $r_i=A^i_{LR} /A^{NC}_{LRtot}$, where
$A^{NC}_{LRtot}$ denotes the total neutral current contribution.
The dotted line gives the Siegert contribution; the long-dashed line is for
the PV d-wave; the short dashed lines give our \lq\lq reasonable  range" for
the anapole effect; and the solid line is for axial vector neutral current
contribution. All the other parameters are the same as in Figure \ref{Fig.8}.
}
\label{Fig.9a}
\end{figure}

\begin{figure}
\epsfxsize=10.0cm
\centerline{\epsffile{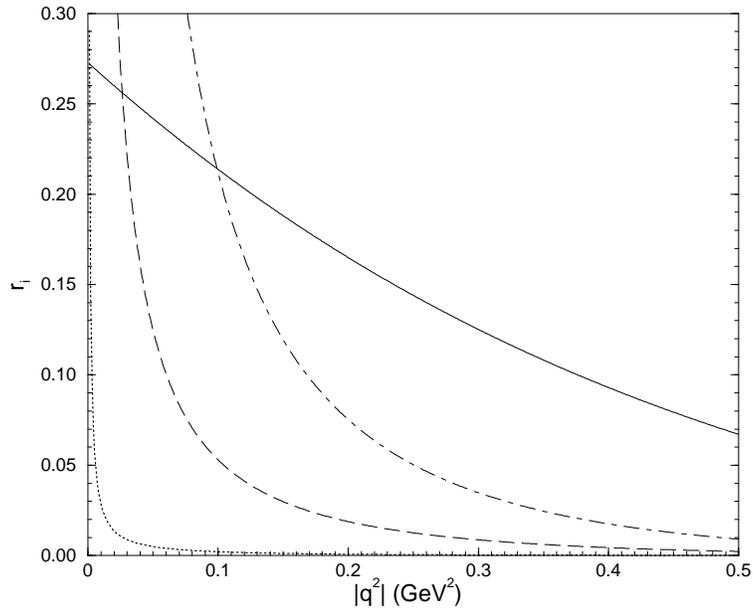}}
\vspace{1cm}
\caption{
Same as Fig. \ref{Fig.9a} but omitting the anapole and PV d-wave curves and
showing
Siegert contribution for several values of the coupling $d_\Delta$. The
dotted,
dashed and dashed-dotted lines are for $d_\Delta= 1 g_\pi,\ 25 g_\pi$ and
$100 g_\pi$ respectively. The solid line is for the axial vector neutral
current
contribution. All the other parameters are the same as in Figure \ref{Fig.9a}.
}
\label{Fig.9b}
\end{figure}

For the purpose of analyzing prospective measurements, it is also useful to
consider the
contributions to the total asymmetry generated by the various ${\cal
O}(\alpha G_F)$
effects. In Figs. (\ref{Fig.9a},\ref{Fig.9b}), we plot the ratios
\begin{equation}
\label{eq:asyratio}
 {\alrd[\Delta^\pi_{(3)}(i)]\over\alrd({\mbox{NC-tot}})}\ \ \ ,
\end{equation}
where  $\alrd({\mbox{NC-tot}})$ is the total neutral current contribution
to the asymmetry
and $i$ denotes the Siegert, anapole, and d-wave contributions. In Fig.
\ref{Fig.9a}, we show the
band generated by the anapole, where the limits are
determined by the ranges
in Table \ref{tab6}. For simplicity, we show the Siegert contribution for
only the single case:
$d_\Delta=25 g_\pi$, where the effective coupling $d_\Delta$
contains both the counterterm and loop effects, noting that
$d_\Delta$ is dominated by $d_\Delta^{CT}$. In Fig. \ref{Fig.9b}, we
give the variation of the Siegert contribution for a range of
$d_\Delta$ values,
where this range is essentially determined by the range for
$d_\Delta^{CT}$ given in
Table \ref{tab4}.
%%%%%%%%%%%%%%%%%%%%%%%%%%%%%%%%%%%% Figure 9a
%%%%%%%%%%%%%%%%%%%%%%%%%%%%%%%%%%%%%%%%%%%%%%%%%%

%%%%%%%%%%%%%%%%%%%%%%%%%%%%%%%%%%%%%% Figure 9b
%%%%%%%%%%%%%%%%%%%%%%%%%%%%%%%%%%%%%%%%%%%%%%%%555
\begin{table}
\begin{center}~
\begin{tabular}{|c||c|c|}\hline
Source & $\RAd({\mbox{best}})$ & $\RAd({\mbox{range}})$
\\\hline\hline
One-quark (SM)  & $-0.54$ & -\\
 Siegert $(+)$ & $0.21$ & $0.02\to 0.85$  \\
 Siegert $(-)$ & $-0.21$ & $-0.85 \to -0.02$  \\
 Anapole  & $0.04$ & $ -0.09\to 0.21$  \\
 d-wave & 0.0006 & $ -0.003 \to 0.003$ \\
\hline
Total $(+)$ & $-0.29$ &$-0.61\to 0.52$  \\
Total $(-)$ &$-0.71$ & $-1.48\to -0.35$  \\
\hline
\end{tabular}
\end{center}
%\vspace{0.5cm}
\caption{\label{tab6} One-quark Standard Model (SM) and many-quark anapole
and Siegert's contributions to $V(A)\times A(N)$ radiative corrections.
Values are computed in the on-shell scheme using
$Q^2=0.1$ (GeV$/c)^2$ and $q_0+W-M=0.6$ GeV. The plus and minus signs
correspond to the positive and negative values for $d^{CT}_\Delta$.}
\end{table}

From the plots in Figs. (\ref{Fig.9a},\ref{Fig.9b}), we observe
that the uncertainty associated with the anapole
and d-wave terms can be as much as $\sim 25\%$ of the nominal axial NC
contribution. The
uncertainty associated with the Siegert contribution is even more
pronounced. For
$Q^2\simle 0.1$ (GeV$/c)^2$,  this uncertainty is $\pm 100\%$ of the axial
NC contribution,
decreasing to $\simle 15\%$ at $Q^2=0.5$ (GeV$/c)^2$. Evidently, in order
to perform a
meaningful determination of the $C_i^A(Q^2)$, one must also determine the
size of the
Siegert contribution. Since the $Q^2$ variation of the latter can be as
large as that
associated with the $C_i^A(Q^2)$ for $0.1 \simle Q^2\simle 0.5$
(GeV$/c)^2$, one may not be able to
rely solely on the $Q^2$-dependence of the asymmetry in this regime
in order to disentangle the various effects.

Rather, in order to separate the Siegert contribution from the other axial
terms,
one would ideally
measure $\alrd$ in a regime where the Siegert term dominates the asymmetry.
%%%%%%%%%%%%%%%%%%%%%%%%%%%%%%%5 Figure 10
\begin{figure}
\epsfxsize=10.0cm
\centerline{\epsffile{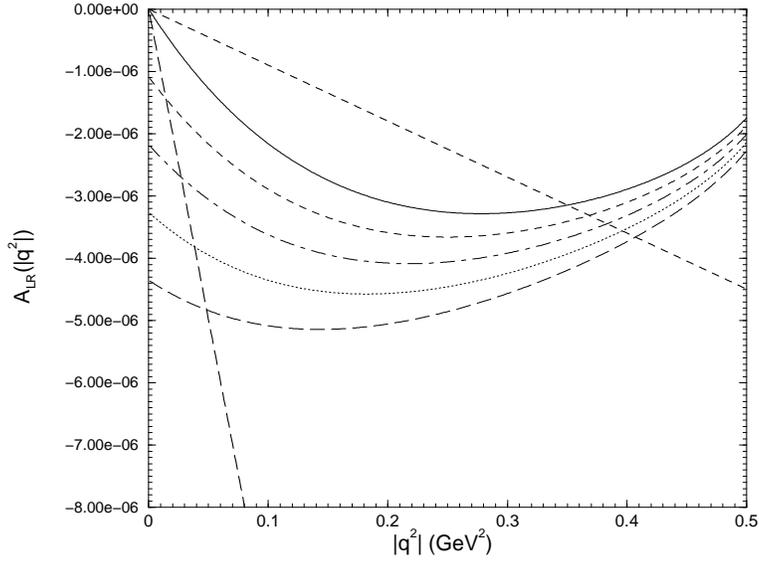}}
\vspace{1cm}
\caption{
Asymmetry components as a function of $|q^2|$ and beam energy $0.424$ GeV.
Except
for $d_\Delta$, all the parameters are taken from the central values of the
table
(\ref{tab5}). The bold long-dashed (dashed) line is for
$A_{LR}(\Delta_{(1)}^\pi)$ ($A_{LR}(\Delta_{(2)}^\pi)$). The  solid,
dashed-dotted, dotted and dashed lines are for $A_{LR}(\Delta_{(3)}^\pi)$
at $d_\Delta= 0, 25, 75$ and $100\ g_\pi$.
}
\label{Fig.10}
\end{figure}
%%%%%%%%%%%%%%%%%%%%%%%%%%%%%%%%%%%%%%%%%%%%%%
As shown in Fig.
\ref{Fig.10}, the Siegert contribution can become as large as the leading,
$\Delta^\pi_{(1)}$
contribution for $Q^2\simle 0.05$ (GeV$/c)^2$. To estimate the experimental
kinematics
optimal for a determination of $d_\Delta$ in this regime, we plot in Fig.
\ref{Fig.11} the total
asymmetry for low-$Q^2$. To set the scale, we use the benchmark feasibility
estimates of
ref. \cite{muk}, based on the  experimental conditions in Table \ref{tab77}.

\begin{table}
\begin{center}~
\begin{tabular}{|c||c|}\hline
Experimental Parameter & Benchmark Value
\\\hline\hline
Luminosity ${\cal L}$ & $2\times 10^{38}\ {\mbox{cm}}^{-2} {\mbox{s}}^{-1}$
\\
Running time  $T$ &   1000 hours  \\
Solid angle $\Delta\Omega$ & 20 msr \\
Electron polarization $P_e$ & 100\%\\
\hline
\end{tabular}
\end{center}
%\vspace{0.5cm}
\caption{\label{tab77}
Possible experimental conditions for $\alrd$ measurement.}
\end{table}

From the figure of merit computed in ref. \cite{muk}, one obtains a
prospective
statistical accuracy of $\sim 27\%$ at $E=400$ MeV, $\theta=180^\circ$ and
$Q^2=0.054$ (GeV$/c)^2$. A measurement with such precision would barely
resolve
the effect of $d_\Delta=\pm 100 g_\pi$. Doubling the beam energy
and going to more forward angles ({\em e.g.} $\theta=20^\circ$),
while keeping
$Q^2$ essentially the same, would reduce the statistical uncertainty to
roughly
5\% . At this level, one would be able to resolve the effect of
$d_\Delta$
having roughly the size of our \lq\lq best value". 
%%%%%%%%%%%%%%%%%%%%%%%%%%%%%%%%%%%Figure 11
\begin{figure}
\epsfxsize=15.0cm
\centerline{\epsffile{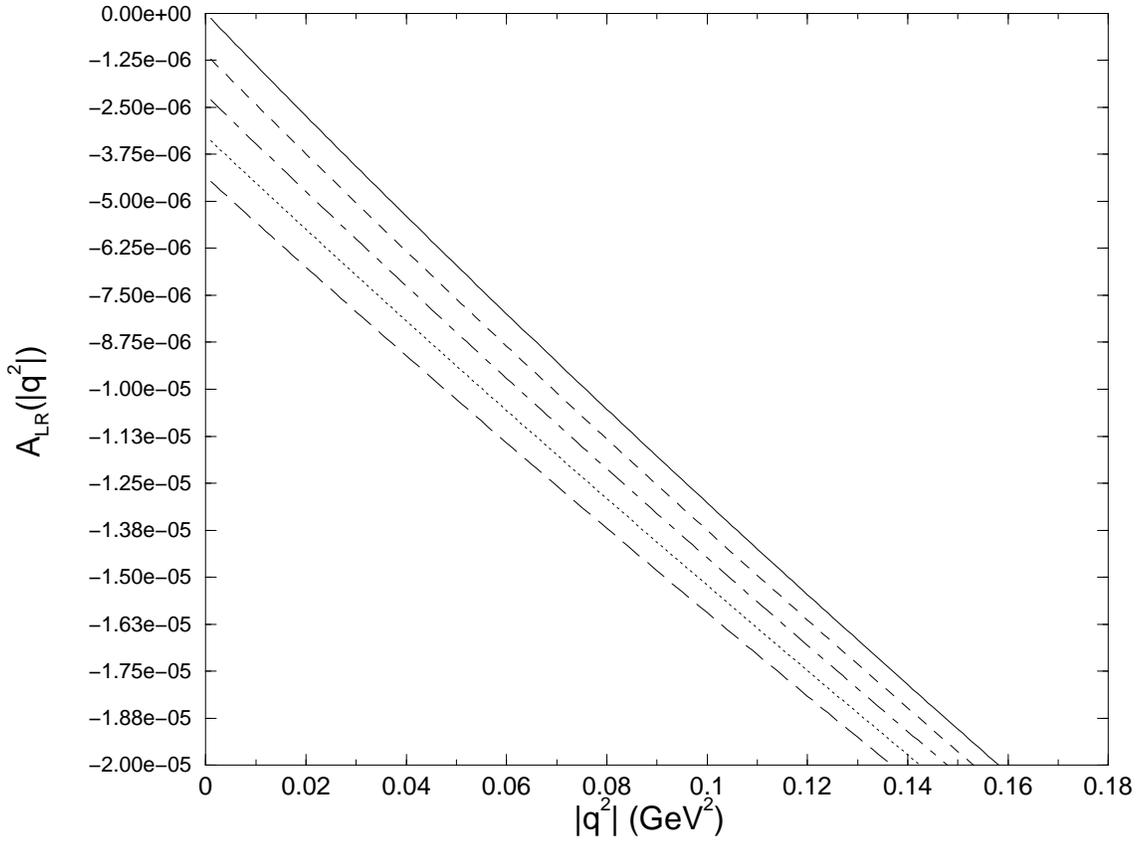}}
\vspace{1cm}
\caption{Total asymmetry at small $|q^2|$ for several $d_\Delta$.
The couplings are at central values of table (\ref{tab5}).
The lines for $d_\Delta= 0, 25, 75$ and $100\ g_\pi$ are the solid, dashed,
dashed-dotted, dotted and long-dashed line.
}
\label{Fig.11}
\end{figure}
%%%%%%%%%%%%%%%%%%%%%%%%%%%%%%%%%%%%%%%%%%%%%%%%%5
More generally, a
forward angle
($\theta\simle 20^\circ$) measurement for $E\sim 1$ GeV appears to offer
the most
promising possibility for determining $d_\Delta$. Such a
measurement would
have two benefits: (a) providing a test in the $\Delta S=0$ channel
of the
mechanism proposed to explain the violation of Hara's theorem in the
$\Delta S=1$ hyperon
radiative decays, and (b)  helping constrain the $d_\Delta$-related
uncertainty in
an extraction of the $C_i^A(Q^2)$ for $Q^2\simge 0.1$ (GeV$/c)^2$.
\section{Conclusions}
\label{sec10}
%\setcounter{equation}{0}
%%%%%%%%%%%%%%%%%%%%%%%%%%%%%%%%%%%%%%%%%%%%%%%%%%%%%%%%%%%%%%%%%%%%

Parity-violation in the weak interaction has become an important tool for
probing
novel aspects of hadron and nuclear structure. At present, an extensive
program of
of PV electron scattering experiments to determine the strange-quark vector
form
factors of the nucleon is underway at MIT-Bates, Jefferson Lab, and
Mainz\cite{pvexpts}. A measurement of the neutron radius of $^{208}$Pb is
planned for
the future at Jefferson Lab\cite{prex}, and measurements of non-leptonic PV
observables are being developed at Los Alamos, NIST, and Jefferson
Lab\cite{hadpvexpts}. In the present study, we have discussed the application
of PV
electron scattering to study the $N\to\Delta$ transition, which holds
significant
interest for our understanding of the low-lying
$qqq$ spectrum. We have argued that:

\begin{itemize}

\item [(i)] The ${\cal O}(\alpha G_F)$ contributions to the axial vector
$N\to \Delta$ response generate a significant contribution to the PV
asymmetry. One
must, therefore, take these effects into consideration when interpreting any
measurement of the asymmetry.

\item [(ii)] A substantial fraction of the ${\cal O}(\alpha G_F)$
contributions
arise from weak interactions among quarks. A particularly interesting \lq\lq
many-quark" contribution of this nature involves the PV $\gamma N\Delta$
electric
dipole coupling, $d_\Delta$, whose presence leads to a non-vanishing
asymmetry at the
photon point.

\item [(iii)] A determination of $d_\Delta$ via, {\em e.g.}, a low-$Q^2$
asymmetry
measurement, would both sharpen the interpretation of a planned Jefferson
Lab
PV $\Delta$ electroexcitation experiment and shed light on the dynamics of
mesonic and radiative hyperon weak decays. Indeed, one may conceivably
discover whether the
anomalously large violation of QCD symmetries observed in the latter are
simply a
peculiarity of the $\Delta S=1$ channel or a more general feature of
low-energy
hadronic weak interactions. At the same time, knowledge of $d_\Delta$ would
allow one
to place new constraints on the axial transition form factors $C_i^A(Q^2)$
from
PV asymmetry measurements taken over a modest kinematic range.

\item [(iv)] Experimental results for the $\Delta S=1$ decays suggest that
the PV
$N\to\Delta$ asymmetry generated by $d_\Delta$ could be large, approaching a
few
$\times 10^{-6}$ as $Q^2\to 0$. Measurement of an asymmetry having this
magnitude using forward angle kinematics at existing medium energy
facilities appears
to lie within the realm of feasibility.

\end{itemize}

More generally, the subject of hadronic effects in electroweak radiative
corrections
has taken on added interest recently in light of new measurements of the
muon
anomalous magnetic moment \cite{muon} and backward angle PV elastic $ep$ and
quasielastic $ed$ scattering \cite{sample}. The results in both cases differ
from
Standard Model predictions, with implications resting on the degree to which
one can
compute hadronic contributions to radiative processes. The interpretation of
future
precision measurements, including determination of the asymmetry parameter
in
neutron $\beta$-decay and the rate for neutrinoless $\beta\beta$-decay, will
demand a
similar degree of confidence in theoretical calculations of higher-order,
hadronic
electroweak effects. Thus, any insight which one might derive from studies
in other
contexts would represent a welcome contribution. To this end, a comparison
of PV
electroexcitation of the $\Delta$ with the corresponding neutral current
$\nu$-induced $\Delta$-excitation would be particularly interesting, as the
latter
process is free from the large ${\cal O}(\alpha G_F)$ hadronic effects
entering
PV electroexcitation \cite{mike,mjrm}.

% All the references are located in here
%\input{biblio.tex}

% appendices: we put them after the references in what ever order
% we prefer

\appendix

\chapter{Fitting Parameters of the Four Fermion Twist Four Contributions to the SFs}
\label{a1}
\begin{figure}  
\centerline{\psfig{figure=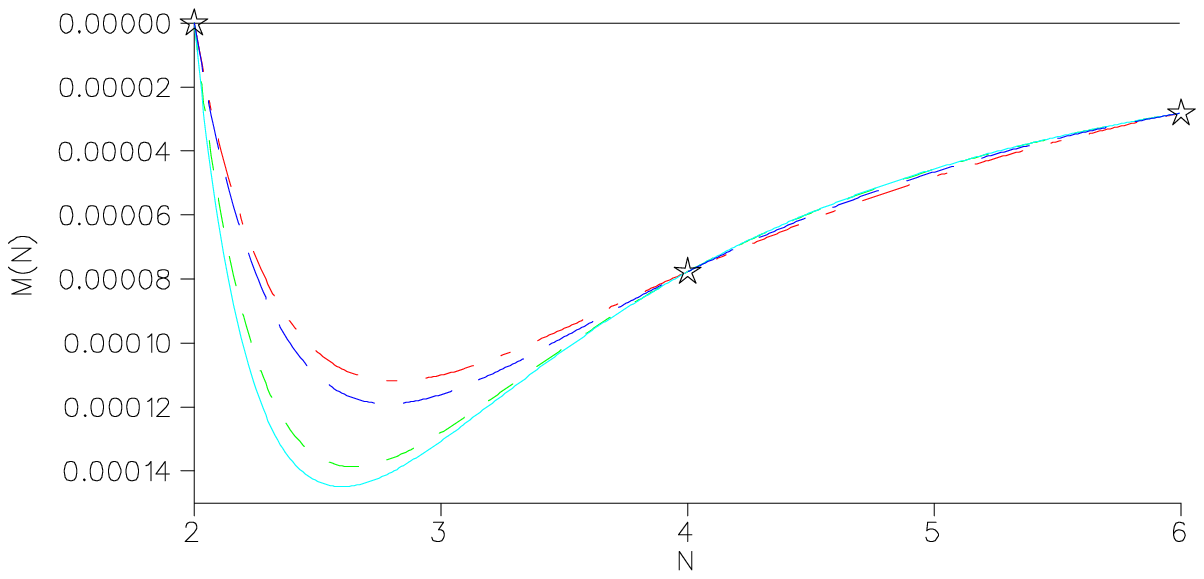}}
\caption{\label{f11} Comparison of the different fits for the moment function $M(N)$ relative to the $F_1^{EM}(x_B)$ SF ($U_1 $ contribution). The solid line 
is relative to the set of parameters in eq. (\ref{e50}), 
the long dashed line is relative to the set of parameters in eq. (\ref{e51}), the dot-dashed line is relative to the set of parameters in eq. (\ref{e52}),
and the short dashed line is relative to the set of parameters in eq. (\ref{e53}). The three moments obtained from the bag model are represented by stars.}
\end{figure}

\begin{figure}
\centerline{\psfig{figure=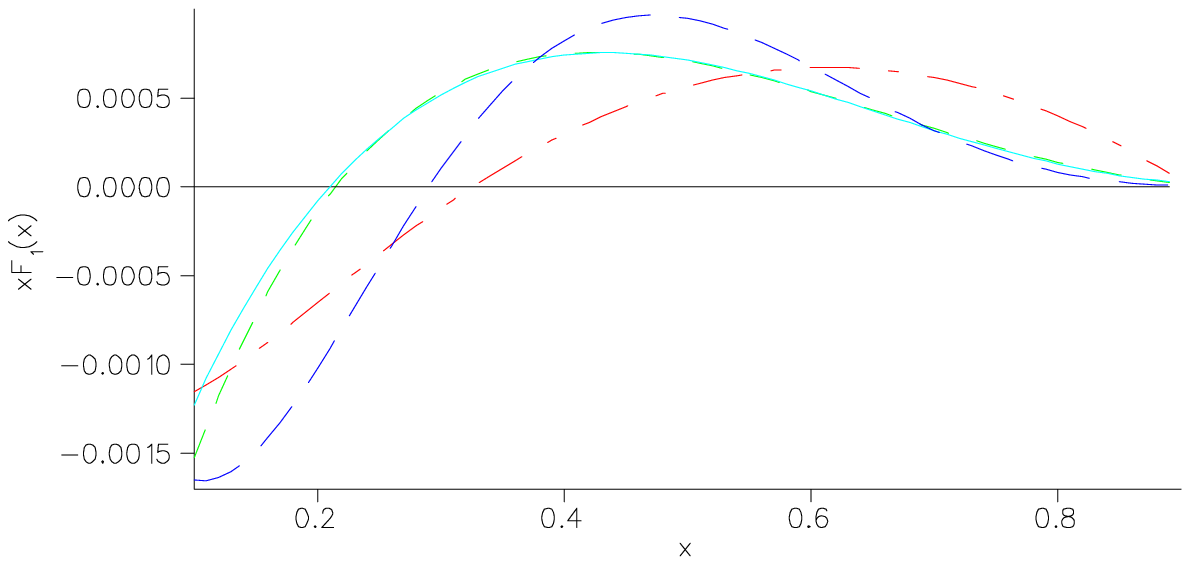}}
\caption{\label{f12}Reconstruction of the SF $x_BF_1^{EM}(x_B)$ through the use of the inverse Mellin transform technique ($U_1$ contribution). 
The type of line used for each 
parametrization is the same as in Fig.\ref{f11}.}
\end{figure}

\begin{figure}[h]  
\centerline{\psfig{figure=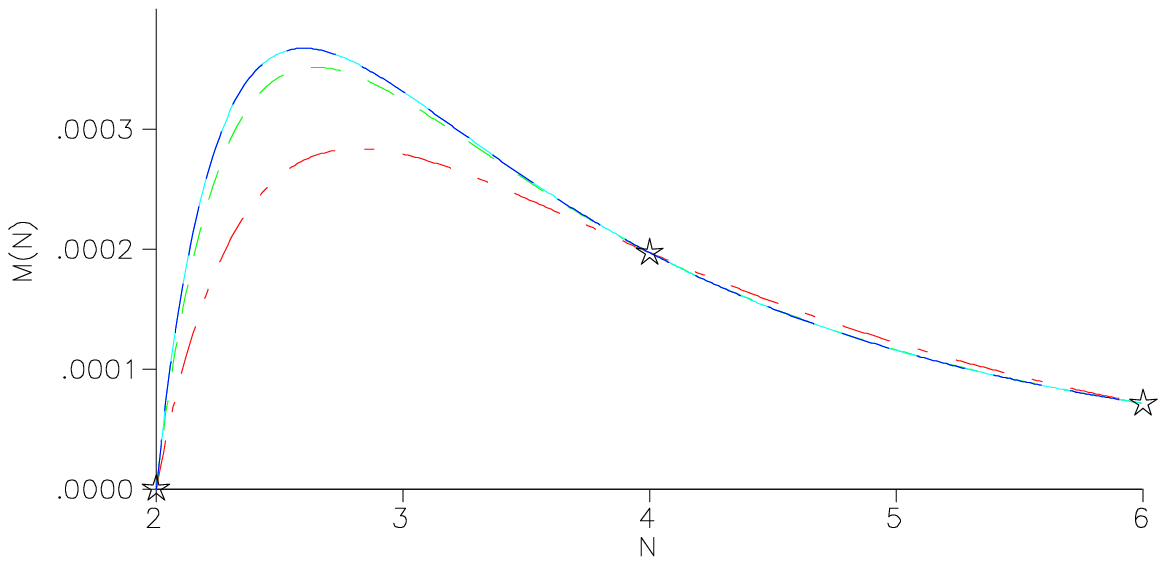}}
\caption{\label{f13} Comparison of the different fits for the moment function $M(N)$ relative to the $F_1^{EMNC}(x_B)$ SF  ($U_1$ contribution). 
The solid line is 
relative to the 
set of parameters in eq. (\ref{e54}), 
the long dashed line is relative to the set of parameters in eq. (\ref{e55}), the dot-dashed line is relative to the set of parameters in eq. (\ref{e56}),
and the short dashed line is relative to the set of parameters in eq. (\ref{e57}). The three moments obtained from the bag model are represented by stars.}
\end{figure}

\begin{figure}[h]  
\centerline{\psfig{figure=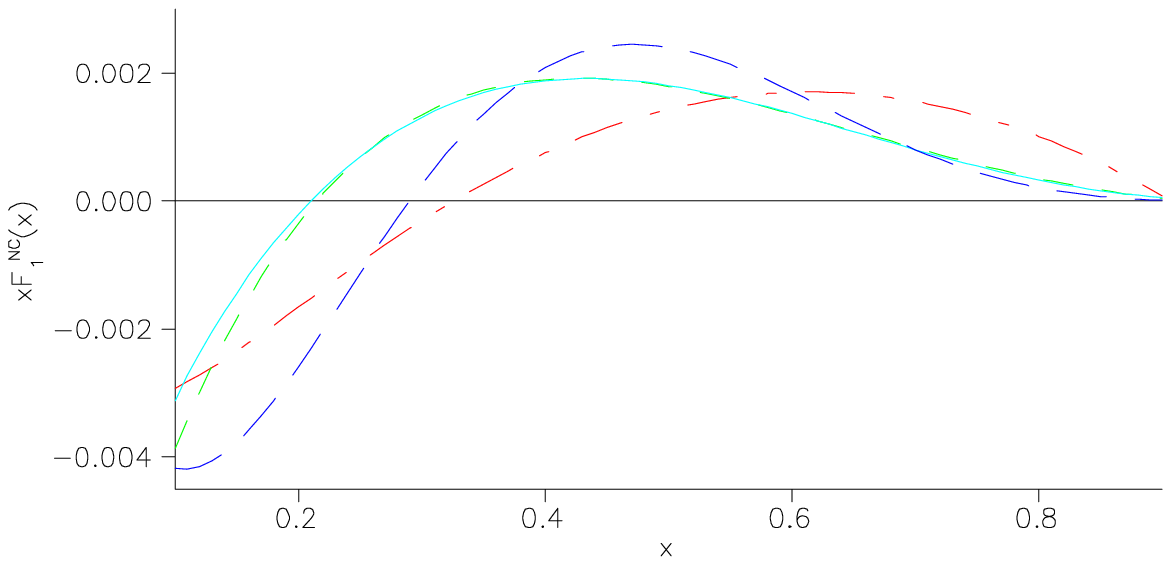}}
\caption{\label{f14} Reconstruction of the SF $x_BF_1^{EMNC}(x_B)$ through the use of the inverse Mellin transform technique  ($U_1$ contribution). 
The type of line used for each 
parametrization is the same as in Fig.\ref{f13}.}
\end{figure}

\begin{figure}[h]  
\centerline{\psfig{figure=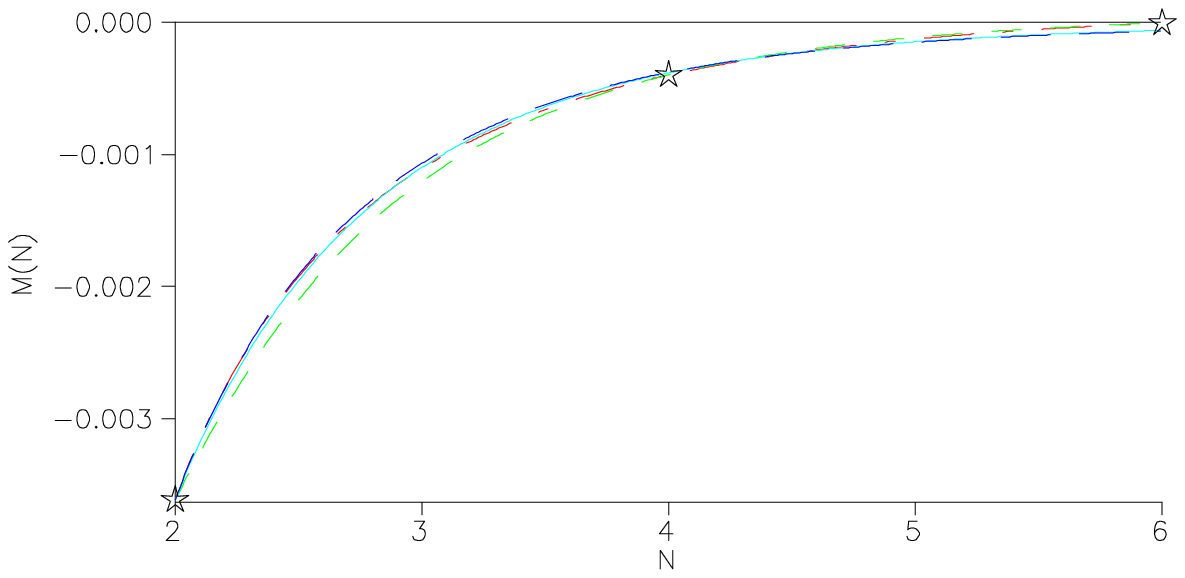}}
\caption{\label{f15} Comparison of the different fits for the moment function $M(N)$ relative to the $F_1^{EMNC}(x_B)$ SF  ($U_2$ contribution). 
The solid line is relative to the 
set of parameters in eq. (\ref{e58}), 
the long dashed line is relative to the set of parameters in eq. (\ref{e59}), the dot-dashed line is relative to the set of parameters in eq. (\ref{e60}),
and the short dashed line is relative to the set of parameters in eq. (\ref{e61}). The three moments obtained from the bag model are represented by stars.}
\end{figure}

\begin{figure}[h]  
\centerline{\psfig{figure=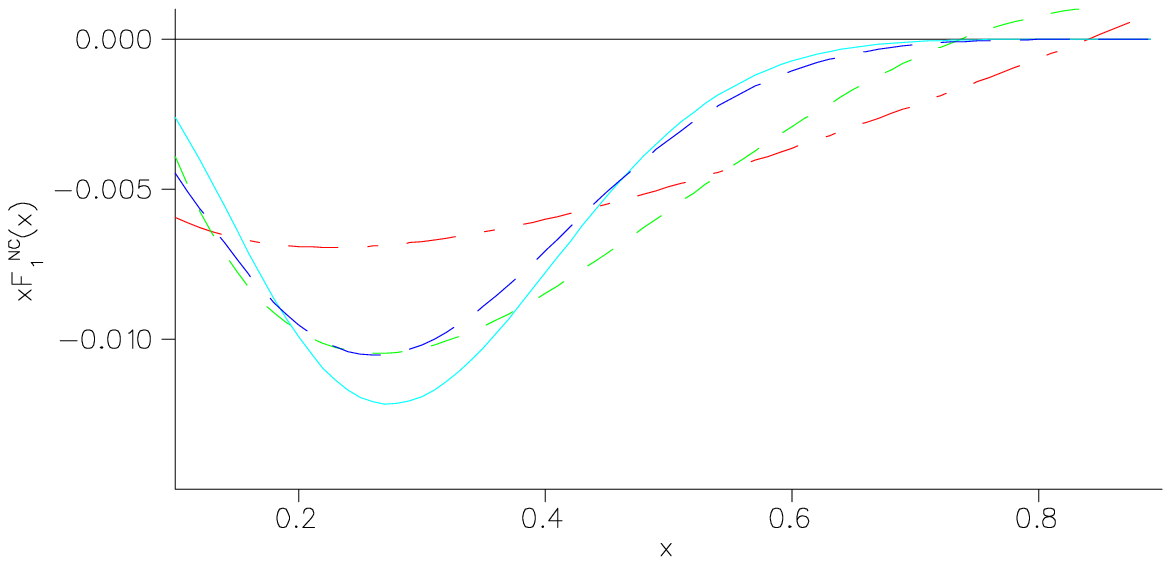}}
\caption{\label{f16} Reconstruction of the SF $x_BF_1^{EMNC}(x_B)$ through the use of the inverse Mellin transform technique  ($U_2$ contribution). 
The type of line used for each 
parametrization is the same as in Fig.\ref{f15}.}
\end{figure}

\begin{figure}[h]  
\centerline{\psfig{figure=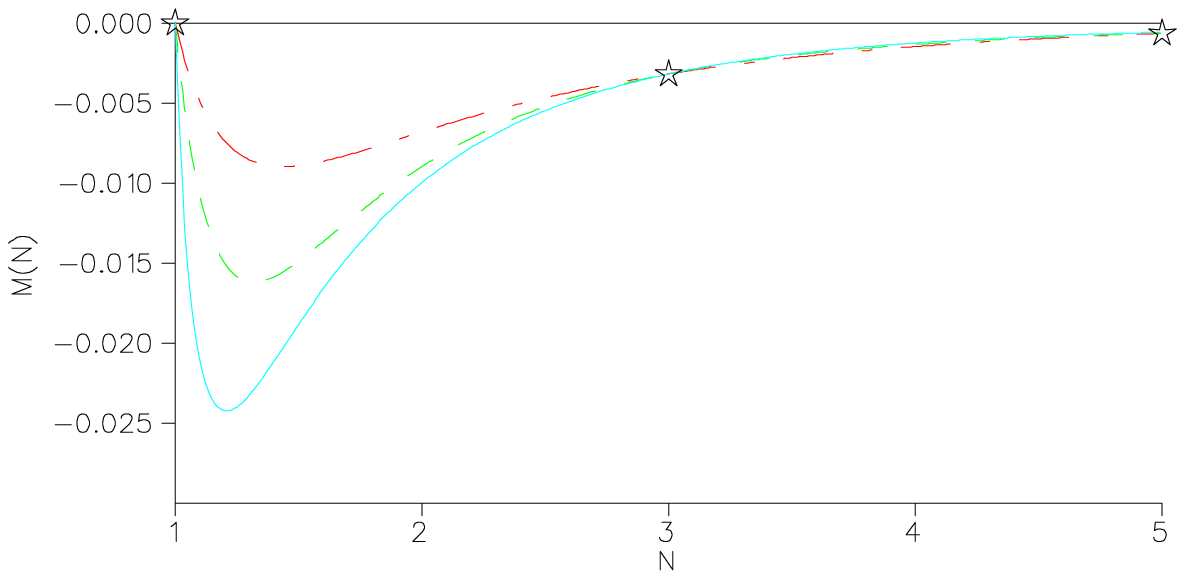}}
\caption{\label{f17} Comparison of the different fits for the moment function $M(N)$ relative to the $F_3^{EMNC}(x_B)$ SF ($U_1$ contribution).
 The solid line is relative
 to the set of parameters in eq. (\ref{e62}), 
the dashed line is relative to the set of parameters in eq. (\ref{e63}) and  the dot-dashed line is relative to the set of parameters in eq. (\ref{e64}). 
The three moments 
obtained from the bag model are represented by stars.  }
\end{figure}

\begin{figure}[h]  
\centerline{\psfig{figure=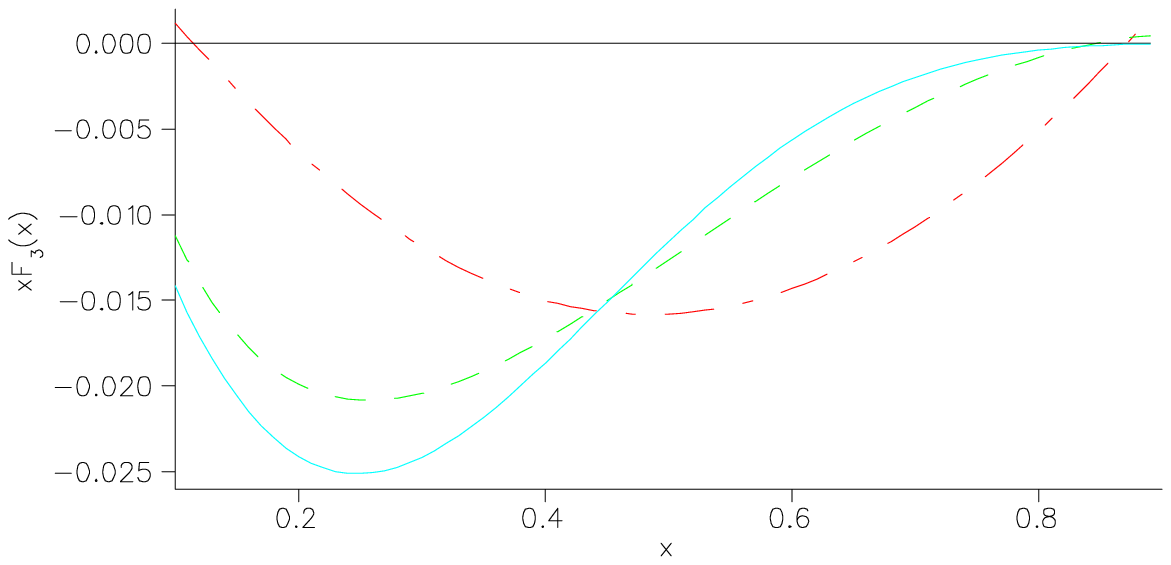}}
\caption{\label{f18} Reconstruction of the SF $x_BF_3^{EMNC}(x_B)$ through the use of the inverse Mellin transform technique ($U_1$ contribution). 
The type of line used for each 
parametrization is the same as in Fig.\ref{f17}.}
\end{figure}

In this appendix we present the results for the different parametrizations used to fit the moments of the SFs. 
The results for  to contibution to $F_1^{EM}(x_B)$ arising from the operator $U_1$ are:\\
\bga \label{e50}
&&\!\!\!\!\!\!\!\!\!\!\!\!a=36.4\times 10^{-6}\,\,\,\,\, b=-1.153\,\,\,\,\,c=2.797\,\,\,\,\,d=3.478\,\,\,\,\,e=6.118\nn\\
\ea 
 and 
\bga \label{e51}
&&\!\!\!\!\!\!\!\!\!\!\!\!a=754.472\times 10^{-6}\,\,\,\,\, b=.038\,\,\,\,\,c=4.317\,\,\,\,\,d=2.815\,\,\,\,\,e=2.181\nn\\
\ea 
 for a parametrization of  type  (\ref{e25}) and 
\bga \label{e52}
a_1=-9.00 \times 10^{-6}\,\,\,\,\, a_2= 100.08\times 10^{-6}\,\,\,\,\, a_3 =-164.16\times 10^{-6}&&\nn\\
\ea 
\bga \label{e53}
a_2=-7.92 \times 10^{-6}\,\,\,\,\, a_3= 231.84\times 10^{-6}\,\,\,\,\, a_4 =-432.00\times 10^{-6}&&\nn\\
\ea 

in the case of a parametrization of type \ref{e26}.  Notice   that also in this case
we found two sets of parameters $\{a,b,c,d,e \}$ which equally fit the moments of the SF. In Fig.\ref{f11} we have plotted the momentum function $M(N)$ and
in Fig.\ref{f12} we have plotted the corresponding contribution to $x_B F_1^{EM}(x_B)$. 
The contribution   of the operator $U_1$ to $F_1^{EMNC}(x_B)$ gave as a result the following set of parameters
\bga \label{e54}
&&\!\!\!\!\!\!\!\!\!\!\!\!a=-9.631\times 10^{-4}\,\,\,\,\, b=-1.153\,\,\,\,\,c=2.797\,\,\,\,\,d=3.478\,\,\,\,\,e=6.118\nn\\
\ea 
\bga \label{e55}
&&\!\!\!\!\!\!\!\!\!\!\!\!a=-199.621\times 10^{-4}\,\,\,\,\, b=.038\,\,\,\,\,c=4.317\,\,\,\,\,d=2.815\,\,\,\,\,e=2.181\nn\\
\ea 
\bga \label{e56}
a_1=2.381 \times 10^{-4}\,\,\,\,\, a_2= -26.480\times 10^{-4}\,\,\,\,\, a_3 =43.434\times 10^{-4}&&\nn\\
\ea 
\bga \label{e57}
a_2=2.096 \times 10^{-4}\,\,\,\,\, a_3= -61.341\times 10^{-4}\,\,\,\,\, a_4 =114.300\times 10^{-4}&&\nn\\
\ea 
The graphs of $M(N)$ and $x_BF_1^{EMNC}(x_B)$ are in Figs.\ref{f13} and \ref{f14} respectively.
The parameters for the contribution of the operator $U_2$  to the SF $F_1^{EMNC}(x_B)$ are
\bga \label{e58}
&&\!\!\!\!\!\!\!\!\!\!\!\!a=302.360\times 10^{-4}\,\,\,\,\, b=1.545\,\,\,\,\,c=9.362\,\,\,\,\,d=265.331\,\,\,\,\,e=3.006\nn\\
\ea 
 and 
\bga  \label{e59}
&&\!\!\!\!\!\!\!\!\!\!\!\!a=77.404\times 10^{-4}\,\,\,\,\, b=-.112\,\,\,\,\,c=7.477\,\,\,\,\,d=5.447\,\,\,\,\,e=179.207\nn\\
\ea 
\bga \label{e60}
a_1=-8.709 \times 10^{-4}\,\,\,\,\, a_2= 48.282\times 10^{-4}\,\,\,\,\, a_3 =36.746\times 10^{-4}&&\nn\\
\ea 
\bga  \label{e61}
a_2=-59.569 \times 10^{-4}\,\,\,\,\, a_3= 433.076\times 10^{-4}\,\,\,\,\, a_4 =-431.403\times 10^{-4}&&\nn\\
\ea 
 The graphs relative to $M(N)$ and $x_BF_1^{EMNC}(x_B)$ are plot in Figs.\ref{f15} and \ref{f16}.
Finally the parameters relative to the contribution of the operator $U_1$ to the SF $F_3^{NC}(x_B)$ are
\bga \label{e62}
&&\!\!\!\!\!\!\!\!\!\!\!\!a=-15.522\times 10^{-4}\,\,\,\,\, b=-.868\,\,\,\,\,c=4.517\,\,\,\,\,d=35.106\,\,\,\,\,e=45.197\nn\\
\ea 
\bga \label{e63}
a_1=17.213 \times 10^{-4}\,\,\,\,\, a_2= -132.571\times 10^{-4}\,\,\,\,\, a_3 =115.358\times 10^{-4}&&\nn\\
\ea 
\bga \label{e64}
a_2=22.344 \times 10^{-4}\,\,\,\,\, a_3=- 280.536\times 10^{-6}\,\,\,\,\, a_4 =258.191\times 10^{-6}&&\nn\\
\ea 
and the graphs for $M(N)$ and $x_BF_3^{NC}(x_B)$ are plotted in Figs.\ref{f17} and \ref{f18}.
\newpage

\chapter{Spin-Flavor-Color Factors}
\label{a2}
In this appendix we are going to present  the spin-flavor-color factors for the different contributions to SFs. Let's first consider the spin-flavor part.
 If we call $P_q^\sigma=q^\dag_\sigma q_\sigma$ the projector operator
 onto flavor $q$ 
and spin $\sigma$, by using eq. (\ref{e20}) we have
\bga \label{e35}
&&\la P|P_u^{1/2}P_u^{1/2}| P \ra=3 \,\,\,\,\,\,\,\,\,\, \la P|P_u^{-1/2}P_u^{-1/2}| P \ra=\frac{1}{3}\nn \\
&&\la P| P_u^{1/2}P_u^{-1/2}| P \ra=\la P|P_u^{-1/2}P_u^{1/2}| P \ra=\frac{1}{3} \nn \\
&&  \la P|P_u^{1/2}P_d^{1/2}| P \ra=\la P|P_d^{1/2}P_u^{1/2}| P \ra=\frac{1}{3}\nn\\
&&\la P| P_u^{1/2}P_d^{-1/2}| P \ra=\la P|P_d^{-1/2}P_u^{1/2}| P \ra=\frac{4}{3} \nn \\
&&\la P|P_u^{-1/2}P_d^{1/2}| P \ra=\la P|P_d^{1/2}P_u^{-1/2}| P \ra
=\frac{1}{3}\nn \\
&&\la P|P_u^{-1/2}P_d^{-1/2}| P \ra=\la P|P_d^{-1/2}P_u^{-1/2}| P \ra=0\nn \\
&&\la P|P_d^{1/2}P_d^{-1/2}| P \ra=\la P|P_d^{-1/2}P_d^{1/2}| P \ra=0\nn \\
&&\la P| P_d^{1/2}P_d^{1/2}| P \ra=\frac{1}{3}\,\,\,\,\,\,\,\,\,\,\la P|P_d^{-1/2}P_d^{-1/2}| P \ra=\frac{2}{3}\nn\\
\ea
For the color factors, we have
\begin{table}
\caption{Color factors for the opertor  $\la P | u^\dag \lambda_a u u^\dag \lambda^a u | P \ra $}
\centerline{\begin{tabular}{|r|c|}
\multicolumn{2}{c}{}\\
\hline
$\lambda^a$   &$\la P | u^\dag \lambda_a u u^\dag \lambda^a u | P \ra $\\
\hline
$ \lambda^1$ &$-\frac{2}{3}$\\
$ \lambda^2$ &$-\frac{2}{3}$\\
$ \lambda^3$ &$+\frac{2}{3}$\\
$ \lambda^4$ &$-\frac{2}{3}$\\
$ \lambda^5$ &$-\frac{2}{3}$\\
$ \lambda^6$ &$-\frac{2}{3}$\\
$ \lambda^7$ &$-\frac{2}{3}$\\
$ \lambda^8$ &$+\frac{2}{3}$\\
\hline
\end{tabular}}
\label{t36}
 \end{table}
\begin{table}
\caption{Color factors for the opertor   $\la P | d^\dag \lambda_a d d^\dag \lambda^a d | P \ra $}
\centerline{\begin{tabular}{|r|c|}
\multicolumn{2}{c}{}\\
\hline
$\lambda^a$ & $\la P | d^\dag \lambda_a d d^\dag \lambda^a d | P \ra $\\
\hline
$ \lambda^1$ &$0$\\
$ \lambda^2$ &$0$\\
$ \lambda^3$ &$\frac{2}{3}$\\
$ \lambda^4$ &$0$\\
$ \lambda^5$ &$0$\\
$ \lambda^6$ &$0$\\
$ \lambda^7$ &$0$\\
$ \lambda^8$ &$\frac{2}{3}$\\
\hline
\end{tabular}}
\label{t37}
 \end{table}
\begin{table}
\caption{Color factors for the opertor $\la P | d^\dag \lambda_a d u^\dag \lambda^a u | P \ra $}
\centerline{\begin{tabular}{|r|c|}
\multicolumn{2}{c}{}\\
\hline
$\lambda^a$ & $\la P | d^\dag \lambda_a d u^\dag \lambda^a u | P \ra $\\
\hline
$ \lambda^1$ &$-\frac{2}{3}$\\
$ \lambda^2$ &$-\frac{2}{3}$\\
$ \lambda^3$ &$-\frac{2}{3}$\\
$ \lambda^4$ &$-\frac{2}{3}$\\
$ \lambda^5$ &$-\frac{2}{3}$\\
$ \lambda^6$ &$-\frac{2}{3}$\\
$ \lambda^7$ &$-\frac{2}{3}$\\
$ \lambda^8$ &$-\frac{2}{3}$\\
\hline
\end{tabular}}
\label{t38}
 \end{table}

\chapter{Moments Formulae for the Four Fermion Contributions}
\label{a3}
In this appendix I am going to present the rest of the formulae derived in Sec. \ref{section_fourfermion}  to compute the moments of the four fermion operators $U_1$ and $U_2$. 
\section{Useful Formulae I}
Proceeding in the same way that lead to Eq. (\ref{e4}) we find, for the other three $\Delta$ functions
\bga \label{e5}
&&\int dx\, dy\,dz\int dx_B  x_B^{N}\Delta(x,z,y,x_B) \nn \\
&\times&\int \frac{d\lambda}{2\pi}\frac{d\mu}{2 \pi}
\frac{d\nu}{2\pi}e^{i\lambda x}e^{i\mu (y-x)}e^{i\nu(z-y)} f_1(\nu) f_2(\mu) f_3(\lambda)\nn\\
&=&\sum_{j=1}^{N-1}\sum_{k=0}^{j-1} \sum_{m=0}^{j-k-1}\sum_{p=0}^{j-1-m}i^{N-2}{j-1-m\choose p}\nn\\
 &\times&{j-k-1\choose m}  (n\cdot\partial)^mf_1(0)(n\cdot\partial)^pf_2(0)(n\cdot\partial)^{N-2-p-m}f_3(0)\nn \\
\ea
and
\bga \label{e6}
&&\int dx\, dy\,dz\int dx_B  x_B^{N}\Delta(y-x,z,y,x_B) \nn \\
&\times&\int \frac{d\lambda}{2\pi}\frac{d\mu}{2 \pi}
\frac{d\nu}{2\pi}e^{i\lambda x}e^{i\mu (y-x)}e^{i\nu(z-y)} f_1(\nu) f_2(\mu) f_3(\lambda)\nn\\
&=&\sum_{j=1}^{N-1}\sum_{k=0}^{j-1}  \sum_{t=0}^{j-k-1}\sum_{m=0}^t\sum_{p=0}^{j-1-m}i^{N-2} (-1)^{t}{j-k-1\choose t}\nn \\
 &\times& {t \choose m} {j-1-m\choose p}(n\cdot\partial)^mf_1(0)(n\cdot\partial)^pf_2(0)(n\cdot\partial)^{N-2-p-m}f_3(0)\nn \\
\ea
and
\bga \label{e7}
&&\int dx\, dy\,dz\int dx_B  x_B^{N}\Delta(x,y-z,y,x_B) \nn \\
&\times&\int \frac{d\lambda}{2\pi}\frac{d\mu}{2 \pi}
\frac{d\nu}{2\pi}e^{i\lambda x}e^{i\mu (y-x)}e^{i\nu(z-y)} f_1(\nu) f_2(\mu) f_3(\lambda)\nn\\
&=&\sum_{j=1}^{N-1}\sum_{k=0}^{j-1}  \sum_{t=0}^{N-j-1}\sum_{m=0}^{j-k-1}\sum_{p=0}^{N-2-t-m}i^{N-2} (-1)^{t}
{N-j-1\choose t}{j-k-1\choose m}\nn\\
 &\times&  {N-2-t-m\choose p} (n\cdot\partial)^mf_1(0)(n\cdot\partial)^pf_2(0)(n\cdot\partial)^{N-2-p-m}f_3(0)\nn \\
\ea
\section{Useful Formulae II}
In Sec. \ref{section_fourfermion} we computed the contribution of the operator $U_2$ to the moments of the SFs (see Eq. (\ref{e170})). The same calculation can be easily extended to the operator  $U_1$ by noticing that  Eq. (\ref{e18}) would now 
read
\bga \label{e171}
&& (n\cdot\partial)^p \bar{q}_{\sigma } (x)n \sla
(n\cdot\partial)^qq_{\sigma }(x)\nn \\
&=&\frac{N_m^2}{M}\Big[ (n\cdot\partial)^p (j_0(E|\mathbf{x}|)e^{iEt} ) (n\cdot\partial)^m (j_0(E|\mathbf{x}|)e^{-iEt} )\nn \\
&-&i  (n\cdot\partial)^m (j_0(E|\mathbf{x}|)e^{-iEt} )(n\cdot\partial)^p (j_1(E|\mathbf{x}|) \frac{z}{|\mathbf{x}|}e^{iEt}  )\nn \\
&+&i  (n\cdot\partial)^p (j_0(E|\mathbf{x}|)
e^{iEt} )(n\cdot\partial)^m (j_1(E|\mathbf{x}|)  \frac{z}{|\mathbf{x}|}e^{-iEt} ) \nn \\
&+&(n\cdot\partial)^p (j_1(E|\mathbf{x}|) \frac{z}{|\mathbf{x}|}e^{iEt}) (n\cdot\partial)^m (j_1(E|\mathbf{x}|) \frac{z}{|\mathbf{x}|}
e^{-iEt} )\nn \\
&+& (x^2+y^2)(n\cdot\partial)^p (\frac{j_1(E|\mathbf{x}|) }{|\mathbf{x}|}e^{iEt}) (n\cdot\partial)^m ( \frac{j_1(E|\mathbf{x}|)}{|\mathbf{x}|}
e^{-iEt} )
\Big]\nn \\
\ea
Notice that in this case there is not a spin dependent factor $(-1)^\sigma$ and notice the change of sign in the last factor in the previous relation compared to Eq. (\ref{e18}). Using Eq. (\ref{e171}) we get the operator $U_1$
\bga 
&&\int dx_B \,x_B^N \Delta(x,y,z,x_B) \int dxdydz U_1(x,y,z) \nn \\
&=& k_{\sigma \sigma'} \frac{N_m^2}{M} \sum_{j=1}^{N-1}\sum_{k=0}^{j-1} \sum_{m=0}^{j-k-1}\sum_{p=0}^{j-1-m}
i^{N-2}{j-1-m\choose p}{j-k-1\choose m}\nn\\
 &\times&\int d\mathbf{x}\Big[  (j_0(E|\mathbf{x}|)e^{iEt} ) (n\cdot\partial)^m (j_0(E|\mathbf{x}|)e^{-iEt} )\nn \\
&-&i  (n\cdot\partial)^m (j_0(E|\mathbf{x}|)e^{-iEt} ) (j_1(E|\mathbf{x}|) \frac{z}{|\mathbf{x}|}e^{iEt}  )\nn \\
&+&i   (j_0(E|\mathbf{x}|)
e^{iEt} )(n\cdot\partial)^m (j_1(E|\mathbf{x}|)  \frac{z}{|\mathbf{x}|}e^{-iEt} ) \nn \\
&+& (j_1(E|\mathbf{x}|) \frac{z}{|\mathbf{x}|}e^{iEt}) (n\cdot\partial)^m (j_1(E|\mathbf{x}|) \frac{z}{|\mathbf{x}|}
e^{-iEt} )\nn \\
&+& (x^2+y^2) (\frac{j_1(E|\mathbf{x}|) }{|\mathbf{x}|}e^{iEt}) (n\cdot\partial)^m ( \frac{j_1(E|\mathbf{x}|)}{|\mathbf{x}|}
e^{-iEt} )\Big]\nn \\
&\times&\Big[ (n\cdot\partial)^p (j_0(E|\mathbf{x}|)e^{iEt} ) (n\cdot\partial)^q (j_0(E|\mathbf{x}|)e^{-iEt} )\nn \\
&-&i  (n\cdot\partial)^q (j_0(E|\mathbf{x}|)e^{-iEt} )(n\cdot\partial)^p (j_1(E|\mathbf{x}|) \frac{z}{|\mathbf{x}|}e^{iEt}  )\nn \\
&+&i  (n\cdot\partial)^p (j_0(E|\mathbf{x}|)
e^{iEt} )(n\cdot\partial)^q (j_1(E|\mathbf{x}|)  \frac{z}{|\mathbf{x}|}e^{-iEt} ) \nn \\
&+&(n\cdot\partial)^p (j_1(E|\mathbf{x}|) \frac{z}{|\mathbf{x}|}e^{iEt}) (n\cdot\partial)^q (j_1(E|\mathbf{x}|) \frac{z}{|\mathbf{x}|}
e^{-iEt} )\nn \\
&+& (x^2+y^2)(n\cdot\partial)^p (\frac{j_1(E|\mathbf{x}|) }{|\mathbf{x}|}e^{iEt}) (n\cdot\partial)^q ( \frac{j_1(E|\mathbf{x}|)}{|\mathbf{x}|}
e^{-iEt} )
\Big]
 \nn \\
\ea
The rest of the contributions for the operaors $U_1$ and $U_2$ arising from the other $\Delta$ functions  can be easily  computed by using Eqs. (\ref{e5}-\ref{e7}) and Eqs. (\ref{e170}) and (\ref{e171}).

\chapter{Bremsstrahlung Tensors}\label{app:tensors}
In this appendix we presents the results for the leptonic tensors (the electromagnetic and the interference), and also the contractions of the leptonic tensors with the hadronic ones.
The electromagnetic leptonic tensor has the following form
\bga  
L^{\mu\nu}_{em}&=& -8\frac{k\sd k_2 k\sd q +k_1\sd k_2(m^2+k\sd k_2)}{k_1\sd k_2 (k\sd k_2)^2}k_2^{\mu}k_2^{\nu}-\frac{1}{(k_1\sd k_2)^2(k\sd k_2)^2}\nn \\
&&\Big\{ 8 k^{\mu}k^{\nu}\Big((m^2-Q^2+2 k_2\sd q -2k\sd    k_2-2 k \sd q) (k\sd k_2 )^2 \nn \\
&+& 2 k_1 \sd k_2 k\sd k_2 (-m^2-k_2 \sd q + k \sd k_2+ k \sd q) + (k_1\sd k_2)^2 m^2\Big)\nn \\
&+& 4(k_2^{\mu}k^{\nu}+k^{\mu} k_2^{\nu}) \Big((2m^2+k \sd k_2)( k_1\sd k_2)^2\nn \\
&-& k\sd k_2 k_1\sd k_2(2m^2+k_2 \sd q-2k \sd q) + (k\sd k_2)^2(k\sd k_2 -k_2 \sd q) \Big)\nn \\
&-&4 g^{\mu\nu}\Big[\Big ((k\sd k_2)^2 +(k\sd k_2 +k\sd q) m^2 -k_2 \sd q (m^2 +k\sd k_2)\Big) (k_1\sd k_2)^2 \nn \\
&+&k\sd k_2 k_1\sd k_2 \Big(-k\sd k_2 Q^2 +2(-k_2 \sd q +k\sd k_2 +k\sd q) (k\sd q-m^2)\Big ) \nn \\
& +& (k\sd k_2)^2 \Big((k\sd k_2)^2 +(m^2-2 k\sd q ) k\sd k_2 -k_2 \sd q (m^2 +k\sd k_2 - 2k\sd q)\nn \\
&+&
k\sd q (m^2-Q^2 -2 k\sd q) \Big)\Big]\Big\}\,,
\ea
while, for the interference one, we get
\bga 
L^{\mu\nu}_{nc}&=&-\frac{1}{(k_1\sd k_2)^2k\sd k_2}\Big[2 g_V^e m \Big(k^{\mu}s^{\nu}(k_1\sd k_2 -k\sd k_2)(k_1\sd k_2+k_2\sd q -k\sd k_2)\nn \\
&-&k_2^{\mu}s^{\nu}[(k_1\sd k_2)^2+(k_2\sd q-2 k\sd k_2) k_1\sd k_2 +(k_2\sd q -k\sd k_2) k\sd k_2]\Big)\Big]\nn \\
&-&\frac{1}{(k_1\sd k_2)^2(k\sd k_2)^2}\Big\{2g_A^e m \Big[\Big(-(k\sd k_2-2m^2)(k_1\sd k_2)^2\nn \\
&+& k\sd k_2 k_1 \sd k_2[ 4 (-m^2+k\sd k_2+k\sd q)-3 k_2 \sd q]+(k\sd k_2)^2[3 k_2\sd q
\nn \\
&-&3 k\sd k_2 
+2 (m^2-Q^2 -2k\sd q)]\Big) s^{\mu}k^{\nu}+\Big( (k\sd k_2 -2 m^2) (k_1\sd k_2)^2\nn \\
&+&k\sd k_2 k_1 \sd k_2 (2m^2 + k_2 \sd q -2 k\sd q ) + (k_2\sd q-k\sd k_2) (k\sd k_2)^2 \Big) s^{\mu}k_2^{\nu}\nn \\
& -&2 g_A^e m g^{\mu\nu}\Big( [-k_2\sd q k_2\sd s +(k\sd k_2 +k\sd q) k_2\sd s\nn \\
&+&(m^2-k\sd k_2 )q\sd s ] (k_1\sd k_2)^2+k\sd k_2 [-k_2\sd s (Q^2+k\sd q)\nn \\
& +&(-2 m^2-2 k_2\sd q +3 k \sd  k_2 +2 k\sd q ) q\sd s] k_1 \sd k_2 +(k\sd k_2)^2 [k_2\sd s k\sd k_2\nn \\
&-&k_2 \sd q (k_2\sd s - 2 q\sd s ) +(m^2-Q^2 -2 k\sd k_2 -2 k \sd q) q\sd s]\Big) \Big]\nn \\
&+&\frac{4 g_V^e m}{k_1\sd k_2 (k\sd k_2)^2}\Big( (k_1\sd k_2 k_2 \sd s) k^{\mu} k_2^{\nu} 
+ [k_1\sd k_2 k_2\sd s + k \sd k_2 (q\sd s \nn \\
&-& k_2\sd s)] k _2^{\mu} k ^{\nu}-(k_2\sd s) (k_1\sd k_2-k\sd k_2)k^{\mu} k^{\nu}\nn \\
& -&(k_1\sd k_2 k_2\sd s +k\sd k_2 q\sd s)k_2^{\mu}k_2^{\nu}\Big)\nn\\
&+& \frac{2 i g_V^e m }{(k_1\sd k_2)^2 k\sd k_2}\Big[(k_1\sd k_2-k\sd k_2)(\epsilon^{\alpha \beta \delta \nu  }k_2^{\mu}-\mu \leftrightarrow \nu )k_{\alpha }q_{\beta }s_{\delta }
\nn \\
&-& [(k_1\sd k_2  -k\sd k_2) k^ \mu +2 (k\sd k_2) k_2^\mu ](\epsilon^{\alpha \beta \delta \nu}  -\mu \leftrightarrow \nu)k_{2\,\alpha }q_{\beta }s_{\delta }\nn \\
&+&(k_1\sd k_2-k\sd k_2)^2\epsilon^{\alpha \beta \mu\nu}k_\alpha s_\beta -\Big( ( k_1\sd k_2)^2 +(k_2\sd q -k\sd q) k_1\sd k_2\nn \\
& -& k\sd k_2 (k_2 \sd q -Q^2 +k\sd k_2 - k\sd q) \Big)\epsilon^{\alpha \beta \mu\nu}k_{2\, \alpha} s_\beta
\Big]\nn \\
 &-&\frac{2 i g_V^e m }{(k_1\sd k_2)^2 (k\sd k_2)^2}\Big[ k_2\sd s (k_1\sd k_2 -k\sd k_2 )^2 \epsilon^{\alpha \beta \mu\nu}k_{  \alpha} q_\beta \nn \\
&-&(k_1\sd k_2-k\sd k_2 )\Big((k_1\sd k_2+2 k\sd k_2) k_2 \sd s +k\sd k_2 q\sd s \Big)\epsilon^{\alpha \beta \mu\nu}k_{2\, \alpha} q_\beta \nn \\
&+& \Big( (k\sd k_2-m^2) (k_1\sd k_2)^2+k\sd k_2 k_1 \sd k_2 (2 m^2 +k_2 \sd q -3 k\sd k_2\nn \\
& -&2 k\sd q )+ (k\sd k_2)^2(-m^2+Q^2-k_2\sd q +2 k \sd k_2 +2 k \sd q)\Big)\epsilon^{\alpha \beta \mu\nu}q_{  \alpha} s_\beta
\Big]\,.\nn \\
\ea
For both tensors we dropped terms proportional to $q_\mu$ and $q_\nu$, because gauge invariance  implies  that $q_\mu W^{\mu\nu}=q_\nu W^{\mu\nu}=0$.

The contraction of the electromagnetic leptonic and hadronic tensors is
\bga\label{e218}
W_{\mu\nu}^{em}L^{\mu\nu}_{em}&=& \frac{1}{(k_1\sd k_2)^2 (k\sd k_2)^2}\Big\{ 8 \Big( [m^4+2k\sd q m^2 +(k\sd k_2)^2\nn \\
&-&2k_2\sd q(m^2+k\sd k_2) ] (k_1\sd k_2)^2 +k\sd k_2 [2(k\sd q-m^2)\nn\\
&& (m^2  
+2 k\sd q) +k_2\sd q (2m^2+k\sd k_2-4k\sd q) \nn \\
&+&2k\sd k_2 (k\sd q -Q^2)] k_1\sd k_2 +(k\sd k_2)^2 [ m^4+(k\sd k_2)^2 -4 (k\sd q)^2 \nn\\
&-&k_2\sd q k\sd k_2+4 k_2\sd q k\sd q -4 k\sd k_2 k\sd q -(m^2
 + 2 k\sd q) Q^2 ]\Big) W_1^{em}\nn \\
& -&4 \Big[2 \Big([ k_1\sd k_2 -k\sd k_2][(k_1\sd k_2-k\sd k_2) m^2 +2k\sd k_2(-k_2\sd q\nn \\
&+&k\sd k_2+k\sd q)]-(k\sd k_2)^2 Q^2 \Big)E^2 -4k_1\sd k_2 (k_1\sd k_2 \nn \\
&-&k\sd k_2) k_0 m^2 E -2 k\sd k_2 k_0((k_1\sd k_2)^2 -k_2\sd q k_1 \sd k_2 +2 k\sd q k_1\sd k_2\nn \\
&+&(k\sd k_2)^2-k_2\sd q k\sd k_2  ) E +(k_1\sd k_2)^2 [(2 k_0^2 -2k_2 \sd q \nn \\
&+&k \sd k_2+k\sd q )m^2 +k\sd k_2 (2k_0^2 -k_2\sd q 
+k\sd k_2)]\nn \\
& +&(k\sd k_2)^2 [(k\sd k_2)^2 +(m^2-2 k\sd q) k\sd k_2-k_2\sd q (m^2+k\sd k_2\nn \\
& -& 2 k\sd q )+k\sd q (m^2 -2 k\sd q -Q^2 ) ] + k_1\sd k_2 k\sd k_2 [-2(k\sd k_2 \nn \\
&+&k\sd q ) m^2 +2 k\sd q (k_0^2 +k\sd k_2 +k\sd q  )-2k_2\sd q (k\sd q-m^2 ) \nn \\
&-&k\sd k_2 Q^2 ]\Big]W_2^{em} \Big\}\,,
\ea
while the contraction of the interference leptonic tensor with the hadronic one is
\bga\label{e219}
W_{\mu\nu}^{nc}L^{\mu\nu}_{nc}&=& \frac{1}{(k_1\sd k_2)^2(k\sd k_2)^2}\Big\{ 8g_a m \Big[ \Big(-2k_2 \sd q k_2\sd s +(k\sd k_2 +2k\sd q ) \nn \\
&&k_2\sd s +2(m^2-k\sd k_2)q\sd s \Big) (k_1\sd k_2)^2+k\sd k_2
 \Big( -4q\sd s m^2\nn \\
&-&3k_2\sd s k\sd q-2k_2\sd sQ^2+k_2\sd q (k_2\sd s -4 q\sd s )+5 k\sd k_2 q\sd s 
\nn \\
&+&4k\sd q q\sd s \Big)k_1\sd k_2+(k\sd k_2)^2\Big( k_2\sd s k\sd k_2 -k_2\sd q (k_2\sd s-4q\sd s) \nn \\
&+&2(m^2-2 k\sd k_2 -2 k\sd q -Q^2 ) q\sd s\Big)\Big]W_1^{nc} \nn \\
&-& 4 g_a \Big[ 2k_1\sd k_2 k_2\sd s (k_1\sd k_2 -k\sd k_2 ) m E^2-2  \Big([2 k_2\sd s k_0 m \nn \\
&+&\sqrt{E^2-m^2}(k\sd k_2 -m^2)] (k_1 \sd k_2)^2 + k\sd k_2 [2m^2\sqrt{E^2-m^2} \nn \\
& -& k_2\sd s k_0 m +k_0 m q\sd s+2 (k_2 \sd q  -3 k\sd k_2    -2k\sd q   )\sem ] k_1\sd k_2\nn \\
& -&(k\sd k_2)^2 \sem (m^2 + 2 k_2\sd q -2 k\sd k_2 -2 k\sd q -Q^2 )  \Big)E\nn \\
& +& k_1 \sd k_2 k\sd k_2 \Big( -2 q\sd s m^3 +2 k_0 m^2 \sem \nn \\
&+&k_2\sd s (-k\sd q -Q^2 ) m +q\sd s m [ 2 k_0^2 -2 k_2\sd q +3 k\sd k_2 +2k\sd q]\nn \\
&+& 2 k_0 \sem [k_2\sd q -k\sd k_2 -k\sd q ] \Big)+(k_1\sd k_2)^2 \Big( -k_2\sd q k_2\sd s m \nn \\
&+&k_2\sd s (2k_0^2 +k\sd k_2 +k\sd q )m -(k\sd k_2 -m^2 ) (m q\sd s\nn \\
& -&2 k_0 \sem )\Big)+ (k\sd k_2)^2\Big(q\sd  s m(m^2 -2 k\sd k_2 -2 k\sd q - Q^2 ) \nn \\
&+& k_2\sd s k \sd  k_2 m +k_2 \sd q (2 \sem k_0 -k_2 \sd s m + 2 m q\sd s )\nn \\
& -& 2 k \sd k_2 k_0 \sem  \Big) \Big] W_2^{nc} \nn \\
&+&8g_e \Big[ m k_1\sd k_2 k_2\sd s \Big([ E (k\sd k_2 -k_1\sd k_2 )+(k_1 \sd k_2 +k\sd k_2 ) k_0 ]Q^2\nn \\
& -& (E-E')[k_1\sd k_2 (k_2\sd q -k\sd q ) +k\sd k_2 (k_2 \sd q +k \sd q) ] \Big) \nn \\
&+&\sem \Big( -(k\sd k_2)^2 (Q^2)^2 +\{k\sd k_2 [(k_1\sd k_2 -2 k\sd k_2 ) (k_1 \sd k_2 + k_2 \sd q \nn \\
&-&  k \sd k_2  ) - 2 (k_1 \sd k_2 -k\sd k_2 ) k\sd q ] - (k_1 \sd k_2 -k \sd k_2 )^2 m^2  \} Q^2 +(k_1\sd k_2\nn \\
& +&k_2 \sd q - k\sd k_2 ) k\sd k_2 [k_1\sd k_2 (k_2 \sd q - k \sd q  ) +k\sd k_2 (k_2 \sd q + k \sd q )] \Big) \nn \\
&+& m \Big( E Q^2 (k \sd k_2)^2 -E' Q^2 (k \sd k_2 )^2 + k_0 [k_1\sd k_2 Q^2 -(k_1\sd k_2
 + k_2 \sd q \nn \\
&-& k\sd k_2 )(k_1 \sd k_2 + k \sd k_2 )] k\sd k_2 + E (k_1 \sd k_2 - k\sd k_2 ) \nn \\
&& [(k_1 \sd k_2 -k \sd k_2 )m^2 + k \sd  k_2 (-k_2 \sd q + k \sd k_2 + 2 k \sd q )] \nn \\
&+& E' (k_1 \sd k_2 - k \sd k_2 ) \{k_1 \sd k_2 (k\sd k_2 -m^2 ) +k \sd  k_2 [m^2 +2 k_2 \sd q \nn \\
&-& 2 (k \sd k_2 + k \sd q)] \}\Big) q\sd s \Big]W_3^{nc}\Big\}\,.
\ea 


\begin{thebibliography}{99}
\bibitem{CJ01b} The operator in Eq. (\ref{eq:Siegert2}) was independently
written down by the authors of Ref. \cite{CJ01}.
\end{thebibliography}
\end{document}